\input epsf.tex
\epsfclipon
 \documentclass[11pt,a4paper,amsart]{article}
 \usepackage{jheppub}
\newtheorem{proposition}{Proposition}
\newtheorem{lemma}{Lemma}
\usepackage{hyperref}

\usepackage{mathtools}
\usepackage{color}

\usepackage{epsfig}
\usepackage{epstopdf}
\usepackage{epsf}
\input{epsf.sty}
\usepackage{graphicx,amsmath,amssymb}
\usepackage{graphicx}
\usepackage{epsfig,multicol}
\usepackage{multirow}
\allowdisplaybreaks

\usepackage{tikz} 

\usepackage{bbm,bm,amsmath,amssymb}
 
\newcommand{\Su}{\color{blue}}
\newcommand{\red}{\color{red}}
\usepackage[normalem]{ulem}

\usepackage{comment}
\def\be{\begin{equation}}
\def\ee{\end{equation}}
\def\figs/B{B}
\def\bea{\begin{eqnarray}}
\def\eea{\end{eqnarray}}
\def\bg{\begin{eqnarray}}
\def\nd{\end{eqnarray}}
\def\sin{{\rm sin}}
\def\cos{{\rm cos}}

\def\log{{\rm log}}

\def\be{\begin{equation}}
\def\ee{\end{equation}}

\def\doi{http://doi.org}

\def\m{\mathrm{m}}

\usepackage{physics}
\usepackage{tikz}
\usetikzlibrary{arrows.meta}
\usetikzlibrary{angles,quotes} % for pic
\usetikzlibrary{decorations.markings,arrows.meta}
\tikzset{>=latex} % for LaTeX arrow head

\tikzset{
  midarr/.style={decoration={markings,mark=at position #1 with {\arrow{stealth}}},postaction={decorate}},
  midarr/.default=0.5
}
\colorlet{xcol}{blue!70!black}

\title{Resurgence of a de Sitter Glauber-Sudarshan State: Nodal Diagrams and Borel Resummation}
\author{{\Su Suddhasattwa Brahma}$^{1}$, {\Su Keshav Dasgupta}$^{2}$, 
{\Su Mir-Mehedi Faruk}$^{2, 3, 4}$, {\Su Bohdan Kulinich}$^{2}$, {\Su Viraj Meruliya}$^{2}$, {\Su Brent Pym}$^{5}$ and {\Su Radu Tatar}$^{6}$\\
	\vskip.03in
	${}^1$ Higgs Centre for Theoretical Physics, School of Physics \& 
	Astronomy, University of Edinburgh, Edinburgh, EH9 3FD, United Kingdom\\
	${}^2$ Department of Physics, McGill University, Montr\'{e}al, Qu\'{e}bec, H3A 2T8, Canada \\
	${}^3$ Institute for Theoretical Physics, University of Amsterdam, Science Park 904, 1090 GL Amsterdam, The Netherlands\\
	${}^4$ Delta Institute for Theoretical Physics, Science Park 904, PO Box 94485, 1090 GL Amsterdam, Netherlands\\
	${}^5$ Department of Mathematics and Statistics, McGill University,
	Montr\'{e}al, Qu\'{e}bec, H3A 3K6, Canada\\
	${}^6$ Department of Mathematical Sciences,
University of Liverpool,  Liverpool, L69 7ZL, United Kingdom \\
	{\tt suddhasattwa.brahma@gmail.com, keshav@hep.physics.mcgill.ca}
	
{\tt mir.faruk, bohdan.kulinich, viraj.meruliya@mail.mcgill.ca, brent.pym@mcgill.ca, Radu.Tatar@Liverpool.ac.uk}}

\date{\today}

\abstract{We show that an explicit construction of a four-dimensional 
de Sitter space may be performed using a diagrammatic approach via nodal diagrams emanating from the path integral representation of the Glauber-Sudarshan state. Sum of these diagrams typically leads to an asymptotic series of Gevrey kind which can then be Borel resummed, thus reproducing the non-perturbative structure of the system. Our analysis shows that four-dimensional de Sitter space is not only possible in string theory 
 overcoming the no-go and the swampland criteria $-$ albeit as a Glauber-Sudarshan state $-$ but it may also be non-perturbatively stable within a controlled temporal domain. Somewhat consistently, the Borel resummation of the Gevrey series provides strong hint towards the positivity of the cosmological constant.} 

\begin{document}

\maketitle

\newpage

\hskip2.2in {\it Waiting a million years, just for us.}

\hskip2in $-$ Peter Weir's ``{\Su Picnic at Hanging Rock}" (1975)

\section{Introduction and summary}	
\label{sec:intro}

Perturbative expansions, which were long thought to be useful techniques to address and analyze physical phenomena, are no longer considered reliable in the light of many recent remarkable results that questioned these computations \cite{unsal}. In fact the issue is not new. It was pointed out by Dyson \cite{dyson} long ago that perturbative QED cannot be the right way to study how electrons and photons interact, despite the massive success of such a point of view. The reasoning, as explained in \cite{dyson}, is simple and generic: if a quantity, computed as some expectation value, makes no sense when the {\it sign} of the coupling constant is reversed, the perturbative theory {\it cannot be convergent}!
Once we lose convergence, clearly no prediction can be made and mathematically we can term this as having zero radius of convergence. With zero radius of convergence the only reliable data is the tree-level results, which is of course devoid of any interesting details. On the other hand, we have very little technical machinery to study anything, especially for subjects like string and M theories, beyond the perturbation theory. Could this mean that we give up the present state of the computational techniques, or is there a reliable way out of this conundrum?

The answer was given more than a century ago by 
\'Emile Borel \cite{borelE} who provided a powerful way to resum an asymptotic (or divergent) series. In retrospect, we now know that the asymptotic growth of any quantum series hides a deeper fact, namely, that the series should correctly be expressed non-perturbatively. The so-called perturbative limits appear only as an approximation and only up-to some orders in the expansion parameters (which are typically the coupling constants). Reliable results can therefore only be got non-perturbatively, although the saving grace here is that the perturbative viewpoint  breaks down generally at very high loop orders $-$ for example perturbative QED results are valid till $137e$-th loop $-$ because the Feynman diagrams show factorial growths.

A similar fate is shared by string theory: string perturbation theory is generically asymptotic, but the growth of the string diagrams may be much more complicated than the Feynman diagrams in field theories. Early success of the application of this idea in string theory came from a proposal by Shenker that such asymptotic growths of string diagrams should imply the existence of branes whose tensions go as $g_s^{-1}$ and {\it not} $g_s^{-2}$ \cite{shenker}, where $g_s$ is the string coupling. (The latter were already known to exist as standard solitonic states in the Neveu-Schwarz sector of string theory.) This eventually led to the discovery of the D-branes by Polchinski \cite{polchinski}. 

String theory is not supergravity, not even approximately, but at low energies supergravity is known to emulate various aspects of string theory. The question however is how {\it low} is the low-energy limit required for our purpose. Clearly one of the essential requirements for the low-energy limit is the absence of massive stringy modes: they should all be integrated out, including of course all the massive Kaluza-Klein (KK) degrees of freedom. Even for such a situation, the growth of the supergravity loops may not be simple, implying that we need to explore 
divergent growths of the loops that are beyond the simpler ones studied by Borel. Fortunately, there is in fact a massive literature on the subject that analyses these scenarios, the prominent ones being the ones studied by Gevrey \cite{gevreyorig} and Mittag-Leffler \cite{mittag}, all developed a century ago!

Unfortunately however, despite these prominent developments in mathematics, the progress in the field of physics related to the applications of Feynman diagrams has somehow steered clear of these. The reason, as mentioned earlier, is that the asymptotic nature is only visible at very high loop orders, so essentially perturbation theory suffices. But what if we need to extract results that depend on {\it all orders} in the perturbation theory? This is where the so-called first {\it limitation} of the perturbation theory begins to emerge \cite{unsal}.

There is a way out of it of course by resorting to Borel resummation as discussed above and in many recent works including \cite{unsal, marino, Emelin}, but the main point is what we emphasized right at the beginning: perturbation theory is not sufficient to study Quantum Field
Theories. Resurgence comes to our rescue by replacing the original asymptotic series by an integral function through Borel transformation. As we will show later on, the singularities of this function on the Borel plane can be avoided by deforming the contour of integration, and cancelling the resulting non-perturbative ambiguities by adding instanton corrections\footnote{In QFT, momentum integrals can also give additional factorial growth which can only be resummed using ``renormalons" -- objects which, unlike instantons, do not have any semiclassical interpretation on the Borel plane.} to the original series. The trans-series is a result of adding these instanton corrections to the original perturbative expansion, thus rendering the whole expression meaningful and well-behaved.

\subsection{The puzzle with de Sitter space in string theory \label{sec1.1}}

The situation with the realization of a four-dimensional de Sitter space in string theory $-$ here we will take type IIB string theory $-$ in light of what we said above, is more complicated. Recall that one of the key advantage we get is from the usage of the supergravity limit which, in turn, depends on our ability to successfully integrate out all the high energy modes in the theory. One of the early attempts to realize de Sitter space in type IIB theory under some controlled laboratory, is the KKLT scenario \cite{KKLT}. Unfortunately, questions have been raised on the validity of the scenario, most prominently from the advocates of the swampland idea \cite{swampland} who point out that the KKLT construction cannot have a well defined effective field theory (EFT) 
description\footnote{For a more balanced view on the KKLT construction and the swampland criteria, see for example
\cite{beno, bergshoeff, sothi, donof, evan, kochru, russo, heliudson, dologoto, kalush} and references therein.}. 

The issue however is more puzzling. Even before we question on the existence of an EFT, we should ask what does EFT entail here. EFT here would mean that the theory should be described only using the low energy degrees of freedom that we somehow got by integrating out the massive stringy degrees of freedom (including the ones on the branes)\footnote{A more accurate statement would be that an EFT description exists when, to any given order in ${g^a_s\over {\rm M}^b_p}$, there are {\it finite} number of operators. Here $(a, b) \in \mathbb{Z}_+$ and we have 
denoted the eleven-dimensional Planck mass as ${\rm M}_p$ and the four-dimensional one as ${\rm M}_{\rm Pl}$. When $(a, b)$ are negative then the system should be viewed non-perturbatively where similar finiteness of the number of operators is required for an EFT description. This aspect has been described in much details in \cite{desitter2}.}. Sadly this is where the situation gets a bit tricky. The small fluctuations over a de Sitter background are given by wave-functions that have {\it time-dependent} frequencies. To allow for an EFT description, {\it \`a la} Wilsonian, we have to integrate out the high energy modes. However if the frequencies are time-dependent, such integration procedure makes no sense and therefore there is no easy way to write an EFT at low energies! (For more details on the consequences from other patches of de Sitter, the readers may refer to \cite{coherbeta2}.) One may try to censor the modes with 
wave-lengths smaller than ${\rm M}_{\rm Pl}^{-1}$, by imposing a temporal bound on the inflationary evolution $-$ the so-called trans-Planckian cosmic censorship (TCC) conjecture \cite{tcc} $-$ but the fundamental issue still remains: there appears to be no simple way to define a Wilsonian effective action\footnote{To handle such systems in which there is no globally well-defined time-like Killing vector, and consequently no energy conservation law, one has to resort to open Quantum Field Theories (somewhat along the lines of Feynman-Vernon \cite{vernon} formalism for dissipative systems) which take into account the entanglement between system and environment degrees of freedom \cite{balasub}, going beyond Wilsonian EFTs (see \cite{brahma1, colas1, burgess1}). The problem however is that, in string and M theories, it is not {\it a priori} clear how one would go about doing this.
Alternatively, one might also envision a {\it waterfall} like scenario where the depleted IR modes are quickly replaced by red-shifted UV modes. But since the temporal growth of the frequencies are not linear, there is no guarantee that an efficient integrating-out procedure would ever work here.}. 

The above issues are not the only hurdles. Since string theory is not supergravity, the evolution of the massive stringy degrees of freedom over a temporally varying background like de Sitter\footnote{We don't consider the static patch because of the issues mentioned above and also in \cite{coherbeta2}.} itself is highly non-trivial and there is no guarantee that the masses of the stringy modes will remain time-independent. If the temporal behavior of the massive string modes show similar behavior as the red-shifted UV modes discussed above, then it is not even clear whether one may define a nice supergravity limit of the system. Introduction of D-branes and NS-branes would further complicate the scenario, including additional complications resulting from the uncancelled zero point energies of the bosonic and fermionic degrees of freedom due to supersymmetry breaking at the vacuum level.

All in all it seems asking for de Sitter as a vacuum solution of string theory appears futile. Is there then a possibility of realising de Sitter solution without sacrificing the low energy {\it supergravity} limit $-$ that helped us to cancel the zero-point energies $-$ but allow a non-supersymmetric temporally varying background and yet retain the nice Wilsonian effective action at low energies?

The answer turned out to be surprisingly yes if we view de Sitter space as a Glauber-Sudarshan state\footnote{Or as a {\it coherent state}, but the distinction is important. Coherent state is typically defined by using a displacement operator on a {\it free} vacuum. Since string theory (or M-theory uplift that we use here) doesn't have a simple free vacuum due to non-zero $g_s$, the displacement operator has to act on an {\it interacting} vacuum. The state resulting from such an operation is called a {\it Glauber-Sudarshan} state to distinguish it from the usual coherent state. See  \cite{coherbeta, coherbeta2} for details on this.} over a supersymmetric Minkowski background in string theory. The idea isn't exactly new, see for example \cite{dvali} for a realization in four space-time dimensions\footnote{The key difference is that the analyses in \cite{dvali} are performed using a coherent state and not using a Glauber-Sudarshan state. Alternatively, for a realization of de Sitter space as a resonance in four-dimensions see for example \cite{maltz}.}, but as far as we are aware of, it was first realized in string theory in \cite{coherbeta, coherbeta2}. Such a construction automatically takes care of all the issues discussed above including the swampland \cite{swampland} and the no-go \cite{GMN} criteria. In particular Wilsonian effective action {\it is} possible because the frequencies (or the massive KK and the stringy modes) that we are integrating out are described over a (static) supersymmetric Minkowski background. The Glauber-Sudarshan state breaks supersymmetry spontaneously (meaning that the vacuum remains supersymmetric, thus cancelling the zero-point energies). The cosmological constant is then given by the non-perturbative effects as we shall see soon.

Once we construct such a Glauber-Sudarshan state, the space-time metric (including the metric components of the internal space) may be computed by taking expectation values of the graviton operators over such a state. The expectation values have to be computed off-shell completely, {\it i.e.} as path-integrals by including all possible interactions, because the construction of the Glauber-Sudarshan state involves {\it interacting} vacua and not the free ones as described earlier. An interesting conundrum now appears: since we can only know the complete form of the perturbative interactions $-$ at least formally $-$ as an infinite series in polynomial powers of the fields and their derivatives, the knowledge of the full non-perturbative effects would in principle be very hard. How should we then proceed? This is where the two concepts that we described earlier, namely the {\it asymptotic nature} of the perturbative series and its {\it factorial growth}, become immensely useful. 

The expectation value of the graviton operator\footnote{We take a simple model in eleven-dimensions with three scalar fields, each representing a component of the graviton, a component of the three-form flux and a component of the fermionic condensate respectively.\label{256split}}, expressed as a series in powers of the coupling constants, turns out to be asymptotic with a factorial growth of Gevrey kind \cite{gevreyorig}. The asymptotic nature of the Gevrey series clearly implies the necessity of the non-perturbative effects which, in turn, may now be included in by Borel resumming the series. Thus our original failure of including the full non-perturbative effects has now been reverted to triumph\footnote{Note however that the Borel resummation of any asymptotic series of a given expectation value directly provides the non-perturbative corrections that would make the series convergent. It doesn't however tell us how a given Lagrangian should be modified unless we compute the expectation value of the action itself. The latter would appear once we study the Schwinger-Dyson equation as in \eqref{toula} where we could encounter the expectation value of the action itself as the second equation in \eqref{watcher}. See section \ref{sec4.6.1} and footnote \ref{tasty} for more details.}  rather miraculously by the powerful computational machinery of Borel resummation!

But this is not the only advantage we get by Borel resumming the Gevrey series. The non-perturbative nature of the Borel resummed series has another satisfying feature because of its connection to the cosmological constant (details of which will become clearer in sections \ref{sec4.5} and \ref{sec4.6}): we can now justify the oft-quoted fact that the cosmological constant can 
{\it only} be realized non-perturbatively. There are no perturbative limits of the above expression, showing that no perturbative computations can ever give us the form of the cosmological constant! Such a conclusion matches rather well with the computations performed in \cite{desitter2} wherein a formally exact expression of the cosmological constant was presented in terms of the non-perturbative quantum effects\footnote{It is instructive at this stage to compare the cosmological constant that we get by Borel resumming a Gevrey series and the one we got from the {\it bulk} equations in \cite{desitter2}. The analysis in \cite{desitter2} explicitly requires non-perturbative effects and non-local counter-terms to be inserted to the EOMs. One possible equivalence between the two computations 
appear from the usage of the Schwinger-Dyson's equations that in some sense replace the bulk EOMs with the ones related to the expectation values. This means there should be a deeper connection between the four disparate facts: expectation value of the metric over the Glauber-Sudarshan state, Gevrey series, Borel resummation and the Schwinger-Dyson's equations. We will come back to this in section \ref{sec4.6.1}.}.

\subsection{Brief summary and organization of the paper \label{sec1.2}}

The paper is broadly organized into three sections. In section \ref{BA095} we study the dynamics in the configuration space of the Glauber-Sudarshan state. In section \ref{sec3} we introduce the {\Su nodal diagrams}, the set of diagrams that succinctly capture the dynamics of one-point and higher points functions in a {\it shifted} interacting vacua. The analysis is most easily performed using path-integrals and we dedicate the section in a careful evaluation of the path-integrals using three different scalar fields $-$ one representative sample from a set of 44 gravitons, 84 flux components and 128 Rarita-Schwinger fermions (for the latter we take a condensate so as to restrict ourselves to bosonic path integrals only). 
And in section \ref{borel} we do a detailed study of the asymptotic series resulting from summing up the nodal diagrams from the previous section. The asymptotic series turns out to be of the {\Su Gevrey-$\alpha$} kind with 
$\alpha$ being one less than the total number of fields in the spectra.

A more elaborate summary of the paper is as follows. Section \ref{BA095} is further subdivided into two sections, sections \ref{sec2.1} and \ref{sec2.2}. In the first sub-section we discuss why we expect de Sitter space to be represented by a Glauber-Sudarshan state, and the second sub-section we study the dynamics of a corresponding interacting theory from the configuration space point of view. Such an analysis, which is distinct from the analysis from the target space point of view, is much more useful here when we are dealing with excited states in QFT (such as the Glauber-Sudarshan state). For example, it helps us to quantify the temporal evolution of the Glauber-Sudarshan state much more easily and in-turn lends us a way to apply the powerful machinery of the path-integral formalism. 

In section \ref{sec3} we use this insight to study the nodal diagrams using the path integrals with shifted vacua. Section \ref{lincoln22} is a warm-up section where a qualitative analysis is performed first before going towards a more detailed quantitative analysis in section \ref{sec3.2}. This section is further sub-divided into seven sub-sections starting from 
sub-section \ref{sec3.2.1}, where we do the tree-level analysis,  till sub-section \ref{3.2.6}, where we combine all the nodal diagrams for the three fields. The study of the nodal diagrams for the second and the third fields $-$ which are the representative sample from 84 flux components and 
fermionic condensate respectively $-$ are performed in sub-sections \ref{sec3.3.2} and \ref{sec3.2.4} respectively. The analysis therein does not involve the source field, and therefore easier to work out. The subtlety however appears when we impose the momentum conservation on the third field, and the amplitudes therein take a different, more complicated form as shown in sub-section \ref{sec3.2.4}. In section \ref{sec3.2.5} we introduce the source field which, in our case forms a representative sample from the set of 44 gravitons (we don't worry about the tensor structures here). Expectedly, the amplitudes of the nodal diagrams are very different from the other two set of nodal diagrams as the source wave-function now appears in the amplitudes themselves. This leads to spatial and temporal dependences of the expectation value  of the first field over the Glauber-Sudarshan state in eleven-dimensional space-time. Two interluding sub-sections, namely \ref{sec3.2.3} and \ref{sec3.2.f}, are introduced to clarify couple of subtleties regarding the UV/IR mixing and the differences between the nodal and the Feynman diagrams respectively, before we combine all the nodal diagrams in sub-section \ref{3.2.6}.

The combined series of the nodal diagrams from the three set of fields 
turns out to be asymptotic in nature thus ruling out perturbation theory. We dedicate the last section, {\it i.e.} section \ref{borel}, for a detailed study of the consequence of Borel resumming the asymptotic series. Unfortunately, the situation at hand turns out to be more subtle as a naive Borel resummation leads to erroneous conclusions. We discuss the pitfalls of such naive summation techniques in section \ref{tagcorfu}. In section \ref{sec4.2} we show that the errors appear once we carefully evaluate the nodal diagrams themselves: the factorial growths that one would naively expect from these diagrams actually do not appear, but appear from the combinatoric factors from the other two set of fields. This means, if we take $1+\alpha$ number of fields, the factorial growth at order ${\rm N}$ would be at least $({\rm N}!)^\alpha$. Such factorial growths are classified by the {\Su Gevrey-$\alpha$} series and we elaborate on this in section \ref{sec4.3}. In sub-section \ref{sec4.3.1} we provide some details on the mathematical aspects of Gevrey series and in sub-section \ref{sec4.3.2} we discuss the consequence of Borel resummation of the Gevrey-$\alpha$ series. The Borel resummation of the Gevrey-$\alpha$ series leads to new subtleties on the Borel plane. Section \ref{sec4.4} is dedicated to elaborating all the subtleties. The first one appears from the so-called {\it off-shell} contributions to the Glauber-Sudarshan state. This is studied in sub-section \ref{sec4.4.1}. These off-shell contributions may be introduced directly using the source wave-function or by modifying the behavior of the Glauber-Sudarshan state. The former approach is discussed in sub-section \ref{sec4.4.1} whereas the latter approach in discussed in sub-section 
\ref{sec4.4.2}. Both approaches lead to similar answers thus showing the self-consistency of the underlying picture. 

Unfortunately these are not the only subtleties. As with usual field theory computations, the denominator of the path-integral is responsible for 
decoupling the vacuum bubbles. A somewhat similar story happens for the nodal diagrams too: the denominator of the path integral removes all diagrams where the source does not couple to the interacting parts of the Lagrangian. In sub-section \ref{sec4.4.3} we show how this removes all the {\it dominant} nodal diagrams leaving the NLO diagrams that are suppressed by ${\cal O}\left({1\over {\rm V}}\right)$, where ${\rm V}$ is the volume of space-time which in-turn is related to the IR cut-off discussed earlier in sub-section \ref{sec3.3.2}. 

The ${\cal O}\left({1\over {\rm V}}\right)$ suppression of the NLO nodal diagrams leads to another subtlety because they {\it shift} the original tree-level that we studied in sub-section \ref{sec3.2.1}. As elaborated in 
sub-section \ref{sec4.4.4}, such a shift clashes somewhat with the de Sitter no-go theorems, unless we demand the vanishing of such a term. This vanishing leads to a precise prediction of the behavior of the Glauber-Sudarshan state that turns out to be surprisingly consistent with the full non-perturbative structure coming from Borel resumming the Gevrey-$\alpha$ series. We also discuss the alternative off-shell contributions from sub-section \ref{sec4.4.2} in the new light and again find consistent picture by demanding the vanishing of the shift in the tree-level diagram.

All our above analysis leads to a specific prediction for the four-dimensional cosmological constant $\Lambda$. In section \ref{sec4.5} we provide a precise mathematical proof to demonstrate the {\it positivity} of the expression for $\Lambda$. Such a proof justifies that a four-dimensional de Sitter space may indeed appear as a Glauber-Sudarshan state in full string theory overcoming the no-go and the swampland criteria. 

The last section {\it i.e.} section \ref{sec4.6} is dedicated to clarifying couple more subtleties in sub-section \ref{sec4.6.1} and elaborating on the two new predictions of our work, namely (a) the prediction for the form of the cosmological constant in sub-section \ref{sec4.6.2} and (b) the prediction for the form of the Glauber-Sudarshan state that represents the four-dimensional de Sitter space in sub-section \ref{sec4.6.3}. We conclude with some elaborations on future directions in section \ref{disco}.

\subsection{A note on the notations and conventions used \label{sec1.3}}

In this paper, although most of the computations will be done using the path-integral formalism $-$ where operators are not necessary $-$ we would still need to distinguish between operators and their eigenvalues especially in sections \ref{sec2.2} and \ref{sec4.6}. The convention we follow in those two sections is as follows. In section \ref{sec2.2} all operators will be denoted using bold-faced symbols unless mentioned otherwise, whereas in section \ref{sec4.6} (or elsewhere if any) the operators will be denoted by hatted bold-faced symbols. In the remaining sections, bold-faced symbols will {\it not} represent operators as we will be using path-integrals formalism throughout. As an example, ${\bf S}_{\rm int}$ and 
$\widehat{\bf S}_{\rm int}$ will represent operators in sections \ref{sec2.2} and \ref{sec4.6} respectively. In the remaining parts of the paper we will mostly not use the hatted symbol, so a bold-faced symbol like
${\bf S}_{\rm int}$ will simply represent an interacting action and not an operator associated to it.  

Another convention is for the dimensions of the fields. As is popular in string theory, most of the fields will be taken to be dimensionless unless mentioned otherwise. We will also take the usual convention of $\hbar = c = 1$, but explicitly show the ${\rm M}_p$ dimensions of the action and other related objects till at least \eqref{angelslune}. After which, to avoid un-necessary clutter, we will keep ${\rm M}_p = 1$. This would imply, for example, that the momenta $k_i$, the volume ${\rm V}$ etc. are dimensionless (see footnote \ref{yaya} for an exact implementation of this). Generally such a convention doesn't affect any of the result, but becomes very useful when we try to compare {\it large} and {\it small} values of various tunable parameters. We will face this in section \ref{sec4.6} wherein our choice of making these parameters dimensionless would work rather well when we want to give meanings to large and small values.

Finally, we shall define the number of fields in our model to be $1 + \alpha$, and we will take a conservative approach with $\alpha = 2$, so there will be three scalar fields representing one of the 44 graviton components, one of the 84 three-form field components and one fermionic condensate respectively. We will denote the corresponding Glauber-Sudarshan states using the standard notation (used in \cite{coherbeta, coherbeta2}) as $(\alpha(k), \beta(k), \gamma(k))$, and as $(\overline\alpha(k), \overline\beta(k), \overline\gamma(k))$ for the volume-suppressed ones for the three fields respectively. Hopefully there should be no source of confusion between $\alpha, \alpha(k)$ and $\overline\alpha(k)$.

\section{Dynamics in the configuration space of the Glauber-Sudarshan state \label{BA095}}

As with any quantum field theories, with or without gravity, the most efficient way to analyze the interactions is to address them from the configuration space point of view. Such viewpoint is especially useful when we are dealing with excited states in QFT (or string theory), where simple applications of the Feynman rules are not enough to capture the full story. We will soon see examples of these where new diagrammatic approach via the  so-called 
{\it nodal diagrams} will allow us to tackle the dynamics of Glauber-Sudarshan states more efficiently. The Glauber-Sudarshan state, as developed in \cite{coherbeta} and \cite{coherbeta2}, appears to be the {\it only} way by which four-dimensional de Sitter space may exist in string theory overcoming the no-go, the swampland and other criteria. However before delving into this, let us analyze few basic structures of these states pertaining to the four-dimensional de Sitter space.

\subsection{Existence of de Sitter space as a Glauber-Sudarshan state \label{sec2.1}}

Let us remind ourselves why and how should we express a four-dimensional de Sitter space in type IIB theory as a Glauber-Sudarshan state. The analysis is detailed and most of it has appeared in \cite{coherbeta} and \cite{coherbeta2}, so we will be brief here. The aim is to express the following type IIB background:
\bg\label{rihan}
ds^2 = {1\over \Lambda t^2 {\rm H}^2(y)}\left(-dt^2 + \sum_{i = 1}^2 dx_i^2 + dx_3^2\right) + {\rm H}^2(y) g_{mn}(y, t) dy^m dy^n, \nd
as a Glauber-Sudarshan state over a supersymmetric Minkowski vacuum, where 
${\rm H}(y)$ (not to be confused with the Hubble constant ${\rm H}$ or the Hermite polynomial ${\bf H}_n$) is the warp-factor, $\Lambda$ is the cosmological constant and $t$ is the conformal time. Note that we have chosen a specific slicing of de Sitter, here it is the {\it flat slicing}, but the analysis as shown in \cite{coherbeta, coherbeta2} is independent of the de Sitter slicings\footnote{Due to various issues, as explained in sections 2.4 and 2.5 of \cite{coherbeta2}, a {\it static patch} of de Sitter or other patches related to the static patch, does not serve as a good embedding to study de Sitter space for the types of problems that we want to resolve. Static patch is sometimes 
useful, and thus extensively used, in analyzing the thermodynamical properties of a de Sitter space but as discussed in \cite{coherbeta} this is not a necessary requirement.}. As argued in \cite{desitter2} the background 
\eqref{rihan} can only exist in the presence of time-dependent fluxes, sources and quantum corrections $-$ the latter include the whole slew of 
non-perturbative, non-local and topological corrections. If we regard \eqref{rihan} as a {\it vacuum} solution in IIB then fluctuations over such a background creates time-dependent frequencies which, having non-linear temporal growths, create problems in constructing a Wilsonian effective action as discussed earlier in section \ref{sec1.1}. In fact, even with linear temporal growths of the frequencies, there doesn't exist a simple way to integrate out degrees of freedom to write an effective theory at low energies. Additionally, 
\eqref{rihan} breaks supersymmetry explicitly and therefore there is no way to cancel (or renormalize) the zero point energies, implying that they would contribute to the cosmological constant leading to the oft mentioned cosmological constant problem. 

There are other issues stemming from the temporal dependences of the degrees of freedom, and as concluded in \cite{coherbeta, coherbeta2}, it appears that the only way out of these conundrums is to view the de Sitter space as a Glauber-Sudarshan state ({\it i.e.} as a coherent state). In fact the complete background, that includes the metric configuration 
\eqref{rihan} as well as the axio-dilaton, fluxes etc., should be viewed as a Glauber-Sudarshan state over a supersymmetric Minkowski background with time-independent fluxes, axio-dilaton etc.

Due to various technical issues, as detailed in \cite{coherbeta, coherbeta2}, it is harder to construct directly such a background in the IIB side. Therefore one can dualize this background to M-theory and therein one can easily construct the Glauber-Sudarshan state as shown in 
\cite{coherbeta, coherbeta2}. The IIB Glauber-Sudarshan state then can be got by shrinking the M-theory torus to zero size. Interestingly, the M-theory uplift of the background \eqref{rihan} takes the following form:
\bg\label{carrie}
ds^2 = g_s^{-{8\over 3}}\left(-dt^2 + \sum_{i = 1}^2 dx_i^2\right) + 
g_s^{-{2\over 3}} {\rm H}^2(y) g_{mn}(y, t) dy^m dy^n + g_s^{4\over 3}~ \eta_{ab} dw^a dw^b, \nd
where $g_s$ is the dual type IIA coupling which takes the form 
$g_s = \sqrt{\Lambda} \vert t\vert {\rm H}(y)$, and $\eta_{ab}$ is the flat metric on the toroidal direction of M-theory when we take vanishing axio-dilaton in the IIB side (which is the constant coupling limit of F-theory). This could be easily generalized to arbitrary axio-dilaton (with both spatial and temporal dependences), but this will not change the {\it form} of the M-theory metric (except that the toroidal direction will no longer be a square torus), or any of the conclusions that we advocated in \cite{desitter2, coherbeta, coherbeta2}. This is of course one of the profound advantage that we gain by uplifting the system to M-theory (the other being the existence of a low-energy action at far IR). Our claim in 
\cite{coherbeta} and \cite{coherbeta2} is that the metric \eqref{carrie} along-with the temporal degrees of freedom supporting the background can be realized as a Glauber-Sudarshan state over the following solitonic background:

{\scriptsize
\bg\label{sissys}
ds^2 = \left({1\over \sqrt{\Lambda} a_o {\rm H}(y)}\right)^{8\over 3} \left(-dt^2 + \sum_{i = 1}^2 dx_i^2\right) + 
\left({{\rm H}^{3}(y)\over \sqrt{\Lambda} a_o {\rm H}(y)}\right)^{2\over 3} \widetilde{g}_{mn}(y) dy^m dy^n + 
\left(\sqrt{\Lambda} a_o {\rm H}(y)\right)^{4\over 3}~ \eta_{ab} dw^a dw^b, \nonumber\\ \nd}
with time-independent fluxes and other degrees of freedom. Note that we have written the warped Minkowski $2 + 1$ direction in a suggestive way by putting a constant dimensionless factor $\sqrt{\Lambda} a_o$ and the internal metric to be $\widetilde{g}_{mn}(y)$. This way of writing \eqref{sissys} would tell us that the Glauber-Sudarshan state 
representation for \eqref{carrie} would satisfy the boundary condition, which is of course one of the necessary condition for realizing a coherent state over a vacuum configuration. For the analysis in this paper we will keep ${\rm H}(y) =$ constant, so that 
$g_s = \sqrt{\Lambda} \vert t \vert$. With a non-trivial embedding of de Sitter the temporal dependence is captured by ${g_s\over {\rm H}(y)
{\rm H}_o({\bf x})} \equiv {g_s\over {\rm HH}_o}$ where ${\rm H}_o({\bf x})$ is related to the embedding (see section 2 of \cite{coherbeta2}), implying that the temporal dependences may be traded with the $g_s$ dependences.  

The construction of a de Sitter Glauber-Sudarshan state over the Minkowski background \eqref{sissys} is highly non-trivial because we are in-principle dealing with an interacting vacuum. Shifting such a vacuum with an appropriate displacement operator creates the necessary Glauber-Sudarshan state. The expectation values of the metric operators over such a state 
create the configuration \eqref{carrie} from such a state. Computing the expectation values using path integrals would automatically insert all possible quantum corrections, and therefore our first goal would be to see how the quantum effects are realized from the configuration space, and how they act on simple eigenstates.

\subsection{Dynamics in the configuration space and quantum effects \label{sec2.2}}

One of our first aim is analyze how any given quantum series 
influences the dynamics in the configuration space, and in particular, the dynamics of the corresponding Glauber-Sudarshan state. The analysis will be slightly technical, so we will start with a simple example of a real scalar field theory in four-dimensions with polynomial and derivative interactions. Such a theory is endowed with the following action:

\begin{figure}[h]
\centering
\vskip-1in
\begin{tabular}{c}
\includegraphics[width=6in]{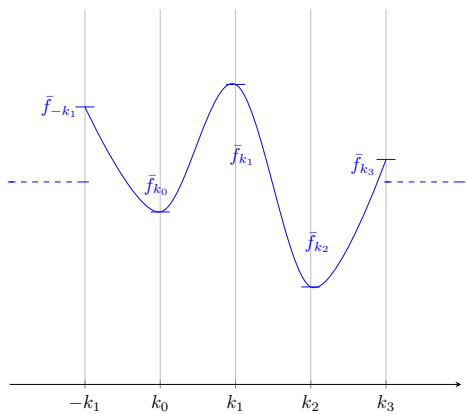}
\end{tabular}
\vskip-4.5in
\caption[]{A delta function state \eqref{warduluk} in the configuration state parametrized by discrete set of momenta $\pm{\bf k}_i$. This leads to a {\it classical} field configuration at a given instant of time, that becomes quantum almost immediately.}
\label{delta1}
\end{figure}

\begin{figure}[h]
\centering
\vskip-1in
\begin{tabular}{c}
\includegraphics[width=6in]{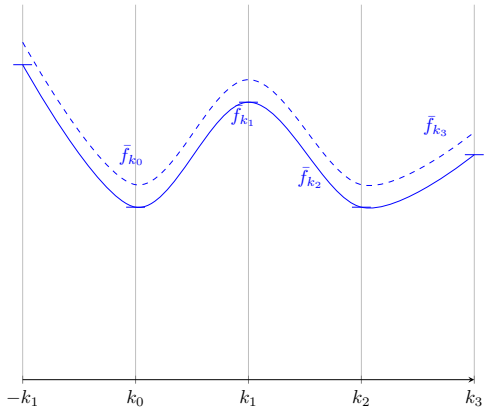}
\end{tabular}
\vskip-4.4in
\caption[]{Deformation of a delta function state via the operator ${\bf S}_{\rm int}(\{{\bf f}_{{\bf k}_n}\})$ at a given instant of time. As shown in \eqref{parker}, the state \eqref{warduluk} is an eigenstate of the operator
${\bf S}_{\rm int}(\{{\bf f}_{{\bf k}_n}\})$ and therefore the deformation shown by the dotted lines can be easily quantified.}
\label{delta2}
\end{figure}

\bg\label{sabbath}
{\rm S}[{\varphi}] \equiv {\rm S}_{\rm kin} + {\rm S}_{\rm int} = \sum_{n, m, p} {1\over {\rm M}_p^{2m - 4}}\int d^4x \sqrt{-{\bf g}_4} ~c_{nmp} \varphi^n \left({\bf D}\cdot {\bf D}\right)^m \varphi^{p}, \nd
where $c_{nmp}$ and $\varphi$ are dimensionless coupling constants and dimensionless field respectively, ${\bf D}$ is the covariant derivative defined using a four-dimensional metric, and $(m, n, p) \in \mathbb{Z}_+$. If we ignore the curved space-time, ${\bf D}$ will become the standard derivative and the system will simplify. For our example, we will resort to the simpler case. The dimensional counting will tell us that for $m > 2$ most of the terms will be suppressed by powers of ${\rm M}_p$ unless $\varphi$ is a localized field, but we will not worry too much about that either. To allow for a canonical kinetic term, we will take $c_{111} \equiv 1$ and we can interpret \eqref{sabbath} as the Wilsonian effective action at a given scale. In the presence of a solitonic background, we can express $\varphi(x)$ for $x \equiv ({\bf x}, t)$ in the following standard way:
\bg\label{telephone}
\varphi(x) = {1\over {\rm M}^3_p}\int d^3{\bf k} f_{\bf k}(t) \psi_{\bf k}({\bf x}) \approx {1\over {\rm M}^3_p{\rm V}} \sum_n e^{-i{\bf k}_n. {\bf x}} 
f_{{\bf k}_n}(t) , \nd
where $\psi_{\bf k}({\bf x}) \approx e^{-i{\bf k}.{\bf x}}$ are the fluctuating modes over a solitonic background $\varphi_0({\bf x})$, $k^i_{n} = {2\pi n^i\over {\rm L}}$ with $i = 1, 2, 3$ and ${\rm V} = {\rm L}^3$. (Appropriate powers of ${\rm M}_p$'s are inserted in to keep $\varphi$ and its Fourier transform dimensionless.) The decomposition \eqref{telephone} is typically off-shell, and for the usage in path-integral we will continue with this. As discussed in \cite{coherbeta}, such an off-shell decomposition fits well with the temporal evolution of a Glauber-Sudarshan state (or in fact with any quantum state created over the solitonic configuration).  
However before discussing the Glauber-Sudarshan state, let us start with the following simpler state in the configuration space for the scalar field in the box:
\bg\label{warduluk}
\vert \overline{f} \rangle \equiv \prod_n \vert \overline{f}_{{\bf k}_{n}}\rangle, \nd
which is in fact a delta-function state in the configuration space as shown in {\Su Figure} \ref{delta1}. Although such a state leads to a purely {\it classical} state in space-time at a given instant of time, it doesn't survive long and almost immediately deforms to a highly quantum state. (See also {\Su Figure} \ref{delta100} for a space-time delta function state.) Nevertheless, as we shall see, a state like \eqref{warduluk} will be immensely useful in analyzing the dynamics of our Glauber-Sudarshan state. The usefulness of 
\eqref{warduluk} lies in the fact that it is an {\it eigenstate} of the field \eqref{telephone} once we convert it into an operator. This is because an operator ${\bf S}_{\rm int}$ from \eqref{sabbath} acts on \eqref{warduluk} as:

{\footnotesize
\bg\label{parker}
&&{\bf S}_{\rm int} \vert \overline{f}\rangle \equiv  \sum_{n, m, p} {(-1)^m c_{nmp}\over {\rm V}^{n + p} {\rm M}_p^{2m' - 4}} 
 \sum_{\{n_i, n_j\}} 
 \prod_{(i, j) = (1, 1)}^{(n, p)} \delta^3\left(\sum_{i = 1}^n {\bf k}_{n_i}  + \sum_{j = 1}^p {\bf k}_{n_j}\right) 
\left(\sum_{l = 1}^n {\bf k}_{n_l}\right)^{2m}{\bf f}_{{\bf k}_{n_i}} {\bf f}_{{\bf k}_{n_j}}
\prod_s \vert \overline{f}_{{\bf k}_{s}}\rangle\nonumber\\
&~~~~ &=  \sum_{n, m, p} {(-1)^m c_{nmp}\over {\rm V}^{n + p} 
{\rm M}_p^{2m' - 4}} 
 \sum_{\{n_i, n_j\}} 
 \prod_{(i, j) = (1, 1)}^{(n, p)} \delta^3\left(\sum_{i = 1}^n {\bf k}_{n_i}  + \sum_{j = 1}^p {\bf k}_{n_j}\right) 
\left(\sum_{l = 1}^n {\bf k}_{n_l}\right)^{2m}\prod_{q = i}^j\delta(s - n_q)\overline{f}_{{\bf k}_{n_q}}
\prod_s \vert \overline{f}_{{\bf k}_{s}}\rangle, \nonumber\\ \nd}
at any given instant of time, with bold-faced ${\bf S}_{\rm int}$ and ${\bf f}_{{\bf k}_{n_i}}$ representing the 
{\it operators}; $m' \equiv m + {1\over 2}(n + p)$; and $\overline{f}_{{\bf k}_{n_q}}$ as the corresponding eigenvalues. The above equation \eqref{parker} is an eigenvalue equation and therefore the action of ${\bf S}_{\rm int}$ simply deforms the delta function state \eqref{warduluk} at a given instant of time as shown in {\Su Figure \ref{delta2}}. However this isn't true once time dependences are switched on, because the operator ${\bf S}_{\rm int}$ now decomposes as:
\bg\label{welkomeden}
{\bf S}_{\rm int} \equiv {\rm M}_p\int dt\left({\bf S}^{(1)}_{\rm int}(\{{\bf f}_{{\bf k}_n}\}) + {\bf S}^{(2)}_{\rm int}(\{\dot{\bf f}_{{\bf k}_n}\}, \{{\bf f}_{{\bf k}_m}\})\right), \nd
and because of the second term, which is a function of $\dot{\bf f}_{{\bf k}_n}$, the delta-function state 
\eqref{warduluk} no longer remains an eigenstate of the full operator ${\bf S}_{\rm int}$ in \eqref{welkomeden} (although it is still an eigenstate\footnote{In other words 
${\bf S}^{(1)}_{\rm int}(\{{\bf f}_{{\bf k}_n}\})\vert \overline{f} \rangle = 
{\rm S}^{(1)}_{\rm int}(\{\overline{f}_{{\bf k}_n}\})\vert \overline{f} \rangle$, but this is not the case with 
${\bf S}^{(2)}_{\rm int}(\{\dot{\bf f}_{{\bf k}_n}\}, \{{\bf f}_{{\bf k}_m}\})$.}
 of ${\bf S}^{(1)}_{\rm int}(\{{\bf f}_{{\bf k}_n}\})$). Of course this is what we expect and it is only because of the second term in \eqref{welkomeden} that the path integral formalism makes sense. Additionally, 
the choice of ${\bf S}_{\rm int}$ is useful because it is directly related to the interacting Hamiltonian ${\bf H}_{\rm int}$ via
 ${\rm M}_p\int dt~{\bf H}_{\rm int} \equiv - {\bf S}_{\rm int}$. The temporal evolution of any state is now governed by the standard formula: 

{\scriptsize
 \bg\label{hermanos} 
 {\bf U}[\dot{\bf f}, {\bf f}] &\equiv& {\rm exp}\left[-i{\rm M}_p\int \left({\bf H}_0 - {\bf S}^{(2)}_{\rm int}(\{\dot{\bf f}_{{\bf k}_n}\}, \{{\bf f}_{{\bf k}_m}\})\right) dt + i{\rm M}_p\int {\bf S}^{(1)}_{\rm int}(\{{\bf f}_{{\bf k}_n}\}) dt\right] \\
 & = & {\rm exp}\left[-i{\rm M}_p\int \left({\bf H}^{(2)}_0(\{\dot{\bf f}_{{\bf k}_n}\}, \{{\bf f}_{{\bf k}_m}\}) - {\bf S}^{(2)}_{\rm int}(\{\dot{\bf f}_{{\bf k}_n}\}, \{{\bf f}_{{\bf k}_m}\})\right) dt - i{\rm M}_p\int \left({\bf H}^{(1)}_0(\{{\bf f}_{{\bf k}_n}\}) 
 - {\bf S}^{(1)}_{\rm int}(\{{\bf f}_{{\bf k}_n}\})\right) dt\right], \nonumber \nd}
  where we have decomposed ${\bf H}_0$, the free Hamiltonian ({\it i.e.} the one proportional to 
 $c_{111} \equiv 1$ in \eqref{sabbath} and we keep it dimensionless), as ${\bf H}_0 \equiv   {\bf H}^{(1)}_0(\{{\bf f}_{{\bf k}_n}\}) 
 + {\bf H}^{(2)}_0(\{\dot{\bf f}_{{\bf k}_n}\}, \{{\bf f}_{{\bf k}_m}\})$. This decomposition, which splits 
 ${\bf U}[\dot{\bf f}, {\bf f}]$ into two pieces that depend on $\dot{\bf f}_{{\bf k}_n}$ and ${\bf f}_{{\bf k}_n}$, is useful because the part that depends only on ${\bf f}_{{\bf k}_n}$ would simply deform the state 
 \eqref{warduluk} as shown in {\Su Figure \ref{delta1}}. We can make this more explicit by re-writing 
 \eqref{hermanos} in the following suggestive way that illustrates the behavior of ${\bf U}[\dot{\bf f}, {\bf f}]$ more clearly:
 
 \begin{figure}[h]
\centering
\vskip-1in
\begin{tabular}{c}
\includegraphics[width=6in]{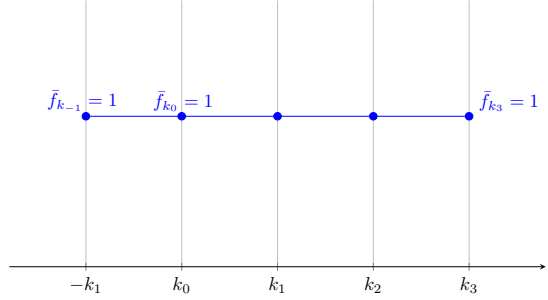}
\end{tabular}
\vskip-5in
\caption[]{A delta function state \eqref{warduluk} in the configuration state that leads to a delta function state in space-time. The spreading of this state in the space-time and in the configuration space will be correlated.}
\label{delta100}
\end{figure}

{\footnotesize
\bg\label{miagoth}
{\bf U}[\dot{\bf f}, {\bf f}] = {\rm exp}\left[-i{\rm M}_p\int {\bf H}_1 dt - \sum_{n = 1}^\infty {1\over n(n +1)}\int_0^1 dh 
\Big(\mathbb{I} - {\rm exp}\left(\mathbb{L}_{\bf A}\right){\rm exp}\left(h \mathbb{L}_{\bf B}\right)\Big)^n{\bf B}\right] {\rm exp}\left(-i{\rm M}_p\int {\bf H}_2 dt\right), \nonumber\\ \nd}
which would act on the ket \eqref{warduluk} by two exponential factors, where we have defined the action of $\mathbb{L}_{\bf A}$ on an operator ${\bf B}$ as $\mathbb{L}_{\bf A} {\bf B} \equiv [{\bf A}, {\bf B}]$. This clearly means ${\rm exp}\left(\mathbb{L}_{\bf B}\right) {\bf B} = {\bf B}$. The other operators appearing  in 
\eqref{miagoth} are defined in the following way:

{\footnotesize
\bg\label{miagoth2}
&& {\bf A} \equiv -i{\rm M}_p\int {\bf H}_1(\{\dot{\bf f}_{{\bf k}_n}\}, \{{\bf f}_{{\bf k}_m}\}) dt - i{\rm M}_p\int {\bf H}_2(\{{\bf f}_{{\bf k}_m}\}) dt, ~~~ {\bf B} \equiv i{\rm M}_p\int {\bf H}_2(\{{\bf f}_{{\bf k}_m}\}) dt \nonumber\\
&&{\bf H}_2 \equiv {\bf H}^{(1)}_0(\{{\bf f}_{{\bf k}_n}\}) - {\bf S}^{(1)}_{\rm int}(\{{\bf f}_{{\bf k}_n}\}), ~~~~
{\bf H}_1 \equiv {\bf H}^{(2)}_0(\{\dot{\bf f}_{{\bf k}_n}\}, \{{\bf f}_{{\bf k}_m}\}) - {\bf S}^{(2)}_{\rm int}(\{\dot{\bf f}_{{\bf k}_n}\}, \{{\bf f}_{{\bf k}_m}\}), \nd}
with the further assumption that integrals (or summations) over the momenta are taken into account for each of the individual operators ${\bf H}_1$ and ${\bf H}_2$. With the two exponential forms in \eqref{miagoth} it is easy to see how the evolution operator ${\bf U}[\dot{\bf f}, {\bf f}]$ now acts on the delta-function state \eqref{warduluk} in the configuration space: if the first exponential was not there, the second exponential, expressed in terms of ${\bf H}_2$, would have simply shifted the state \eqref{warduluk} in the configuration space as shown in {\Su Figure \ref{delta2}} (the dotted line in the figure would have shifted with time). The reason, as mentioned earlier, is because the state \eqref{warduluk} remains as eigenstate of the second exponential factor. However this state in {\it not} an eigenstate of any of the operators $\{\dot{\bf f}_{{\bf k}_n}\}$, and since:
\bg\label{britsnow}
\mathbb{L}_{\bf A} {\bf B} = {\rm M}^2_p\int dt dt'\left[{\bf H}_1(t), {\bf H}_2(t')\right], \nd
the state \eqref{warduluk} cannot be an eigenstate of the first exponential operator in \eqref{miagoth}. Thus because of the action of the first exponential operator, the state \eqref{warduluk} expands all over the configuration space, instead of just deforming as in {\Su Figure \ref{delta2}}, leading simultaneously to all possible configurations of the scalar field in space-time.  

The decompositions \eqref{welkomeden} and \eqref{miagoth2} make our analysis easier once we go to the M-theory perturbative corrections, for example as in eq. (4.81) of \cite{coherbeta2}: all the generalized curvature and G-flux components that are not expressed using derivatives of the temporal coordinate will contribute to ${\bf S}_{\rm int}^{(1)}$, and the rest will contribute to ${\bf S}_{\rm int}^{(2)}$. There is however some subtlety in our choice of the delta-function state \eqref{warduluk} now because we have metric, G-flux and fermionic components. We will quantify this as we go along. The Glauber-Sudarshan state at any given instant of time may be written as\footnote{We are ignoring a factor of the configuration space volume that appears when we convert the integral into discrete sum.}:
\bg\label{thar}
\vert {\sigma}\rangle \equiv \prod_{\bf k} \left(\sum_{f_{\bf k}} \Psi_\Omega^{({\sigma}_{\bf k})}\left(f_{\bf k}\right) \vert f_{\bf k}\rangle\right), \nd
where $ \Psi_\Omega^{({\sigma}_{\bf k})}\left(f_{\bf k}\right)$ is the wave-function of the Glauber-Sudarshan state for a given mode ${\bf k}$ in ten space-dimensions (recall that $k = (k_0, {\bf k})$), by shifting the interacting vacuum $\vert\Omega\rangle$ in the configuration space. The shifting parameter 
$\bar{\sigma}_{\bf k}$ and other parameters appearing in \eqref{thar} may be defined in the following 
way\footnote{A slight change in notation from the scalar field theory example to the one in M-theory. Bold-faced letters will continue to denote warped fields, and the operators will henceforth be denoted by {hats}, {\it i.e.} 
$\widehat{\bf g}_{\rm CD}$ will denote the operator corresponding to the field ${\bf g}_{\rm CD}$, unless mentioned otherwise. Similar distinctions will go for the three-form field and the Rarita-Schwinger fermion.}:

{\footnotesize
\bg\label{bggracie}
f_{\bf k} \equiv \Big(\{{\bf g}_{\rm AB}({\bf k})\}, \{{\bf C}_{\rm ABE}({\bf k})\}, \{\psi_{\rm B}({\bf k})\}\Big), ~~~
{\sigma}_{\bf k} \equiv 
\Big(\{{\alpha}_{\rm AB}({\bf k})\}, \{{\beta}_{\rm ABE}({\bf k})\}, 
\{{\gamma}_{\rm B}({\bf k})\}\Big), \nd}
where $({\rm A, B, ..}) \in {\bf R}^{2, 1} \times {\cal M}_4 \times {\cal M}_2 \times {\mathbb{T}^2\over {\cal G}}$, with $({\alpha}, {\beta})$ being bosonic and ${\gamma}$ being fermionic. As mentioned in \cite{coherbeta2, coherbeta}, the Fourier transform of the shift $\sigma$ {\it does not} give the classical de Sitter background due to the subtlety of wave-function renormalization. This is discussed briefly in section 6 of \cite{coherbeta2} and we shall provide some more details on it soon.

\begin{figure}[h]
\centering
\vskip-1in
\begin{tabular}{c}
\includegraphics[width=6in]{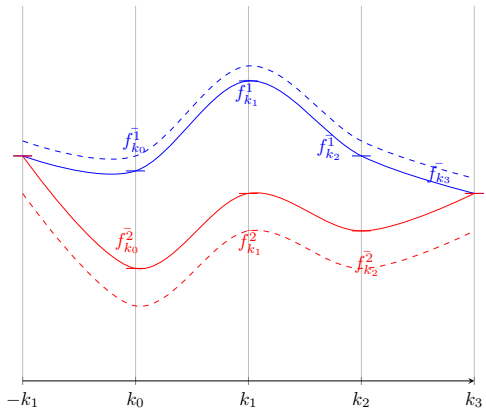}
\end{tabular}
\vskip-4.4in
\caption[]{Two delta function states with coincident points in the configuration space can deform differently
via the operator ${\bf S}_{\rm int}(\{{\bf f}_{{\bf k}_n}\})$ at a given instant of time. Again each of these deformations may be quantified.}
\label{delta3}
\end{figure}

Let us now see how the full interacting action \eqref{sabbath} acts on the Glauber-Sudarshan state 
\eqref{thar}. This is where our work on the simpler scalar field theory becomes useful.
Comparing \eqref{thar} to \eqref{warduluk} we see that the action of ${\bf S}^{(1)}_{\rm int}$ and 
${\bf S}^{(2)}_{\rm int}$, or more appropriately the evolution operator \eqref{miagoth}, on \eqref{thar} 
may be quantified by the following steps.

\vskip.1in

\noindent $\bullet$ Express \eqref{thar} as a linear combination of an infinite number of delta-function states 
\eqref{warduluk}. All these states are further modulated by the corresponding products of the Glauber-Sudarshan wave-functions. 

\vskip.1in

\noindent $\bullet$ The action of ${\bf S}_{\rm int}^{(1)}$ will now {\it deform} each of these delta-function states, as each of these delta-functions states are individually eigenstates of the interacting Hamiltonian $-{\bf S}^{(1)}_{\rm int}$. Since the eigenvalues are expected to be different, as shown in {\Su Figure \ref{delta3}}, these deformations would all be generically different. Furthermore, these deformations may be computed in the temporal domain where the type IIA string coupling $g_s < 1$. 

\vskip.1in

\noindent $\bullet$ The action of ${\bf S}_{\rm int}^{(2)}$ (as well as ${\bf H}_1, 
\mathbb{L}_{{\bf A}}$ and ${\bf B}$), now spreads each of these delta-function states in the full configuration space. When we sum-up these evolutions in the temporal domain where the type IIA coupling $g_s < 1$, it reproduces precisely the temporal evolution of our Glauber-Sudarshan state under the full perturbative quantum corrections. One may similarly add the non-perturbative as well as the non-local quantum corrections and study the temporal evolution in a similar fashion. How this may be achieved by Borel resummation will be discussed soon.

\vskip.1in

\noindent There is one puzzle that needs to be clarified before we move on, and has to do the expression of $\mathbb{D}(\sigma)$ in terms of 
$\mathbb{D}_0(\sigma)$. One possible connection is given by the following relation \cite{coherbeta}:
\bg\label{hyurino}
{\mathbb{D}}(\sigma, t) = \lim_{{\rm T} \to \infty(1-i\epsilon)} {\mathbb{D}}_0(\sigma, t)~{\rm exp}\left(i{\rm M}_p\int_{-{\rm T}}^t 
dt~{\bf H}_{\rm int}\right), \nd
which should be viewed as an operator relation. Note the appearance of $t$ in the definition of ${\mathbb{D}}_0(\sigma, t)$, which is the displacement operator for the free theory\footnote{Recall that we have taken the non-unitary part of the displacement operator \cite{coherbeta}.}. Turning \eqref{hyurino} around would imply either of the two things: one, the temporal dependence of 
${\mathbb{D}}(\sigma, t)$ cancels the temporal dependence from the exponential piece; or two, the temporal dependences do not cancel. The former would imply\footnote{Here $\widehat{\bf g}_{\mu\nu}$ denotes the operator corresponding to the metric component. We also define the warped metric ${\bf g}_{\mu\nu} \equiv \langle \widehat{\bf g}_{\mu\nu}\rangle_{\sigma}$ to be consistent with both \eqref{bggracie} as well as the notations used in \cite{coherbeta, coherbeta2}. See also section \ref{sec1.3} for details on the conventions used in the paper.}:
\bg\label{yuhana}
\langle\alpha \vert ~\widehat{\bf g}_{\mu\nu}({\bf x}, y, z; t)~\vert \alpha\rangle = {\eta_{\mu\nu}\over {\rm H}^{2/3}(y)} +
{\bf Re}\left({1\over {\rm M}^9_p}\int{d^{10}{\bf k} \over 2\omega^{(\psi)}_{\bf k}}~
{\alpha}^{(\psi)}_{\mu\nu}({\bf k}, t) \psi_{\bf k}({\bf x}, y, z)\right), \nd
which reproduces the classical de Sitter background when 
$\alpha^{(\psi)}_{\mu\nu} = \alpha^{({\rm c};\psi)}_{\mu\nu}$ and ${\rm H}(y)$ is the warp-factor from \eqref{rihan}\footnote{A slight change of notation from \cite{coherbeta, coherbeta2}: the Glauber-Sudarshan state is denoted by $\vert\sigma\rangle$ and the coherent state by $\vert\sigma^{({\rm c})}\rangle$ here compared to $\vert\overline\sigma\rangle$ and $\vert\sigma\rangle$ respectively in \cite{coherbeta, coherbeta2} as we shall reserve the notation $\vert\overline\sigma\rangle$ to denote a volume suppressed Glauber-Sudarshan state (see \eqref{nehaprit}).\label{bldride}}. This is the free-field result and as such the quantum corrections are {\it not} taken into account. On the other hand, the latter choice with un-cancelled temporal dependences would only make sense if the displacement of the interacting vacuum would match the displacements $\overline\sigma$ of the free vacua (for all momenta ${\bf k}$) at a {\it given instant of time}. But since both the interacting vacuum $\vert\Omega \rangle$ and 
the displacement operators ${\mathbb{D}}(\sigma, t)$ are changing with time, a classical result like \eqref{yuhana} seems 
unlikely\footnote{Plus they would be forbidden by the no-go theorems \cite{GMN}. Such a constraint is borne out of a hitherto unexplored deeper connection between the Schwinger-Dyson's equations that control the dynamics in the {\it bulk} and the expectation values of the metric operators over the Glauber-Sudarshan state. Elucidating this connection is beyond the scope of the present work and will be dealt separately elsewhere. \label{dee}}, and therefore relation like \eqref{hyurino} needs to be carefully implemented in the path-integral so that all the quantum corrections may be taken into account.

One can be a bit more quantitative here. The displacement operator for the fully interacting theory is naturally a complicated object as it is in general constructed out of the creation and the annihilation operators for all modes and all degrees of freedom, including the metric, fluxes and the fermions (see discussion in \cite{coherbeta}). If we consider, for simplicity, an interacting scalar field theory of the form \eqref{sabbath}, then a possible candidate for the displacement operator in Fourier space may be written as:
\bg\label{milenej}
{\rm log}~{\mathbb{D}}(\sigma(k), z) = \sum_{n,m,p} {\mathbb{D}}_{nmp}(k) \equiv \sum_{n,m,p} z_{nmp}
~k^{2n}
\big(\sigma^{(m,p)}(k) {\widetilde\varphi}^{\ast p}(k) + {\rm c.c}\big), \nd
where $\sigma^{(m, p)}(k)$ are generic functions of $k$;  $z_{nmp}$ are dimensionful coefficients with $n \ge 0, m \ge 1$, 
$p\ge 1$, $z_{011} \equiv 1$ and $\sigma^{(1, 1)}(k) \equiv \sigma(k)$. This is already pretty involved, but we can study it term by term and see whether any simplification results. (Note that terms with $z_{nmp} \equiv z_{np} \delta_{nm}, ~\forall p$ are suppressed by powers of ${\rm M}_p$.) The first term of the form:
\bg\label{ijacob}
{\cal D}_{11}(k) \equiv \sum_n {\mathbb{D}}_{n11}(k) =
\sum\limits_n z_{n11}~k^{2n}\big(\sigma(k) {\widetilde\varphi}^\ast(k)
+ \sigma^\ast(k)\widetilde\varphi(k)\big), \nd may be easily simplified by absorbing $\sum\limits_n z_{n11}k^{2n}$ in the definition of $\sigma(k)$. (Alternatively, we can use this freedom to make $z_{n11} = 0$, for $n \ge 1$.) The second term, of the form:
\bg\label{irene1}
{\cal D}_2(k) \equiv \sum_{n,m} {\mathbb{D}}_{nm2}(k) =
\sum_{n, m}z_{nm2}~ k^{2n}\big(\sigma^{(m, 2)}(k) {\widetilde\varphi}^{\ast 2}(k) + \sigma^{\ast(m, 2)}(k) \widetilde\varphi^2(k)\big),\nd is more subtle because its presence will change the propagator of the theory unless $\sigma(k)$ itself is proportional to $k^2$. Since such a choice of $\sigma(k)$ will {\it not} lead to the de Sitter state that we want, we will take $z_{nm2} = 0$. The third term onwards, namely for $m \ge 1, p \ge 1$, renormalizes the
interaction terms in \eqref{sabbath} implying that it is only the first term that can actually change the dynamics of the theory\footnote{This is a bit of a subtle point, so let us elaborate it in some details. For start we will take a scalar field $\varphi(x, y, z)$ whose Fourier components are given by $\widetilde\varphi(k) \equiv \widetilde\varphi_k$, and define the corresponding Glauber-Sudarshan state as $\sigma^{(m, p)}(k) \equiv (-\sigma_k)^m$. If we now choose $z_{nmp}$ in \eqref{milenej} in such a way that every term of ${\bf S}_{\rm tot} \equiv {\bf S}_{\rm kin} + {\bf S}_{\rm int}$ is shifted accordingly, then the expectation value of $\varphi$ over the Glauber-Sudarshan state is given by the following path-integral expression (we will elaborate more on this in the next section):

{\tiny
\bg\label{mouny}
\langle\varphi\rangle_{\sigma} &= &{\int\prod\limits_k d\widetilde\varphi_k~{\rm exp}\left[-{a_k\over {\rm M_p^9V}}\left(\widetilde\varphi_k - {{\rm M_p^2}\sigma_k\over a_k}\right)^2\right]\left(1 + \sum\limits_{k'}\mathbb{F}\left[\left\{\widetilde\varphi_{k'} - {{\rm M_p^2}\sigma_{k'}\over a_{k'}}\right\}\right]\right) ~{1\over {\rm M_p^{11}V}}\sum\limits_{k''} \widetilde\varphi_{k''}\psi_{{\bf k}''}({\bf x}, y, z)e^{-ik''_0t}
\over
\int\prod\limits_k d\widetilde\varphi_k~{\rm exp}\left[-{a_k\over {\rm M_p^9V}}\left(\widetilde\varphi_k - {{\rm M_p^2}\sigma_k\over a_k}\right)^2\right]\left(1 + \sum\limits_{k'}\mathbb{F}\left[\left\{\widetilde\varphi_{k'} - {{\rm M_p^2}\sigma_{k'}\over a_{k'}}\right\}\right]\right)} \nonumber\\
& = & {\bf Re}\left({1\over {\rm M}_p^9}\int d^{11}k ~{\sigma_k\over a_k}~
\psi_{\bf k}({\bf x}, y, z) ~e^{-ik_0 t}\right) + 
{\int {\cal D}\varphi~{\rm exp}\left[-i{\bf S}_{\rm tot}(\varphi)\right]~
\varphi(x, y, z) \over \int {\cal D}\varphi~{\rm exp}\left[-i{\bf S}_{\rm tot}(\varphi)\right]}, \nonumber\nd}
where $\mathbb{F}[\varphi] = {\rm exp}\left[-i{\bf S}_{\rm int}\right] - 1$
and $a_k = {ik^2}$ is the propagator. The first is the tree-level answer from \eqref{yuhana} (we are ignoring the solitonic contribution) and the second 
term is the one-point function over the usual vacuum. Both are problematic: if the second term vanishes, then we are only getting the tree-level answer which would clash with the no-go theorems \cite{GMN}. On the other hand, if the second term doesn't vanish, it will create uncancelled tadpoles making the underlying analysis inconsistent.\label{maisonmey}}. This then tells us that there is a simple choice of displacement operator for the interacting theory that takes the form: 
\bg\label{lucliu}
{\mathbb{D}}(\sigma)\vert\Omega\rangle \equiv  
~\prod_k {\mathbb{D}}(\sigma(k), 0)\vert\Omega\rangle
= ~\prod_k {\rm exp}\left(
 {{\cal D}_{11}(k) \over \sum\limits_n z_{n11}~k^{2n}}\right) \vert\Omega\rangle, \nd
on an interacting vacuum $\vert\Omega\rangle$ which suffices for all practical computations, with ${\cal D}_{11}$ as in \eqref{ijacob} and discretized momenta (we are ignoring volume factor that will be inserted in \eqref{lincoln}). The reason is simple: if we take a more complicated form like \eqref{milenej}, then the equation relating\footnote{For details on this, see section 6.1 of \cite{coherbeta2} and footnote \ref{bldride}.} $\sigma(k)$ with $\sigma^{({\rm c})}(k)$ will be more complicated, resulting in a more involved wave-function renormalization (in addition to the issues pointed out in footnote \ref{maisonmey}), but the underlying physics will not change. Thus while the choice 
\eqref{hyurino} with cancelling temporal dependence leads to a coherent state in the free theory, the choice \eqref{lucliu} on an interacting vacuum will lead to the required Glauber-Sudarshan state. 

The final analysis from the aforementioned point of view then matches exactly with the analysis that we discussed in section 6 of \cite{coherbeta2}. However there is yet one more point that remains to be elaborated on and has to do with the expectation value of the metric operator in the presence of all possible corrections from an equivalent 
quantum series like
\eqref{sabbath} (or as in \eqref{angelslune} later) and the non-perturbative/non-local ones. This is where 
path integral formulation becomes immensely useful, and we turn to it next.

\section{Path integral analysis, nodal diagrams and quantum corrections \label{sec3}}

The path integral formulation of the Glauber-Sudarshan state has been discussed earlier in \cite{coherbeta, coherbeta2}, but those analysis 
have mostly been relegated to the free part of the action (with small inputs from the interaction parts). Here we want to rectify the situation by including all possible quantum corrections, much along the lines of 
say eq. (4.81) of \cite{coherbeta2}. However
because of the sheer complexity of the quantum series like the ones in \cite{coherbeta2}, including the non-perturbative/non-local ones, it is not practical to carry out this computation (although this could in principle be done). In the following therefore we will simplify this computation by resorting to a set-up with scalar degrees of freedom. We will take three scalar fields 
$(\varphi_1, \varphi_2, \varphi_3)$ corresponding to the three set of degrees of freedom 
$(\{{\bf g}_{\mu\nu}\}, \{{\bf C}_{\rm ABD}\}, \{{\bf \Psi}_{\rm A}\})$, where $(\mu, \nu) \in {\bf R}^{2, 1}$ and 
$({\rm A, B}) \in {\bf R}^{2, 1} \times {\cal M}_4 \times {\cal M}_2 \times {\mathbb{T}^2 \over {\cal G}}$. The path-integral structure that we are looking for here may be represented by:
\bg\label{wilsonrata}
\langle \varphi_1\rangle_{\sigma} \equiv {\int {\cal D}\varphi_1 {\cal D}\varphi_2 {\cal D}\varphi_3
~e^{i{\bf S}_{\rm tot}}~ \mathbb{D}^\dagger({\alpha}, {\beta}, {\gamma}) \varphi_1(x, y, z)
\mathbb{D}({\alpha}, {\beta}, {\gamma}) \over 
\int {\cal D}\varphi_1 {\cal D}\varphi_2 {\cal D}\varphi_3
~e^{i{\bf S}_{\rm tot}} ~\mathbb{D}^\dagger({\alpha}, {\beta}, {\gamma}) 
\mathbb{D}({\alpha}, {\beta}, {\gamma})}, \nd 
where ${\sigma} = ({\alpha}, {\beta}, {\gamma})$ is associated with 
$(\{{\bf g}_{\mu\nu}\}, \{{\bf C}_{\rm ABD}\}, \{{\bf \Psi}_{\rm A}\})$ degrees of freedom (see also 
\cite{coherbeta, coherbeta2}), $\mathbb{D}(\sigma)$ is a non-unitary displacement operator, {\it i.e.} $\mathbb{D}^\dagger(\sigma) \mathbb{D}(\sigma) \ne 
\mathbb{D}(\sigma) \mathbb{D}^\dagger(\sigma) \ne 1$ 
(see 
\cite{coherbeta} for details),  and the total action ${\bf S}_{\rm tot} \equiv {\bf S}_{\rm kin} + {\bf S}_{\rm int} + 
{\bf S}_{\rm ghost} + {\bf S}_{\rm gf}$ where the perturbative part of ${\bf S}_{\rm int}$ comes from an interaction term like eq. (4.81) in \cite{coherbeta2} and ${\bf S}_{\rm gf}$ is the gauge-fixing term. For the case in point here, {\it i.e.} \eqref{wilsonrata}, and as mentioned above, we will take a slightly simpler version\footnote{One might wonder if further simplification of \eqref{wilsonrata} is possible. For example a possibility would be to replace the denominator of 
\eqref{wilsonrata} by taking $\langle\Omega\vert\Omega\rangle$ instead of $\langle{\sigma}\vert{\sigma}\rangle$ in the following way:
\bg\label{shaystone}
\langle\varphi_1\rangle_{\sigma} = {\langle\sigma\vert \varphi_1\vert\sigma\rangle\over \langle\Omega\vert\Omega\rangle} =
{\int {\cal D}\varphi_1 {\cal D}\varphi_2 {\cal D}\varphi_3
~e^{i{\bf S}_{\rm tot}}~ \mathbb{D}^\dagger({\alpha}, {\beta}, {\gamma}) \varphi_1(x, y, z)
\mathbb{D}({\alpha}, {\beta}, {\gamma}) \over 
\int {\cal D}\varphi_1 {\cal D}\varphi_2 {\cal D}\varphi_3
~e^{i{\bf S}_{\rm tot}}}. \nd
While the above formula for the average will keep our computations simple, in the sense that we do not have to evaluate the denominator separately (and just expand it perturbatively), the above formalism cannot provide the full story. In section \ref{sec4.4} we will show the difference once the denominator is carefully introduced. Meanwhile we will continue with the evaluation of the numerator of the path integral. \label{gertrude}}
from the one in \cite{coherbeta2} where the various degrees of freedom are represented by {\it real} scalar degrees of freedom $(\varphi_1, \varphi_2, \varphi_3)$. Such a mapping works well for the metric and the three-form degrees of freedom but one might be worried about the replacement of Rarita-Schwinger fermion with a scalar degree of freedom. In the language of generalized metric and the three-form fields, this may not be an issue if we consider fermionic condensate. Following this strategy, a generic form for ${\bf S}_{\rm int}$ $-$ compared to what we had in \eqref{sabbath} $ - $ may be represented in the following way:
\bg\label{angelslune}
{\bf S}_{\rm int} &\equiv & \int d^{11} x \sum_{n, ..., s} c_{nmpqrs} \partial^n \varphi_1^q(x) \partial^m \varphi_2^r(x) 
\partial^p\varphi_3^s(x)\\
& = &\sum_{n, .., s} \int \prod_{i = 1}^q d^{11} \hat{k}_i \prod_{j = 1}^r d^{11} \hat{l}_j \prod_{t = 1}^{s - 1} d^{11} \hat{f}_t~c_{nmpqrs} 
(-1)^{(n + m + 3p)/2} \left(\sum_{i = 1}^q \hat{k}_i\right)^n \left(\sum_{j = 1}^r \hat{l}_j\right)^m\nonumber\\
&\times & \left(\sum_{i = 1}^q \hat{k}_i + \sum_{j = 1}^ r \hat{l}_j\right)^p
 \prod_{i = 1}^q \widetilde{\varphi}_1(\hat{k}_i) \prod_{j = 1}^r \widetilde{\varphi}_2(\hat{l}_j) \prod_{t = 1}^{s - 1} \widetilde{\varphi}_3(\hat{f}_t)
\widetilde\varphi^\ast_3\left(\sum_{i = 1}^q \hat{k}_i + \sum_{j = 1}^ r \hat{l}_j + \sum_{t = 1}^{s-1} \hat{f}_t\right), \nonumber \nd
where $n + m + p \in 2\mathbb{Z}_+$ and they are appropriately arranged so that the system is Lorentz invariant. The raising and lowering are done using the flat metric because we want to do a representative computations with the aforementioned identifications of the scalar degrees of freedom with the degrees of freedom in say eq. (4.81) of \cite{coherbeta2}. We also expect $(q, r, s) \in \mathbb{Z}_+$, and $c_{nmpqrs}$ to be dimensionful coefficients.
Putting everything together, the numerator of the path-integral in \eqref{wilsonrata} may be re-written as:

{\footnotesize
\bg\label{lincoln}
 {\rm Num}\left[\langle\varphi_1\rangle_{{\sigma}}\right] &\equiv &
\prod_{i, j, t}\int d\left({\bf Re}~\widetilde{\varphi}_1(k_i)\right) d\left({\bf Im}~\widetilde{\varphi}_1(k_i)\right) d\left({\bf Re}~\widetilde{\varphi}_2(l_j)\right) d\left({\bf Im}~\widetilde{\varphi}_2(l_j)\right)
\nonumber\\
&\times & d\left({\bf Re}~\widetilde{\varphi}_3(f_t)\right) d\left({\bf Im}~\widetilde{\varphi}_3(f_t)\right)
\cdot {1\over {\rm V}} \sum_{i} \psi_{{\bf k}''_i}({\bf x}, y, z) e^{-ik''_{i,0} t} 
\left({\bf Re}~\widetilde{\varphi}_1(k''_i) + i {\bf Im}~\widetilde{\varphi}_1(k''_i)\right) \nonumber\\
& \times & {\rm exp}\Big[-{1\over {\rm V}}\Big(\sum_{i} \vert \alpha(k'_i)\vert^2 + \sum_{j} \vert \beta(l'_j)\vert^2 +
\sum_{t} \vert \gamma(f'_t)\vert^2\Big)\Big]\nonumber\\
&\times & {\rm exp}\Big[{2\over {\rm V}}\sum_{i, j, t}\Big({\bf Re}~\alpha(k'_i)
~{\bf Re}~\widetilde{\varphi}_1(k'_i) + {\bf Im}~\alpha(k'_i)
~{\bf Im}~\widetilde{\varphi}_1(k'_i) + {\bf Re}~\beta(l'_j)
~{\bf Re}~\widetilde{\varphi}_2(l'_j) \nonumber\\
& + & {\bf Im}~\beta(l'_j)
~{\bf Im}~\widetilde{\varphi}_2(l'_j) + {\bf Re}~\gamma(f'_t)
~{\bf Re}~\widetilde{\varphi}_3(f'_t) + {\bf Im}~\gamma(f'_t)
~{\bf Im}~\widetilde{\varphi}_3(f'_t) + ...\Big)\Big] \nonumber\\ 
& \times & {\rm exp}\Big[{i\over {\rm V}} \Big(\sum_{i} k_i^2 \vert \widetilde\varphi_1(k_i)\vert^2 + 
\sum_{j} l_j^2 \vert \widetilde\varphi_2(l_j)\vert^2 + 
\sum_{t} f_t^2 \vert \widetilde\varphi_3(f_t)\vert^2\Big) + i {\bf S}_{\rm sol} + i{\bf S}_{\rm top}\Big] \nonumber\\
&\times & {\rm exp}\Big[ \sum_{n,..,s}{i\over {\rm V}^u} \sum_{\{u_i, v_j, w_t\}} c_{nmpqrs} (-1)^{(n + m + 3p)/2}
\left(\sum_{i = 1}^q {k}_{u_i}\right)^n \left(\sum_{j = 1}^r {l}_{v_j}\right)^m
\left(\sum_{i = 1}^q {k}_{u_i} + \sum_{j = 1}^ r {l}_{v_j}\right)^p\nonumber\\
& \times & 
\prod_{i, j, t = 1}^{q, r, s-1}\Big({\bf Re}~\widetilde{\varphi}_1(k_{u_i}) + i {\bf Im}~\widetilde{\varphi}_1(k_{u_i})\Big) 
\Big({\bf Re}~\widetilde{\varphi}_2(l_{v_j}) + i {\bf Im}~\widetilde{\varphi}_2(l_{v_j})\Big) 
\Big({\bf Re}~\widetilde{\varphi}_3(f_{w_t}) + i {\bf Im}~\widetilde{\varphi}_3(f_{w_t})\Big) 
\nonumber\\
&\times &
\Big({\bf Re}~\widetilde{\varphi}_3\Big(\sum_{i = 1}^q {k}_{u_i} + \sum_{j = 1}^ r {l}_{v_j} + \sum_{t = 1}^{s-1}{f}_{w_t}\Big) - i {\bf Im}~\widetilde{\varphi}_3\Big(\sum_{i = 1}^q {k}_{u_i} + \sum_{j = 1}^ r {l}_{v_j} + \sum_{t = 1}^{s-1}{f}_{w_t}\Big)\Big)\Big], \nd}
%&\times & {\rm exp}\Big[{\cal O}\Big({\rm exp}({\bf S}_{\rm loc})\Big) + %{\cal O}\Big({\rm exp}({\bf S}_{\rm nloc})\Big)
%+ {\cal O}\Big({\rm exp}({\rm exp}({\bf S}_{\rm ferm}))\Big)\Big], \nd} 
where $u = q + r + s - 1$ and the dotted terms in the fifth line comes only from the determinant of the metric (if any) as there are {\it no} additional contributions to $\mathbb{D}(\sigma)$ because of our choice \eqref{lucliu} instead of \eqref{milenej}. (See also footnote \ref{maisonmey}.) The reality conditions in \eqref{milenej} and \eqref{ijacob} have been chosen to provide a relative plus sign between the real and the imaginary 
components of the products of $\sigma(k)$ and $\widetilde\varphi(k)$ as shown in the fourth and the fifth lines of \eqref{lincoln}. This is also borne out from an explicit computation in \cite{coherbeta}. For the convenience of the computations we can keep various factors {\it dimensionless} in the following way. The fields $\varphi_i(x)$ for $i = 1, 2, 3$ are taken to be dimensionless, so to keep the Fourier transforms $\widetilde{\varphi}_i(k)$ also dimensionless, we have to insert appropriate powers of ${\rm M}_p$ in the definition of the Fourier transforms. We can also keep $(\alpha(k), \beta(k), \gamma(k))$ dimensionless if we take the momentum factors $k$ to be dimensionless. 
The coefficients $c_{nmpqrs}$ are generically dimensionful coefficients but we can make them dimensionless, just as we did for the $c_{nmp}$ coefficients in \eqref{sabbath}, by inserting appropriate powers of ${\rm M}_p$. Finally it is the volume factor ${\rm V}$ that is dimensionful, but in the presence of ${\rm M}_p$ the whole set of terms can be made dimensionless and therefore exponentiated. {\Su All of these may be easily achieved, and in turn help to avoid unnecessary clutter, by making ${\rm M}_p \equiv 1$ unless mentioned otherwise}\footnote{Once we take the solitonic Minkowski background into account, there appears an additional scale (excluding the UV cut-off $\Lambda$) compared to ${\rm M}_p$ $-$ the scale at which new degrees of freedom enter $-$ in the problem. This is $\hat\mu$, the scale associated with the size of the internal compact manifold. Clearly we expect $\hat\mu < {\rm M}_p$ with the KK degrees of freedom integrated out. The hierarchy of scales being $\Lambda >> {\rm M}_p \ge \hat\mu > \mu > k_{\rm IR}$, where $\mu$ is the energy scale of the experiment and $k_{\rm IR}$ is the IR cut-off. Additionally $\Lambda$ and $k_{\rm IR}$ will be related by the UV/IR correspondence that we shall discuss in section \ref{sec3.2.3}. We can now use 
$\hat\mu$ to make both the momenta $k_i$ and the coupling constants $c_{nmpqrs}$ dimensionless via ${k_i\over \hat\mu} \equiv \hat{k}_i$ and $\hat{g} \equiv c_{nmpqrs}\left({\hat\mu\over {\rm M}_p}\right)^{n + m + p}$, where $g \equiv c_{nmpqrs}$ is the coupling constant that we add by hand here (in string theory $g = 1$, and the coupling constant is completely determined by $\hat\mu$ and ${\rm M}_p$). This
 keeps the volume ${\rm V}$ dimensionless via ${\rm M}^{11}_p {\rm V} \equiv \hat{\rm V}$, but introduces an additional factor of $\left({\hat\mu\over {\rm M}_p}\right)^2$ as the coefficients of the kinetic terms on the sixth line of \eqref{lincoln}. To avoid all the un-necessary clutter we will henceforth assume ${\hat\mu\over {\rm M}_p} \equiv 1$ which will make $\hat{k}_i \equiv k_i, \hat{\rm V} \equiv {\rm V}$ and $\hat{g} \equiv g$ all {\it dimensionless}, as emphasized earlier. Such an assignment will be particularly useful when we define the Hermite polynomials in terms of the dimensionless momenta in \eqref{greek}, or raise $g$ and ${\rm V}$ to arbitrary powers. Another thing to note is that all the coupling constants are given in terms of $\hat\mu$ and ${\rm M}_p$, but no string coupling $g_s$ appears. This is intentional as we want to restrict ourselves to a simpler set-up in M-theory. \label{yaya}}.
${\bf S}_{\rm sol}$  and 
${\bf S}_{\rm top}$ are respectively the actions of the solitonic background, which is basically the time-independent warped Minkowski background, and the topological terms that we studied in \cite{DRS} and elaborated further in \cite{coherbeta}.  Note also the distributions of the eleven-dimensional {\it discrete} momenta $(k_i, k'_i, k''_i, k_{u_i})$ for $i = 1, 2, 3, ..., \infty$ over the various terms of the path-integral. (Similar assignments are for the $(l_i, ...)$ and $(f_i, ...)$ momenta.) These assignments of momenta\footnote{The arrangement of momenta in \eqref{angelslune} as well as in \eqref{lincoln} follow from our choice of modes over the solitonic background which is $\psi_{\bf k}({\bf x}, y, z) e^{-ik_0 t}$ with 
$\psi_{\bf k}({\bf x}, y, z) \equiv {\rm exp}\left(i{\bf k}.{\rm X}\right)
{\bf G}_{\bf k}({\rm X})$ and ${\rm X} = ({\bf x}, y, z)$. For the Minkowski background that we took, ${\bf G}_{\bf k}({\rm X})$ approaches 1, allowing us to impose:
\bg\label{catulin}
\int d^{11}x~\prod_{i = 1}^q \psi_{{\bf k}_{u_i}}({\rm X})
\prod_{j = 1}^r \psi_{{\bf l}_{v_j}}({\rm X}) 
\prod_{p = 1}^s \psi_{{\bf f}_{w_p}}({\rm X}) e^{-i(k_{0,u_i}+ l_{0,v_j} + 
f_{0,w_p})t} = \delta^{11}\left( \sum_{i = 1}^q k_{u_i} + 
\sum_{j = 1}^r l_{v_j} + \sum_{t = 1}^s f_{w_s}\right), \nonumber \nd
which we used to simplify the expression in \eqref{angelslune} and 
\eqref{lincoln}. We have also defined $k_{u_i} = ({\bf k}_{u_i}, k_{0, u_i})$, $l_{v_j} = ({\bf l}_{v_j}, l_{0, v_j})$ and $f_{w_p} = ({\bf f}_{w_p}, f_{0, w_p})$ as discrete momenta with $(u_i, v_j, w_p) \in \mathbb{Z}_+$. \label{fisshu}}
have to carefully match-up when we perform the integrals in \eqref{lincoln}, otherwise we cannot get non-zero answer from \eqref{lincoln}.

\subsection{First look at the path-integral \eqref{lincoln} \label{lincoln22}}

\noindent The propagators of the three scalars are respectively ${1\over k_i^2}, {1\over l_i^2}$ and ${1\over f_i^2}$, which are because of our choice of gauge-fixing
${\bf S}_{\rm gf}$ mentioned earlier. The interaction terms, parametrized by 
$c_{nmpqrs}$, with $(n, .., s) \in \mathbb{Z}_+$  and $(q, r, s) > (2, 2, 2)$ make the path-integral rather hard to compute but we can make a few simplifying steps to ease the process. First, would be to give some dimensions to the scalar fields so that the $c_{nmpqrs}$ terms may be perturbatively expanded. Secondly, 
 we will integrate over {\it one} set of momenta to elucidate the process. Finally, a few symbolic manipulations like:
\bg\label{neveC}
&&{\bf Re}~\widetilde{\varphi}_1(k_1) \equiv y_1, ~~~ {\bf Re}~\widetilde{\varphi}_1(k'_1) \equiv y'_1, ~~~
{\bf Re}~\widetilde{\varphi}_1(k''_1) \equiv y''_1\nonumber\\  
&&{\bf Im}~\widetilde{\varphi}_1(k_1) \equiv y_2, ~~~ {\bf Im}~\widetilde{\varphi}_1(k'_1) \equiv y'_2, ~~~
{\bf Im}~\widetilde{\varphi}_1(k''_1) \equiv y''_2, \nd
with similar replacements $(z_i, \omega_i)$ for $(\widetilde{\varphi}_2(l_1), \widetilde{\varphi}_3(f_1))$ will further ease the process of the computations. Note however the absence of the ghost contributions to the path-integral in \eqref{lincoln}. There are two ways to reconcile this. One, since the system is abelian, the ghost would typically decouple much like what we have in QED. Two, we can always choose a gauge fixing condition that would decouple the ghosts at least for the metric modes ${\bf g}_{\mu\nu}$ where 
$(\mu, \nu) \in {\bf R}^{2, 1}$. A hint that the ghost sector decouples can be seen from our study of 
Schwinger-Dyson's equations in \cite{coherbeta, coherbeta2}, which in retrospect makes sense if we demand that the Glauber-Sudarshan state reproduces the de Sitter background in the temporal domain where the IIA coupling $g_s < 1$. For one set of momentum modes, the path-integral \eqref{lincoln} takes the simplified form:

{\footnotesize 
\bg\label{mortuarymey}
{\rm Num}\left[\langle\varphi_1\rangle_{{\sigma}}\right](k_1, l_1, f_1) &=& {1\over {\rm V}} \int 
dy_1 dy_2 dz_1 dz_2 d\omega_1 d\omega_2 ~\psi_{{\bf k}''_1}({\bf x}, y, z) e^{-ik''_0 t}(y_1'' + i y_2'')\\
&\times & {\rm exp}\Big[ {2\over {\rm V}} \Big(\alpha_1' y_1' + \alpha_2' y_2' + 
\beta_1' z_1' + \beta_2' z_2' + \gamma_1' \omega_1' 
+ \gamma_2' \omega_2' + ...\Big)+ i{\bf S}_{\rm sol} \Big]\nonumber\\
& \times & {\rm exp}\Big[ i^{t_1}{\rm V}^{t_2}\tilde{c}_{nmpqrs}a_1^{n/2} b_1^{m/2} \tilde{c}_1^{p}(y_1 + y_2)^q(z_1 + iz_2)^r (\omega_1 + i\omega_2)^{s-1}(\hat{\omega}_1 - i\hat{\omega}_2) + ..\Big] \nonumber\\
&\times &{\rm exp}\Big[-{1\over {\rm V}}\left(|\alpha'_1|^2 + |\beta'_1|^2 
+ |\gamma'_1|^2\right) -
 a_1(y_1^2 + y_2^2) - b_1(z_1^2 + z_2^2) - c_1(\omega_1^2 + \omega_2^2)+ ..\Big],
\nonumber\nd}
 where we have made one crucial assumption for the momenta in the interaction sector:  $k_{u_1} = k_{u_2} = ... = 
 k_{u_q} \equiv k_1$ with similar identifications for $l_{v_j}$ for $j = 1, .., r$ with $l_1$ and 
 $f_{w_t}$ for $t = 1, .., s-1$ with $f_1$.  This will however make $f_{w_s} = -qk_1 - rl_1 - (s-1)f_1$ from momentum conservation. In the language of \eqref{lincoln}, we are taking the special case where 
 all but one of the scalar modes for the three scalar fields are stacked against the first set of their respective Gaussian integrals. Because of the shifted vacuum structure (coming from $\mathbb{D}(\sigma)$) there is no restriction on $(q, r, s)$ to be even integers. The other parameters are defined in the following way:
 $\alpha'_1 \equiv \alpha(k_1')$ with similar definitions for $(\beta'_1, \gamma'_1)$; and 
 $(a_1, b_1, c_1) = -{i\over {\rm V}}(k_1^2, l_1^2, f_1^2)$ with:
 \bg\label{casin529}
 &&\tilde{c}_{nmpqrs} = q^n r^m c_{nmpqrs}, ~~~~~ \tilde{c}_1 = q\sqrt{a_1} + r\sqrt{b_1}, \nonumber\\
 && t_1 = 1 + 3(n + m + 2p), ~~~~ t_2 = 1 - q - r- s + {1\over 2}(n + m + p) \nonumber\\
 && \hat{\omega}_1 = {\bf Re}~\widetilde\varphi_3(qk_1 + rl_1 + (s-1)f_1), ~~~
 \hat{\omega}_2 = {\bf Im}~\widetilde\varphi_3(qk_1 + rl_1 + (s-1)f_1), \nd
 where $(k_1, l_1, f_1)$ could in-principle have either signs. However one might worry, because of our choice of $f_{w_s}$ above, we are no longer restricted to one set of momentum modes for the three scalar fields. But if we choose momenta $(k_1, l_1, f_1)$ in such a 
way that:
\bg\label{lenuray} 
 sf_1 + qk_1 + rl_1 = 0, \nd
 where $(s, q, r) \in \mathbb{Z}_+$, then $f_{w_s} = f_1$ and $\hat{\omega}_1 = \omega_1$ and 
 $\hat\omega_2 = - \omega_2$. This way the path-integral structure in \eqref{mortuarymey} will indeed be restricted to one set of momentum modes. Finally,  the dotted terms in the second and the third lines of 
 \eqref{mortuarymey} have the same meaning as in 
 \eqref{lincoln} except that they are now defined for a single set of modes. 
 
 Unfortunately even with a single set of momentum modes, the above path-integral is hard to  do unless the interaction terms are expanded perturbatively. As mentioned earlier this is possible if $(y_i, z_i, w_i)$ etc. are given some ${\rm M}_p$ dimensions (instead of taking them dimensionless). That being said, we will however indulge in yet another simplification: ignore the complex parts of the Fourier modes. With all these the analysis becomes reasonably manageable, and the ratio of the numerator and the denominator for a given set of momentum modes takes the following form:
 
 {\footnotesize
 \bg\label{bgracie}
 {{\rm Num}\over {\rm Den}} = {-{\alpha_1\over {\rm V}a_1} + \sum\limits_{\{h_j\}} \mathbb{A}(h_1, h_2, h_3) \left(-\alpha_1\right)^{q + 1 - h_1} \left(-\beta_1\right)^{r  - h_2}
 \left(-\gamma_1\right)^{s  - h_3} + {\cal O}(c^2_{nmpqrs}) \over 
 1 + \sum\limits_{\{h_j\}} \mathbb{B}(h_1, h_2, h_3) \left(-\alpha_1\right)^{q - h_1} \left(-\beta_1\right)^{r  - h_2} \left(-\gamma_1\right)^{s  - h_3} + {\cal O}(c^2_{nmpqrs})}, 
 \nd} 
 where $h_i \in 2\mathbb{Z}_+$, and we expect $\mathbb{B}(h_1, h_2, h_3)$ to be related to $\mathbb{A}(h_1, h_2, h_3)$ via the standard Gaussian identity
 $\mathbb{B}(h_1, h_2, h_3)\left(q + 1\right)! = q!~a_1{\rm V}\mathbb{A}(h_1, h_2, h_3)$. We can also determine $\mathbb{A}(h_1, h_2, h_3)$ by appropriately Taylor expanding over the parameters in the following way:
 
 {\footnotesize
 \bg\label{foreign}
\mathbb{A}(h_1, h_2, h_3) = 8\pi (-i)^n {\rm V}^{f_4}c_{nmpqrs}\prod_{i = 1}^3 (h_i -1)!!~a_1^{f_1} b_1^{f_2} c_1^{f_3} 
\left(\begin{matrix}q + 1 \\ h_1 \\ \end{matrix}\right)
\left(\begin{matrix}r \\ h_2 \\ \end{matrix}\right)
\left(\begin{matrix} s \\ h_3 \\ \end{matrix}\right), \nd} 
where $n$ is an integer determined from the parameter of the model, ${\rm V}$ is the volume and $(a_1, b_1, c_1)$ are the propagators. Note the appearance of ${\rm V}^{-1}$ in \eqref{bgracie} which is a signature that only one momentum mode is taken into account. The other parameters are defined in the following way:
{\bg\label{thickand}
&& f_4 = v - q - r - s + \sum_{j = 1}^3 h_j - 1, ~~~~~
v = {1\over 2}(n + m + p) - 1 \\
&&f_1 = {n\over 2} - q + {h_1\over 2} - {1\over 2}, 
~~ f_2 = {m\over 2} - r + {h_2\over 2} - {1\over 2},
~~ f_3 = {p\over 2} - s + {h_3\over 2} - {1\over 2}. \nonumber \nd}
The question that we ask now is what happens when we sum over, not one, but all the modes. This is where the system becomes more involved and in the following we turn to a detailed elaboration. 
 
\vskip.1in

\subsection{A more elaborate analysis of the path-integral \eqref{lincoln} and nodal diagrams \label{sec3.2}}

\noindent Our above analysis with one set of momentum modes was a useful toy example but it raised a few questions. In the ratio \eqref{bgracie} we see that appearances of ${\beta}$ and ${\gamma}$, which one would have expected. However the puzzle now is what happens when we sum over all momenta? Do the ${\beta}(l_i)$ and ${\gamma}(f_t)$ integrate out in some way? Additionally we made the choice \eqref{casin529} to simplify the ensuing analysis. Could this be relaxed? In the following we will try to elaborate on this. We will find that there is a diagrammatic way to analyze the path-integral \eqref{lincoln} order by order in the coupling and momenta. However to simplify the analysis we will again resort to the case where we ignore the complex parts of the various fields\footnote{We don't lose any information from such restrictions. Taking real Fourier components with only positive momenta means that any field component may be expressed as:
\bg\label{kiara}
\varphi({\rm X}, t) \equiv \int_0^\infty d^{11}k ~\widetilde\varphi(k) ~{\bf Re}\left(\psi_{\bf k}({\rm X})e^{-ik_0t}\right), \nonumber \nd
where ${\rm X} = ({\bf x}, y, z)$ is the set of spatial coordinates. We could have also integrated in the regime $k \in [-\infty, +\infty]$ where the reality of the field would have translated into taking a Fourier component that is an even function of $k$. The latter involves taking both positive and negative momenta.\label{redskix3}}. With this, the numerator of the path-integral takes the following form:

{\footnotesize
\bg\label{fontini}
{\rm Num}[\langle\varphi_1\rangle_{\overline\sigma}] & = & \mathbb{C}(\overline\alpha_i, a_j, {\rm V}, ..)
\int d\widetilde\varphi_1(k_1) 
~{\rm exp}\Big[-a_1\Big(\widetilde\varphi_1(k_1) + {\overline\alpha_1\over a_1}\Big)^2\Big]
~d\widetilde\varphi_1(k_2) 
~{\rm exp}\Big[-a_2\Big(\widetilde\varphi_1(k_2) + {\overline\alpha_2\over a_2}\Big)^2\Big] ...... \nonumber\\
&&~~\times  
d\widetilde\varphi_2(l_1) 
~{\rm exp}\Big[-b_1\Big(\widetilde\varphi_2(l_1) + {\overline\beta_1\over b_1}\Big)^2\Big]
~d\widetilde\varphi_2(l_2) 
~{\rm exp}\Big[-b_2\Big(\widetilde\varphi_2(l_2) + {\overline\beta_2\over b_2}\Big)^2\Big]  
~d\widetilde\varphi_2(l_3)  ...... \nonumber\\
&& ~~ \times 
d\widetilde\varphi_3(f_1) 
~{\rm exp}\Big[-c_1\Big(\widetilde\varphi_3(f_1) + {\overline\gamma_1\over c_1}\Big)^2\Big]
~d\widetilde\varphi_3(f_2) 
~{\rm exp}\Big[-c_2\Big(\widetilde\varphi_3(f_2) + {\overline\gamma_2\over c_2}\Big)^2\Big]   
~d\widetilde\varphi_3(f_3) ...... \nonumber\\
&& ~~ \times \Big(\widetilde\varphi_1(k_1)\psi_{{\bf k}_1}({\bf x}, y, z)e^{-ik_{0, 1}t} + 
\widetilde\varphi_1(k_2)\psi_{{\bf k}_2}({\bf x}, y, z)e^{-ik_{0, 2}t} + 
\widetilde\varphi_1(k_3)\psi_{{\bf k}_3}({\bf x}, y, z)e^{-ik_{0, 3}t} + .. \Big)\nonumber\\
&&~~ \times\Big(1 + {i}\sum_{{\cal S}}{\rm V}^{-u} c_{nmpqrs} (-1)^{v}
\Big(\begin{matrix} p \\ e_o \\ \end{matrix}\Big)(k_{u_1} + k_{u_2} + ... + k_{u_q})^{n + e_o}(l_{v_1} + l_{v_2} + ... + l_{v_r})^{m + p - e_o}\nonumber\\
&&~~ \times \widetilde\varphi_1(k_{u_1})\widetilde\varphi_1(k_{u_2})... \widetilde\varphi_1(k_{u_q}) \widetilde\varphi_2(l_{v_1})
\widetilde\varphi_2(l_{v_2})... \widetilde\varphi_2(l_{v_r}) \widetilde\varphi_3(f_{w_1}) \widetilde\varphi_3(f_{w_2}) ... \widetilde\varphi_3(f_{w_{s-1}})
\widetilde\varphi_3(f_{w_s}) \nonumber\\ 
&& ~~ + {\cal O}(c^2_{nmpqrs})\Big)\cdot{1\over {\rm V}}, \nd} 
%{\rm exp}\Big[{\cal O}\Big({\rm exp}({\bf S}_{\rm loc})\Big) + {\cal %O}\Big({\rm exp}({\bf S}_{\rm nloc})\Big)
%+ {\cal O}\Big({\rm exp}({\rm exp}({\bf S}_{\rm ferm}))\Big)\Big], 
%\nd}
where ${\cal S}$ is the set ${\cal S} \equiv \left(\{u_i\}, \{v_j\}, \{w_t\}, n, m, ..., s, e_o\right)$, 
$\Big(\begin{matrix} p \\ e_o \\ \end{matrix}\Big) = {p!\over e_o! (p - e_o)!}$ (note the sum over $e_o$ in \eqref{fontini}), 
$v \equiv {1\over 2}(n + m + 3p)$, $u \equiv q + r + s - 1$, and we have defined the other variables in the following way:
\bg\label{nehaprit}
&&(a_j, b_j, c_j) \equiv (a(k_j), b(l_j), c(f_j)) =  -{i\over {\rm V}}(k_j^2, l_j^2, f_j^2) \nonumber\\
&& (\overline{\alpha}_i, \overline{\beta}_i, \overline{\gamma}_i) \equiv (\overline{\alpha}(k_i), \overline{\beta}(l_i), \overline{\gamma}(f_i)) = {1\over {\rm V}}({\alpha}(k_i), 
{\beta}(l_i), {\gamma}(f_i)), \nd
such that ${\overline{\alpha}_i\over a_i} = {\overline\alpha(k_i)\over a(k_i)}$ is independent of ${\rm V}$ {\it i.e.} henceforth both $a(k)$ and $\overline\alpha(k)$ have inverse volume dependences (similarly for the other ratios) and we can denote the Glauber-Sudarshan state by $\vert\overline\sigma\rangle$ and the corresponding displacement operator by $\mathbb{D}(\overline\sigma)$ as alluded to in footnote \ref{bldride}. Note that the definition \eqref{nehaprit} can allow us to even go to the Euclidean space\footnote{As is well-known, path-integral with an UV cut-off is best expressed in the Euclidean formalism. For us this is easy to implement by changing $(a_j, b_j, c_j)$ to their Euclidean values and then shift the Euclidean vacua using $(\overline\alpha_i,
\overline\beta_i, \overline\gamma_i)$. This way no $i = \sqrt{-1}$ would appear in the definitions. However, since these are symbolic manipulations, we will not worry too much about them here.} from here. $\mathbb{C}(\overline\alpha_i, a_j, {\rm V}, ..)$ is the coefficient that depends on $(\overline\alpha_i,
\overline\beta_i, \overline\gamma_i), (a_j, b_j, c_j)$ and ${\rm V}$ and it decouples once we divide 
\eqref{fontini} with the denominator of the path-integral. Thus we will also not worry too much on the explicit form for $\mathbb{C}(\overline\alpha_i, a_j, {\rm V}, ..)$.

The shifted vacuum structures of the three scalar fields make the analysis non-trivial, but as we shall see, the path-integral structure can be decomposed as collections of certain {\it nodal} diagrams. These diagrams stem from how the various momenta in the interaction sector are summed over. This is non-trivial because, although most momenta are {\it aligned} with their corresponding measures, we still have a set of momenta
$(k_{u_i}, l_{v_j}, f_{w_t})$ which can take any values.  Thus it appears that the best way to deal with the scenario is to take various cases and discuss their contributions. In the following we will study some of these cases in details.

\subsubsection{Contributions at the tree level \label{sec3.2.1}}

\noindent The tree level contributions come from \eqref{fontini} when we take vanishing coupling constants, {\it i.e.} we take $c_{nmpqrs} = 0$. As such this makes sense, at least for the perturbative corrections in
\eqref{lincoln}, when the interaction terms are sub-dominant in the limit ${\rm M}_p \to \infty$ and $g_s << 1$ because M-theory doesn't have any other  adjustable parameters\footnote{Note that this $g_s$ is the string coupling over the warped Minkowski background and therefore should not be confused with the 
${g_s\over {\rm HH}_o}$ that we used earlier in the Schwinger-Dyson's equations or in the analysis of the the quantum series
in say eq. (4.81) of \cite{coherbeta2} and the scaling relation in eq. (4.82) therein.}. In this limit we can express 
\eqref{fontini} in the following way:

\pgfmathsetmacro\MathAxis{height("$\vcenter{}$")}
\bg\label{luna1}
\begin{tikzpicture}[baseline={(0, 0cm-\MathAxis pt)},every node/.style={ball  color=white, circle,  minimum size=25pt}]
    \draw (-1.5, 0) -- (0, 0) node{i}  -- (1.5, 0);
    \filldraw [black] (-1.5,0) circle (1pt);
    \filldraw [black] (1.5,0) circle (1pt);
\end{tikzpicture}
\; =  \prod_{j \ne i} 
\int d\widetilde\varphi_1(k_j) 
~{\rm exp}\Big[-a_j\Big(\widetilde\varphi_1(k_j) + {\overline\alpha_j\over a_j}\Big)^2\Big]\\
\times \prod_{u} 
\int d\widetilde\varphi_2(l_u) 
~{\rm exp}\Big[-b_u\Big(\widetilde\varphi_2(l_u) + {\overline\beta_u\over b_u}\Big)^2\Big] \nonumber\\
~\times \prod_{v} 
\int d\widetilde\varphi_3(f_v) 
~{\rm exp}\Big[-c_v\Big(\widetilde\varphi_3(f_v) + {\overline\gamma_v\over c_v}\Big)^2\Big] \nonumber\\
\times  \int d\widetilde\varphi_1(k_i) 
~{\rm exp}\Big[-a_i\Big(\widetilde\varphi_1(k_i) + {\overline\alpha_i\over a_i}\Big)^2\Big]
\widetilde\varphi_1(k_i)\psi_{{\bf k}_i}({\bf x}, y, z)e^{-ik_{0, i}t}\nonumber\\
 = -\prod_{j} \Big({\pi_j\over a_j}\Big)^{1/2} 
\prod_{u} \Big({\pi_u\over b_u}\Big)^{1/2} \prod_{v} \Big({\pi_v\over c_v}\Big)^{1/2} 
\Big({\overline\alpha_i\over a_i}\Big)\psi_{{\bf k}_i}({\bf x}, y, z)e^{-ik_{0, i}t} \nonumber \nd 
where $\pi_j = \pi_u = \pi_v = \pi$ and the overall minus sign is due to our choice of convention for the displacement operator $\mathbb{D}(\overline\sigma)$. The above diagram is only for the momentum mode
$k_i$. We can clearly sum over all the $k_i$ modes to get the full answer, but before we do this we can draw an equivalent tree-level diagram for the denominator. We can represent this as:

\bg\label{luna2}  
\begin{tikzpicture}
\draw (-1,0) -- (1,0);
\filldraw [black] (-1,0) circle (1pt);
\filldraw [black] (1,0) circle (1pt);
\end{tikzpicture}
\; = \prod_{j} \Big({\pi_j\over a_j}\Big)^{1/2} 
\prod_{u} \Big({\pi_u\over b_u}\Big)^{1/2} \prod_{v} \Big({\pi_v\over c_v}\Big)^{1/2}, \nd 
which could in principle be written in a more condensed form in the limit when all the momenta become continuous. In the continuum limit, ${\rm V} \to \infty$, and therefore we can sum over all $i$ from 
\eqref{luna1}, to get the following result:

\bg\label{luna3}
\frac{1}{V} \sum_{i=1}^{\infty} \;
   \begin{tikzpicture}[baseline={(0, 0cm-\MathAxis pt)}, vertex/.style={anchor=base, ball color=white, circle,  minimum size=25pt}]
    \draw (-1.2, 0) -- (0, 0) node [vertex] {i}  -- (1.2, 0);
    \filldraw [black] (-1.2,0) circle (1pt);
    \filldraw [black] (1.2,0) circle (1pt);
\end{tikzpicture} 
\; = \; 
\begin{tikzpicture}[baseline={(0, 0cm-\MathAxis pt)}]
\draw (-1,0) -- (1,0);
\filldraw [black] (-1,0) circle (1pt);
\filldraw [black] (1,0) circle (1pt);
\end{tikzpicture}
\otimes 
\left(-\int d^{11}k ~{\overline\alpha(k)\over a(k)}~\psi_{{\bf k}}({\bf x}, y, z)e^{-ik_{0}t}\right), \nd
which of course matches exactly with the result that we had in \cite{coherbeta, coherbeta2} when we take the ratio between \eqref{luna3} and \eqref{luna2} because 
$a(k) \propto k^2 = (k_0 + \omega_{\bf k})(k_0 - \omega_{\bf k})$ and the residue at the pole appears to reproduce the de Sitter metric when we replace the scalar field $\varphi_1$ by the metric degrees of 
freedom\footnote{As will become clear later, the tree-level result in fact does not give the de Sitter metric as one might have expected. The difference lies in the wave-function renormalization factor. How is this connected to the asymptotic nature of the perturbation theory, or to the Borel resummation will be the subject of the coming sections.}. This matches with the classical result from \eqref{yuhana} once we add the solitonic configuration. The question however is what happens when we switch on the interactions, here parametrized by non-zero 
$c_{nmpqrs}$. This is where the system becomes non-trivial, and we will begin by analyzing first the scalar fields $\varphi_2$ and $\varphi_3$ before addressing the main scalar degree of freedom $\varphi_1$.

\subsubsection{Contributions from the $\varphi_2$ fields \label{sec3.3.2}}

The scalar field $\varphi_2$, as mentioned earlier, is a representative field for the three-form flux components in M-theory (where we suppress the tensorial indices). The three-form fields have 84 massless degrees of freedom, so in principle we should take $\varphi_{2i}$ with $i = 1, ..., 84$ components. This will make the analysis substantially more involved, so in the following we will suffice with only {\it one} component, but with arbitrary copies (represented by $r$ in the coupling constant $c_{nmpqrs}$
from \eqref{lincoln}). Each of these copies (or fields) can have different momenta, which should then be integrated over. In the following we discuss a few cases.

\vskip.2in

\noindent{\bf Case 1: $l_{v_1} = l_{v_2} = l_{v_3} = ..... = l_{v_r} \equiv l_i$ for the $\varphi_2$ fields}

\vskip.2in

\noindent For the scalar degree of freedom $\varphi_2$ let us first consider the case where all the discrete momentum modes $l_{v_j}$ take the same values of $l_i$. Such a choice is possible because we are summing over the set $\{v_j\}$ in \eqref{fontini}. (Note that this does {\it not} imply that the discrete momenta $l_i$ are all equivalent!) Also since the $l_{v_j}$ momentum modes are separate from the $k_{u_i}$ or the 
$f_{w_t}$ momentum modes, we can keep fixed the other momenta and study the cases associated with the $l_{v_j}$ modes. (There is some subtlety with the $f_{w_t}$ modes due to momentum conservation that we shall elaborate soon.) For such a scenario, we can have the following nodal diagram that succinctly captures the value of the path-integral:

\bg\label{luna4}
\begin{tikzpicture}[baseline={(0, 0cm-\MathAxis pt)}, thick, 
main/.style = {draw, circle, fill=black, minimum size=12pt},
dot/.style={inner sep=0pt,fill=black, circle, minimum size=3pt},
dots/.style={inner sep=0pt, fill=black, circle, minimum size=2pt}]
  \node[dot] (1) at (-1.5, 0) {};
  \node[dot] (2) at (0, 0) {};
  \node[main] (3) at (60:1.5) {};
  \node[main] (4) at (30:1.5) {};
  \node[main] (5) at (0:1.5) {};
  \node[main] (6) at (-30:1.5) {};
  \node[main] (7) at (-65:1.5) {};
  \node[dot] (b1) at (-35:1.5) {};
  \node[dots] (d1) at (-40:1.3) {};
  \node[dots] (d2) at (-50:1.3) {};
  \node[dots] (d3) at (-60:1.3) {};
\draw (1) -- node[midway, above right, pos=0.2] {$r l_i$} (2);
\draw (2) -- node[midway, left, pos=0.7] {$l_i$} (3);
\draw (2) -- node[midway, left, pos=0.8] {$l_i$} (4);
\draw (2) -- node[midway, above right, sloped, pos=0.7] {$l_i$} (5);
\draw (2) -- node[midway, right, pos=0.7] {$l_i$} (6);
\draw (2) -- node[midway, right, pos=0.7] {$l_i$} (7);
\end{tikzpicture}
\; &&= \prod_{j \ne i} 
\int d\widetilde\varphi_2(l_j) 
~{\rm exp}\Big[-b_j\Big(\widetilde\varphi_2(l_j) + {\overline\beta_j\over b_j}\Big)^2\Big]\\
&& \times \int d\widetilde\varphi_2(l_i) 
~{\rm exp}\Big[-b_i\Big(\widetilde\varphi_2(l_i) + {\overline\beta_i\over b_i}\Big)^2\Big] ~
(rl_i)^{m+p-e_o}\widetilde\varphi_2^r(l_i)\nonumber\\
&&=  \prod_{j} \Big({\pi_j\over b_j}\Big)^{1/2} 
\sum_{e_p \in 2\mathbb{Z}_+}^r \left(\begin{matrix} r \\ e_p \\ \end{matrix}\right) 
\left(-{\overline\beta_i\over b_i}\right)^{r - e_p} {(rl_i)^{m+p-e_o}(e_p - 1)!!\over (2b_i)^{e_p/2}}, \nonumber\nd
where $\pi_j = \pi$, the repeated indices are not summed over and $e_p \in 2\mathbb{Z}_+$ to keep the integral non-vanishing. However this doesn't put any constraint on $r$ itself: it can be an even or an odd integer. If $r$ is an odd integer then the summation is till $(r - 1)$. (There is also a sum over $e_o$ from \eqref{lincoln} that we will insert at the very end.) The nested sum is of course for a given choice of $l_i$. As before, we can sum over all the discrete momenta $l_i$, and in the limit ${\rm V} \to \infty$, 
the summation will turn into an integral. The result we get is:

{\footnotesize
\bg\label{luna5}
\frac{1}{\rm V} \sum_{i=1}^{\infty} \;
\begin{tikzpicture}[baseline={(0, 0cm-\MathAxis pt)}, thick,
main/.style = {draw, circle, fill=black, minimum size=12pt},
dot/.style={inner sep=0pt,fill=black, circle, minimum size=3pt},
dots/.style={inner sep=0pt, fill=black, circle, minimum size=2pt}] 
  \node[dot] (1) at (-1.5, 0) {};
  \node[dot] (2) at (0, 0) {};
  \node[main] (3) at (60:1.5) {};
  \node[main] (4) at (30:1.5) {};
  \node[main] (5) at (0:1.5) {};
  \node[main] (6) at (-30:1.5) {};
  \node[main] (7) at(-65:1.5) {};
  \node[dots] (d1) at (-40:1.3) {};
  \node[dots] (d2) at (-50:1.3) {};
  \node[dots] (d3) at (-60:1.3) {};
  
\draw (1) -- node[midway, above right, pos=0.2] {$r l_i$} (2);
\draw (2) -- node[midway, above left, pos=0.7] {$l_i$} (3);
\draw (2) -- node[midway, above left, pos=0.9] {$l_i$} (4);
\draw (2) -- node[midway, above right, sloped, pos=0.7] {$l_i$} (5);
\draw (2) -- node[midway, above right, pos=0.7] {$l_i$} (6);
\draw (2) -- node[midway, right, pos=0.7] {$l_i$} (7);
\end{tikzpicture} 
\; = \prod_{j} \Big({\pi_j\over b_j}\Big)^{1/2} 
\int d^{11} l ~(rl)^{m+p-e_o} \sum_{e_p \in 2\mathbb{Z}_+}^r \left(\begin{matrix} r \\ e_p \\ \end{matrix}\right) 
\left(-{\overline\beta(l)\over b(l)}\right)^{r - e_p} {(e_p - 1)!!\over \left(2b(l)\right)^{e_p/2}},\nonumber\\ \nd }
where notice two things: one, we have extracted a factor of ${\rm V}^{-1}$ from ${\rm V}^{-u}$ in \eqref{lincoln}; and two, we have kept some part outside the integral with a product structure in the discrete form. The latter is because when we take the other two set of momentum modes, namely $k_i$ and $f_j$, we expect a piece like \eqref{luna2} to separate out. Such a piece will {\it cancel} out from the denominator of the path-integral, thus leaving an integral structure like \eqref{luna5}. For the former, there would be an extra ${\rm V}^{-r+1}$ suppression factor from ${\rm V}^{-u}$ in \eqref{lincoln} once we convert the summation to an integral. We will ignore this in the subsequent discussion and only insert it later when we combine all the nodal diagrams in section \ref{3.2.6}.

\vskip.2in

\noindent{\bf Case 2: $l_{v_1} = l_{v_2} = l_{v_3} = .... = l_{v_{r-1}} = l_i, l_{v_r} = l_j, i \ne j$ for the
$\varphi_2$ fields}

\vskip.2in

\noindent The two diagrams in \eqref{luna4} and \eqref{luna5} were our first foray beyond the tree-level and into the quantum regime. Therein we took the simplest case where all the momenta take the same values. Here we go into a slightly more non-trivial case where all but one of the momentum modes take the same values. As we shall see, this itself will lead to different answers from what we got earlier.

The choice of $(l_i, l_j)$ for the $l_{v_k}$ momenta is interesting because now both $i$ and $j$ can vary. So when we sum over, just as we did for {\bf Case 2} above, we will have to sum over both $i$ and $j$. To see how this may work out, let us start with the case where $l_i = l_1$ and $l_j = l_2$. The nodal diagram for this case then yields:

{\footnotesize
\bg\label{luna6}
&& \hskip.5in{ 
\begin{tikzpicture}[baseline={(0, 0cm-\MathAxis pt)}, thick,
main/.style = {draw, circle, fill=black, minimum size=12pt},
dot/.style={inner sep=0pt,fill=black, circle, minimum size=4pt},
dots/.style={inner sep=0pt, fill=black, circle, minimum size=2pt}]
  \node[main] (a) at (-3.2, 0) {};
  \node[dot] (1) at (-1.7, 0) {};
  \node[dot] (2) at (0, 0) {};
  \node[main] (3) at (60:1.5) {};
  \node[main] (4) at (30:1.5) {};
  \node[main] (5) at (0:1.5) {};
  \node[main] (6) at (-30:1.5) {};
  \node[main] (7) at(-65:1.5) {};
  \node[dots] (d1) at (-40:1.3) {};
  \node[dots] (d2) at (-50:1.3) {};
  \node[dots] (d3) at (-60:1.3) {};
\draw (a) -- node[midway, above right, pos=0.2] {$l_2$} (1);
\draw (1) -- node[midway, above right, pos=0] {$(r-1)\; l_1$} (2);
\draw (2) -- node[midway, left, pos=0.7] {$l_1$} (3);
\draw (2) -- node[midway, left, pos=0.8] {$l_1$} (4);
\draw (2) -- node[midway, above right, sloped, pos=0.7] {$l_1$} (5);
\draw (2) -- node[midway, right, pos=0.7] {$l_1$} (6);
\draw (2) -- node[midway, right, pos=0.7] {$l_1$} (7);
\end{tikzpicture} } \nonumber
\\
=&& \prod_{j} \Big({\pi_j\over b_j}\Big)^{1/2}  \sum_{e_1, e_2}r(rl_1-l_1)^{e_1} l_2^{m_1}
\left(\begin{matrix} m_1 + e_1 \\ e_1\\ \end{matrix}\right) \left(\begin{matrix} r-1\\ e_2\\ 
\end{matrix}\right)\left(-{\overline\beta_1\over b_1}\right)^{r_1} \left(-{\overline\beta_2\over b_2}\right)
{(e_2 - 1)!!\over (2b_1)^{e_2/2}},  \nd} 
where $m_1 \equiv m + p - e_o - e_1, r_1 \equiv r - 1 -e_2$ and $e_2 \in 2\mathbb{Z}_+$ otherwise integrals would vanish. Again, this doesn't put any constraints on either $m, p$ or $r$. The extra factor of $r$ is from combinatoric. Note that the sum is over all 
$e_o, e_1$ and $e_2$, although we haven't explicitly shown the sum over $e_o$. This will be inserted 
in later.

It is now easy to see that, for a given choice of $l_i = l_1$, we can choose all values of $l_j$ starting with $l_2$. Similarly, when $l_i = l_2$, $l_j$ will take all values except $l_2$. This shows that the value of the nodal diagram for all $i$ and $j$, with $i \ne j$, will have a nested integral structure of the following form:

{\footnotesize
\bg\label{luna7}
&&
\hskip.5in{\frac{1}{{\rm V}^2} \sum_{i, j} \quad
\begin{tikzpicture}[thick, baseline={(0, 0cm-\MathAxis pt)},
main/.style = {draw, circle, fill=black, minimum size=12pt},
dot/.style={inner sep=0pt,fill=black, circle, minimum size=4pt},
dots/.style={inner sep=0pt, fill=black, circle, minimum size=2pt}]
  \node[main] (a) at (-3.2, 0) {};
  \node[dot] (1) at (-1.7, 0) {};
  \node[dot] (2) at (0, 0) {};
  \node[main] (3) at (60:1.5) {};
  \node[main] (4) at (30:1.5) {};
  \node[main] (5) at (0:1.5) {};
  \node[main] (6) at (-30:1.5) {};
  \node[main] (7) at(-65:1.5) {};
  \node[dots] (d1) at (-40:1.3) {};
  \node[dots] (d2) at (-50:1.3) {};
  \node[dots] (d3) at (-60:1.3) {};
\draw (a) -- node[midway, above right, pos=0.2] {$l_j$} (1);
\draw (1) -- node[midway, above right, pos=0.1] {$(r-1)\; l_i$} (2);
\draw (2) -- node[midway, left, pos=0.7] {$l_i$} (3);
\draw (2) -- node[midway, left, pos=0.8] {$l_i$} (4);
\draw (2) -- node[midway, above right, sloped, pos=0.7] {$l_i$} (5);
\draw (2) -- node[midway, right, pos=0.7] {$l_i$} (6);
\draw (2) -- node[midway, right, pos=0.7] {$l_i$} (7);
\end{tikzpicture}} \nonumber
\\
&& =  \prod_{k} \Big({\pi_k\over b_k}\Big)^{1/2} 
\sum_{e_1, e_2}r(r - 1)^{e_1} 
\left(\begin{matrix} m_1 + e_1 \\ e_1\\ \end{matrix}\right) \left(\begin{matrix} r-1\\ e_2\\ 
\end{matrix}\right) {(e_2 - 1)!!\over 2^{e_2/2}}
\\
&& \times \Bigg[\int_0^\infty d^{11}l ~{l^{e_1}\over b^{e_2/2}(l)} \left(-{\overline\beta(l)\over 
b(l)}\right)^{r_1} \int_l^\infty d^{11} l' ~{l^{'m_1} \overline\beta(l') \over b(l')}
+ \int_0^\infty d^{11} l ~{l^{m_1} \overline\beta(l) \over b(l)} 
\int_l^\infty d^{11}l' ~{l^{'e_1}\over b^{e_2/2}(l')} \left(-{\overline\beta(l')\over 
b(l')}\right)^{r_1}\Bigg],\nonumber \nd}
where for simplicity we have only taken the positive momentum modes. The negative frequency modes will simply add to the existing structure so we don't lose any important information by taking this assumption. Furthermore $b(-l) = b(l)$ and $\overline\beta(-l) = \overline\beta^\ast(l) = \overline\beta(l)$ because we have consistently ignored the imaginary parts of the Fourier modes, which is again consistent with our choice of positive frequencies. The other parameters, namely $(m_1, r_1)$, are defined as before.

Note two things. One, the first nested integral in \eqref{luna7} could be viewed as an integral over the top half of a triangular region of a two eleven-dimensional 
$(l, l')$ spaces. Similarly the second nested integral of \eqref{luna7} could be mapped to the lower half of the triangular region of the 
$(l, l')$ space. This means we could possibly combine the two nested integrals to one integral over the full $(l, l')$ space as:

{\footnotesize
\bg\label{lafyre}
&&\hskip.5in{\frac{1}{{\rm V}^2} \sum_{i, j} \quad
\begin{tikzpicture}[thick, baseline={(0, 0cm-\MathAxis pt)},
main/.style = {draw, circle, fill=black, minimum size=12pt},
dot/.style={inner sep=0pt,fill=black, circle, minimum size=4pt},
dots/.style={inner sep=0pt, fill=black, circle, minimum size=2pt}]
  \node[main] (a) at (-3.2, 0) {};
  \node[dot] (1) at (-1.7, 0) {};
  \node[dot] (2) at (0, 0) {};
  \node[main] (3) at (60:1.5) {};
  \node[main] (4) at (30:1.5) {};
  \node[main] (5) at (0:1.5) {};
  \node[main] (6) at (-30:1.5) {};
  \node[main] (7) at(-65:1.5) {};
  \node[dots] (d1) at (-40:1.3) {};
  \node[dots] (d2) at (-50:1.3) {};
  \node[dots] (d3) at (-60:1.3) {};
\draw (a) -- node[midway, above right, pos=0.2] {$l_j$} (1);
\draw (1) -- node[midway, above right, pos=0.1] {$(r-1)\; l_i$} (2);
\draw (2) -- node[midway, left, pos=0.7] {$l_i$} (3);
\draw (2) -- node[midway, left, pos=0.8] {$l_i$} (4);
\draw (2) -- node[midway, above right, sloped, pos=0.7] {$l_i$} (5);
\draw (2) -- node[midway, right, pos=0.7] {$l_i$} (6);
\draw (2) -- node[midway, right, pos=0.7] {$l_i$} (7);
\end{tikzpicture}} \nonumber\\
&= &\prod_{k} \Big({\pi_k\over b_k}\Big)^{1/2} 
\sum_{e_1, e_2}r\hat{r}^{e_1} 
\left(\begin{matrix} m_e \\ e_1\\ \end{matrix}\right) \left(\begin{matrix} \hat{r}\\ e_2\\ 
\end{matrix}\right)
{(e_2 - 1)!!\over 2^{e_2/2}}
\int_0^\infty d^{11}l \int_0^\infty d^{11}l'~{l^{e_1}\over b^{e_2/2}(l)} \left(-{\overline\beta(l)\over 
b(l)}\right)^{r_1}{l^{'m_1} \overline\beta(l') \over b(l')}, \nonumber\\ \nd}
where $\hat{r} \equiv r - 1$ and $m_e \equiv m_1 + e_1$. One may easily see that the domain of the integrations have changed.
Interestingly, while \eqref{lafyre} is a possible simplification of 
\eqref{luna7}, it hides the $2!$ growth of \eqref{luna7} from 
\eqref{luna5}\footnote{Further simplifications cannot happen. For example one might also wonder if, by including \eqref{luna5}, one could convert the nested integrals in \eqref{luna7} from 0 to $\infty$. This is in general subtle because of the presence of shifted Gaussians: the growths are {\it not} linear as some parts of the integrals vanish as one goes from even to odd integrands. Additionally \eqref{luna7} is a double integral (in $11d$ variables), plus there are differences in powers of momenta: $l^{e_1}$ for one part and $l^{m_1}$ for the other. A careful comparison will easily reveal that our case by case study is probably the most efficient way to handle the path integral \eqref{fontini}.}. 

And two, in the nested integral structure, we have 
integrated from $l$ to $\infty$. The lower limit of the integral, which is $l$, should be clear from our above discussion. However one might be concerned by the upper limit of the integral because of the Wilsonian effective action, the integral should be till scale $\mu \propto {\rm M}_p$. 
This can be reconciled by going to the limit where ${\rm M}_p \to \infty$. Thus the $\infty$ appearing in the upper limits of the nested integrals are therefore the symbolic way to express the energy scale of the system. Such an approach helps us to resolve any further issues coming from allowing positive powers of 
$l$ in the integrals. For example, if $m' > 2$, there might be possibilities that the nodal diagrams have UV divergences. But since the UV of the nodal diagrams are all controlled by $\mu \propto {\rm M}_p$, none of the nodal diagrams blow up, and we always get finite answers from them.

On the other hand, the lower limits of the outermost integrals are kept at zero so one might worry about IR issues. This in general should not be much of a concern and we should be able to control the IR issues somewhat along the lines of the IR issues studied for gravitational theories with UV/IR mixing. Moreover the compactness of the internal eight-manifold will tell us that internal  momenta could go as low as the inverse sizes of the cycles\footnote{For example local orthogonal one-cycles. The non-K\"ahlerity of the internal eight-manifold could even support these one-cycles globally, although here we are only concerned about the local ones.} in the internal eight-manifold over the warped Minkowski background. The time-independence of the internal manifold can serve as IR cut-off for at least the internal momenta.
This is of course another advantage we gain by realizing the de Sitter space (including the internal manifold) as a Glauber-Sudarshan state
over the supersymmetric Minkowski background.

Another point to note here is the appearance of $\overline\beta(l)$ {\it inside} the nested integrals 
in \eqref{luna7} (and also \eqref{luna5}). This means the equations connecting $\sigma = (\alpha, \beta, \gamma)$ with $\overline\sigma = (\overline\alpha, \overline\beta, \overline\gamma)$ could become slightly non-trivial. We already anticipated this in \cite{coherbeta2}, and here we will demonstrate this in more details. The good thing is clearly the quantitative control that we have over such computations even in the presence of all possible quantum corrections. This will be demonstrated further below as we tackle more complicated nodal diagrams.

\vskip.2in

\noindent{\bf Case 3: $l_{v_1} = l_{v_2} = l_{v_3} = ... l_{v_{r-2}} = l_i, l_{v_{r-1}} \ne l_i, 
l_{v_{r}} \ne l_i$ for the $\varphi_2$ fields}

\vskip.2in

\noindent Our next interesting case may be split into two set of nodal diagrams: one with 
$l_{v_{r-1}} = l_{v_r}$ and the other with $l_{v_{r-1}} \ne l_{v_r}$. For the first case we can consider 
$l_i = l_1$ and $l_{v_{r-1}} = l_{v_r} = l_2$ and then change both $l_1$ and $l_2$ to add all these diagrams together. With $(l_1, l_2)$ the nodal diagram gives us:

{\footnotesize
\bg\label{luna8}
&& 
\centerline{\begin{tikzpicture}[thick,
main/.style = {draw, circle, fill=black, minimum size=12pt},
dot/.style={inner sep=0pt,fill=black, circle, minimum size=4pt},
dots/.style={inner sep=0pt, fill=black, circle, minimum size=2pt}]
  \node[dot] (a) at (-3.2, 0) {};
  \node[dot] (1) at (-1.7, 0) {};
  \node[dot] (2) at (0, 0) {};
  \node[main] (3) at (60:1.5) {};
  \node[main] (4) at (30:1.5) {};
  \node[main] (5) at (0:1.5) {};
  \node[main] (6) at (-30:1.5) {};
  \node[main] (7) at (-65:1.5) {};
  \node[main] (a1) at (-4.4, 1) {};
  \node[main] (a2) at (-4.4, -1) {};
  \node[dots] (d1) at (-40:1.3) {};
  \node[dots] (d2) at (-50:1.3) {};
  \node[dots] (d3) at (-60:1.3) {};
\draw (a1) -- node[midway, above right, pos=0.2] {$l_2$} (a);
\draw (a2) -- node[midway, below right, pos=0.2] {$l_2$} (a);
\draw (a) -- node[midway, above right, pos=0.2] {$2\; l_2$} (1);
\draw (1) -- node[midway, above right, pos=0.1] {$(r-2)\; l_1$} (2);
\draw (2) -- node[midway, left, pos=0.7] {$l_1$} (3);
\draw (2) -- node[midway, left, pos=0.8] {$l_1$} (4);
\draw (2) -- node[midway, above right, sloped, pos=0.7] {$l_1$} (5);
\draw (2) -- node[midway, right, pos=0.7] {$l_1$} (6);
\draw (2) -- node[midway, right, pos=0.7] {$l_1$} (7);
\end{tikzpicture}} 
\nonumber\\
&&= \prod_{j} \Big({\pi_j\over b_j}\Big)^{1/2} 
\sum_{e_3, e_4}r(r-1)(rl_1 - 2l_1)^{e_3}(2l_2)^{m_2} 
\left(\begin{matrix} m_2 + e_3 \\ e_3\\ \end{matrix}\right) \left(\begin{matrix} r-2\\ e_4\\ 
\end{matrix}\right)\left(-{\overline\beta_1\over b_1}\right)^{r_2} \left({1\over 4b_2} + {\overline\beta^2_2\over 2b^2_2}\right)
{(e_4 - 1)!!\over (2b_1)^{e_4/2}}, \nonumber\\ \nd}
where $m_2 = m + p - e_o - e_3$ and $r_2 = r - 2 - e_4$, where 
$(e_3, e_4) \in (\mathbb{Z}_+, \mathbb{Z}_+)$. Note the similarity and the difference with \eqref{luna6}: the structure is similar but the terms with $b_2$ is more involved, plus the combinatoric factor is different. This is expected because of the equal split of the
$l_2$ momentum modes in the nodal diagram. This is clearly not the case with unequal split, as may be seen from the following nodal diagram:

{\footnotesize
\bg\label{luna9}
&&
\centerline{\begin{tikzpicture}[thick,
main/.style = {draw, circle, fill=black, minimum size=12pt},
dot/.style={inner sep=0pt,fill=black, circle, minimum size=4pt},
dots/.style={inner sep=0pt, fill=black, circle, minimum size=2pt}]
  \node[dot] (a) at (-3.2, 0) {};
  \node[dot] (1) at (-1.7, 0) {};
  \node[dot] (2) at (0, 0) {};
  \node[main] (3) at (60:1.5) {};
  \node[main] (4) at (30:1.5) {};
  \node[main] (5) at (0:1.5) {};
  \node[main] (6) at (-30:1.5) {};
  \node[main] (7) at (-65:1.5) {};
  \node[main] (a1) at (-4.4, 1) {};
  \node[main] (a2) at (-4.4, -1) {};
  \node[dots] (d1) at (-40:1.3) {};
  \node[dots] (d2) at (-50:1.3) {};
  \node[dots] (d3) at (-60:1.3) {};
\draw (a1) -- node[midway, above right, pos=0.2] {$l_3$} (a);
\draw (a2) -- node[midway, below right, pos=0.2] {$l_2$} (a);
\draw (a) -- node[midway, above right, pos=0.2] {$l_2 + l_3$} (1);
\draw (1) -- node[midway, above right, pos=0] {$(r-2)\; l_1$} (2);
\draw (2) -- node[midway, left, pos=0.7] {$l_1$} (3);
\draw (2) -- node[midway, left, pos=0.8] {$l_1$} (4);
\draw (2) -- node[midway, above right, sloped, pos=0.7] {$l_1$} (5);
\draw (2) -- node[midway, right, pos=0.7] {$l_1$} (6);
\draw (2) -- node[midway, right, pos=0.7] {$l_1$} (7);
\end{tikzpicture}} 
\nonumber\\
&& = \prod_{j} \Big({\pi_j\over b_j}\Big)^{1/2} 
\sum_{\{e_u\}} a_r (rl_1 - 2l_1)^{e_5} l_2^{e_6} l_3^{m_3} 
\left(\begin{matrix} m_2 + e_3 \\ e_5\\ \end{matrix}\right)
\left(\begin{matrix} m_3 + e_6 \\ e_6\\ \end{matrix}\right) \left(\begin{matrix} r - 2 \\ e_7\\ \end{matrix}\right)
\left(-{\overline\beta_1\over b_1}\right)^{r_3}\left({\overline\beta_2 
\overline\beta_3 \over b_2 b_3}\right)
{(e_7 - 1)!!\over (2b_1)^{e_7/2}}, \nonumber\\ \nd}
where $\{e_u\} = (e_5, e_6, e_7)$, $a_r = r(r-1)$, $m_2$ is defined as above, $m_3 = m + p - e_o - e_5 - e_6$ and 
$r_3 = r - 2 - e_7$ with $(e_5, e_6, e_7) \in \mathbb{Z}_+$. With these we are ready to sum over all the possible permutations of the lines in the nodal diagrams. We start with \eqref{luna8}. Summing over all such diagrams, again with $i \ne j$, gives us the following value:

{\footnotesize
\bg\label{luna10}
&&
\hskip.4in{\frac{1}{{\rm V}^2} \sum_{i, j} \; \; 
\begin{tikzpicture}[thick, baseline={(0, 0cm-\MathAxis pt)},
main/.style = {draw, circle, fill=black, minimum size=12pt},
dot/.style={inner sep=0pt,fill=black, circle, minimum size=4pt},
dots/.style={inner sep=0pt, fill=black, circle, minimum size=2pt}]
  \node[dot] (a) at (-3.2, 0) {};
  \node[dot] (1) at (-1.7, 0) {};
  \node[dot] (2) at (0, 0) {};
  \node[main] (3) at (60:1.5) {};
  \node[main] (4) at (30:1.5) {};
  \node[main] (5) at (0:1.5) {};
  \node[main] (6) at (-30:1.5) {};
  \node[main] (7) at (-65:1.5) {};
  \node[main] (a1) at (-4.4, 1) {};
  \node[main] (a2) at (-4.4, -1) {};
  \node[dots] (d1) at (-40:1.3) {};
  \node[dots] (d2) at (-50:1.3) {};
  \node[dots] (d3) at (-60:1.3) {};
\draw (a1) -- node[midway, above right, pos=0.2] {$l_j$} (a);
\draw (a2) -- node[midway, below right, pos=0.2] {$l_j$} (a);
\draw (a) -- node[midway, above right, pos=0.2] {$2\; l_j$} (1);
\draw (1) -- node[midway, above right, pos=0] {$(r-2)\; l_i$} (2);
\draw (2) -- node[midway, left, pos=0.7] {$l_i$} (3);
\draw (2) -- node[midway, left, pos=0.8] {$l_i$} (4);
\draw (2) -- node[midway, above right, sloped, pos=0.7] {$l_i$} (5);
\draw (2) -- node[midway, right, pos=0.7] {$l_i$} (6);
\draw (2) -- node[midway, right, pos=0.7] {$l_i$} (7);
\end{tikzpicture}}
\nonumber\\
&& = \prod_{k} \Big({\pi_k\over b_k}\Big)^{1/2} 
\sum_{e_3, e_4}a_r(r - 2)^{e_3}(2l_2)^{m_2} 
\left(\begin{matrix} m_2 + e_3 \\ e_3\\ \end{matrix}\right) \left(\begin{matrix} r-2\\ e_4\\ 
\end{matrix}\right) {(e_4 - 1)!!\over 2^{e_4/2 - m_2 + 1}}\\
&& \times \Bigg[\int_0^\infty d^{11} l~{l^{e_3}\over (b(l))^{e_4/2}} 
\left(-{\overline\beta(l)\over b(l)}\right)^{r_2} \int_l^\infty d^{11}l'~
l^{'m_2}\left({1\over 2b(l')} + {\overline\beta^2(l')\over b^2(l')}\right)
\nonumber\\ 
&& + \int_0^\infty d^{11} l~
l^{m_2}\left({1\over 2b(l)} + {\overline\beta^2(l)\over b^2(l)}\right)
\int_l^\infty d^{11} l'~{l^{'e_3}\over (b(l'))^{e_4/2}} 
\left(-{\overline\beta(l')\over b(l')}\right)^{r_2}\Bigg]\nonumber\\ 
&& = \prod_{k} \Big({\pi_k\over b_k}\Big)^{1/2} 
\sum_{e_3, e_4}a_r(r - 2)^{e_3}(2l_2)^{m_2} 
\left(\begin{matrix} m_2 + e_3 \\ e_3\\ \end{matrix}\right) \left(\begin{matrix} r-2\\ e_4\\ 
\end{matrix}\right) {(e_4 - 1)!!\over 2^{e_4/2 - m_2 + 1}}\nonumber\\
&&\times \int_0^\infty d^{11} l~\int_0^\infty d^{11}l'~{l^{e_3}\over (b(l))^{e_4/2}} 
\left(-{\overline\beta(l)\over b(l)}\right)^{r_2} 
l^{'m_2}\left({1\over 2b(l')} + {\overline\beta^2(l')\over b^2(l')}\right)
\nonumber\nd}
where $a_r = r(r-1)$ and expectedly, the nested structure as in \eqref{luna7} appears again which may be combined together under one roof. Note however the difference from \eqref{luna7}: there are terms independent of $\overline\beta(l)$ or $\overline\beta(l')$ inside the integrals. In fact for $e_4 = r - 2$ there would be terms completely independent of 
$\overline\beta(l)$ and would only depend on powers of momentum and the propagator $b(l)$. On the other hand, summing over all possible permutations of \eqref{luna9} is more non-trivial because now three different momenta appear. As an example consider the following nodal diagram where $r-2$ legs have $l_3$ momenta, and out of the remaining two legs, one has momentum $l_1$ and the other has momentum $l_j$ where 
$j \ne (1, 3)$. We then sum over all $j$ to get the following value:

{\footnotesize
\bg\label{luna11}
&& 
\frac{1}{\rm V} \sum_{j} \; \; 
\begin{tikzpicture}[thick, baseline={(0, 0cm-\MathAxis pt)},
main/.style = {draw, circle, fill=black, minimum size=12pt},
dot/.style={inner sep=0pt,fill=black, circle, minimum size=4pt},
dots/.style={inner sep=0pt, fill=black, circle, minimum size=2pt}]
  \node[dot] (a) at (-3.2, 0) {};
  \node[dot] (1) at (-1.7, 0) {};
  \node[dot] (2) at (0, 0) {};
  \node[main] (3) at (60:1.5) {};
  \node[main] (4) at (30:1.5) {};
  \node[main] (5) at (0:1.5) {};
  \node[main] (6) at (-30:1.5) {};
  \node[main] (7) at (-65:1.5) {};
  \node[main] (a1) at (-4.4, 1) {};
  \node[main] (a2) at (-4.4, -1) {};
  \node[dots] (d1) at (-40:1.3) {};
  \node[dots] (d2) at (-50:1.3) {};
  \node[dots] (d3) at (-60:1.3) {};
\draw (a1) -- node[midway, above right, pos=0.2] {$l_1$} (a);
\draw (a2) -- node[midway, below right, pos=0.2] {$l_j$} (a);
\draw (a) -- node[midway, above right, pos=0] {$l_1 + l_j$} (1);
\draw (1) -- node[midway, above right, pos=0] {$(r-2)\; l_3$} (2);
\draw (2) -- node[midway, left, pos=0.7] {$l_3$} (3);
\draw (2) -- node[midway, left, pos=0.8] {$l_3$} (4);
\draw (2) -- node[midway, above right, sloped, pos=0.7] {$l_3$} (5);
\draw (2) -- node[midway, right, pos=0.7] {$l_3$} (6);
\draw (2) -- node[midway, right, pos=0.7] {$l_3$} (7);
\end{tikzpicture} 
\nonumber\\
= && \prod_{k} \Big({\pi_k\over b_k}\Big)^{1/2} 
\sum_{\{e_u\}} a_r (r - 2)^{e_5}
\left(\begin{matrix} m_2 + e_3 \\ e_5\\ \end{matrix}\right)
\left(\begin{matrix} m_3 + e_6 \\ e_6\\ \end{matrix}\right) \left(\begin{matrix} r - 2 \\ e_7\\ \end{matrix}\right)
{(e_7 - 1)!!\over 2^{e_7/2}}\nonumber\\
&& \times ~ {l_3^{e_5}\over b_3^{e_7/2}}
\left(-{\overline\beta_3\over b_3}\right)^{r_3}\left[{\overline\beta_1 l_1^{e_6}\over b_1}\left(\int_{l_1}^\infty d^{11} l'~{\overline\beta(l')l^{'m_3}\over b(l')} - 
\int_{l_2}^{l_3} d^{11} l'~{\overline\beta(l')l^{'m_3}\over b(l')}\right)\right], \nd}
with $(r_3, m_3, a_r, \{e_u\})$ taking the same values as in \eqref{luna9}. The crucial thing to note here is the appearance of a relative minus sign between two integrals. The limits of the second integral is from one specific value of the momentum $-$ here it is $l_2$ $-$ to the {\it next} allowed value of the momentum $-$ which is $l_3$. In the discrete sum approach, this particular value of the momentum is absent in the sum over $l_j$. Since the appearance, or the absence thereof, comes with zero measure, we expect:
\bg\label{dcmika}
\int_{l_n}^{l_{n+1}} d^{11} l~{\overline\beta(l)l^{m_3}\over b(l)} 
~ \to ~ 0, \nd
in the infinite volume limit where ${\rm V} \to \infty$, for two consecutive momenta in the discrete sum. Thus in the integral picture of \eqref{luna11} we can ignore the second integral and retain only the first one. Putting \eqref{luna11} and \eqref{dcmika} together, we can sum over all the nodal diagrams with $(r-2)$ legs of momenta $l_3$ to get:

{\footnotesize
\bg\label{luna12}
&&
\frac{1}{\rm V} \sum_{j}
\begin{tikzpicture}[thick, baseline={(0, 0cm-\MathAxis pt)},
main/.style = {draw, circle, fill=black, minimum size=12pt},
dot/.style={inner sep=0pt,fill=black, circle, minimum size=4pt},
dots/.style={inner sep=0pt, fill=black, circle, minimum size=2pt}]
  \node[dot] (a) at (-3, 0) {};
  \node[dot] (1) at (-1.7, 0) {};
  \node[dot] (2) at (0, 0) {};
  \node[main] (3) at (60:1.5) {};
  \node[main] (4) at (30:1.5) {};
  \node[main] (5) at (0:1.5) {};
  \node[main] (6) at (-30:1.5) {};
  \node[main] (7) at (-65:1.5) {};
  \node[main] (a1) at (-4.2, 1) {};
  \node[main] (a2) at (-4.2, -1) {};
  \node[dots] (d1) at (-40:1.3) {};
  \node[dots] (d2) at (-50:1.3) {};
  \node[dots] (d3) at (-60:1.3) {}; 
\draw (a1) -- node[midway, above right, pos=0.2] {$l_1$} (a);
\draw (a2) -- node[midway, below right, pos=0.2] {$l_j$} (a);
\draw (a) -- node[midway, above right, pos=0] {$l_1 + l_j$} (1);
\draw (1) -- node[midway, above right, pos=0] {$(r-2)\; l_3$} (2);
\draw (2) -- node[midway, left, pos=0.7] {$l_3$} (3);
\draw (2) -- node[midway, left, pos=0.8] {$l_3$} (4);
\draw (2) -- node[midway, above right, sloped, pos=0.7] {$l_3$} (5);
\draw (2) -- node[midway, right, pos=0.7] {$l_3$} (6);
\draw (2) -- node[midway, right, pos=0.7] {$l_3$} (7);
\end{tikzpicture} 
 \; + \;
\frac{1}{\rm V} \sum_{j}
\begin{tikzpicture}[thick, baseline={(0, 0cm-\MathAxis pt)},
main/.style = {draw, circle, fill=black, minimum size=12pt},
dot/.style={inner sep=0pt,fill=black, circle, minimum size=4pt},
dots/.style={inner sep=0pt, fill=black, circle, minimum size=2pt}]
  \node[dot] (a) at (-3, 0) {};
  \node[dot] (1) at (-1.7, 0) {};
  \node[dot] (2) at (0, 0) {};
  \node[main] (3) at (60:1.5) {};
  \node[main] (4) at (30:1.5) {};
  \node[main] (5) at (0:1.5) {};
  \node[main] (6) at (-30:1.5) {};
  \node[main] (7) at (-65:1.5) {};
  \node[main] (a1) at (-4.2, 1) {};
  \node[main] (a2) at (-4.2, -1) {};
  \node[dots] (d1) at (-40:1.3) {};
  \node[dots] (d2) at (-50:1.3) {};
  \node[dots] (d3) at (-60:1.3) {};  
\draw (a1) -- node[midway, above right, pos=0.2] {$l_2$} (a);
\draw (a2) -- node[midway, below right, pos=0.2] {$l_j$} (a);
\draw (a) -- node[midway, above right, pos=0] {$l_2 + l_j$} (1);
\draw (1) -- node[midway, above right, pos=0] {$(r-2)\; l_3$} (2);
\draw (2) -- node[midway, left, pos=0.7] {$l_3$} (3);
\draw (2) -- node[midway, left, pos=0.8] {$l_3$} (4);
\draw (2) -- node[midway, above right, sloped, pos=0.7] {$l_3$} (5);
\draw (2) -- node[midway, right, pos=0.7] {$l_3$} (6);
\draw (2) -- node[midway, right, pos=0.7] {$l_3$} (7);
\end{tikzpicture}
\nonumber \\
+&& \quad
\frac{1}{\rm V} \sum_{j} \;\;
\begin{tikzpicture}[thick, baseline={(0, 0cm-\MathAxis pt)},
main/.style = {draw, circle, fill=black, minimum size=12pt},
dot/.style={inner sep=0pt,fill=black, circle, minimum size=4pt},
dots/.style={inner sep=0pt, fill=black, circle, minimum size=2pt}]
  \node[dot] (a) at (-3.2, 0) {};
  \node[dot] (1) at (-1.7, 0) {};
  \node[dot] (2) at (0, 0) {};
  \node[main] (3) at (60:1.5) {};
  \node[main] (4) at (30:1.5) {};
  \node[main] (5) at (0:1.5) {};
  \node[main] (6) at (-30:1.5) {};
  \node[main] (7) at (-65:1.5) {};
  \node[main] (a1) at (-4.4, 1) {};
  \node[main] (a2) at (-4.4, -1) {};
  \node[dots] (d1) at (-40:1.3) {};
  \node[dots] (d2) at (-50:1.3) {};
  \node[dots] (d3) at (-60:1.3) {};  
\draw (a1) -- node[midway, above right, pos=0.2] {$l_4$} (a);
\draw (a2) -- node[midway, below right, pos=0.2] {$l_j$} (a);
\draw (a) -- node[midway, above right, pos=0] {$l_4 + l_j$} (1);
\draw (1) -- node[midway, above right, pos=0] {$(r-2)\; l_3$} (2);
\draw (2) -- node[midway, left, pos=0.7] {$l_3$} (3);
\draw (2) -- node[midway, left, pos=0.8] {$l_3$} (4);
\draw (2) -- node[midway, above right, sloped, pos=0.7] {$l_3$} (5);
\draw (2) -- node[midway, right, pos=0.7] {$l_3$} (6);
\draw (2) -- node[midway, right, pos=0.7] {$l_3$} (7);
\end{tikzpicture} 
\quad + \quad ......
\nonumber\\
= && \prod_{k} \Big({\pi_k\over b_k}\Big)^{1/2} 
\sum_{e_5, e_6, e_7}r(r-1)(r - 2)^{e_5}
\left(\begin{matrix} m_2 + e_3 \\ e_5\\ \end{matrix}\right)
\left(\begin{matrix} m_3 + e_6 \\ e_6\\ \end{matrix}\right) \left(\begin{matrix} r - 2 \\ e_7\\ \end{matrix}\right)
{(e_7 - 1)!!\over 2^{e_7/2}}\nonumber\\
&& \times 
\Bigg[{\overline\beta_1 \over b_1} \left(l_1^{e_6}\int_{l_1}^\infty d^{11} l'~{\overline\beta(l')l^{'m_3}\over b(l')} + l_1^{m_3}\int_{l_1}^\infty d^{11} l'~{\overline\beta(l')l^{'e_6}\over b(l')}\right) \nonumber\\
&& +~~
{\overline\beta_2 \over b_2} \left(l_2^{e_6}\int_{l_2}^\infty d^{11} l'~{\overline\beta(l')l^{'m_3}\over b(l')} + l_2^{m_3}\int_{l_2}^\infty d^{11} l'~{\overline\beta(l')l^{'e_6}\over b(l')}\right)\nonumber\\
&& +~~ {\overline\beta_4 \over b_4} \left(l_4^{e_6}\int_{l_4}^\infty d^{11} l'~{\overline\beta(l')l^{'m_3}\over b(l')} + l_4^{m_3}\int_{l_4}^\infty d^{11} l'~{\overline\beta(l')l^{'e_6}\over b(l')}\right) \nonumber\\
&& +~~
{\overline\beta_5 \over b_5} \left(l_5^{e_6}\int_{l_5}^\infty d^{11} l'~{\overline\beta(l')l^{'m_3}\over b(l')} + l_5^{m_3}\int_{l_5}^\infty d^{11} l'~{\overline\beta(l')l^{'e_6}\over b(l')}\right) + ..... 
\Bigg]~ {l_3^{e_5}\over b_3^{e_7/2}}
\left(-{\overline\beta_3\over b_3}\right)^{r_3}, \nd}
where note the absence of $\overline\beta_3$ inside the square bracket, as one might have expected. There are also sum over $m_i$ as well as over 
$(r, r_i)$ etc., but we will insert them later. Similar series as in 
\eqref{luna12} can be built with $(r-2)$ legs of momenta $l_i$, and we can sum them all over for $i \ne j \ne k$. The result is straightforward, albeit a bit more complicated:

{\footnotesize
\bg\label{luna13}
&&\hskip1.2in{\frac{1}{{\rm V}^3} \sum_{i,j,k} \; \;
\begin{tikzpicture}[thick,  baseline={(0, 0cm-\MathAxis pt)},
main/.style = {draw, circle, fill=black, minimum size=12pt},
dot/.style={inner sep=0pt,fill=black, circle, minimum size=4pt},
dots/.style={inner sep=0pt, fill=black, circle, minimum size=2pt}]
  \node[dot] (a) at (-3.2, 0) {};
  \node[dot] (1) at (-1.7, 0) {};
  \node[dot] (2) at (0, 0) {};
  \node[main] (3) at (60:1.5) {};
  \node[main] (4) at (30:1.5) {};
  \node[main] (5) at (0:1.5) {};
  \node[main] (6) at (-30:1.5) {};
  \node[main] (7) at (-65:1.5) {};
  \node[main] (a1) at (-4.4, 1) {};
  \node[main] (a2) at (-4.4, -1) {};
  \node[dots] (d1) at (-40:1.3) {};
  \node[dots] (d2) at (-50:1.3) {};
  \node[dots] (d3) at (-60:1.3) {};
\draw (a1) -- node[midway, above right, pos=0.2] {$l_j$} (a);
\draw (a2) -- node[midway, below right, pos=0.2] {$l_k$} (a);
\draw (a) -- node[midway, above right, pos=0.1] {$l_j + l_k$} (1);
\draw (1) -- node[midway, above right, pos=0] {$(r-2)\; l_i$} (2);
\draw (2) -- node[midway, left, pos=0.7] {$l_i$} (3);
\draw (2) -- node[midway, left, pos=0.8] {$l_i$} (4);
\draw (2) -- node[midway, above right, sloped, pos=0.7] {$l_i$} (5);
\draw (2) -- node[midway, right, pos=0.7] {$l_i$} (6);
\draw (2) -- node[midway, right, pos=0.7] {$l_i$} (7);
\end{tikzpicture}}
\nonumber \\
&& =  \prod_{g} \Big({\pi_g\over b_g}\Big)^{1/2} 
\sum_{e_5, e_6, e_7} r(r-1)(r - 2)^{e_5}
\left(\begin{matrix} m_2 + e_3 \\ e_5\\ \end{matrix}\right)
\left(\begin{matrix} m_3 + e_6 \\ e_6\\ \end{matrix}\right) \left(\begin{matrix} r - 2 \\ e_7\\ \end{matrix}\right)
{(e_7 - 1)!!\over 2^{e_7/2}}\\
&& \times \Bigg[\int_0^\infty d^{11} l~
{\overline\beta(l)l^{e_6} \over b(l)}
\int_{l}^\infty d^{11} l' ~ {l^{'e_5}\over b^{e_7/2}(l')}
\left(-{\overline\beta(l')\over b(l')}\right)^{r_3}
\int_{l'}^\infty d^{11} l''~{\overline\beta(l'')l^{''m_3}\over b(l'')}
\nonumber\\
&& + ~~ \int_0^\infty d^{11} l~{\overline\beta(l)l^{m_3}\over b(l)}
\int_{l}^\infty d^{11} l' ~ {l^{'e_5}\over b^{e_7/2}(l')}
\left(-{\overline\beta(l')\over b(l')}\right)^{r_3}
\int_{l'}^\infty d^{11} l''~
{\overline\beta(l'')l^{''e_6} \over b(l'')} \nonumber\\
&& + ~ \int_0^\infty d^{11} l~
{\overline\beta(l) \over b(l)} \left(l^{e_6}\int_{l}^\infty d^{11} l'~{\overline\beta(l')l^{'m_3}\over b(l')} + l^{m_3}\int_{l}^\infty d^{11} l'~{\overline\beta(l')l^{'e_6}\over b(l')}\right)
\int_{l'}^\infty d^{11} l'' ~ {l^{''e_5}\over b^{e_7/2}(l'')}
\left(-{\overline\beta(l'')\over b(l'')}\right)^{r_3}\nonumber\\
&& + ~\int_0^\infty d^{11} l ~ {l^{e_5}\over b^{e_7/2}(l)}
\left(-{\overline\beta(l)\over b(l)}\right)^{r_3}
\int_l^\infty d^{11} l'~
{\overline\beta(l') \over b(l')} \left(l^{'e_6}\int_{l'}^\infty d^{11} l''~{\overline\beta(l'')l^{''m_3}\over b(l'')} + l^{'m_3}\int_{l'}^\infty d^{11} l''~{\overline\beta(l'')l^{''e_6}\over b(l'')}\right)
\Bigg]\nonumber\\
&& =  \prod_{g} \Big({\pi_g\over b_g}\Big)^{1/2} 
\sum_{\{e_i\}} r_1r_2^{e_5} (-1)^{r_3} {\cal A}_{\{e_i\}}
\int_0^\infty d^{11} l\int_{0}^\infty d^{11} l' 
\int_{0}^\infty d^{11} l''~
{\overline\beta(l)l^{e_6} \over b(l)}
 {l^{'e_5}\over b^{e_7/2}(l')}
\left({\overline\beta(l')\over b(l')}\right)^{r_3}
{\overline\beta(l'')l^{''m_3}\over b(l'')}\nonumber
\nd}
where ${\cal A}_{\{e_i\}}$ denote the combinatorial factor ${\cal A}_{\{e_i\}} \equiv \Big(\begin{matrix} m_2 + e_3 \\ e_5\\ \end{matrix}\Big)
\Big(\begin{matrix} m_3 + e_6 \\ e_6\\ \end{matrix}\Big) \Big(\begin{matrix} r - 2 \\ e_7\\ \end{matrix}\Big){(e_7 - 1)!!\over 2^{e_7/2}}$ with 
$\{e_i\} = (e_5, e_6, e_7)$, 
$r_1 \equiv r(r-1)$, $r_2 \equiv r - 2$; and 
we see a multiple nested structure appearing compared to what we had earlier. They can all be combined together under one roof with triple integrals. Again the upper limits of the integrals should be till the energy scale $\mu \propto {\rm M}_p$, and we have taken ${\rm M}_p \to \infty$ to simplify the Gaussian integrals. We can go to more complicated nodal diagram like:

\vskip.1in

\bg\label{luna14}
{1\over {\rm V}^{c_o}} \sum_{i, j, ...} 
\begin{tikzpicture}[thick, baseline={(0, 0cm-\MathAxis pt)},
main/.style = {draw, circle, fill=black, minimum size=12pt},
dot/.style={inner sep=0pt,fill=black, circle, minimum size=4pt},
dots/.style={inner sep=0pt, fill=black, circle, minimum size=2pt},
mdot/.style={inner sep=0pt, fill=black, circle}] 
  \node[mdot] (b) at (-5, 0) {};
  \node[mdot] (b1) at (-5, 2) {};
  \node[mdot] (b2) at (-5, -2.25) {};
  \node[dot] (b11) at (-5.8, 2) {};
  \node[dot] (b22) at (-5.8, -2.25) {};
  \node[main] (b111) at (-6.8, 2.5) {};
  \node[main] (b112) at (-6.8, 1.5) {};
  \node[main] (b221) at (-6.8, -1.5) {};
  \node[main] (b222) at (-7, -2) {};
  \node[main] (b223) at (-7, -2.5) {};
  \node[main] (b224) at (-6.8, -3) {};
  \node[dot] (a) at (-5.8, 0) {};
  \node[dot] (1) at (-1.7, 0) {};
  \node[dot] (2) at (0, 0) {};
  \node[main] (3) at (60:1.5) {};
  \node[main] (4) at (30:1.5) {};
  \node[main] (5) at (0:1.5) {};
  \node[main] (6) at (-30:1.5) {};
  \node[main] (7) at (-65:1.5) {};
  \node[main] (a1) at (172: 6.6) {};
  \node[main] (a2) at (188: 6.6) {};
  \node[main] (a0) at (180: 6.6) {};
  \node[dots] (d1) at (-40:1.3) {};
  \node[dots] (d2) at (-50:1.3) {};
  \node[dots] (d3) at (-60:1.3) {};
  \node[dots] (d40) at (-5, -0.5) {};
  \node[dots] (d41) at (-5, -1) {};
  \node[dots] (d42) at (-5, -1.3) {};
  \node[dots] (d43) at (-5, -1.5) {};
  \node[dots] (d44) at (-5, -1.7) {};
  \node[dots] (d45) at (-5, -2) {};

\draw (a) -- node [ above right, pos=-0.3] {$3 l_k$} (b);
\draw (b1) -- node {} (b);
\draw (b1) -- node [midway, above right, pos=0.9] {$2\; l_j$} (b11);
\draw (b2) -- node [midway, above right, pos=1] {$4\; l_m$} (b22);
\draw (b11) -- node [midway, above right, pos=0.5] {$l_j$} (b111);
\draw (b11) -- node [midway, below right, pos=0.5] {$l_j$} (b112);
\draw (b22) -- node [midway, above right, pos=0.5] {$l_m$} (b221);
\draw (b22) -- node {} (b222);
\draw (b22) -- node {} (b223);
\draw (b22) -- node [midway, below right, pos=0.5] {$l_m$} (b224);
\draw (a1) -- node[midway, above right, pos=0.2] {$l_k$} (a);
\draw (a2) -- node[midway, below right, pos=0.2] {$l_k$} (a);
\draw (a0) -- node  {} (a);
\draw (b) -- node[midway, below right, pos=0] {$2 l_j + 3 l_k + 4 l_m + ...$} (1);
\draw (1) -- node[midway, above right, pos=0] {$(r-c)\; l_i$} (2);
\draw (2) -- node[midway, left, pos=0.7] {$l_i$} (3);
\draw (2) -- node[midway, left, pos=0.8] {$l_i$} (4);
\draw (2) -- node[midway, above right, sloped, pos=0.7] {$l_i$} (5);
\draw (2) -- node[midway, right, pos=0.7] {$l_i$} (6);
\draw (2) -- node[midway, right, pos=0.7] {$l_i$} (7);
\end{tikzpicture} 
\nd
\vskip.1in

\noindent where $c = 2 + 3 + 4 + ...$, $c_o$ is the order of the nested integrals, and the distribution of momenta follow similar pattern as before. This 
would lead to an even more complicated multiple integral structure, which we leave here as an exercise for the reader. 

\subsubsection{Interlude 1: UV/IR mixing and IR cut-off in nodal 
diagrams \label{sec3.2.3}}

We briefly alluded to IR cut-off on the nodal diagrams by referring to the UV/IR mixing. Since this is important, let us take a short interlude to discuss this issue in some details starting with the UV/IR mixing and then how this imposes an IR cut-off. 

UV/IR mixing in gravity is often invoked as a solution to hierarchy problems in physics \cite{uvir}. For instance, the cosmological constant problem can be understood as the {\it mismatch} between the naive value calculated for the vacuum energy from theoretical computations and what is measured for the late-time acceleration of the universe. However, invoking a relationship between the UV and IR cutoff for the effective field theory can resolve this problem. Let us elaborate the story further in the following. 

One of the key problem in the usual formulation of non-supersymmetric field theories is the appearance of the uncontrolled vacuum energies from both the bosonic and the fermionic sectors. In the simple case, 
the loop corrections to the vacuum energy density (assuming a massless, \textit{i.e.} $m \ll \Lambda_{\rm UV}$, scalar field) go as:
\begin{eqnarray}
V_0 \sim \Lambda_{\rm UV}^4\,.
\end{eqnarray}
In general, $\Lambda_{\rm UV}$ can be brought closer to  ${\rm M}_{\rm Pl}$ for a quantum theory of gravity. (We use $\Lambda$ and ${\rm M}_p$ respectively in eleven-dimensions.) On the other hand, we know that the observed value of the cosmological constant is only as big as $V_{\rm obs} \sim ({\rm meV})^4$. This leads to the much discussed $10^{120}$ orders of magnitude mismatch for the cosmological constant problem. 

The way UV/IR mixing fixes this is to include non-local effects coming from gravity into the calculation. The main idea is that the EFT, for a given region of space characterized by a length scale $\ell_{\rm IR}$, cannot be valid to any arbitrary UV cutoff $\Lambda_{\rm UV}$. Due to the UV/IR mixing from gravitational dynamics of sufficiently dense systems, the UV cutoff is inherently related to the IR one, and vice versa. A heuristic way to think about this is to consider only loop corrections to the vacuum energy in this region of space $\ell_{\rm IR}$ which are not high enough to make the entire region collapse into a black hole. In other words, given an IR cutoff $\Lambda_{\rm IR} = 1/\ell_{\rm IR}$, we can only have vacuum fluctuations of energy density which are controlled by the Schwarzschild radius of the region of space $\ell_{\rm IR}$,  ${\rm R}_{\rm S} = 2{\rm G M}$, \textit{i.e.}
\begin{eqnarray}
\Lambda_{\rm UV}^4 \sim \delta \rho \leq \left( \frac{{\rm M}_{\rm Pl}}{\ell_{\rm IR}}\right)^2\,.
\end{eqnarray}
Using the Friedmann equation, it can be shown that given the current Hubble scale as the IR cutoff leads to a UV cutoff such that the vacuum energy contribution is indeed of the order $({\rm meV})^4$. Similar arguments have also led to solving the electroweak hierarchy problems and predictions for beyond Standard Model physics.

Our detour for the above discussion is to emphasize the following. Since we are describing how a four-dimensional de Sitter space is constructed as a Glauber-Sudarshan state in string theory, we are necessarily dealing with a gravitational system which must account for UV/IR mixing\footnote{The Glauber-Sudarshan state is realized over a supersymmetric Minkowski background so one might wonder if the UV/IR mixing can be imposed here at all. The answer is that there are strong gravitational effects (due to time-independent fluxes, sources and quantum corrections) that make the system a warped Minkowski space with a compact curved non-K\"ahler internal manifold (albeit time-independent). The reason for choosing such a background as against a simpler background of the form ${\cal M}_{11} = \mathbb{R}^{2, 1} \times {\bf T}^8$, where ${\bf T}^8$ is an eight-dimensional toroidal space, is because of the Dine-Seiberg runaway problem \cite{dineseiberg}. Due to the runaway, the toroidal internal space will decompatify thus making the $2+1$ dimensional Newton's constant zero. To avoid this we will need fluxes and quantum effects to stabilize the moduli which, in turn, will convert the internal toroidal manifold to a non-K\"ahler space (that may or may not even be complex). Because of these gravitational dynamics there is a maximum energy fluctuation which is allowed beyond which we will have the entire region collapse into a black hole, thus implying the UV/IR mixing. \label{vmyerss}}. Since we know that our path integrals are valid up-to a UV cutoff $\Lambda_{\rm UV}$, we can turn around the above argument to infer that UV/IR mixing will lead to an infrared cutoff for the action. We do not need to make the relationship between these two cutoffs explicit, as they will not play a crucial role in our calculation. Nevertheless, since our theory is only well-defined for a Planckian scale UV-cutoff, we must invoke an IR cutoff to account for the non-local gravitational interactions. As will become evident later on, an IR cutoff is also necessary for subsequent calculations in our model.

\subsubsection{Contributions from the $\varphi_3$ fields \label{sec3.2.4}}

\noindent Our above exercises were all about $\varphi_2$ field which, in the language of eq. (4.81) of \cite{coherbeta2}, is the sector with G-fluxes. We would now like to quantify the scenario with $\varphi_3$ field. This field maps to the fermionic condensate, which is the form it appears in our path-integral \eqref{lincoln}, or its simplified version  \eqref{fontini}\footnote{As an example, the low energy dynamics of QCD is generically captured by hadronic or mesonic condensates that form the dynamical degrees of freedom at far IR. In our analysis the RG flows do not necessarily lead to confinement, but we are using similar condensates to study the IR theory. This is for technical simplification and for avoiding path-integrals with Rarita-Schwinger fermions. As mentioned earlier, a more elaborate analysis may be done with these fermionic degrees of freedom but we will not do it here. We will also not discuss other fermionic (and bosonic) degrees of freedom on M-branes, and leave them for future works.}. There are 128 gravitino degrees of freedom, and the fermionic condensate combines them together to provide the $\varphi_3$ field. We can then take $s$ copies of them with different momenta (plus an additional constraint on the {\it total} momentum) and study a few cases below.

Before moving ahead, let us clarify some subtleties associated with the 
$\varphi_3$ path integral.
The path integral structure for the $\varphi_3$ field looks very similar to the sector with $\varphi_2$ field, so the analysis should follow similar pattern as before. However there is one key difference: because of the momentum conservation, the momentum $f_{w_s}$ is not independent and is determined by the momenta $k_{u_i}, l_{v_j}$ and $f_{w_t}$ as:
\bg\label{headhunters}
-f_{w_s} = \sum_{i = 1}^q k_{u_i} + \sum_{j = 1}^r l_{v_j} + 
\sum_{t = 1}^{s-1} f_{w_t}. \nd
The minus sign is not very crucial because $\widetilde{\varphi}_3(-f_{w_s}) = \widetilde{\varphi}^\ast_3(f_{w_s}) = \widetilde{\varphi}_3(f_{w_s})$, as we have taken real fields. (This was also one of the reason for ignoring the negative momenta.) The equation \eqref{headhunters} means that the
$s$ momenta cannot be all equal thus ruling out nodal diagrams of the form 
\eqref{luna4} and \eqref{luna5} for the $\varphi_3$ field.

The choice of momentum for $\widetilde\varphi_3(f_{w_s})$ will then crucially depend on the choice of momenta $k_{u_i}, l_{v_j}$ and $f_{w_s}$ up to the limits prescribed in \eqref{headhunters}. For example with equal distribution of momenta, $-f_{w_s} = qk_i + rl_j + (s-1) f_k$ for a given choice of $(q, r, s) \in \mathbb{Z}_+$ and $(i, j, k) \in \mathbb{Z}_+$. In general we expect:

{\footnotesize
\bg\label{snelmad}
{1\over f^{2n_o}_{w_s}} = \sum_{n_1, e_{n_1}} (-1)^{n_1} 
\left(\sum_{i = 1}^q k_{u_i}\right)^{e_{n_1}}
\left(\sum_{j = 1}^r l_{v_j}\right)^{n_1 - e_{n_1}}
\left(\begin{matrix} n_1 \\ e_{n_1}\\ \end{matrix}\right)
{2n_o(2n_0 + 1)... (2n_o + n_1 - 1)\over (n_1)! \left(\sum\limits_{t = 1}^{s-1} 
f_{w_t}\right)^{2n_o + n_1}}, \nd}
where $(n_o, n_1, e_{n_1}) \in \mathbb{Z}_+$. Note that the effect of \eqref{snelmad} on \eqref{lincoln} is to simply change $p \to p + n_1$, or alternatively, change both $p \to p + n_1$ and $e_o \to e_0 + e_{n_1}$ in \eqref{fontini}, along with the usage of the identity: 
\bg\label{blofessr}
\left(\begin{matrix} p + n_1 \\ e_v\\ \end{matrix}\right) = \sum_{e_o, e_{n_1}}
\left(\begin{matrix} p \\ e_{o}\\ \end{matrix}\right)
\left(\begin{matrix} n_1 \\ e_{n_1}\\ \end{matrix}\right)~\delta(e_o + e_{n_1} - e_v).\nd 
Such an approach will tell how to introduce the effects of the momentum conservation in the $\varphi_3$ sector in the nodal diagrams: simply change the powers of the $k_{u_i}$ and the $l_{v_j}$ series appearing in 
\eqref{fontini} and sum over them. However \eqref{snelmad} is valid if we can impose the following condition:
\bg\label{perfume}
\zeta \equiv \sum_{t = 1}^{s-1} f_{w_t} > \sum_{i = 1}^q k_{u_i} + 
\sum_{j = 1}^r l_{v_j}, \nd
which is possible because the momenta are all lower than ${\rm M}_p$, and we can go to the limit where $f_{w_t}$ momenta are all larger than either 
$k_{u_i}$ or $l_{v_j}$ momenta. (What happens when we go to the regime where \eqref{perfume} is not valid will be discussed later.)
This would have {\it almost} done the job, if not for the subtlety of the appearance of $\overline\gamma(f_{w_s})  \in \{\overline\gamma_1, \overline\gamma_2, ...., \overline\gamma_\infty\}$.  In the limit \eqref{perfume} we can approximate any powers of $\overline\gamma(f_{w_s})$ as:

{\footnotesize
\bg\label{casin18}
\overline\gamma^{n_o}(f_{w_s}) = \sum_{\{n_k, e_{n_k}\}}\prod_{k = 2}^{n_o + 2}{(-1)^{n_k}\over (n_k)!} 
\Bigg(\begin{matrix} n_k \\ e_{n_k}\\ \end{matrix}\Bigg)
\Bigg(\sum_{i = 1}^q k_{u_i}\Bigg)^{e_{n_k}}\Bigg(\sum_{j = 1}^r l_{v_j}\Bigg)^{n_k - e_{n_k}} 
{\partial^{n_k}\overline{\gamma}(-\zeta)\over \partial\zeta^{n_k}}, \nd}
where $(n_k, e_{n_k}) \in \mathbb{Z}_+$. For large values of $n_k$, the above series is suppressed by $(e_{n_k})!(n_k - e_{n_k})!$. Similarly the overall suppression factor for the series ${\overline\gamma^{n_o}(f_{w_s})\over f^{2n_o}_{w_s}}$ becomes:
\bg\label{jun11kabab}
(e_{n_1})!(e_{n_k})!(n_1 - e_{n_1})!(n_k - e_{n_k})!\zeta^{2n_o + n_1}, \nd
which in the limit \eqref{perfume} should result in convergence. However we shouldn't be too worried about the convergence because there are also suppression factors coming from 
${\overline\gamma^{n_{ot}}(f_{w_t})\over f^{2n_{ot}}_{w_t}}$ for 
$t \in [1, s-1]$ and $n_{ot} \in \mathbb{Z}_+$, including the ones from the
denominator of the path-integral \eqref{wilsonrata}. Combining \eqref{snelmad} and \eqref{casin18} then tells us that the momentum conservation in the $\varphi_3$ sector then changes \eqref{lincoln}, or more appropriately \eqref{fontini}, by:

{\footnotesize
\bg\label{tagjap}
p ~ \to ~ p + n_1 + n_k, ~~ e_o ~ \to ~ e_o + e_{n_1} + e_{n_k},
~~{\rm and ~inclusion ~ of}~~ {1\over \zeta^{2n_o + n_1}}
{\partial^{n_k}\overline{\gamma}(-\zeta)\over \partial\zeta^{n_k}}, \nd}
in the nodal diagrams where $k \ge 2$ with no extra factors of ${\rm V}$ from \eqref{nehaprit}. Because of the summation over 
all $(n_1, n_k)$, one should be careful in appropriately inserting the 
derivatives from \eqref{tagjap}. In the following we will study few examples to illustrate this point.

\vskip.2in

\noindent{\bf Case 1: $f_{w_1} = f_{w_2} = f_{w_3} = .... = f_{w_{s-1}} = f_i, f_{w_s} \ne f_i$ for the $\varphi_3$ fields}

\vskip.2in

\noindent Our first example is the one with equal values for all the $f_{w_t}$ momenta for $1 \le t \le s-1$. Clearly now $f_{w_s}$ will depend on what values do $k_{u_i}$ and $l_{v_j}$ momenta take, resulting in nodal diagrams depending on our earlier choices. Let us then take $k_{u_1} = ... = k_{u_q} = k_i$ and $l_{v_1} = ... = l_{v_r} = l_j$ $-$ the latter leading to the nodal diagram \eqref{luna4} $-$ giving us $-f_{w_s} = qk_i + rl_j + (s-1) f_k$ as mentioned earlier. The value of the nodal diagrams for all $f_i$ 
then becomes:

{\footnotesize
\bg\label{luna15}
&& \hskip1.2in{
\frac{1}{{\rm V}} \sum_{i} \quad
\begin{tikzpicture}[thick, baseline={(0, 0cm-\MathAxis pt)},
main/.style = {draw, circle, fill=black, minimum size=12pt},
dot/.style={inner sep=0pt,fill=black, circle, minimum size=4pt},
dots/.style={inner sep=0pt, fill=black, circle, minimum size=2pt}]
  \node[main] (a) at (-3.2, 0) {};
  \node[dot] (1) at (-1.7, 0) {};
  \node[dot] (2) at (0, 0) {};
  \node[main] (3) at (60:1.5) {};
  \node[main] (4) at (30:1.5) {};
  \node[main] (5) at (0:1.5) {};
  \node[main] (6) at (-30:1.5) {};
  \node[main] (7) at(-65:1.5) {};
  \node[dots] (d1) at (-40:1.3) {};
  \node[dots] (d2) at (-50:1.3) {};
  \node[dots] (d3) at (-60:1.3) {};
\draw (a) [dashed] -- node[midway, above right, pos=0.2] {$f_{\omega_s}$} (1);
\draw (1) -- node[midway, above right, pos=0] {$(s-1)\; f_i$} (2);
\draw (2) -- node[midway, left, pos=0.7] {$f_i$} (3);
\draw (2) -- node[midway, left, pos=0.8] {$f_i$} (4);
\draw (2) -- node[midway, above right, sloped, pos=0.7] {$f_i$} (5);
\draw (2) -- node[midway, right, pos=0.7] {$f_i$} (6);
\draw (2) -- node[midway, right, pos=0.7] {$f_i$} (7);
\end{tikzpicture} }
\nonumber\\
&& = \prod_{t = 1}^s \left({\pi_t\over c_t}\right)^{1/2} 
\sum_{e_s \in 2\mathbb{Z}_+} \left(\begin{matrix} s - 1 \\ e_s\\ \end{matrix}\right) {(e_s - 1)!!\over 2^{e_s/2}(s-1)^{n_1 + n_k + 2}} 
\int d^{11} f~{(-\overline\gamma(f))^{s - 1 -e_s} \over f^{2s - e_s + n_1}}
\cdot {\partial^{n_k}\overline\gamma(-(s-1)f)\over \partial f^{n_k}},\nonumber\\ \nd}
where we have to further sum over $(n_1, n_k)$ for $k \ge 2$ as evident from \eqref{snelmad} and \eqref{casin18}. This summation will depend on how the other nodal diagrams with $qk_i$ and $rl_j$ arranged. The dotted line in the nodal diagram indicates that the momentum $f_{w_s}$ is not independent and depends on other momenta as mentioned above. Due to all these subtleties, the value of the nodal diagram in \eqref{luna15} is quite different from what we encountered earlier. We have also taken $c(f) \propto f^2$ $-$ henceforth ignoring factors of ${\rm V}$ from \eqref{nehaprit} $-$ to simplify the integral structure in \eqref{luna15}. 

\vskip.2in

\noindent{\bf Case 2: $f_{w_1} = .... = f_{w_{s-2}} = f_g, f_{w_{s-1}} = f_h, f_{w_s} \ne (f_g, f_h)$ for the $\varphi_3$ fields}

\vskip.2in

\noindent Our next case is when we distribute the independent momenta in two set: one set with $(s-2)$ momenta taking values $f_g$ and the other set with one momentum taking the value $f_h$ with $f_g \ne f_h$. On the other hand, we will continue with the simpler case of $k_{u_i} = k_i$ and 
$l_{v_j} = l_j$. This will make $-f_{w_s} = qk_i + rl_j + f_h + (s-2)f_g$ so we will have to be a bit more careful in inserting this in the path-integral \eqref{fontini}. The nodal diagram for this case then takes the following value:

{\footnotesize
\bg\label{luna16}
&& \hskip1.2in{
\frac{1}{{\rm V^2}} \sum_{g, h} \quad
\begin{tikzpicture}[thick, baseline={(0, 0cm-\MathAxis pt)},
main/.style = {draw, circle, fill=black, minimum size=12pt},
dot/.style={inner sep=0pt,fill=black, circle, minimum size=4pt},
dots/.style={inner sep=0pt, fill=black, circle, minimum size=2pt}]
  \node[main] (a) at (-4, 0) {};
  \node[dot] (1) at (-2, 0) {};
  \node[main] (11) at (-2, 2) {};
  \node[dot] (2) at (0, 0) {};
  \node[main] (3) at (60:1.5) {};
  \node[main] (4) at (30:1.5) {};
  \node[main] (5) at (0:1.5) {};
  \node[main] (6) at (-30:1.5) {};
  \node[main] (7) at(-65:1.5) {};
  \node[dots] (d1) at (-40:1.3) {};
  \node[dots] (d2) at (-50:1.3) {};
  \node[dots] (d3) at (-60:1.3) {};
\draw (a) [dashed] -- node[midway, above right, pos=0.2] {$f_{\omega_s}$} (1);
\draw (1) -- node[midway, above right, pos=0] {$(s-2)\; f_g$} (2);
\draw (1) -- node[midway, right, pos=0.8] {$f_h$} (11);
\draw (2) -- node[midway, left, pos=0.7] {$f_g$} (3);
\draw (2) -- node[midway, left, pos=0.8] {$f_g$} (4);
\draw (2) -- node[midway, above right, sloped, pos=0.7] {$f_g$} (5);
\draw (2) -- node[midway, right, pos=0.7] {$f_g$} (6);
\draw (2) -- node[midway, right, pos=0.7] {$f_g$} (7);
\end{tikzpicture}}
\nonumber\\
&& =  \prod_{t = 1}^s \left({\pi_t\over c_t}\right)^{1/2} 
\sum_{e_s \in 2\mathbb{Z}_+} \left(\begin{matrix} s - 2 \\ e_s\\ \end{matrix}\right) {(s-1)(e_s - 1)!!\over 2^{e_s/2}}\\
&& \times \Bigg[\int_0^\infty d^{11}f~{\left(-\overline\gamma(f)\right)^{s-2-e_s}\over \left(c(f)\right)^{s-2-e_s/2}} \int_f^\infty d^{11}f'~{\overline\gamma(f')\over c(f')((s-2)f + f')^{n_1 + 2}}
\Bigg({1\over s-2}~{\partial \over \partial f} + {\partial \over \partial f'}\Bigg)^{n_k} \overline\gamma\left(-(s-2) f - f'\right) \nonumber\\
&& + \int_0^\infty d^{11} f~{\overline\gamma(f)\over c(f)} \int_{f}^\infty 
d^{11}f'~ {\left(-\overline\gamma(f')\right)^{s-2-e_s}\over \left(c(f')\right)^{s-2-e_s/2}\left((s-2)f'+f\right)^{n_1+2}}
\Bigg({1\over s-2}~{\partial \over \partial f'} + {\partial \over \partial f}\Bigg)^{n_k} \overline\gamma\left(-(s-2) f' - f\right)\Bigg] \nonumber\\
&& =  \prod_{t = 1}^s \left({\pi_t\over c_t}\right)^{1/2} 
\sum_{e_s \in 2\mathbb{Z}_+} \left(\begin{matrix} s - 2 \\ e_s\\ \end{matrix}\right) {(s-1)(e_s - 1)!!\over 2^{e_s/2}}\nonumber\\
&& \times\int_0^\infty d^{11}f~\int_0^\infty d^{11}f'~{\overline\gamma(f')\left(-\overline\gamma(f)\right)^{s-2-e_s}\over c(f')\left(c(f)\right)^{s-2-e_s/2}((s-2)f + f')^{n_1 + 2}}
\Bigg({1\over s-2}~{\partial \over \partial f} + {\partial \over \partial f'}\Bigg)^{n_k} \overline\gamma\left(-(s-2) f - f'\right) \nonumber
\nd}
where appropriate summation over $(n_1, n_k)$ with $k \ge 2$ will have to be performed in the end once other nodal diagrams with momenta $(k_i, l_j)$ are inserted in. As expected, the combined multiple integral structure is more involved than what we had earlier as it involves both $(f, f')$ momenta and their partial derivatives.

One might also wonder what would happen if we change the choices for 
$(k_{u_i}, l_{v_j})$ momenta but keep the above choice for $f_{w_k}$ momenta? Could we, for example, continue to use the result of 
\eqref{luna16}. The answer is yes because with  different choices for 
$(k_{u_i}, l_{v_j})$ momenta, the nodal diagrams would automatically change due to our aforementioned condition \eqref{tagjap}, and the system will continue to allow for an approximate product structure as long as we are within the constraint \eqref{perfume}.

\vskip.2in

\noindent{\bf Case 3: $f_{w_1} = .. = f_{w_{s-3}} = f_g, f_{w_{s-2}} = f_h, f_{w_{s-1}} = f_m, f_{w_s} \ne (f_g, f_h, f_m)$ for the $\varphi_3$ fields}

\vskip.2in
\noindent This case is a little more complicated from the case 4 studied earlier because of the additional constraint from \eqref{tagjap}, so we will divide the nodal diagrams into two classes: one with $f_m = f_h \ne f_g$ and the other with $f_m \ne f_h \ne f_g$. The former is surprisingly straightforward, and the nodal diagram for this case gives us the following value:

{\footnotesize
\bg\label{luna17}
&& \hskip1.2in{
\frac{1}{{\rm V^2}} \sum_{g, h} \quad
\begin{tikzpicture}[thick, baseline={(0, 0cm-\MathAxis pt)},
main/.style = {draw, circle, fill=black, minimum size=12pt},
dot/.style={inner sep=0pt,fill=black, circle, minimum size=4pt},
dots/.style={inner sep=0pt, fill=black, circle, minimum size=2pt}]
  \node[main] (a) at (-4, 0) {};
  \node[dot] (1) at (-2, 0) {};
  \node[dot] (11) at (-2, 1.5) {};
  \node[main] (111) at (-2.9, 2.3) {};
  \node[main] (112) at (-1.1, 2.3) {};
  \node[dot] (2) at (0, 0) {};
  \node[main] (3) at (60:1.5) {};
  \node[main] (4) at (30:1.5) {};
  \node[main] (5) at (0:1.5) {};
  \node[main] (6) at (-30:1.5) {};
  \node[main] (7) at(-65:1.5) {};
  \node[dots] (d1) at (-40:1.3) {};
  \node[dots] (d2) at (-50:1.3) {};
  \node[dots] (d3) at (-60:1.3) {};
\draw (a) [dashed] -- node[midway, above right, pos=0.2] {$f_{\omega_s}$} (1);
\draw (1) -- node[midway, above right, pos=0] {$(s-3)\; f_g$} (2);
\draw (1) -- node[midway, right, pos=0.8] {$2 f_h$} (11);
\draw (2) -- node[midway, left, pos=0.7] {$f_g$} (3);
\draw (2) -- node[midway, left, pos=0.8] {$f_g$} (4);
\draw (2) -- node[midway, above right, sloped, pos=0.7] {$f_g$} (5);
\draw (2) -- node[midway, right, pos=0.7] {$f_g$} (6);
\draw (2) -- node[midway, right, pos=0.7] {$f_g$} (7);
\draw (11) -- node[midway, right, pos=0.5] {$f_h$} (112);
\draw (11) -- node[midway, left, pos=0.5] {$f_h$} (111);
\end{tikzpicture} }
\nonumber\\
&& =  \prod_{t = 1}^s \left({\pi_t\over c_t}\right)^{1/2} 
\sum_{e_s \in 2\mathbb{Z}_+} \left(\begin{matrix} s - 3 \\ e_s\\ \end{matrix}\right) {(s-1)(s-2)(e_s - 1)!!\over 2^{e_s/2 + 1}}\\
&& \times \Bigg[\int_0^\infty d^{11}f~{\left(-\overline\gamma(f)\right)^{s_o}\over \left(c(f)\right)^{s'_o}} \int_f^\infty d^{11}f'
~{c(f') + 2\overline\gamma^2(f')\over 2c^2(f')((s-3)f + f')^{n_1 + 2}}
\Bigg({1\over s-3}~{\partial \over \partial f} + {1\over 2} {\partial \over \partial f'}\Bigg)^{n_k} \overline\gamma\left(-(s-3) f - f'\right) \nonumber\\
&& + \int_0^\infty d^{11} f~\left({c(f) + 2\overline\gamma^2(f)\over 2c^2(f)}\right)\int_f^\infty d^{11}f'~{\left(-\overline\gamma(f')\right)^{s_o} \left(c(f')\right)^{-s'_o}\over \left((s-3)f'+2f\right)^{n_1 + 2}}
\Bigg({1\over s-3}~{\partial \over \partial f'} + {1\over 2} {\partial \over \partial f}\Bigg)^{n_k} \overline\gamma\left(-(s-3) f' - f\right)
\Bigg]\nonumber\\
&& =  \prod_{t = 1}^s \left({\pi_t\over c_t}\right)^{1/2} 
\sum_{e_s \in 2\mathbb{Z}_+} \left(\begin{matrix} s - 3 \\ e_s\\ \end{matrix}\right) {(s-1)(s-2)(e_s - 1)!!\over 2^{e_s/2 + 1}}\nonumber\\
&&\times \int_0^\infty d^{11} f~\int_0^\infty d^{11}f'~{\left(-\overline\gamma(f')\right)^{s_o} \left(c(f')\right)^{-s'_o}\over \left((s-3)f'+2f\right)^{n_1 + 2}}
\left({c(f) + 2\overline\gamma^2(f)\over 2c^2(f)}\right)
\Bigg({1\over s-3}~{\partial \over \partial f'} + {1\over 2} {\partial \over \partial f}\Bigg)^{n_k} \overline\gamma\left(-(s-3) f' - f\right) \nonumber \nd}
where $s_o = s - 3 - e_s, s_o' = s_0 + e_s/2$ and as before we have to sum over $(n_1, n_k)$ once we combine the other nodal diagrams. One should also compare the internal derivative structure in the above nodal diagram with the one from \eqref{luna16}: the difference in the relative factors comes from the momentum distribution. Once we change the momentum distribution to 
$f_h \ne f_g \ne f_m$, the story becomes more involved and the nodal diagram for this case takes the following value:

{\footnotesize
\bg\label{luna18}
&& \hskip1.2in{
\frac{1}{{\rm V^3}} \sum_{g, h, m} \quad
\begin{tikzpicture}[thick, baseline={(0, 0cm-\MathAxis pt)},
main/.style = {draw, circle, fill=black, minimum size=12pt},
dot/.style={inner sep=0pt,fill=black, circle, minimum size=4pt},
dots/.style={inner sep=0pt, fill=black, circle, minimum size=2pt}]
  \node[main] (a) at (-4, 0) {};
  \node[dot] (1) at (-2, 0) {};
  \node[dot] (11) at (-2, 1.5) {};
  \node[main] (111) at (-2.9, 2.3) {};
  \node[main] (112) at (-1.1, 2.3) {};
  \node[dot] (2) at (0, 0) {};
  \node[main] (3) at (60:1.5) {};
  \node[main] (4) at (30:1.5) {};
  \node[main] (5) at (0:1.5) {};
  \node[main] (6) at (-30:1.5) {};
  \node[main] (7) at(-65:1.5) {};
  \node[dots] (d1) at (-40:1.3) {};
  \node[dots] (d2) at (-50:1.3) {};
  \node[dots] (d3) at (-60:1.3) {};
\draw (a) [dashed] -- node[midway, above right, pos=0.2] {$f_{\omega_s}$} (1);
\draw (1) -- node[midway, above right, pos=0] {$(s-3)\; f_g$} (2);
\draw (1) -- node[midway, right, pos=0.8] {$ f_{h} + f_{m}$} (11);
\draw (2) -- node[midway, left, pos=0.7] {$f_g$} (3);
\draw (2) -- node[midway, left, pos=0.8] {$f_g$} (4);
\draw (2) -- node[midway, above right, sloped, pos=0.7] {$f_g$} (5);
\draw (2) -- node[midway, right, pos=0.7] {$f_g$} (6);
\draw (2) -- node[midway, right, pos=0.7] {$f_g$} (7);
\draw (11) -- node[midway, right, pos=0.5] {$f_m$} (112);
\draw (11) -- node[midway, left, pos=0.5] {$f_h$} (111);
\end{tikzpicture} }
\nonumber\\
&& = \prod_{t = 1}^s \left({\pi_t\over c_t}\right)^{1/2} 
\sum_{e_s \in 2\mathbb{Z}_+} \left(\begin{matrix} s - 3 \\ e_s\\ \end{matrix}\right) {(s-1)(s-2)(e_s - 1)!!\over 2^{e_s/2 - 1}}\\
&& \times \Bigg[\int_0^\infty d^{11}f~{\left(-\overline\gamma(f)\right)^{s_o}\over \left(c(f)\right)^{s'_o}} \int_f^\infty d^{11}f'~{\overline\gamma(f')\over c(f')} \int_{f'}^\infty d^{11} f''~{\overline\gamma(f'')\over 
c(f'')\Big((s-3)f + f' + f''\Big)^{n_1+2}}
\left({1\over s-3}~{\partial\over \partial f} + {\partial\over \partial f'} + {\partial\over \partial f''}\right)^{n_k}\nonumber\\
&& \times ~\overline\gamma(-(s-3)f - f' - f'') ~+~ 
\int_0^\infty d^{11}f~{\overline\gamma(f)\over c(f)}\int_f^\infty d^{11}f'~{\overline\gamma(f')\over c(f')}
\int_{f'}^\infty d^{11}f''~{\left(-\overline\gamma(f'')\right)^{s_o}\over \left(c(f'')\right)^{s'_o}\Big((s-3)f'' + f' + f\Big)^{n_1+2}}\nonumber\\
&&\times~\left({1\over s-3}~{\partial\over \partial f''} + {\partial\over \partial f'} + {\partial\over \partial f}\right)^{n_k}
\overline\gamma(-(s-3)f'' - f' - f)
+ \int_0^\infty d^{11}f~{\overline\gamma(f)\over c(f)} \int_f^\infty d^{11}f'~{\left(-\overline\gamma(f')\right)^{s_o}\over \left(c(f')\right)^{s'_o}} \nonumber\\
&& \times ~ \int_{f'}^\infty d^{11} f''~{\overline\gamma(f'')\over 
c(f'')\Big((s-3)f' + f + f''\Big)^{n_1+2}}
\left({1\over s-3}~{\partial\over \partial f'} + {\partial\over \partial f} + {\partial\over \partial f''}\right)^{n_k} \overline\gamma(-(s-3)f' - f - f'')\Bigg]\nonumber\\
&& = \prod_{t = 1}^s \left({\pi_t\over c_t}\right)^{1/2} 
\sum_{e_s \in 2\mathbb{Z}_+} \left(\begin{matrix} s - 3 \\ e_s\\ \end{matrix}\right) {(s-1)(s-2)(e_s - 1)!!\over 2^{e_s/2 - 1}}
\int_0^\infty d^{11}f~\int_0^\infty d^{11}f'~\int_{0}^\infty d^{11} f''\nonumber\\
&& \times ~{\overline\gamma(f')\overline\gamma(f'')\left(-\overline\gamma(f)\right)^{s_o}\over c(f') c(f'')\left(c(f)\right)^{s'_o}
\Big((s-3)f + f' + f''\Big)^{n_1+2}}
\left({1\over s-3}~{\partial\over \partial f} + {\partial\over \partial f'} + {\partial\over \partial f''}\right)^{n_k} \overline\gamma(-(s-3)f - f' - f'')
\nonumber \nd}
with the expected triple integral structure. $(s_o, s'_o)$ are defined as above and again, we will have to sum over $(n_1, n_k)$ at the end. The story could now be extended to more complicated nodal diagrams of the form:

\bg\label{luna19} 
\frac{1}{{\rm V^{c_0}}} \sum_{g, h, ...} \quad
\begin{tikzpicture}[thick, baseline={(0, 0cm-\MathAxis pt)},
main/.style = {draw, circle, fill=black, minimum size=12pt},
dot/.style={inner sep=0pt,fill=black, circle, minimum size=4pt},
dots/.style={inner sep=0pt, fill=black, circle, minimum size=2pt}]
  \node[main] (a) at (-7, 0) {};
  \node[dot] (1) at (-3, 0) {};
  \node[dot] (11) at (-3, 2) {};
  \node[dot] (1D) at (-3, -1.7) {};
  \node[main] (1D0) at (-3, -3.5) {};
  \node[main] (1D1) at (- 2.4, -3.2) {};
  \node[main] (1D2) at (-1.8, -2.8) {};
  \node[main] (1D3) at (-3.6, -3.2) {};
  \node[main] (1D4) at (-4.2, -2.8) {};
  \node[dots] (1Dd1) at (-3.7, -2.5) {};
  \node[dots] (1Dd2) at (-3.5, -2.6) {};
  \node[dot] (11R) at (-1, 2) {};
  \node[dot] (11L) at (-6, 2) {};
  \node[dots] (11L2) at (-4.5, 2) {};
  \node[dots] (11L1) at (-3.5, 2) {};
  \node[dots] (11Ld1) at (-3.7, 2) {};
  \node[dots] (11Ld1) at (-3.7, 2) {};
  \node[dots] (11Ld1) at (-4, 2) {};
  \node[dots] (11Ld1) at (-4.3, 2) {};
  \node[dot] (11RU) at (-1, 3.5) {};
  \node[dot] (11U) at (-3, 3) {};
  \node[dot] (11LU) at (-6, 3.5) {};   
  \node[main] (11RU1) at (-2, 4.5) {};
  \node[main] (11RU2) at (0, 4.5) {};
  \node[main] (11U1) at (-4, 3.5) {};
  \node[main] (11U2) at (-3, 3.7) {};
  \node[main] (11U3) at (-2, 3.5 ) {};
  \node[main] (11LU1) at (-6, 5) {};
  \node[main] (11LU2) at (-7.5 , 4) {};
  \node[main] (11LU3) at (-4.5, 4) {};
  \node[main] (11LU4) at (-5, 4.5) {};
  \node[main] (11LU5) at (-7, 4.5) {};
  \node[dots] (11LUd2) at (-6.8, 4.2) {};
  \node[dots] (11LUd3) at (-7, 4) {};
  \node[dot] (2) at (0, 0) {};
  \node[main] (3) at (60:1.5) {};
  \node[main] (4) at (30:1.5) {};
  \node[main] (5) at (0:1.5) {};
  \node[main] (6) at (-30:1.5) {};
  \node[main] (7) at(-65:1.5) {};
  \node[dots] (d1) at (-40:1.3) {};
  \node[dots] (d2) at (-50:1.3) {};
  \node[dots] (d3) at (-60:1.3) {};
\draw (a) [dashed] -- node[midway, above right, pos=0.2] {$f_{\omega_s}$} (1);
\draw (1) -- node[midway, above right, pos=0] {$(s-c)\; f_g$} (2);
\draw (1) -- node[midway, right, pos=0.8] {$ 2f_{h} + .. + rf_l$} (11);
\draw (2) -- node[midway, left, pos=0.7] {${}^{f_g}$} (3);
\draw (2) -- node[midway, left, pos=0.8] {${}^{f_g}$} (4);
\draw (2) -- node[midway, above right, sloped, pos=0.7] {${}^{f_g}$} (5);
\draw (2) -- node[midway, right, pos=0.7] {${}^{f_g}$} (6);
\draw (2) -- node[midway, right, pos=0.7] {${}^{f_g}$} (7);
\draw (11) -- node {} (11R);
\draw (11) -- node {} (11R);
\draw (11) -- node {} (11L1);
\draw (11L2) -- node {} (11L);
\draw (11R) -- node [midway, right, pos=0.8] {$2 f_h$} (11RU);
\draw (11) -- node [midway, right, pos=0.5] {$3 f_m$} (11U);
\draw (11L) -- node [midway, left, pos=0.8] {$r f_l$} (11LU);
\draw (11RU) -- node [midway, right, pos=0.8] {${}_{f_h}$} (11RU1);
\draw (11RU) -- node [midway, right, pos=0.7] {${}_{f_h}$} (11RU2);
\draw (11U) -- node {} (11U1);
\draw (11U) -- node [midway, left, pos=0.8] {${}_{f_m}$} (11U2);
\draw (11U) -- node [midway, left, pos=0.8] {${}_{f_m}$} (11U3);
\draw (11LU) -- node [midway, right, pos=0.8] {${}_{f_l}$} (11LU1);
\draw (11LU) -- node [midway, left, pos=0.5] {${}_{f_l}$} (11LU2);
\draw (11LU) -- node [midway, right, pos=0.8] {} (11LU3);
\draw (11LU) -- node {} (11LU4);
\draw (11LU) -- node [midway, right, pos=0.8] {${}_{f_l}$} (11LU5);
\draw (1) -- node [midway, right, pos=0.5] {$q f_n$} (1D);
\draw (1D) -- node {} (1D0);
\draw (1D) -- node  [midway, right, pos=0.7] {${}_{f_n}$}  (1D1);
\draw (1D) -- node [midway, right, pos=0.7] {$ {}_{f_n}$} (1D2);
\draw (1D) -- node {} (1D3);
\draw (1D) -- node [midway, left, pos=0.5] {${}_{f_n}$}   (1D4);
\end{tikzpicture}
 \nd
where $c = 1 + 2 + 3 + ... + r + q$, and $c_o$ is the order of the nested diagram which, in turn, is related to the number of distinct {\it branches}
in the nodal diagram. The analysis of above diagram is a little more non-trivial from what we had earlier but is nevertheless straightforward. We will however not try it here and, as before, leave it as an exercise for the diligent readers.

Before going to the study of $\varphi_1$ fields, let us clarify one issue related to the lower limits of the outermost integrals in the nodal diagrams \eqref{luna15} till \eqref{luna18}. While the lower limits of the nested integrals in the aforementioned diagrams are clear, the lower limits
of the outermost integrals $-$ which are zeroes $-$ would seem to clash with the condition \eqref{perfume} and \eqref{casin18}. We are specifically looking for the energy scale where:
\bg\label{capitaineQ}
\zeta \equiv \sum_{t = 1}^{s-1} f_{w_t} ~ \le ~ (q + r) \mu, \nd
where $\mu << {\rm M}_p$ (or $\mu << 1$ because we took ${\rm M}_p \equiv 1$) is the typical energy scale for the Wilsonian effective action \eqref{lincoln}. One saving grace here is the summation structure of $\zeta$: we can allow almost all the $(s-2)$ momenta to take any values as long as {\it one} momentum is bounded from below by 
$(q+r)\mu$. Of course one could fathom an easy fix to the problem by imposing an IR cutoff $\kappa_{\rm IR}$ to all the integrals, and replace:
\bg\label{katuholm}
f_{w_t} \rightarrow \kappa_{\rm IR} + f_{w_t}, ~~~
k_{u_i} \rightarrow \kappa_{\rm IR} + k_{u_i}, ~~~
l_{v_j} \rightarrow \kappa_{\rm IR} + l_{v_j}, \nd
in all the nodal diagrams, or simply change the lower limits of the outermost integrals of the nodal diagrams. The replacement \eqref{katuholm} is perfectly consistent with the UV/IR mixing that we discussed earlier
and clearly helps us for the regime \eqref{capitaineQ}. Additionally, imposing an IR cutoff in the Minkowski spacetime works well here as elaborated in footnote \ref{vmyerss}. Not only that, the IR cut-off is in fact {\it necessary} for the system to make sense and therefore
 we will continue using the condition \eqref{tagjap} in all the nodal diagrams for the $\varphi_3$ field.

\subsubsection{Contributions from the $\varphi_1$ fields \label{sec3.2.5}}

With all the above discussion for the $(\varphi_2, \varphi_3)$ fields, we are now ready to tackle the $\varphi_1$ field. This field is related to the graviton components, which means there are 44 massless degrees of freedom. Thus in principle we should take 44 scalar fields (with suppressed tensor indices) of the form $\varphi_{1i}$ with $i = 1, ...., 44$. For simplicity however, and as we discussed earlier, we will take multiple copies of {\it one} component of the scalar field with different momenta. In the following we study the path integral analysis for a few cases with different momenta.

\vskip.2in

\noindent{\bf Case 1: $k_{u_1} = k_{u_2} = k_{u_3} = .. = k_{u_q}$ and $k_i$ for the $\varphi_1$ fields}

\vskip.2in

\noindent  As before, we will start with the simplest case where all the $k_{u_i}$ momenta take the same value, namely $k_i$ in \eqref{fontini}. There are a few subtleties now. First, the presence of the extra field $\varphi_1$ in the outer leg of the nodal diagram (see for example case 1) means that there are two possible cases now: one, with the field momenta aligned with the momenta of the $q$ fields, and two, with the field momenta not aligned with the $q$ fields. Secondly, the powers of both the $k_{u_i}$ momenta as well as the $l_{v_j}$ momenta have to be changed as in \eqref{tagjap} to account for the momentum conservation. With the aligned field momenta, the nodal diagram takes the following value:

{\footnotesize
\bg\label{luna20}
&& \hskip1.2in{ \frac{1}{\rm V} \sum_i \;
\begin{tikzpicture}[baseline={(0, 0cm-\MathAxis pt)}, thick,
ball/.style={ball  color=white, circle,  minimum size=25pt},
main/.style = {draw, circle, fill=black, minimum size=12pt},
dot/.style={inner sep=0pt,fill=black, circle, minimum size=3pt},
dots/.style={inner sep=0pt, fill=black, circle, minimum size=2pt}]
  \node[ball] (1) at (-1.5, 0) {i};
  \node[dot] (2) at (0, 0) {};
  \node[main] (3) at (60:1.5) {};
  \node[main] (4) at (30:1.5) {};
  \node[main] (5) at (0:1.5) {};
  \node[main] (6) at (-30:1.5) {};
  \node[main] (7) at (-65:1.5) {};
  \node[dot] (b1) at (-35:1.5) {};
  \node[dots] (d1) at (-40:1.3) {};
  \node[dots] (d2) at (-50:1.3) {};
  \node[dots] (d3) at (-60:1.3) {};
\draw (1) -- node[midway, above right, pos=0.2] {$q k_i$} (2);
\draw (2) -- node[midway, left, pos=0.7] {$k_i$} (3);
\draw (2) -- node[midway, left, pos=0.8] {$k_i$} (4);
\draw (2) -- node[midway, above right, sloped, pos=0.7] {$k_i$} (5);
\draw (2) -- node[midway, right, pos=0.7] {$k_i$} (6);
\draw (2) -- node[midway, right, pos=0.7] {$k_i$} (7);
\end{tikzpicture} }
\nonumber\\
&& = \prod_{l = 1}^q \left({\pi_l\over a_l}\right)^{1/2} 
\sum_{e_q \in 2\mathbb{Z}_+} \left(\begin{matrix} q + 1 \\ e_q\\ \end{matrix}\right) {(e_q - 1)!! q^{n_q}\over 2^{e_q/2}}
\int d^{11}k~{k^{n_q}\left(-\overline\alpha(k)\right)^{q+1-e_q}\over 
\left(a(k)\right)^{q + 1 - e_q/2}}~\psi_{\bf k}({\bf x}, y, z) e^{-ik_0 t}, \nd}
where we can take $a(k) \propto k^2$ (ignoring ${\rm V}$ from \eqref{nehaprit} henceforth unless mentioned otherwise) and 
$n_q \equiv n + e_o + e_{n_{q'}}$ with $q' \ge 1$ from \eqref{tagjap}. There is also a sum over $(e_o, e_{n_{q'}})$ which we shall insert at the very end. The above result is the first non-trivial deviation from the 
tree-level result \eqref{luna1}, and points towards the series corrections in powers of $\overline\alpha(k)$ as shown in section 6.1 of \cite{coherbeta2}. On the other hand, for the non-aligned field momenta, the nodal diagram gives us:

{\footnotesize
\bg\label{luna21}
&&  \hskip1.2in{ \frac{1}{{\rm V}^2} \sum_{i,j} \; 
\begin{tikzpicture}[baseline={(0, 0cm-\MathAxis pt)}, thick, 
ball/.style={ball  color=white, circle,  minimum size=25pt},
main/.style = {draw, circle, fill=black, minimum size=12pt},
dot/.style={inner sep=0pt,fill=black, circle, minimum size=3pt},
dots/.style={inner sep=0pt, fill=black, circle, minimum size=2pt}]
  \node[ball] (1) at (-1.5, 0) {i};
  \node[dot] (2) at (0, 0) {};
  \node[main] (3) at (60:1.5) {};
  \node[main] (4) at (30:1.5) {};
  \node[main] (5) at (0:1.5) {};
  \node[main] (6) at (-30:1.5) {};
  \node[main] (7) at (-65:1.5) {};
  \node[dot] (b1) at (-35:1.5) {};
  \node[dots] (d1) at (-40:1.3) {};
  \node[dots] (d2) at (-50:1.3) {};
  \node[dots] (d3) at (-60:1.3) {};
\draw (1) -- node[midway, above right, pos=0.2] {$q k_j$} (2);
\draw (2) -- node[midway, left, pos=0.7] {$k_j$} (3);
\draw (2) -- node[midway, left, pos=0.8] {$k_j$} (4);
\draw (2) -- node[midway, above right, sloped, pos=0.7] {$k_j$} (5);
\draw (2) -- node[midway, right, pos=0.7] {$k_j$} (6);
\draw (2) -- node[midway, right, pos=0.7] {$k_j$} (7);
\end{tikzpicture} }
\nonumber\\
&& = -\prod_{l = 1}^q \left({\pi_l\over a_l}\right)^{1/2} 
\sum_{e_q \in 2\mathbb{Z}_+} \left(\begin{matrix} q  \\ e_q\\ \end{matrix}\right) {(e_q - 1)!! q^{n_q}\over 2^{e_q/2}}\\
&&\times\Bigg[\int d^{11}k~{\overline\alpha(k)\over k^2}~\psi_{\bf k}({\rm X})e^{-ik_0t} \int_k^\infty d^{11}k'~{\left(-\overline\alpha(k')\right)^{q_1}\over k^{'q_2}} 
+ \int d^{11}k~{\left(-\overline\alpha(k)\right)^{q_1}\over k^{q_2}}
\int_k^\infty d^{11}k'~{\overline\alpha(k')\over k'^2}~\psi_{\bf k'}({\rm X})e^{-ik'_0t}\Bigg], \nonumber \nd}
where $q_1 \equiv q - e_q$, $q_2 \equiv 2q - e_q - n_q$ with $n_q$ as defined above\footnote{There are extra multiplicative factors
of ${\rm V}$ from \eqref{nehaprit} once we compare $a(k)$ with $k^2$ but, as mentioned earlier, these are not important for the illustrative examples that we present here. They do become important when we compare the relative suppression factor of one diagram over another. As such we will insert them later when we combine all the amplitudes.}, and ${\rm X}\equiv ({\bf x}, y, z)$. The nested integral structure of the second term is interesting: it involves an integral over the wave-function $\psi_{\bf k'}({\rm X})e^{-ik'_0t}$ leading to possible temporal dependence. Such temporal dependence may be traded with the 
${g_s\over {\rm HH}_o}$ dependence where $g_s$ is the string coupling in the dual IIA side for a de Sitter space (thus is {\it not} the $g_s$ used for the Minkowski background). This will have some important consequences, and we will elaborate further on it below when we go to more involved nodal diagrams.

\vskip.2in

\noindent{\bf Case 2: $k_{u_1} = k_{u_2} = k_{u_3} = .... = k_{u_{q-2}} = k_i$, $(k_{u_{q-1}}, k_{u_q}) \ne k_i$ for the $\varphi_1$ fields}

\vskip.2in

\noindent The case with two momenta unequal to the rest of the $(q-2)$ momenta is a bit more subtle, again because of the momentum carried by the external leg. There are a few possibilities now: {\Su one}, when the external momenta match with the internal $(q-2)$ momenta but differ from the other two internal momenta (which are kept equal); {\Su two}, when the external momenta match with the internal two (equal) momenta but differ from the internal $(q-2)$ momenta; {\Su three}, when the external momenta match with one of the two internal momenta, but differ from both $(q-2)$ momenta as well as from the other internal momenta; and {\Su four}, when the external momenta do not match with either of the internal $(q-2)$ momenta or the other two internal momenta (which in turn are kept unequal). The first one is relatively easier, and the nodal diagram gives:

{\footnotesize
\bg \label{luna22}
 && \hskip1.2in{ 
 \frac{1}{{\rm V}^2} \sum_{i,j} \; 
\begin{tikzpicture}[baseline={(0, 0cm-\MathAxis pt)}, thick, 
ball/.style={ball  color=white, circle,  minimum size=25pt},
main/.style = {draw, circle, fill=black, minimum size=12pt},
dot/.style={inner sep=0pt,fill=black, circle, minimum size=3pt},
dots/.style={inner sep=0pt, fill=black, circle, minimum size=2pt}]
  \node[dot] (-1) at (-4.5, 0) {};
  \node[main] (-11) at (-5.5, 1.2) {};
  \node[main] (-12) at (-5.5, -1.2) {};
  \node[ball] (1) at (-2.5, 0) {i};
  \node[dot] (2) at (0, 0) {};
  \node[main] (3) at (60:1.5) {};
  \node[main] (4) at (30:1.5) {};
  \node[main] (5) at (0:1.5) {};
  \node[main] (6) at (-30:1.5) {};
  \node[main] (7) at (-65:1.5) {};
  \node[dot] (b1) at (-35:1.5) {};
  \node[dots] (d1) at (-40:1.3) {};
  \node[dots] (d2) at (-50:1.3) {};
  \node[dots] (d3) at (-60:1.3) {};
\draw (1) -- node[midway, above right, pos=0.8] {$ 2 k_j$} (-1);
\draw (-1) -- node[midway, above right, pos=0.8] {$  k_j$} (-11);
\draw (-1) -- node[midway, above left, pos=0.8] {$  k_j$} (-12);
\draw (1) -- node[midway, above right, pos=0] {$(q - 2)  k_i$} (2);
\draw (2) -- node[midway, left, pos=0.7] {$k_i$} (3);
\draw (2) -- node[midway, left, pos=0.8] {$k_i$} (4);
\draw (2) -- node[midway, above right, sloped, pos=0.7] {$k_i$} (5);
\draw (2) -- node[midway, right, pos=0.7] {$k_i$} (6);
\draw (2) -- node[midway, right, pos=0.7] {$k_i$} (7);
\end{tikzpicture} }
\nonumber\\
&& = \prod_{l = 1}^q \left({\pi_l\over a_l}\right)^{1/2} 
\sum_{e_q \in 2\mathbb{Z}_+} \left(\begin{matrix} q - 1 \\ e_q\\ \end{matrix}\right) {(e_q - 1)!! (q-1)q^{n_q+1}\over 2^{2 + e_q/2}}\\
&& \times \Bigg[\int d^{11}k~{(-\overline\alpha(k))^{q - e_q - 1} \over 
\left(a(k)\right)^{q - 1 - e_q/2}}~\psi_{\bf k}({\bf x}, y, z)e^{-ik_0t} \int_k^\infty 
d^{11}k'~{\Big((q-2)k + 2k'\Big)^{n_q} \over a(k')}\left(1 + {2\overline\alpha^2(k')\over a(k')}\right)\nonumber\\
&& + \int d^{11}k~{1\over a(k)}\left(1 + {2\overline\alpha^2(k)\over a(k)}\right) \int_k^\infty d^{11}k'~{(-\overline\alpha(k'))^{q - e_q - 1} \over \left(a(k')\right)^{q - 1 - e_q/2}}\Big((q-2)k' + 2k\Big)^{n_q}\psi_{\bf k'}({\bf x}, y, z)e^{-ik'_0t}\Bigg],\nonumber \nd}
where $n_q$ is as defined after \eqref{luna20}. Note that now we do {\it not} combine the two nested integrals into a double integral over two equal domains. This is because we want to keep the momentum dependence of the outermost integral explicit so that we can equate this to the Fourier transform of the type of background that we want (see \cite{coherbeta} and \cite{coherbeta2} for details). Interestingly, we see that for $e_q = q - 1$, in the second nested integral structure, there are pieces which are independent of $\overline\alpha(k')$. Similar $(\overline\alpha(k), \overline\alpha(k'))$ independent pieces appear from the second line of \eqref{luna22}. These may be collected together and are expressed as (ignoring factors of ${\rm V}$ from \eqref{nehaprit}): 
\bg\label{luna222}
&& {\cal Z}_1 \equiv \int {d^{11}k\over k^2} \int_k^\mu d^{11}k'~{\Big((q-2)k' + 2k\Big)^{n_q}\over k'^{q-1}}~\psi_{\bf k'}({\bf x}, y, z)e^{-ik'_0 t} \nonumber\\
&& {\cal Z}_2 \equiv  \int d^{11}k~{\psi_{\bf k}({\bf x}, y, z)e^{-ik_0 t}\over k^{q-1}} 
\int_k^\mu d^{11}k'~{\Big((q-2)k + 2k'\Big)^{n_q} \over k'^2}, \nd
where we have explicitly restricted the upper limits of the integrals to $\mu \propto {\rm M}_p$. The first term is interesting because the nested integral will produce a term 
proportional to powers of ${g_s\over {\rm HH}_o}$ once we integrate over 
$dk'_0$ as anticipated earlier and also in \cite{coherbeta, coherbeta2}.
We can make this more concrete by performing the $k'_0$ integral explicitly. The result may be presented succinctly as:
\bg\label{luna223}
{\cal Z}_1 & = & \sum_{c_q, d_q} 2^{n_q - c_q} \Lambda^{d_q + 1/2} (q-2)^{c_q}\left(\begin{matrix} n_q \\ c_q\\ \end{matrix}\right)
\left(\begin{matrix} {c_q - q + 1 - 2d_q\over 2} \\ d_q\\ \end{matrix}\right)\\
&\times & \int d^{11} k~k^{n_q - c_q - 2} \int_{\bf k}^{\vec\mu} d^{10} 
{\bf k}'~(-{\bf k}'^2)^{c_q - q + 1 - 2d_q/2}~ \psi_{\bf k'}({\bf x}, y, z)\nonumber\\
&\times & \left[{\rm x}^n e^{\cal A}\left(\sum_{p = 1}^\infty 
{(a{\rm x})^{2p-2}(-1)^{p+1}\over (n+p) p!} + i
\sum_{p = 1}^\infty {(a{\rm x})^{2p-1}(-1)^{p}\over (n+ 2p) (2p)!}\right) + i\Gamma(n+1)\right]_{k_0/\sqrt{\Lambda}}^{\mu_0/\sqrt{\Lambda}},\nonumber \nd
where $x = (t, {\bf x}), ~\mu = (\mu_0, \vec\mu), ~{\rm x} = {k'_0\over \sqrt{\Lambda}}, ~a = \sqrt{\Lambda} t$ and $n = 2d_q \in 2\mathbb{Z}_+$ with $\Lambda$ being the cosmological constant. We have also defined 
${\cal A} \equiv 2i\pi(n + 1)\lfloor{-{\rm arg}(ia) - {\rm arg}(x) + \pi\over 2\pi}\rfloor$ where $\lfloor z \rfloor$ is the floor function. The above series makes sense when both $(k_0, \mu_0)$ are small, which is of course guaranteed by our Wilsonian effective action. Note also that the series is in powers of $a{\rm x}$, {\it i.e.} in powers of $k_0t$ or $\mu_0 t$. Replacing $t$ by ${g_s\over {\rm HH}_o}$, we get a series in powers of 
${g_s\over {\rm HH}_o}$ independent of $\overline\alpha(k)$ as anticipated earlier.

There are also terms independent of both ${g_s\over {\rm HH}_o}$ and  $\overline\alpha(k)$, and here it is exemplified by the $\Gamma(1 + 2d_q)$
factor (we will not worry too much about the $i$ factor right now, because in the end we will have to extract the real part out of the products of the series of certain nodal diagrams). Another example is the ${\cal Z}_2$ function wherein the nested integral doesn't depend on any temporal or spatial factors\footnote{The integral over $d^{11}k$ is not necessary if we want to follow the identification between $\overline\alpha(k)$ and $\alpha(k)$ as shown in section 6.1 of \cite{coherbeta2}. We will elaborate on this soon.}. This leads to:
\bg\label{luna224}
{\cal Z}_2 & = & \sum_{f_q, g_q} {(-1)^{f_q/2 - g_q}2^{f_q} (q-2)^{n_q - f_q} \over (1 + 2g_q)(8 + f_q - 2g_q)}\left(\begin{matrix} n_q \\ f_q\\ \end{matrix}\right)
\left(\begin{matrix} {f_q - 2\over 2} \\ g_q\\ \end{matrix}\right)\\
&\times & \int d^{11}k~\left(\mu^{1+2g_q}_0 - k^{1+2g_q}_0\right)
\left(({\vec{\mu}}^2)^{4-g_q+f_q/2} - ({\bf k}^2)^{4-g_q + f_q/2}\right)
k^{1-q + n_q -f_q}\psi_{\bf k}({\rm X})e^{-ik_0t}, \nonumber \nd
where ${\rm X} = ({\bf x}, y, z)$, $k = (k_0, {\bf k})$, $(f_q, g_q) \in \mathbb{Z}_+$ and we have ignored factors of $\pi$ associated with the angular integral. Note the dependence on the energy scale $\mu \equiv (\mu_0, \vec\mu)$. The Wilsonian effective action keeps this finite and small, so the integral does not blow up. Thus there is no UV problem. If we keep $q > 1$, one might worry about the IR problem because $f_q \le n_q$ and $1+g_q \le {f_q\over 2}$. This isn't a problem either as elaborated earlier. We will discuss more on the IR issues later.

Our second sub-case is when the external momenta match with the internal two (equal) momenta but differ from the internal $(q-2)$ momenta. This is slightly more non-trivial compared to the earlier one simply because of the placement of the external momenta. Nevertheless this may be exactly evaluated and the nodal diagram gives us:

{\footnotesize
\bg\label{luna23}
&&   \hskip1.2in{ 
\frac{1}{{\rm V}^2} \sum_{i,j} \; 
\begin{tikzpicture}[baseline={(0, 0cm-\MathAxis pt)}, thick, 
ball/.style={ball  color=white, circle,  minimum size=25pt},
main/.style = {draw, circle, fill=black, minimum size=12pt},
dot/.style={inner sep=0pt,fill=black, circle, minimum size=3pt},
dots/.style={inner sep=0pt, fill=black, circle, minimum size=2pt}]
  \node[dot] (-1) at (-4.5, 0) {};
  \node[main] (-11) at (-5.5, 1.2) {};
  \node[main] (-12) at (-5.5, -1.2) {};
  \node[ball] (1) at (-2.5, 0) {i};
  \node[dot] (2) at (0, 0) {};
  \node[main] (3) at (60:1.5) {};
  \node[main] (4) at (30:1.5) {};
  \node[main] (5) at (0:1.5) {};
  \node[main] (6) at (-30:1.5) {};
  \node[main] (7) at (-65:1.5) {};
  \node[dot] (b1) at (-35:1.5) {};
  \node[dots] (d1) at (-40:1.3) {};
  \node[dots] (d2) at (-50:1.3) {};
  \node[dots] (d3) at (-60:1.3) {};
\draw (1) -- node[midway, above right, pos=0.8] {$ 2 k_i$} (-1);
\draw (-1) -- node[midway, above right, pos=0.8] {$  k_i$} (-11);
\draw (-1) -- node[midway, above left, pos=0.8] {$  k_i$} (-12);
\draw (1) -- node[midway, above right, pos=0] {$(q - 2)  k_j$} (2);
\draw (2) -- node[midway, left, pos=0.7] {$k_j$} (3);
\draw (2) -- node[midway, left, pos=0.8] {$k_j$} (4);
\draw (2) -- node[midway, above right, sloped, pos=0.7] {$k_j$} (5);
\draw (2) -- node[midway, right, pos=0.7] {$k_j$} (6);
\draw (2) -- node[midway, right, pos=0.7] {$k_j$} (7);
\end{tikzpicture} }
\nonumber\\
&& = \prod_{l = 1}^q \left({\pi_l\over a_l}\right)^{1/2} 
\sum_{e_q \in 2\mathbb{Z}_+} \left(\begin{matrix} q - 2 \\ e_q\\ \end{matrix}\right) {q(q-1)(e_q - 1)!! \over 2^{2 + e_q/2}}\\
&& \times \Bigg[\int d^{11} k ~{\left(-\overline\alpha(k)\right)^{q-2-e_q}\over \left(a(k)\right)^{q-2-e_q/2}} \int_k^\infty d^{11}k'~{\left((q-2)k + 2k'\right)^{n_q}\over a^3(k')} ~\overline\alpha(k')\left(3a(k') -
2\overline\alpha^2(k')\right) ~\psi_{\bf k'}({\bf x}, y, z) e^{-ik'_0 t}
\nonumber\\
&& + \int d^{11}k~{\overline\alpha(k)\left(3a(k) -
2\overline\alpha^2(k)\right) \over a^3(k)}~\psi_{\bf k}({\bf x}, y, z) e^{-ik_0 t} \int_k^\infty d^{11}k'~{\left((q-2)k' + 2k\right)^{n_q}\over
\left(a(k')\right)^{q-2-e_q/2}}\left(-\overline\alpha(k')\right)^{q-2-e_q}
\Bigg], \nonumber
\nd}
where $n_q = n + e_o + e_{n_p}$ with $p \ge 1$, and we see that the first nested integral does not have a $\overline\alpha$ independent factor so we cannot integrate that piece without prior knowledge of $\overline\alpha$. On the other hand, the second nested integral does have a $\overline\alpha$ independent piece for $e_q = q - 2$. This takes the form:
\bg\label{fajerjee}
{\cal Z}_3 \equiv \sum_{h_q \in \mathbb{Z}_+} 2^{n_q - h_q} (q-2)^{h_q} 
\left(\begin{matrix} n_q \\ h_q\\ \end{matrix}\right)\int_k^\mu 
d^{11}k'~k'^{2 + h_q - q}, \nd
which then contributes to the $k$ dependent pieces that are modulated by the wave-function $\psi_{\bf k}({\bf x}, y, z)e^{-ik_0 t}$.
As before, by taking the upper limits of the integrals to $\mu$ instead of $\infty$ shows that there are no UV issues anywhere.

Our third sub-case is when the external momenta match with one of the two internal momenta, but differ from both $(q-2)$ momenta as well as from the other internal momenta. We can take the three set of momenta as $k_i, k_j$ and $(q-2) k_l$ where $k_i \ne k_j \ne k_l$. The nodal diagram for this case gives us the following value:

{\footnotesize
\bg\label{luna24}
&& \hskip1.2in{ 
 \frac{1}{{\rm V}^3} \sum_{i,j, l} \; 
\begin{tikzpicture}[baseline={(0, 0cm-\MathAxis pt)}, thick, 
ball/.style={ball  color=white, circle,  minimum size=25pt},
main/.style = {draw, circle, fill=black, minimum size=12pt},
dot/.style={inner sep=0pt,fill=black, circle, minimum size=3pt},
dots/.style={inner sep=0pt, fill=black, circle, minimum size=2pt}]
  \node[dot] (-1) at (-4.5, 0) {};
  \node[main] (-11) at (-5.5, 1.2) {};
  \node[main] (-12) at (-5.5, -1.2) {};
  \node[ball] (1) at (-2.5, 0) {i};
  \node[dot] (2) at (0, 0) {};
  \node[main] (3) at (60:1.5) {};
  \node[main] (4) at (30:1.5) {};
  \node[main] (5) at (0:1.5) {};
  \node[main] (6) at (-30:1.5) {};
  \node[main] (7) at (-65:1.5) {};
  \node[dot] (b1) at (-35:1.5) {};
  \node[dots] (d1) at (-40:1.3) {};
  \node[dots] (d2) at (-50:1.3) {};
  \node[dots] (d3) at (-60:1.3) {};
\draw (1) -- node[midway, above right, pos=0.9] {$ k_i + k_j$} (-1);
\draw (-1) -- node[midway, above right, pos=0.8] {$  k_j$} (-11);
\draw (-1) -- node[midway, above left, pos=0.8] {$  k_i$} (-12);
\draw (1) -- node[midway, above right, pos=0] {$(q - 2)  k_l$} (2);
\draw (2) -- node[midway, left, pos=0.7] {$k_l$} (3);
\draw (2) -- node[midway, left, pos=0.8] {$k_l$} (4);
\draw (2) -- node[midway, above right, sloped, pos=0.7] {$k_l$} (5);
\draw (2) -- node[midway, right, pos=0.7] {$k_l$} (6);
\draw (2) -- node[midway, right, pos=0.7] {$k_l$} (7);
\end{tikzpicture} }
\nonumber\\
&& \prod_{l = 1}^q \left({\pi_l\over a_l}\right)^{1/2} 
\sum_{e_q \in 2\mathbb{Z}_+} \left(\begin{matrix} q - 2 \\ e_q\\ \end{matrix}\right) {q(q-1)(e_q - 1)!! \over 2^{e_q/2}}\\
&& \times \Bigg[ \int d^{11}k~{\left(-\overline\alpha(k)\right)^{q_a}\over \left(a(k)\right)^{q_b}}\int_k^\infty d^{11}k'~{\overline\alpha(k')\over a(k')} \int_{k'}^\infty d^{11}k''
f(k, k', k'') \psi_{\bf k''}({\rm X}) e^{-ik''_0t} \left({1\over 2a(k'')} + {\overline\alpha^2(k'')\over a^2(k'')}\right)\nonumber\\
&& + \int d^{11}k~{\overline\alpha(k)\over a(k)}\int_k^\infty d^{11}k'~{\left(-\overline\alpha(k')\right)^{q_a}\over \left(a(k')\right)^{q_b}} \int_{k'}^\infty d^{11}k''
f(k', k, k'') \psi_{\bf k''}({\rm X}) e^{-ik''_0t} \left({1\over 2a(k'')} + {\overline\alpha^2(k'')\over a^2(k'')}\right)\nonumber\\
&& + \int d^{11}k~{\left(-\overline\alpha(k)\right)^{q_a}\over \left(a(k)\right)^{q_b}}\int_k^\infty d^{11}k'~\left({1\over 2a(k')} + {\overline\alpha^2(k')\over a^2(k')}\right)\psi_{\bf k'}({\rm X}) e^{-ik'_0t}\int_{k'}^\infty d^{11}k''f(k, k', k'')~{\overline\alpha(k'')\over a(k'')}\nonumber\\
&& + \int d^{11}k~{\overline\alpha(k)\over a(k)}\int_k^\infty d^{11}k'~
\left({1\over 2a(k')} + {\overline\alpha^2(k')\over a^2(k')}\right)
\psi_{\bf k'}({\rm X}) e^{-ik'_0t}\int_{k'}^\infty d^{11}k''
f(k', k, k'')
{\left(-\overline\alpha(k'')\right)^{q_a}\over \left(a(k'')\right)^{q_b}} \nonumber\\
&& + \int d^{11}k~\left({1\over 2a(k)} + {\overline\alpha^2(k)\over a^2(k)}\right)
\psi_{\bf k}({\rm X}) e^{-ik_0t} \int_{k}^\infty d^{11}k'~
{\left(-\overline\alpha(k')\right)^{q_a}\over \left(a(k')\right)^{q_b}}
\int_{k'}^\infty d^{11}k''
~f(k', k, k'')~{\overline\alpha(k'')\over a(k'')}\nonumber\\
&& + \int d^{11}k~\left({1\over 2a(k)} + {\overline\alpha^2(k)\over a^2(k)}\right)
\psi_{\bf k}({\rm X}) e^{-ik_0t} \int_{k}^\infty d^{11}k'~{\overline\alpha(k')\over a(k')}\int_{k'}^\infty d^{11}k''
~f(k'', k, k')~{\left(-\overline\alpha(k'')\right)^{q_a}\over \left(a(k'')\right)^{q_b}}\Bigg], \nonumber \nd}
where we see that a much more complicated nested structure appears. We have also defined the variables appearing above as: $q_a \equiv q - 2 - e_q, q_b \equiv q - 2 - {e_q\over 2}$ and $f(k_1, k_2, k_3)$ as an ordered polynomial defined as
$f(k_1, k_2, k_3) \equiv \left((q-2)k_1 + k_2 + k_3\right)^{n_q}$. The other quantities $(n_q, {\rm X})$ are defined as above, namely $n_q = n + e_o + e_p$ with $p \ge 1$ and ${\rm X} = ({\bf x}, y, z)$.

The final sub-case is when the external momenta do not match with either of the internal $(q-2)$ momenta or the other two internal momenta (which in turn are kept unequal). This means all the set of momenta $-$ here we take them as $(k_i, k_j, k_m, k_l)$ $-$ are unequal, {\it i.e.} $k_i \ne k_j \ne k_l \ne k_m$. Such unequal distribution of momenta now leads to a slightly more involved nested integral structure, as seen from the following nodal diagram:

{\footnotesize
\bg\label{luna25}
&& \hskip1.2in{ 
 \frac{1}{{\rm V}^4} \sum_{i,j, l, m} \; 
\begin{tikzpicture}[baseline={(0, 0cm-\MathAxis pt)}, thick, 
ball/.style={ball  color=white, circle,  minimum size=25pt},
main/.style = {draw, circle, fill=black, minimum size=12pt},
dot/.style={inner sep=0pt,fill=black, circle, minimum size=3pt},
dots/.style={inner sep=0pt, fill=black, circle, minimum size=2pt}]
  \node[dot] (-1) at (-4.5, 0) {};
  \node[main] (-11) at (-5.5, 1.2) {};
  \node[main] (-12) at (-5.5, -1.2) {};
  \node[ball] (1) at (-2.5, 0) {i};
  \node[dot] (2) at (0, 0) {};
  \node[main] (3) at (60:1.5) {};
  \node[main] (4) at (30:1.5) {};
  \node[main] (5) at (0:1.5) {};
  \node[main] (6) at (-30:1.5) {};
  \node[main] (7) at (-65:1.5) {};
  \node[dot] (b1) at (-35:1.5) {};
  \node[dots] (d1) at (-40:1.3) {};
  \node[dots] (d2) at (-50:1.3) {};
  \node[dots] (d3) at (-60:1.3) {};
\draw (1) -- node[midway, above right, pos=0.9] {$ k_m + k_j$} (-1);
\draw (-1) -- node[midway, above right, pos=0.8] {$  k_j$} (-11);
\draw (-1) -- node[midway, above left, pos=0.8] {$  k_m$} (-12);
\draw (1) -- node[midway, above right, pos=0] {$(q - 2)  k_l$} (2);
\draw (2) -- node[midway, left, pos=0.7] {$k_l$} (3);
\draw (2) -- node[midway, left, pos=0.8] {$k_l$} (4);
\draw (2) -- node[midway, above right, sloped, pos=0.7] {$k_l$} (5);
\draw (2) -- node[midway, right, pos=0.7] {$k_l$} (6);
\draw (2) -- node[midway, right, pos=0.7] {$k_l$} (7);
\end{tikzpicture}} 
\nonumber\\
&& \prod_{l = 1}^q \left({\pi_l\over a_l}\right)^{1/2} 
\sum_{e_q \in 2\mathbb{Z}_+} \left(\begin{matrix} q - 2 \\ e_q\\ \end{matrix}\right) {q(q-1)(e_q - 1)!! \over 2^{e_q/2 - 1}}\\
&& \times \Bigg[ \int d^{11}k~{\left(-\overline\alpha(k)\right)^{q_a}\over \left(a(k)\right)^{q_b}}\int_{k}^\infty d^{11}k'~{\overline\alpha(k')\over a(k')}\int_{k'}^\infty d^{11}k''~f(k, k', k'')~{\overline\alpha(k'')\over a(k'')}\int_{k''}^\infty d^{11}k'''~\psi_{\bf k'''}({\rm X})e^{-ik'''_0t}{\overline\alpha(k''')\over a(k''')}\nonumber\\
&& + \int d^{11}k~{\left(-\overline\alpha(k)\right)^{q_a}\over \left(a(k)\right)^{q_b}}\int_{k}^\infty d^{11}k'~\psi_{\bf k'}({\rm X})e^{-ik'_0t}~{\overline\alpha(k')\over a(k')}\int_{k'}^\infty d^{11}k''~{\overline\alpha(k'')\over a(k'')}\int_{k''}^\infty d^{11}k'''~f(k, k'', k''')~{\overline\alpha(k''')\over a(k''')}\nonumber\\
&& + \int d^{11}k~{\overline\alpha(k)\over a(k)}\int_k^\infty d^{11}k'~{\overline\alpha(k')\over a(k')}\int_{k'}^\infty d^{11}k''~\psi_{\bf k''}({\rm X})e^{-ik''_0t}~{\overline\alpha(k'')\over a(k'')}\int_{k''}^\infty d^{11}k'''~f(k''', k, k')~{\left(-\overline\alpha(k''')\right)^{q_a}\over \left(a(k''')\right)^{q_b}} \nonumber\\
&& + \int d^{11}k~\psi_{\bf k}({\rm X})e^{-ik_0t}~{\overline\alpha(k)\over a(k)}\int_k^\infty d^{11}k'
~{\overline\alpha(k')\over a(k')}
\int_{k'}^\infty d^{11}k''~{\overline\alpha(k'')\over a(k'')}
\int_{k''}^\infty d^{11}k'''~f(k''', k', k'')~{\left(-\overline\alpha(k''')\right)^{q_a}\over \left(a(k''')\right)^{q_b}}\nonumber\\
&& + \int d^{11}k~{\overline\alpha(k)\over a(k)}\int_{k}^\infty d^{11}k'~{\left(-\overline\alpha(k')\right)^{q_a}\over \left(a(k')\right)^{q_b}}\int_{k'}^\infty d^{11}k''~\psi_{\bf k''}({\rm X})e^{-ik''_0t}~{\overline\alpha(k'')\over a(k'')}\int_{k''}^\infty d^{11}k'''~f(k', k, k''')~{\overline\alpha(k''')\over a(k''')}\nonumber\\
&& + \int d^{11}k~\psi_{\bf k}({\rm X})e^{-ik_0t}~{\overline\alpha(k)\over a(k)}\int_{k}^\infty d^{11}k'~{\left(-\overline\alpha(k')\right)^{q_a}\over \left(a(k')\right)^{q_b}}\int_{k'}^\infty d^{11}k''~{\overline\alpha(k'')\over a(k'')}\int_{k''}^\infty d^{11}k'''~f(k', k'', k''')~{\overline\alpha(k''')\over a(k''')}\nonumber\\
&& + \int d^{11}k~{\overline\alpha(k)\over a(k)}\int_{k}^\infty d^{11}k'~{\left(-\overline\alpha(k')\right)^{q_a}\over \left(a(k')\right)^{q_b}}\int_{k'}^\infty d^{11}k''~\psi_{\bf k'''}({\rm X})e^{-ik'''_0t}~{\overline\alpha(k'')\over a(k'')}\int_{k''}^\infty d^{11}k'''~f(k', k, k''')~{\overline\alpha(k''')\over a(k''')}\nonumber\\
&& + \int d^{11}k~{\overline\alpha(k)\over a(k)}\int_{k}^\infty d^{11}k'~{\overline\alpha(k')\over a(k'')}\int_{k'}^\infty d^{11}k''~f(k'', k, k')~{\left(-\overline\alpha(k'')\right)^{q_a}\over \left(a(k'')\right)^{q_b}}\int_{k''}^\infty d^{11}k'''~\psi_{\bf k'''}({\rm X})e^{-ik'''_0t}~{\overline\alpha(k''')\over a(k''')}\nonumber\\
&& + \int d^{11}k~{\overline\alpha(k)\over a(k)}\int_{k}^\infty d^{11}k'~\psi_{\bf k'}({\rm X})e^{-ik'_0t}~{\overline\alpha(k')\over a(k'')}\int_{k'}^\infty d^{11}k''~{\left(-\overline\alpha(k'')\right)^{q_a}\over \left(a(k'')\right)^{q_b}}\int_{k''}^\infty d^{11}k'''~f(k'', k''', k)~{\overline\alpha(k''')\over a(k''')}\nonumber\\
&& + \int d^{11}k~\psi_{\bf k}({\rm X})e^{-ik_0t}~{\overline\alpha(k)\over a(k)}\int_{k}^\infty d^{11}k'~{\overline\alpha(k')\over a(k'')}\int_{k'}^\infty d^{11}k''~{\left(-\overline\alpha(k'')\right)^{q_a}\over \left(a(k'')\right)^{q_b}}\int_{k''}^\infty d^{11}k'''~f(k'', k''', k')~{\overline\alpha(k''')\over a(k''')}\nonumber\\
&& + \int d^{11}k~{\left(-\overline\alpha(k)\right)^{q_a}\over \left(a(k)\right)^{q_b}}\int_{k}^\infty d^{11}k'~{\overline\alpha(k')\over a(k')}\int_{k'}^\infty d^{11}k''~\psi_{\bf k''}({\rm X})e^{-ik''_0t}~{\overline\alpha(k'')\over a(k'')}\int_{k''}^\infty d^{11}k'''~f(k, k', k''')~{\overline\alpha(k''')\over a(k''')}\nonumber\\
&& + \int d^{11}k~{\overline\alpha(k)\over a(k)}\int_k^\infty d^{11}k'
~\psi_{\bf k'}({\rm X})e^{-ik'_0t}~{\overline\alpha(k')\over a(k')}
\int_{k'}^\infty d^{11}k''~{\overline\alpha(k'')\over a(k'')}
\int_{k''}^\infty d^{11}k'''~f(k''', k, k'')~{\left(-\overline\alpha(k''')\right)^{q_a}\over \left(a(k''')\right)^{q_b}}\Bigg], \nonumber
\nd}
where $(q_a, q_b)$ are as defined above and $f(k_1, k_2, k_3)$ is an ordered polynomial defined as $f(k_1, k_2, k_3) \equiv \left((q -2)k_1 + k_2 + k_3\right)^{n_q}$ with $(n_q, {\rm X})$ as above. Notice that in writing the values of the nodal diagrams from \eqref{luna20} till \eqref{luna25}, the {\it ordering} of the external momentum line is not important. This is precisely what one might have expected and our expressions demonstrate this explicitly. Interestingly, there are also 
$4!$ terms contributing to the amplitude (half of them are identical to the other half). This factorial growth, even to the first order in the coupling constant $c_{nmpqrs}$, may be verified for all other nodal diagrams for all the three fields. As mentioned earlier, there appears no immediate reason to combine all the $4!$ integrals into one multiple integral over equal domains, as we have to eventually equate the amplitudes of the diagrams to the Fourier transform of the background that we want. There is however a different reason to combine the integrals under one roof. The story is subtle and will be elaborated soon.

Note that, compared to \eqref{luna22}, there aren't any terms independent of $\overline\alpha$ that only depends on the wave-function 
$\psi_{\bf k'}({\rm X})e^{-ik'_0 t}$ where $(k'_0, {\bf k'})$ are the set of momentum components inside the nested integral structures. In fact generically, as long as the external momenta do not match with any of the set of internal momenta, there won't be any $\overline\alpha$ independent factors. The minimal set of requirements for $\overline\alpha$ independent pieces are:
\vskip.2in

\noindent $\bullet$ The external momentum line should match with any one of the set of internal momentum lines in the nodal diagram.

\vskip.2in

\noindent $\bullet$ The splitting of $qk_l$ momenta into $c_jk_j$ with 
$1 \le j < q$, such that $c_j \in \mathbb{Z}_+$ with $c_j \ge 2$. 

\vskip.2in

\noindent $\bullet$ If for some $j = r$, $c_r = 1$, then the external momentum line should match with the internal momentum $k_r$.

\vskip.2in

\noindent If the aforementioned conditions are satisfied, then the $\overline{\alpha}$ independent pieces would integrate out to provide the 
${g_s\over {\rm HH}_o}$ terms that we mentioned earlier and also in 
\cite{coherbeta, coherbeta2}. For a fixed value of $(n, m,..., r, s)$ in 
\eqref{fontini} the number of possible nodal diagrams are controlled by 
$q, r$ and $s-1$ as well as the coupling constant $c_{nmpqrs}$ (which here is solely dependent on ${\rm M}_p$). However before we venture towards the applications of all the nodal diagrams drawn above, let us clarify few subtleties related to the diagrams themselves.

\subsubsection{Interlude 2: Feynman diagrams or nodal diagrams? \label{sec3.2.f}}

\noindent All the above analysis of the nodal diagrams were done using the action \eqref{fontini} $-$ instead of \eqref{lincoln} $-$ which in turn relied on the usage of {\it real} fields (namely only the real parts of the Fourier transforms of the real fields). Ignoring the complex parts of the Fourier components might raise some questions because the latter is a crucial ingredient in the implementations of the Feynman diagrams to any quantum processes. 

It is easy to see how and why the complex parts of the Fourier fields\footnote{We will use this nomenclature to refer to the Fourier components of the real fields.} are necessary in the standard formulation of QFT. Consider, for simplicity, a two point function in the vanilla $\lambda \varphi^4$ theory. Inserting this in the path integral would involve an integration over six field components (at various values of the {\it discrete} momenta). Now individually, if we take the momenta to be different, the integral vanishes because of the Gaussian functions.
Non-zero result appears when two field components (one from the vertex and one from the external leg) have the same momenta but are complex conjugates of each other, namely:
\bg\label{nnice}
{\bf Re}~\widetilde{\varphi}(k_1) + i {\bf Im}~\widetilde{\varphi}(k_1), 
~~~~~{\bf Re}~\widetilde{\varphi}(k_1) 
- i {\bf Im}~\widetilde{\varphi}(k_1), \nd
for a given discrete momentum $k_1$. In the language of the Feynman diagrams, we have a {\it line} joining the two fields. On the other hand, for the analysis of the nodal diagrams, we saw that the one-point functions of the various fields components are non-zero. This means there is no longer any {\it a priori} need to consider the complex components of the Fourier fields. Their presence (or absence thereof) are not going to significantly change the basic outcome of the path-integral (except of course increasing the {\it number} of terms considerably). For example squaring a complex Fourier field, would result into:
\bg\label{bivanna}
\left({\bf Re}~\widetilde{\varphi}(k_1)\right)^2 - 
\left({\bf Im}~\widetilde{\varphi}(k_1)\right)^2 + 2i \left({\bf Re}~\widetilde{\varphi}(k_1)\right) \left({\bf Im}~\widetilde{\varphi}(k_1)\right), \nd
which vanishes in the standard Feynman diagram analysis but {\it does not} vanish in the nodal diagram analysis. In fact the nodal diagram analysis would typically lead to both complex  and real pieces, which fits rather well with the fact that this quantity when integrated with the wave-function 
$\psi_{\bf k}({\bf x}, y, z) e^{-ik_{0}t}$ reproduces a real function\footnote{Here for example a quantity related to the space-time metric. The complex part easily appears from the complex part of \eqref{bivanna}. The real part, on the other hand, appears not only from the real part of \eqref{bivanna} but also from the one with a relative plus sign (the latter is of course the standard contribution from the corresponding Feynman diagrams).}.

The aforementioned simple considerations imply that for an analysis with the Glauber-Sudarshan states, the nodal diagrams have better chances of extracting the physics of the underlying phenomena $-$ here being the space-time metric, the background G-fluxes, or the fermionic condensates. Taking the full complex Fourier fields from \eqref{lincoln} would add new terms $-$ both real and imaginary pieces $-$ to the already complicated {\it amplitudes} from the three Fourier field components $(\varphi_1, \varphi_2, \varphi_3)$, but the physics will not change. Because of this we will not pursue this exercise any further here and leave it for the diligent reader. 

There is however one subtle issue that requires considerations here, and is related to the {\it growth} of the number of terms as we go to the higher orders in the coupling constants. We already saw that, to the first order
in the coupling constant $c_{nmpqrs}$, the nodal diagrams for all the three Fourier fields $(\widetilde\varphi_1, \widetilde\varphi_2, \widetilde\varphi_3)$ do show factorial growths for all values of 
$(n, m, ..., s) \in \mathbb{Z}_+$. When we go to higher orders in the coupling constants $c_{nmpqrs}$ we expect the growth to increase in similar ways, which would imply that the series cannot be convergent and can only be asymptotic. This asymptotic nature of the series means that we have to apply Borel resummation technique. How this works explicitly will be elaborated after we analyze the numerator and the denominator of the path integral \eqref{fontini} from the nodal diagrams.

\subsubsection{Combining the nodal diagrams for the $(\varphi_1, \varphi_2, \varphi_3)$ fields \label{3.2.6}}

With all the above analysis we are now ready to combine the nodal diagrams to compute the expectation value $\langle\varphi_1\rangle_{\overline\sigma}$ from the path integral \eqref{fontini} or \eqref{lincoln}. In the following we will analyze both the numerator and the denominator of the path integral, which for our case
takes the following form:

{\footnotesize
\bg\label{2summers}
\langle\varphi_1\rangle_{\overline\sigma}  = \frac{\textrm{Num}}{\textrm{Den}} = 
\frac{ \begin{tikzpicture}[baseline={(0, 0cm-\MathAxis pt)},every node/.style={ball  color=white, circle,  minimum size=10pt}]
    \draw (-0.5, 0) -- (0, 0) node{i}  -- (0.5, 0);
    \filldraw [black] (-0.5,0) circle (1pt);
    \filldraw [black] (0.5,0) circle (1pt);
\end{tikzpicture} \;
+ \sum\limits_{n,...,s} c_{nmpqrs} \mathcal{N}^{(1)}_{nmp}(k; q) \otimes \mathcal{N}^{(2)}_{nmp}(\mu; r) \otimes \mathcal{N}^{(3)}_{nmp} (\mu; s) +
{\cal O}(c^2_{nmpqrs})}{ \begin{tikzpicture} [baseline={(0, 0cm-\MathAxis pt)}]
\draw (-0.5,0) -- (0.5,0);
\filldraw [black] (-0.5,0) circle (1pt);
\filldraw [black] (0.5,0) circle (1pt);
\end{tikzpicture} \;
+ \sum\limits_{n,...,s} c_{nmpqrs} \mathcal{N}^{(1)'}_{nmp} (k; q) \otimes \mathcal{N}^{(2)}_{nmp}(\mu; r) \otimes \mathcal{N}^{(3)}_{nmp} (\mu; s) + {\cal O}(c^2_{nmpqrs})}, \nonumber\\ \nd }
where we have shown the series in powers of the coupling constants 
$c_{nmpqrs}$ for all values of $(n, ..., s)$. The first term of the numerator is the tree-level from \eqref{luna1}. In a similar vein, the first term of the denominator is from \eqref{luna2}. For vanishing coupling constants, {\it i.e} for $c_{nmpqrs} = 0$, the result is clearly the integral part of 
\eqref{luna3}, as anticipated earlier. The quantum contributions are represented by the diagrammatic series $({\cal N}^{(1)}_{nmp}(k; q), {\cal N}^{(1)'}_{nmp}(k; q))$ with momenta $k$ and $({\cal N}^{(2)}_{nmp}(\mu; r), {\cal N}^{(3)}_{nmp}(\mu; s))$ with the momenta integrated to the scale $\mu \propto {\rm M}_p$. Note that 
${\cal N}^{(2)}_{nmp}(\mu; r)$, and ${\cal N}^{(3)}_{nmp}(\mu; s)$ appear on both numerator and the denominator, but only ${\cal N}^{(1)}_{nmp}(k; q)$ appears on the numerator and 
${\cal N}^{(1)'}_{nmp}(k; q)$ appears on the denominator. The series 
${\cal N}^{(1)}_{nmp}(k; q)$, ${\cal N}^{(2)}_{nmp}(\mu; r)$, and ${\cal N}^{(3)}_{nmp}(\mu; s)$ are related to the three fields 
$(\varphi_1, \varphi_2, \varphi_3)$ respectively. Similarly ${\cal N}^{(1)'}_{nmp}(k; q)$ will also be related to $\varphi_1$ field, as will be explained soon. 
Note also that we have imposed a product structure to both 
${\cal N}^{(1)}_{nmp}(k; q) \otimes {\cal N}^{(2)}_{nmp}(\mu; r)\otimes{\cal N}^{(3)}_{nmp}(\mu; s)$ and 
${\cal N}^{(1)'}_{nmp}(k; q)\otimes{\cal N}^{(2)}_{nmp}(\mu; r)\otimes {\cal N}^{(3)}_{nmp}(\mu; s)$. This may be justified in the following way. Let us first fix the momenta of the $r$ copies of $\widetilde\varphi_2$ and $s-1$ copies of $\widetilde\varphi_3$ fields. This will automatically fix $\zeta$ from \eqref{perfume}. Once we choose a specific distribution of the momenta for the $\widetilde\varphi_1$ fields (as in say \eqref{luna20}), the powers of $(\{k_{u_i}\}, \{l_{v_j}\})$ from \eqref{tagjap} will also be fixed. One may now change the momentum distributions, from say \eqref{luna20} till \eqref{luna25}, keeping the 
momenta of the $r$ copies of $\widetilde\varphi_2$ and $s-1$ copies of  $\widetilde\varphi_3$ fixed.
We can go even further by noticing that 
as long as the momenta of the $s - 1$ copies of $\widetilde\varphi_3$ fields are kept fixed, but varying the momenta of both the $q$ copies of $\widetilde\varphi_1$ and the $r$ copies of $\widetilde\varphi_2$ fields, $\zeta$ from \eqref{perfume} will remain unchanged. This means we can individually sum the nodal diagrams of the three fields, as long as we keep track of the momentum distribution and 
$\zeta$ from \eqref{perfume} and \eqref{tagjap}. This simplified picture
 is valid in the limit proposed in \eqref{perfume}, and could in-principle change once the choices of momenta deviate from \eqref{perfume}. In fact a generic construction reveals a more subtle story. To see this, let us consider two representative nodal diagrams from the fields $\varphi_1$ and $\varphi_2$. For concreteness, let us take \eqref{luna20} and \eqref{luna5} with equal choices of momenta respectively. For given values of the two nodal diagrams, we can allow the whole series of the nodal diagrams from $\varphi_3$ field in the following way:
  
  {\footnotesize
 \bg\label{luna26}
 &\left(
 \begin{tikzpicture}[baseline={(0, 0cm-\MathAxis pt)}, thick,
main/.style = {draw, circle, fill=black, minimum size=6pt},
dot/.style={inner sep=0pt,fill=black, circle, minimum size=3pt},
ball/.style={ball  color=white, circle,  minimum size=15pt},
dots/.style={inner sep=0pt,fill=black, circle, minimum size=1pt}] 
  \node[dot] (1) at (0, 0) {};
  \node[dot] (11) at (-1, 0) {};
  \node[ball] (111) at (-2, 0) {i};
  \node[main] (3) at (40:1) {};
  \node[main] (4) at (15:1) {};
  \node[main] (5) at (-15:1) {};
  \node[main] (6) at (-40:1) {};
  \node[dots] (0) at (-25: 0.7) {};
  \node[dots] (00) at (-35: 0.7) {};
\draw (1) -- node[midway, above right, pos=1] {$q k_i$} (11);
\draw (11) -- node[midway, above right, pos=0.8] {} (111);
\draw (1) -- node {}  (3);
\draw (1) -- node {}  (4);
\draw (1) -- node {}  (5);
\draw (1) -- node {}  (6);
\end{tikzpicture} 
\; ; \; (q k_i)^n 
 \right) \; \otimes
 \left(
 \begin{tikzpicture}[baseline={(0, 0cm-\MathAxis pt)}, thick,
main/.style = {draw, circle, fill=black, minimum size=6pt},
dot/.style={inner sep=0pt,fill=black, circle, minimum size=3pt},
dots/.style={inner sep=0pt,fill=black, circle, minimum size=1pt}] 
  \node[dot] (1) at (0, 0) {};
  \node[dot] (11) at (-1, 0) {};
  \node[main] (3) at (40:1) {};
  \node[main] (4) at (15:1) {};
  \node[main] (5) at (-15:1) {};
  \node[main] (6) at (-40:1) {};
  \node[dots] (0) at (-25: 0.7) {};
  \node[dots] (00) at (-35: 0.7) {};
\draw (1) -- node[midway, above right, pos=1] {$r l_j$} (11);
\draw (1) -- node {}  (3);
\draw (1) -- node {}  (4);
\draw (1) -- node {}  (5);
\draw (1) -- node {}  (6);
\end{tikzpicture} \; ; \; (r l_j)^m
\right)
\\
& \otimes \; \left(
\begin{tikzpicture}[baseline={(0, 0cm-\MathAxis pt)}, thick,
main/.style = {draw, circle, fill=black, minimum size=6pt},
dot/.style={inner sep=0pt,fill=black, circle, minimum size=3pt},
dots/.style={inner sep=0pt,fill=black, circle, minimum size=1pt}] 
  \node[dot] (1) at (0, 0) {};
  \node[dot] (11) at (-1.5, 0) {};
  \node[main] (111) at (-2.5, 0) {};
  \node[main] (3) at (40:1) {};
  \node[main] (4) at (15:1) {};
  \node[main] (5) at (-15:1) {};
  \node[main] (6) at (-40:1) {};
  \node[dots] (0) at (-25: 0.7) {};
  \node[dots] (00) at (-35: 0.7) {};
  \node[dots] (01) at (-2, 0) {};
  \node[dots] (02) at (-2, -0.3) {};
\draw (1) -- node[midway, above right, pos=1] {$(s-1) f_l$} (11);
\draw (11) [dashed] -- node[below=2mm] {\tiny $\frac{\bar{\gamma}^{n_0} (-q k_i - r l_j - (s-1) f_l)}{(q k_i + r l_j + (s-1) f_l)^{2 n_0}}$} (111);
\draw (1) -- node {}  (3);
\draw (1) -- node {}  (4);
\draw (1) -- node {}  (5);
\draw (1) -- node {}  (6);
\draw (02)[->] -- node {} (01);
\end{tikzpicture} \; ; \; (q k_i + r l_j)^p 
\; + \; 
\begin{tikzpicture}[baseline={(0, 0cm-\MathAxis pt)}, thick,
main/.style = {draw, circle, fill=black, minimum size=6pt},
dot/.style={inner sep=0pt,fill=black, circle, minimum size=3pt},
dots/.style={inner sep=0pt,fill=black, circle, minimum size=1pt}] 
  \node[dot] (1) at (0, 0) {};
  \node[dot] (11) at (-1.5, 0) {};
  \node[main] (111) at (-2.5, 0) {};
  \node[main] (112) at (-1.5, 1) {};
  \node[main] (3) at (40:1) {};
  \node[main] (4) at (15:1) {};
  \node[main] (5) at (-15:1) {};
  \node[main] (6) at (-40:1) {};
  \node[dots] (0) at (-25: 0.7) {};
  \node[dots] (00) at (-35: 0.7) {};
  \node[dots] (01) at (-2, 0) {};
  \node[dots] (02) at (-2, -0.3) {};
\draw (1) -- node[midway, above right, pos=1] {$(s-2) f_h$} (11);
\draw (11) [dashed] -- node[below=2mm] {\tiny $\frac{\bar{\gamma}^{n_0} (-q k_i - r l_j - f_g - (s-2) f_h)}{(q k_i + r l_j + f_g + (s-2) f_h)^{2 n_0}}$} (111);
\draw (1) -- node {}  (3);
\draw (1) -- node {}  (4);
\draw (1) -- node {}  (5);
\draw (1) -- node {}  (6);
\draw (11) -- node[midway, left, pos=0.8] {$f_g$} (112);
\draw (02)[->] -- node {} (01);
\end{tikzpicture} \; ; \; (q k_i + r l_j)^p \; + ... 
\right) \nonumber
  \nd }
 where we have explicitly shown the choices of momenta for each of the nodal diagrams (which are  appropriately integrated in the explicit amplitudes corresponding to the respective diagrams\footnote{For example in the $\varphi_2$ and $\varphi_3$ sectors they are integrated till $\mu < {\rm M}_p$.}), including the functional form for $\overline\gamma$ and the propagator in the $\varphi_3$ sector. Of course both the propagator and the powers of the momenta in the $\varphi_3$ sector may be traded off using \eqref{lincoln} and \eqref{snelmad}, in the limit \eqref{perfume}, to get the following diagrammatic representation:
 
 {\footnotesize 
 \bg\label{luna27}
 &\sum\limits_{n_1, l_{n_1}, l_0}
 \begin{pmatrix}
n_1\\
 l_{n_1}
\end{pmatrix}
 \begin{pmatrix}
p\\
 l_0
\end{pmatrix}
\left(
 \begin{tikzpicture}[baseline={(0, 0cm-\MathAxis pt)}, thick,
main/.style = {draw, circle, fill=black, minimum size=6pt},
dot/.style={inner sep=0pt,fill=black, circle, minimum size=3pt},
ball/.style={ball  color=white, circle,  minimum size=15pt},
dots/.style={inner sep=0pt,fill=black, circle, minimum size=1pt}] 
  \node[dot] (1) at (0, 0) {};
  \node[dot] (11) at (-1, 0) {};
  \node[ball] (111) at (-2, 0) {i};
  \node[main] (3) at (40:1) {};
  \node[main] (4) at (15:1) {};
  \node[main] (5) at (-15:1) {};
  \node[main] (6) at (-40:1) {};
  \node[dots] (0) at (-25: 0.7) {};
  \node[dots] (00) at (-35: 0.7) {};
\draw (1) -- node[midway, above right, pos=1] {$q k_i$} (11);
\draw (11) -- node[midway, above right, pos=0.8] {} (111);
\draw (1) -- node {}  (3);
\draw (1) -- node {}  (4);
\draw (1) -- node {}  (5);
\draw (1) -- node {}  (6);
\end{tikzpicture} 
\; ;  (q k_i)^{n + l_0 + l_{n_1}} 
 \right)  \otimes
 \left(
 \begin{tikzpicture}[baseline={(0, 0cm-\MathAxis pt)}, thick,
main/.style = {draw, circle, fill=black, minimum size=6pt},
dot/.style={inner sep=0pt,fill=black, circle, minimum size=3pt},
dots/.style={inner sep=0pt,fill=black, circle, minimum size=1pt}] 
  \node[dot] (1) at (0, 0) {};
  \node[dot] (11) at (-0.6, 0) {};
  \node[main] (3) at (40:1) {};
  \node[main] (4) at (15:1) {};
  \node[main] (5) at (-15:1) {};
  \node[main] (6) at (-40:1) {};
  \node[dots] (0) at (-25: 0.7) {};
  \node[dots] (00) at (-35: 0.7) {};
\draw (1) -- node[midway, above right, pos=1] {$r l_j$} (11);
\draw (1) -- node {}  (3);
\draw (1) -- node {}  (4);
\draw (1) -- node {}  (5);
\draw (1) -- node {}  (6);
\end{tikzpicture} \; ;  (r l_j)^{m + p + n_1 - l_0 - l_{n_1}}
\right)
\nonumber\\
& \otimes \; \left(
\begin{tikzpicture}[baseline={(0, 0cm-\MathAxis pt)}, thick,
main/.style = {draw, circle, fill=black, minimum size=6pt},
dot/.style={inner sep=0pt,fill=black, circle, minimum size=3pt},
dots/.style={inner sep=0pt,fill=black, circle, minimum size=1pt}] 
  \node[dot] (1) at (0, 0) {};
  \node[dot] (11) at (-1.5, 0) {};
  \node[main] (111) at (-2.5, 0) {};
  \node[main] (3) at (40:1) {};
  \node[main] (4) at (15:1) {};
  \node[main] (5) at (-15:1) {};
  \node[main] (6) at (-40:1) {};
  \node[dots] (0) at (-25: 0.7) {};
  \node[dots] (00) at (-35: 0.7) {};
  \node[dots] (01) at (-2, 0) {};
  \node[dots] (02) at (-2, -0.3) {};
\draw (1) -- node[midway, above right, pos=1] {$(s-1) f_l$} (11);
\draw (11) [dashed] -- node[below=2mm] {\tiny $\bar{\gamma}^{n_0} (-q k_i - r l_j - (s-1) f_l)$} (111);
\draw (1) -- node {}  (3);
\draw (1) -- node {}  (4);
\draw (1) -- node {}  (5);
\draw (1) -- node {}  (6);
\draw (02)[->] -- node {} (01);
\end{tikzpicture}
\; + \; 
\begin{tikzpicture}[baseline={(0, 0cm-\MathAxis pt)}, thick,
main/.style = {draw, circle, fill=black, minimum size=6pt},
dot/.style={inner sep=0pt,fill=black, circle, minimum size=3pt},
dots/.style={inner sep=0pt,fill=black, circle, minimum size=1pt}] 
  \node[dot] (1) at (0, 0) {};
  \node[dot] (11) at (-1.5, 0) {};
  \node[main] (111) at (-2.5, 0) {};
  \node[main] (112) at (-1.5, 1) {};
  \node[main] (3) at (40:1) {};
  \node[main] (4) at (15:1) {};
  \node[main] (5) at (-15:1) {};
  \node[main] (6) at (-40:1) {};
  \node[dots] (0) at (-25: 0.7) {};
  \node[dots] (00) at (-35: 0.7) {};
  \node[dots] (01) at (-2, 0) {};
  \node[dots] (02) at (-2, -0.3) {};
\draw (1) -- node[midway, above right, pos=1] {$(s-2) f_h$} (11);
\draw (11) [dashed] -- node[below=2mm] {\tiny $\bar{\gamma}^{n_0} (-q k_i - r l_j - f_g - (s-2) f_h)$} (111);
\draw (1) -- node {}  (3);
\draw (1) -- node {}  (4);
\draw (1) -- node {}  (5);
\draw (1) -- node {}  (6);
\draw (11) -- node[midway, left, pos=0.8] {$f_g$} (112);
\draw (02)[->] -- node {} (01);
\end{tikzpicture}  \; + ...... 
\right)
 \nd }
 where the appropriate summation structure is inserted in. So far the story has progressed in the way we had described earlier, but now we notice the presence of the $\overline\gamma$ functions accompanying all the nodal diagrams for the $\varphi_3$. In the limit \eqref{perfume}, if we can expand $\overline\gamma$ as in \eqref{casin18}, then the aforementioned product structure ensues. Generically however \eqref{perfume} cannot be applied and therefore ${\overline\gamma/ f^2_{w_s}}$ knows about the momenta in the $(\varphi_1, \varphi_2)$ sectors. This is also evident from the following diagram:
 
 {\footnotesize 
 \bg\label{luna28}
 &\sum\limits_{l_0}
 \begin{pmatrix}
p\\
 l_0
\end{pmatrix}
\left(
 \begin{tikzpicture}[baseline={(0, 0cm-\MathAxis pt)}, thick,
main/.style = {draw, circle, fill=black, minimum size=6pt},
dot/.style={inner sep=0pt,fill=black, circle, minimum size=3pt},
ball/.style={ball  color=white, circle,  minimum size=15pt},
dots/.style={inner sep=0pt,fill=black, circle, minimum size=1pt}] 
  \node[dot] (1) at (0, 0) {};
  \node[ball] (11) at (-1.5, 0) {i};
  \node[main] (111) at (-2.5, 0) {};
  \node[main] (3) at (40:1) {};
  \node[main] (4) at (15:1) {};
  \node[main] (5) at (-15:1) {};
  \node[main] (6) at (-40:1) {};
  \node[dots] (0) at (-25: 0.7) {};
  \node[dots] (00) at (-35: 0.7) {};
\draw (1) -- node[midway, above right, pos=1] {$(q-1) k_i$} (11);
\draw (11) -- node[midway, above right, pos=0.8] {$k_j$} (111);
\draw (1) -- node {}  (3);
\draw (1) -- node {}  (4);
\draw (1) -- node {}  (5);
\draw (1) -- node {}  (6);
\end{tikzpicture} 
\; ;  ((q - 1) k_i + k_j)^{n + l_0} 
 \right) \otimes
 \left(
 \begin{tikzpicture}[baseline={(0, 0cm-\MathAxis pt)}, thick,
main/.style = {draw, circle, fill=black, minimum size=6pt},
dot/.style={inner sep=0pt,fill=black, circle, minimum size=3pt},
dots/.style={inner sep=0pt,fill=black, circle, minimum size=1pt}] 
  \node[dot] (1) at (0, 0) {};
  \node[dot] (11) at (-0.6, 0) {};
  \node[main] (3) at (40:1) {};
  \node[main] (4) at (15:1) {};
  \node[main] (5) at (-15:1) {};
  \node[main] (6) at (-40:1) {};
  \node[dots] (0) at (-25: 0.7) {};
  \node[dots] (00) at (-35: 0.7) {};
\draw (1) -- node[midway, above right, pos=1] {$r l_a$} (11);
\draw (1) -- node {}  (3);
\draw (1) -- node {}  (4);
\draw (1) -- node {}  (5);
\draw (1) -- node {}  (6);
\end{tikzpicture} \; ;  (r l_a)^{m + p - l_0}
\right) \nonumber\\
& \otimes \; \left(
\begin{tikzpicture}[baseline={(0, 0cm-\MathAxis pt)}, thick,
main/.style = {draw, circle, fill=black, minimum size=6pt},
dot/.style={inner sep=0pt,fill=black, circle, minimum size=3pt},
dots/.style={inner sep=0pt,fill=black, circle, minimum size=1pt}] 
  \node[dot] (1) at (0, 0) {};
  \node[dot] (11) at (-1.5, 0) {};
  \node[main] (111) at (-2.5, 0) {};
  \node[main] (3) at (40:1) {};
  \node[main] (4) at (15:1) {};
  \node[main] (5) at (-15:1) {};
  \node[main] (6) at (-40:1) {};
  \node[dots] (0) at (-25: 0.7) {};
  \node[dots] (00) at (-35: 0.7) {};
  \node[dots] (01) at (-2, 0) {};
  \node[dots] (02) at (-2, -0.3) {};
\draw (1) -- node[midway, above right, pos=1] {$(s-1) f_l$} (11);
\draw (11) [dashed] -- node[below=2mm] {\tiny ${\bar{\gamma}^{n_0} (-(q-1) k_i - k_j - r l_a - (s-1) f_l)\over (k_j + (q-1)k_i + rl_a + (s-1)f_l)^{2n_0}}$} (111);
\draw (1) -- node {}  (3);
\draw (1) -- node {}  (4);
\draw (1) -- node {}  (5);
\draw (1) -- node {}  (6);
\draw (02)[->] -- node {} (01);
\end{tikzpicture}
\; + \; 
\begin{tikzpicture}[baseline={(0, 0cm-\MathAxis pt)}, thick,
main/.style = {draw, circle, fill=black, minimum size=6pt},
dot/.style={inner sep=0pt,fill=black, circle, minimum size=3pt},
dots/.style={inner sep=0pt,fill=black, circle, minimum size=1pt}] 
  \node[dot] (1) at (0, 0) {};
  \node[dot] (11) at (-1.5, 0) {};
  \node[main] (111) at (-2.5, 0) {};
  \node[main] (112) at (-1.5, 1) {};
  \node[main] (3) at (40:1) {};
  \node[main] (4) at (15:1) {};
  \node[main] (5) at (-15:1) {};
  \node[main] (6) at (-40:1) {};
  \node[dots] (0) at (-25: 0.7) {};
  \node[dots] (00) at (-35: 0.7) {};
  \node[dots] (01) at (-2, 0) {};
  \node[dots] (02) at (-2, -0.3) {};
\draw (1) -- node[midway, above right, pos=1] {$(s-2) f_h$} (11);
\draw (11) [dashed] -- node[below=2mm] {\tiny ${\bar{\gamma}^{n_0} (-(q-1) k_i - k_j - r l_a - (s-1) f_l)\over (k_j + (q-1)k_i + rl_a + f_g + (s-2)f_h)^{2n_0}}$} (111);
\draw (1) -- node {}  (3);
\draw (1) -- node {}  (4);
\draw (1) -- node {}  (5);
\draw (1) -- node {}  (6);
\draw (11) -- node[midway, left, pos=0.8] {$f_g$} (112);
\draw (02)[->] -- node {} (01);
\end{tikzpicture} \; + ... 
\right) 
 \nd }
 where the choice of the ${\overline\gamma/ f^2_{w_s}}$ functions are {\it different} from what we had in \eqref{luna26}, implying that changing the nodal diagrams in the $(\varphi_1, \varphi_2)$ sectors changes the entire series of diagrams from the $\varphi_3$ sector. 
 This means, in general, the series of nodal diagrams in the three sectors {\it cannot} have a product structure and the condition \eqref{tagjap} is only valid in the limit \eqref{perfume}. Nevertheless we will impose an approximate product structure to the nodal diagrams from the three sectors
 and express the path integral as in \eqref{2summers} for this section, and analyze the generic picture in section \ref{borel}. 
  Putting everything together therefore, 
the set of nodal diagrams contributing to ${\cal N}^{(1)}_{nmp}(k; q)$ may be denoted by the following series:
\bg\label{luna29}
& \mathcal{N}^{(1)}_{nmp}(k; q) = 
\frac{1}{\rm V} \sum\limits_i \; 
\begin{tikzpicture}[baseline={(0, 0cm-\MathAxis pt)}, thick,
main/.style = {draw, circle, fill=black, minimum size=6pt},
dot/.style={inner sep=0pt,fill=black, circle, minimum size=3pt},
ball/.style={ball  color=white, circle,  minimum size=15pt},
dots/.style={inner sep=0pt,fill=black, circle, minimum size=1pt}] 
  \node[dot] (1) at (0, 0) {};
  \node[ball] (11) at (-1, 0) {i};
  \node[main] (3) at (40:1) {};
  \node[main] (4) at (15:1) {};
  \node[main] (5) at (-15:1) {};
  \node[main] (6) at (-40:1) {};
  \node[dots] (0) at (-25: 0.7) {};
  \node[dots] (00) at (-35: 0.7) {};
\draw (1) -- node[midway, above right, pos=1] {\tiny{$q k_i$}} (11);
\draw (1) -- node {}  (3);
\draw (1) -- node {}  (4);
\draw (1) -- node {}  (5);
\draw (1) -- node {}  (6);
\end{tikzpicture} \; + \; 
\frac{1}{{\rm V}^2} \sum\limits_{i, j} \; 
\begin{tikzpicture}[baseline={(0, 0cm-\MathAxis pt)}, thick,
main/.style = {draw, circle, fill=black, minimum size=6pt},
dot/.style={inner sep=0pt,fill=black, circle, minimum size=3pt},
ball/.style={ball  color=white, circle,  minimum size=15pt},
dots/.style={inner sep=0pt,fill=black, circle, minimum size=1pt}] 
  \node[dot] (1) at (0, 0) {};
  \node[ball] (11) at (-1, 0) {i};
  \node[main] (3) at (40:1) {};
  \node[main] (4) at (15:1) {};
  \node[main] (5) at (-15:1) {};
  \node[main] (6) at (-40:1) {};
  \node[dots] (0) at (-25: 0.7) {};
  \node[dots] (00) at (-35: 0.7) {};
\draw (1) -- node[midway, above right, pos=1] {\tiny{$q k_j$}} (11);
\draw (1) -- node {}  (3);
\draw (1) -- node {}  (4);
\draw (1) -- node {}  (5);
\draw (1) -- node {}  (6);
\end{tikzpicture} 
\\
& \frac{1}{{\rm V}^2} \sum\limits_{i, j} \; 
\begin{tikzpicture}[baseline={(0, 0cm-\MathAxis pt)}, thick,
main/.style = {draw, circle, fill=black, minimum size=6pt},
dot/.style={inner sep=0pt,fill=black, circle, minimum size=3pt},
ball/.style={ball  color=white, circle,  minimum size=15pt},
dots/.style={inner sep=0pt,fill=black, circle, minimum size=1pt}] 
  \node[dot] (1) at (0, 0) {};
  \node[ball] (11) at (-1.5, 0) {i};
  \node[dot] (111) at (-2.4, 0) {};
  \node[main] (1111) at (-2.8, 0.5) {};
  \node[main] (1112) at (-2.8, -0.5) {};
  \node[main] (3) at (40:1) {};
  \node[main] (4) at (15:1) {};
  \node[main] (5) at (-15:1) {};
  \node[main] (6) at (-40:1) {};
  \node[dots] (0) at (-25: 0.7) {};
  \node[dots] (00) at (-35: 0.7) {};
\draw (1) -- node[midway, above right, pos=1] {\tiny{$(q-2) k_i$}} (11);
\draw (11) -- node[midway, above right, pos=1] {\tiny{$2 k_j$}} (111);
\draw (111) -- node {} (1111);
\draw (111) -- node {} (1112);
\draw (1) -- node {}  (3);
\draw (1) -- node {}  (4);
\draw (1) -- node {}  (5);
\draw (1) -- node {}  (6);
\end{tikzpicture} \; + \; 
\frac{1}{{\rm V}^4} \sum\limits_{i, j, l, m} \; 
\begin{tikzpicture}[baseline={(0, 0cm-\MathAxis pt)}, thick,
main/.style = {draw, circle, fill=black, minimum size=6pt},
dot/.style={inner sep=0pt,fill=black, circle, minimum size=3pt},
ball/.style={ball  color=white, circle,  minimum size=15pt},
dots/.style={inner sep=0pt,fill=black, circle, minimum size=1pt}] 
  \node[dot] (1) at (0, 0) {};
  \node[ball] (11) at (-1.5, 0) {i};
  \node[dot] (111) at (-3, 0) {};
  \node[main] (1111) at (-3.5, 0.5) {};
  \node[main] (1112) at (-3.5, -0.5) {};
  \node[main] (3) at (40:1) {};
  \node[main] (4) at (15:1) {};
  \node[main] (5) at (-15:1) {};
  \node[main] (6) at (-40:1) {};
  \node[dots] (0) at (-25: 0.7) {};
  \node[dots] (00) at (-35: 0.7) {};
\draw (1) -- node[midway, above right, pos=1] {\tiny{$(q-2) k_l$}} (11);
\draw (11) -- node[midway, above right, pos=1] {\tiny{$k_m + k_j$}} (111);
\draw (111) -- node {} (1111);
\draw (111) -- node {} (1112);
\draw (1) -- node {}  (3);
\draw (1) -- node {}  (4);
\draw (1) -- node {}  (5);
\draw (1) -- node {}  (6);
\end{tikzpicture} \; + ... \nonumber
\\
&+ \; \frac{1}{{\rm V}^{q+1}} \sum\limits_u \; 
\begin{tikzpicture}[baseline={(0, 0cm-\MathAxis pt)}, thick,
main/.style = {draw, circle, fill=black, minimum size=6pt},
dot/.style={inner sep=0pt,fill=black, circle, minimum size=3pt},
ball/.style={ball  color=white, circle,  minimum size=15pt},
dots/.style={inner sep=0pt,fill=black, circle, minimum size=1pt}] 
  \node[dot] (1) at (0, 0) {};
  \node[ball] (11) at (-1.5, 0) {i};
  \node[dot] (13) at (0, 0.5) {};
  \node[dot] (14) at (0, 1) {};
  \node[dot] (12) at (0, -0.3) {};
  \node[dot] (15) at (0, -1) {};
  \node[main] (131) at (1, 0.5) {};
  \node[main] (141) at (1, 1) {};
  \node[main] (121) at (1, -0.3) {};
  \node[main] (151) at (1, -1) {};
  \node[dots] (01) at (0, -0.3) {};
  \node[dots] (01) at (0, -0.6) {};
  \node[dots] (01) at (0, -0.8) {};
  \node[dots] (01) at (-3, -0.6) {};
  \node[dots] (01) at (-3, -0.8) {}; 
  \node[dot] (111) at (-3, 0) {};
  \node[dot] (113) at (-3, 0.5) {};
  \node[dot] (114) at (-3, 1) {};
  \node[dot] (112) at (-3, -0.3) {};
  \node[dot] (115) at (-3, -1) {};
  \node[main] (1131) at (-4, 0.5) {};
  \node[main] (1141) at (-4, 1) {};
  \node[main] (1121) at (-4, -0.3) {};
  \node[main] (1151) at (-4, -1) {};
\draw (1)  -- node[midway, above right, pos=0.8] {} (11);
\draw (12) -- node[midway, above right, pos=0.2] {} (121);
\draw (13) -- node[midway, above right, pos=0.2] {} (131);
\draw (14) -- node[midway, above right, pos=0] {\tiny{$k_{u_{r+1}}$}} (141);
\draw (15) -- node[midway, below right, pos=0] {\tiny{$k_{u_q}$}} (151);
\draw (11)  -- node[midway, above right, pos=0.8] {} (111);
\draw (112) -- node[midway, above right, pos=0.2] {} (1121);
\draw (113) -- node[midway, above right, pos=1] {\tiny{$k_{u_2}$}} (1131);
\draw (114) -- node[midway, above right, pos=1] {\tiny{$k_{u_1}$}} (1141);
\draw (115) -- node[midway, below right, pos=1] {\tiny{$k_{u_r}$}} (1151);
\draw (1) -- node {} (13);
\draw (13) -- node {} (14);
\draw (1) -- node {} (12);
\draw (111) -- node {} (112);
\draw (112) -- node {} (113);
\draw (113) -- node {} (114);
\end{tikzpicture} \nonumber
\nd
where ${\rm U} \equiv (k_{u_1}, k_{u_2}, ..., k_{u_q}, k_i)$ with the condition that $k_{u_1} \ne k_{u_2} \ne .... \ne k_{u_q} \ne k_i$. Clearly such a diagram contributes $(q + 1)!$ terms to the path integral with $(q+1)$ nested integral structures for each term. Note also, and as mentioned earlier, once we convert the summations in \eqref{luna29} to integrals, the first diagram will be suppressed by 
${\rm V}^{-q}$, the second diagram by ${\rm V}^{q-1}$, and so on till we 
hit the last diagram with all different momenta. This will not be suppressed by any volume factor, implying that it will dominate over the other diagrams (at least from the volume point of view).
In a similar vein ${\cal N}^{(1)'}_{nmp}(k; q)$ may be represented by the following series of diagrams:
\bg\label{luna30}
& \mathcal{N}^{(1) '}_{nmp}(k; q) = 
\frac{1}{\rm V} \sum\limits_i \; 
\begin{tikzpicture}[baseline={(0, 0cm-\MathAxis pt)}, thick,
main/.style = {draw, circle, fill=black, minimum size=6pt},
dot/.style={inner sep=0pt,fill=black, circle, minimum size=3pt},
ball/.style={ball  color=white, circle,  minimum size=15pt},
dots/.style={inner sep=0pt,fill=black, circle, minimum size=1pt}] 
  \node[dot] (1) at (0, 0) {};
  \node[main] (11) at (-1, 0) {};
  \node[main] (3) at (40:1) {};
  \node[main] (4) at (15:1) {};
  \node[main] (5) at (-15:1) {};
  \node[main] (6) at (-40:1) {};
  \node[dots] (0) at (-25: 0.7) {};
  \node[dots] (00) at (-35: 0.7) {};
\draw (1) -- node[midway, above right, pos=1] {\tiny{$q k_i$}} (11);
\draw (1) -- node {}  (3);
\draw (1) -- node {}  (4);
\draw (1) -- node {}  (5);
\draw (1) -- node {}  (6);
\end{tikzpicture} \; + \; 
 \frac{1}{{\rm V}^2} \sum\limits_{i, j} \; 
\begin{tikzpicture}[baseline={(0, 0cm-\MathAxis pt)}, thick,
main/.style = {draw, circle, fill=black, minimum size=6pt},
dot/.style={inner sep=0pt,fill=black, circle, minimum size=3pt},
ball/.style={ball  color=white, circle,  minimum size=15pt},
dots/.style={inner sep=0pt,fill=black, circle, minimum size=1pt}] 
  \node[dot] (1) at (0, 0) {};
  \node[dot] (11) at (-1.5, 0) {};
  \node[dot] (111) at (-2.4, 0) {};
  \node[main] (1111) at (-2.8, 0.5) {};
  \node[main] (1112) at (-2.8, -0.5) {};
  \node[main] (3) at (40:1) {};
  \node[main] (4) at (15:1) {};
  \node[main] (5) at (-15:1) {};
  \node[main] (6) at (-40:1) {};
  \node[dots] (0) at (-25: 0.7) {};
  \node[dots] (00) at (-35: 0.7) {};
\draw (1) -- node[midway, above right, pos=1] {\tiny{$(q-2) k_i$}} (11);
\draw (11) -- node[midway, above right, pos=1] {\tiny{$2 k_j$}} (111);
\draw (111) -- node {} (1111);
\draw (111) -- node {} (1112);
\draw (1) -- node {}  (3);
\draw (1) -- node {}  (4);
\draw (1) -- node {}  (5);
\draw (1) -- node {}  (6);
\end{tikzpicture}
\\
&\; + \; 
\frac{1}{{\rm V}^3} \sum\limits_{m, j, l} \; 
\begin{tikzpicture}[baseline={(0, 0cm-\MathAxis pt)}, thick,
main/.style = {draw, circle, fill=black, minimum size=6pt},
dot/.style={inner sep=0pt,fill=black, circle, minimum size=3pt},
ball/.style={ball  color=white, circle,  minimum size=15pt},
dots/.style={inner sep=0pt,fill=black, circle, minimum size=1pt}] 
  \node[dot] (1) at (0, 0) {};
  \node[dot] (11) at (-1.5, 0) {};
  \node[dot] (111) at (-3, 0) {};
  \node[main] (1111) at (-3.5, 0.5) {};
  \node[main] (1112) at (-3.5, -0.5) {};
  \node[main] (3) at (40:1) {};
  \node[main] (4) at (15:1) {};
  \node[main] (5) at (-15:1) {};
  \node[main] (6) at (-40:1) {};
  \node[dots] (0) at (-25: 0.7) {};
  \node[dots] (00) at (-35: 0.7) {};
\draw (1) -- node[midway, above right, pos=1] {\tiny{$(q-2) k_l$}} (11);
\draw (11) -- node[midway, above right, pos=1] {\tiny{$k_m + k_j$}} (111);
\draw (111) -- node {} (1111);
\draw (111) -- node {} (1112);
\draw (1) -- node {}  (3);
\draw (1) -- node {}  (4);
\draw (1) -- node {}  (5);
\draw (1) -- node {}  (6);
\end{tikzpicture} \; + ... 
+ \; \frac{1}{{\rm V}^{q}} \sum\limits_u \; 
\begin{tikzpicture}[baseline={(0, 0cm-\MathAxis pt)}, thick,
main/.style = {draw, circle, fill=black, minimum size=6pt},
dot/.style={inner sep=0pt,fill=black, circle, minimum size=3pt},
ball/.style={ball  color=white, circle,  minimum size=15pt},
dots/.style={inner sep=0pt,fill=black, circle, minimum size=1pt}] 
  \node[dot] (1) at (0, 0) {};
  \node[dot] (11) at (-1.5, 0) {};
  \node[dot] (13) at (0, 0.5) {};
  \node[dot] (14) at (0, 1) {};
  \node[dot] (12) at (0, -0.3) {};
  \node[dot] (15) at (0, -1) {};
  \node[main] (131) at (1, 0.5) {};
  \node[main] (141) at (1, 1) {};
  \node[main] (121) at (1, -0.3) {};
  \node[main] (151) at (1, -1) {};
  \node[dots] (01) at (0, -0.3) {};
  \node[dots] (01) at (0, -0.6) {};
  \node[dots] (01) at (0, -0.8) {};
  \node[dots] (01) at (-3, -0.6) {};
  \node[dots] (01) at (-3, -0.8) {}; 
  \node[dot] (111) at (-3, 0) {};
  \node[dot] (113) at (-3, 0.5) {};
  \node[dot] (114) at (-3, 1) {};
  \node[dot] (112) at (-3, -0.3) {};
  \node[dot] (115) at (-3, -1) {};
  \node[main] (1131) at (-4, 0.5) {};
  \node[main] (1141) at (-4, 1) {};
  \node[main] (1121) at (-4, -0.3) {};
  \node[main] (1151) at (-4, -1) {};
\draw (1)  -- node[midway, above right, pos=0.8] {} (11);
\draw (12) -- node[midway, above right, pos=0.2] {} (121);
\draw (13) -- node[midway, above right, pos=0.2] {} (131);
\draw (14) -- node[midway, above right, pos=0] {\tiny{$k_{u_{r+1}}$}} (141);
\draw (15) -- node[midway, above right, pos=0] {\tiny{$k_{u_q}$}} (151);
\draw (11)  -- node[midway, above right, pos=0.8] {} (111);
\draw (112) -- node[midway, above right, pos=0.2] {} (1121);
\draw (113) -- node[midway, above right, pos=1] {\tiny{$k_{u_2}$}} (1131);
\draw (114) -- node[midway, above right, pos=1] {\tiny{$k_{u_1}$}} (1141);
\draw (115) -- node[midway, above right, pos=1] {\tiny{$k_{u_r}$}} (1151);
\draw (1) -- node {} (13);
\draw (13) -- node {} (14);
\draw (1) -- node {} (12);
\draw (111) -- node {} (112);
\draw (112) -- node {} (113);
\draw (113) -- node {} (114);
\end{tikzpicture} \nonumber
\nd
where ${\rm U}' \equiv (k_{u_1}, k_{u_2}, ..., k_{u_q})$. Comparing to 
\eqref{luna26}, we see that the series \eqref{luna27} differs by the absence of the {\it source} field $\varphi_1$. Clearly now the biggest contribution would come from the $q!$ terms with $q$ nested integral structures when $k_{u_1} \ne k_{u_2} \ne .... \ne k_{u_q}$. This will also be the most dominant nodal diagram of all the other diagrams in \eqref{luna30}.

The nodal diagrams contributing to ${\cal N}^{(2)}_{nmp}(\mu; r)$ would look similar to the series in \eqref{luna27} from ${\cal N}^{(1)'}_{nmp}(k; q)$. There are however two key differences. One, the $\{l_{v_j}\}$ momenta have different powers 
as evident from \eqref{tagjap}, and two, the outermost integrals in the nested integral structures of the nodal diagrams are all integrated upto 
the energy-scale $\mu \propto {\rm M}_p$. This may be symbolically expressed as:
\bg\label{jul13crisis}
{\rm diag}\left[{\cal N}^{(1)'}_{nmp}(k \to l; q \to r)\right]\Big\vert_0^\mu ~ = ~ {\rm diag}\left[{\cal N}^{(2)}_{nmp}(\mu; r)\right], \nd
which is our way of identifying the series of nodal diagrams from \eqref{luna4} till \eqref{luna14} $-$ and beyond $-$ with the ones from 
\eqref{luna27}. One may easily see that, despite diagrammatical equivalences, the amplitudes differ quite a bit as mentioned above.

The nodal diagrams contributing to ${\cal N}^{(3)}_{nmp}(\mu; s)$ differ from the other three sets because of the momentum constraint as evident from the nodal diagrams \eqref{luna15} to \eqref{luna19}. This however raises the following puzzle. What if we had put the momentum constraint not on $\varphi_3$ field, but on $\varphi_1$ or $\varphi_2$ fields? It is easy to see that the answer should not change. The interaction term, to first order in $c_{nmpqrs}$, that enters the path integral with shifted vacua may be represented by:

{\footnotesize
\bg\label{felojane}
{\cal I} \equiv {1\over {\rm V}^{u'}} \sum_{{\cal S}'} c_{nmpqrs}
\prod_{i = 1}^q\widetilde\varphi_1(k_{u_i})\left(\sum_{i = 1}^q k_{u_i}\right)^n
\prod_{j = 1}^r\widetilde\varphi_2(l_{v_j})\left(\sum_{j = 1}^r l_{v_j}\right)^m
\prod_{t = 1}^s\widetilde\varphi_3(f_{w_t})\left(\sum_{t = 1}^s f_{w_t}\right)^p
\delta^{11}\left({\cal F}_{klf}\right), \nonumber\\ \nd}
which may be inserted in the path integral thus replacing the interaction terms in \eqref{fontini}. The way we have defined \eqref{felojane}, it is the first term in the exponential expansion of ${\rm exp}~{\cal I}$ with dimensionless ${\cal I}$ (recall that $c_{nmpqrs}$ for all $(n, .., s)$ have appropriate inverse ${\rm M}_p$ factors). We have also defined $u' \equiv q + r + s - 1$, 
${\cal S}' \equiv \left(\{u_i\}, \{v_j\}, \{w_t\}, n, ..., s\right)$ and
${\cal F}_{klf}$ in a way that the delta function puts the following 
momentum constraint:
\bg\label{colinssA}
{\cal F}_{klf} \equiv \sum_{i = 1}^q k_{u_i} + \sum_{j = 1}^r l_{v_j}
+ \sum_{t = 1}^s f_{w_t} = 0. \nd
In this language we see that we can choose any one of the three momenta:
$(k_{u_q}, l_{v_r}, f_{w_s})$, express it in terms of all other momenta and insert it in the path integral \eqref{lincoln} or \eqref{fontini}. In our analysis we choose $f_{w_s}$ to be fixed via \eqref{headhunters}, but we could in-principle have chosen any one of the three aforementioned momenta. It is clear that, since $(\varphi_2, \varphi_3)$ typically are {\it internal} fields (meaning the external $\varphi_1$ source interacts with the 
$\varphi_1^p$ term), exchanging $f_{w_s}$ with $l_{v_r}$ would not change anything once some changes in nomenclature, and the derivative terms, 
from \eqref{tagjap} are appropriately inserted in. However if we fix 
$k_{u_q}$ via the momentum conservation, then in \eqref{fontini} we have to make the following changes: $n \leftrightarrow p, p \leftrightarrow n, \{k_{u_i}\} \leftrightarrow \{f_{w_t}\}$, and the inclusion of:
\bg\label{zoeekrav}
{1\over \zeta^{2n_o + n_1}} {\partial^{n_k}\overline\alpha(-\zeta)\over 
\partial\zeta^{n_k}}, ~~~{\rm with}~~~ \zeta \equiv \sum_{i = 1}^{q -1} k_{u_i}, \nd
in the nodal diagrams, 
where $(n_o, n_1) \in \mathbb{Z}_+$ are defined as in \eqref{snelmad} with the aforementioned changes to $(n, p)$ inserted in. This insertion of derivative term will not only make the amplitudes of the nodal diagrams for the $\varphi_1$ sector much more involved, but will also change the amplitudes of the nodal diagrams for the $(\varphi_2, \varphi_3)$ sector. 
On the other hand, from the argument presented using the interaction term \eqref{felojane}, the path integral computed using \eqref{zoeekrav}
{\it cannot} be different from what we had in \eqref{2summers}! Therefore 
we can stick with the {\it easier} choice from \eqref{felojane}, and avoid the complicated route via \eqref{zoeekrav} for the present analysis. In fact, from the equivalence of the path integral for the three possible realizations of the momentum conservation condition, we could propose non-trivial identities between the components of $\overline\sigma \equiv (\overline\alpha(k), \overline\beta(l), \overline\gamma(f))$, in addition to the relations connecting them to the various components of
$\sigma \equiv (\alpha(k), \beta(l), \gamma(f))$ respectively (see section 6.1 of \cite{coherbeta2}). The latter of which will be elaborated below, but before moving ahead, let us list the nodal diagrams contributing to 
${\cal N}^{(3)}_{nmp}(\mu; s)$: 

{\footnotesize
\bg\label{luna31}
& \mathcal{N}^{(3)}_{nmp}(\mu; s) = 
\frac{1}{\rm V} \sum\limits_i \; 
\begin{tikzpicture}[baseline={(0, 0cm-\MathAxis pt)}, thick,
main/.style = {draw, circle, fill=black, minimum size=6pt},
dot/.style={inner sep=0pt,fill=black, circle, minimum size=3pt},
dots/.style={inner sep=0pt,fill=black, circle, minimum size=1pt}] 
  \node[dot] (1) at (0, 0) {};
  \node[dot] (11) at (-1.5, 0) {};
  \node[main] (111) at (-3, 0) {};
  \node[main] (3) at (40:1) {};
  \node[main] (4) at (15:1) {};
  \node[main] (5) at (-15:1) {};
  \node[main] (6) at (-40:1) {};
  \node[dots] (0) at (-25: 0.7) {};
  \node[dots] (00) at (-35: 0.7) {};
\draw (1) -- node[midway, above right, pos=1] {$(s - 1) f_i$} (11);
\draw (11) [dashed] -- node[midway, above right, pos=0.8] {$f_{\omega_s}$} (111);
\draw (1) -- node {}  (3);
\draw (1) -- node {}  (4);
\draw (1) -- node {}  (5);
\draw (1) -- node {}  (6);
\end{tikzpicture} 
\; + \; \frac{1}{{\rm V}^2} \sum\limits_{g, h} \; 
\begin{tikzpicture}[baseline={(0, 0cm-\MathAxis pt)}, thick,
main/.style = {draw, circle, fill=black, minimum size=6pt},
dot/.style={inner sep=0pt,fill=black, circle, minimum size=3pt},
dots/.style={inner sep=0pt,fill=black, circle, minimum size=1pt}] 
  \node[dot] (1) at (0, 0) {};
  \node[dot] (11) at (-1.5, 0) {};
  \node[main] (111) at (-3, 0) {};
  \node[main] (112) at (-1.5, 1) {};
  \node[main] (3) at (40:1) {};
  \node[main] (4) at (15:1) {};
  \node[main] (5) at (-15:1) {};
  \node[main] (6) at (-40:1) {};
  \node[dots] (0) at (-25: 0.7) {};
  \node[dots] (00) at (-35: 0.7) {};
\draw (1) -- node[midway, above right, pos=1] {$(s - 2) f_g$} (11);
\draw (11) [dashed] -- node[midway, above right, pos=0.8] {$f_{\omega_s}$} (111);
\draw (11) -- node[midway, left, pos=0.8] {$f_h$} (112);
\draw (1) -- node {}  (3);
\draw (1) -- node {}  (4);
\draw (1) -- node {}  (5);
\draw (1) -- node {}  (6);
\end{tikzpicture} \nonumber\\
& +~
\frac{1}{{\rm V}^3} \sum\limits_{g, h, m} \; 
\begin{tikzpicture}[baseline={(0, 0cm-\MathAxis pt)}, thick,
main/.style = {draw, circle, fill=black, minimum size=6pt},
dot/.style={inner sep=0pt,fill=black, circle, minimum size=3pt},
dots/.style={inner sep=0pt,fill=black, circle, minimum size=1pt}] 
  \node[dot] (1) at (0, 0) {};
  \node[dot] (11) at (-1.5, 0) {};
  \node[main] (111) at (-3, 0) {};
  \node[dot] (112) at (-1.5, 1) {};
  \node[main] (1121) at (-2, 1.5) {};
  \node[main] (1122) at (-1, 1.5) {};
  \node[main] (3) at (40:1) {};
  \node[main] (4) at (15:1) {};
  \node[main] (5) at (-15:1) {};
  \node[main] (6) at (-40:1) {};
  \node[dots] (0) at (-25: 0.7) {};
  \node[dots] (00) at (-35: 0.7) {};
\draw (1) -- node[midway, above right, pos=1] {$(s - 3) f_g$} (11);
\draw (11) [dashed] -- node[midway, below right, pos=0.8] {$f_{\omega_s}$} (111);
\draw (11) -- node[midway, left, pos=0.8] {$f_h + f_m$} (112);
\draw (112) -- node[midway, left, pos=0.8] {} (1121);
\draw (112) -- node[midway, left, pos=0.8] {} (1122);
\draw (1) -- node {}  (3);
\draw (1) -- node {}  (4);
\draw (1) -- node {}  (5);
\draw (1) -- node {}  (6);
\end{tikzpicture} 
\; +\;  ..... \; +  \; 
\frac{1}{{\rm V}^{s-1}} \sum\limits_{\rm U} \; 
\begin{tikzpicture}[baseline={(0, 0cm-\MathAxis pt)}, thick,
main/.style = {draw, circle, fill=black, minimum size=6pt},
dot/.style={inner sep=0pt,fill=black, circle, minimum size=3pt},
dots/.style={inner sep=0pt,fill=black, circle, minimum size=1pt}] 
  \node[dot] (1) at (0, 0) {};
  \node[main] (11) at (-2, 0) {};
  \node[dot] (13) at (0, 0.8) {};
  \node[dot] (14) at (0, 1.5) {};
  \node[dot] (15) at (0, -1.2) {};
  \node[main] (121) at (1, 0) {};
  \node[main] (131) at (1, 0.8) {};
  \node[main] (141) at (1, 1.5) {};
  \node[main] (151) at (1, -1.2) {};
  \node[dots] (01) at (0, -0.3) {};
  \node[dots] (01) at (0, -0.7) {};
  \node[dots] (01) at (0, -1) {};
  \node[dots] (01) at (0, -0.5) {};
\draw (1) [dashed] -- node[midway, above right, pos=0.8] {$f_{\omega_s}$} (11);
\draw (1) -- node[midway, above right, pos=0.2] {$f_{\omega_3}$} (121);
\draw (13) -- node[midway, above right, pos=0.2] {$f_{\omega_2}$} (131);
\draw (14) -- node[midway, above right, pos=0.2] {$f_{\omega_1}$} (141);
\draw (15) -- node[midway, below right, pos=0] {$f_{\omega_{s-1}}$} (151);
\draw (1) -- node {} (13);
\draw (13) -- node {} (14);
\end{tikzpicture} 
 \nd }
where ${\rm U} \equiv (f_{w_1}, f_{w_2}, ...., f_{w_{s-1}})$ with the inequality condition $f_{w_1} \ne f_{w_2} \ne .... \ne f_{w_{s_1}} \ne f_{w_s}$, the last of which is controlled by the momenta of all other fields via \eqref{headhunters}. This will contribute $(s-1)!$ terms with 
$s-1$ nested integrals with the derivative factor from \eqref{tagjap} and, according to our discussion above, will dominate over all the other diagrams in \eqref{luna31}.

\section{Nodal diagrams, resurgence trans-series and Borel resummation
\label{borel}}

In the previous section and especially in \eqref{2summers} we saw how the path integral analysis of \eqref{lincoln} or \eqref{fontini} may be succinctly arranged as a series of nodal diagrams. We also saw that, to the first order in the coupling constant $c_{nmpqrs}$ and for some fixed values of $(n, m, ..., s)$ the nodal diagrams do show factorial growth. The question is what happens when we go to higher orders in the coupling constant, {\it i.e.} to order $c_{nmpqrs}^{\rm N}$ for ${\rm N} \in \mathbb{Z}_+$, where:
\bg\label{macharrsn}
c^{\rm N}_{nmpqrs} \equiv \prod_{i = 1}^{\rm N} c_{n_im_ip_iq_ir_is_i}, \nd
and $n \equiv (n_1, ..., n_{\rm N})$ are subset of possible values of $n$ which may or may not be equal (similarly for $(m_i,..., s_i)$). This would then be the generic possible ways the coupling constants would appear to the ${\rm N}$-th order. We can also look at the interaction terms for the Fourier fields in the $\varphi_1$ sector, and they take the following 
form\footnote{Note that we can easily generalize the interaction terms from the ones appearing in \eqref{angelslune} to more generic terms of the form by the following replacement in \eqref{lincoln} or \eqref{fontini}:
\bg\label{dani}
\partial^n\varphi^p ~ \to ~ \{\partial^n\}\{\varphi^p\} \equiv 
\sum_{\{n_j\}\{p_j\}} \prod_{j = 1}^k \partial^{n_j} \varphi^{p_j}, \nonumber\nd
where the sum is over the subspace satisfying $\sum\limits_{j = 1}^k n_j = n$ and $\sum\limits_{j = 1}^k p_j = p$ with 
$(n_j, p_j) \in \left(\mathbb{Z}_+, \mathbb{Z}_+\right)$. The splitting of the momenta into various components, respectively accompanying each of the field components,
should not change any of our computations as this is how we have worked out the individual nodal diagrams even in the simpler case of \eqref{angelslune}. Thus generalizing \eqref{angelslune} to the aforementioned interaction would simply mean that we change the powers of the momenta accompanying each field components.\label{loraseri}}:  

{\footnotesize
\bg\label{ivabst2}
\sum_{\{u_i\}} \prod_{i = 1}^q \widetilde\varphi_1(k_{u_i}) 
\left(\sum_{j = 1}^q k_{u_j}\right)^n = 
\sum_{i, j, .., p} {q!\over l_i! l_j!..l_p!}~\underbrace{\widetilde\varphi^{l_i}_1(k_i)
\widetilde\varphi^{l_j}_1(k_j)....\widetilde\varphi^{l_p}_1(k_p)}_{l_i + ..+l_p = q} 
\left(l_ik_i + l_jk_j + ..+ l_pk_p\right)^n, \nd}
where $\{u_i\} = (u_1, ..., u_q)$ and the set of $q$ momenta 
$(k_i, ..., k_p)$ may or may not be equal to each other. It should be clear that $l_i + l_j + ... + l_p = q$, and $(l_i, .., l_p) \in \mathbb{Z}_+$ so they form a set of integer distribution of $q$. The discrete momenta 
$(k_i,..., k_p)$ are arranged keeping in mind that there is an upper cut-off in the nested integrals given by $\mu < {\rm M}_p$. This means the momentum $k$ for the Fourier field $\widetilde\varphi(k)$ may be divided as:
\bg\label{aadams}
k \equiv (k_1, k_2, k_3, ......, k_{{\rm N}_\mu}), \nd
where we expect $k_{{\rm N}_\mu} \equiv \mu$ and $k_1 \equiv \kappa_{\rm IR}$ respectively proportional at most to the inverse sizes of the internal eight-manifold and the eleven-dimensional non-compact space (see footnote \ref{yaya}). Such a discrete distribution, using the Wilsonian scale $\mu$ means that $i = 1, ..., {\rm N}_\mu$ (similarly for $(j, ..., p)$ in the summation structure of \eqref{ivabst2}). Therefore the total number of terms contributing to 
\eqref{ivabst2} is:
\bg\label{feserb}
\sum_{l_i,..,l_p} {q!\over l_i!l_j!...l_p!} ~\delta(l_i + l_j + ...+ l_p - q) = {\rm N}_\mu^q, \nd
which follows from simple combinatoric identities,
and the number of terms that are constructed by all different discrete momenta is 
$\left(\begin{matrix} {\rm N}_\mu \\ q\\ \end{matrix}\right)$. Clearly each of these terms appear $q!$ times in \eqref{ivabst2}. With the three fields
$(\varphi_1, \varphi_2, \varphi_3)$ we expect ${\rm N}^{3q}_\mu$ number of 
terms, out of which:
\bg\label{bic88}
{\rm N}(q, r, s) \equiv \left(\begin{matrix} {\rm N}_\mu \\ q\\ \end{matrix}\right)\left(\begin{matrix} {\rm N}_\mu \\ r\\ \end{matrix}\right)\left(\begin{matrix} {\rm N}_\mu \\ s - 1\\ \end{matrix}\right), \nd
number of terms with different discrete momenta from each sector. Each of these terms come with the combinatoric factor of $q!r!(s-1)!$; and one could get rid of this annoying factor by redefining the coupling constants $c_{nmpqrs} \to {c_{nmpqrs}\over q!r!(s-1)!}$. Such a procedure will put unit factors in front of the terms with all unequal momenta, while suppressing the other terms by $q!$, $r!$ and $(s-1)!$ in the $\varphi_1, \varphi_2$ and $\varphi_3$ sectors respectively. 

Let us now come back to issue that we discussed around \eqref{luna26}, namely the fact that the path integral is not being expressed in terms of products of three set of nodal diagrams. In the following we will show that, for each and every nodal diagram from the $\varphi_1$ sector, {\it all} nodal diagrams from $\varphi_2$ and $\varphi_3$ sectors contribute. The difference from the product structure appears from the fact that the contributions of the nodal diagrams from $\varphi_2$ and $\varphi_3$ sectors are sensitive to the momenta carried by the nodal diagrams from $\varphi_1$ sector. 

We start by fixing one nodal diagram from $\varphi_1$ sector, say for example \eqref{luna22}. To this we can multiply by the amplitude of a nodal diagram from $\varphi_2$ sector, namely \eqref{luna7}. We can now see that, for these two representative choices, all possible nodal diagrams from $\varphi_3$ sector to first order in $c_{nmpqrs}$ appears. This is represented by the following:

{\footnotesize
\bg\label{luna32}
& \sum\limits_{l_0}
\begin{pmatrix}
p \\
l_0
\end{pmatrix}
\left(
\begin{tikzpicture}[baseline={(0, 0cm-\MathAxis pt)}, thick,
main/.style = {draw, circle, fill=black, minimum size=6pt},
dot/.style={inner sep=0pt,fill=black, circle, minimum size=3pt},
ball/.style={ball  color=white, circle,  minimum size=15pt},
dots/.style={inner sep=0pt,fill=black, circle, minimum size=1pt}] 
  \node[dot] (1) at (0, 0) {};
  \node[ball] (11) at (-1.5, 0) {i};
  \node[dot] (111) at (-2.4, 0) {};
  \node[main] (1111) at (-2.8, 0.5) {};
  \node[main] (1112) at (-2.8, -0.5) {};
  \node[main] (3) at (40:1) {};
  \node[main] (4) at (15:1) {};
  \node[main] (5) at (-15:1) {};
  \node[main] (6) at (-40:1) {};
  \node[dots] (0) at (-25: 0.7) {};
  \node[dots] (00) at (-35: 0.7) {};
\draw (1) -- node[midway, above right, pos=1] {\tiny{$(q-2) k_i$}} (11);
\draw (11) -- node[midway, above right, pos=1] {\tiny{$2 k_j$}} (111);
\draw (111) -- node {} (1111);
\draw (111) -- node {} (1112);
\draw (1) -- node {}  (3);
\draw (1) -- node {}  (4);
\draw (1) -- node {}  (5);
\draw (1) -- node {}  (6);
\end{tikzpicture}  \; ; \; {\rm S}_{ij}^{n+l_0} (q)
\right) \; \otimes 
\left( 
\begin{tikzpicture}[baseline={(0, 0cm-\MathAxis pt)}, thick,
main/.style = {draw, circle, fill=black, minimum size=6pt},
dot/.style={inner sep=0pt,fill=black, circle, minimum size=3pt}] 
  \node[dot] (1) at (0, 0) {};
  \node[dot] (11) at (-1.5, 0) {};
  \node[main] (111) at (-3, 0) {};
  \node[main] (3) at (40:1) {};
  \node[main] (4) at (15:1) {};
  \node[main] (5) at (-15:1) {};
  \node[main] (6) at (-40:1) {};
  \node[dot] (0) at (-25: 0.7) {};
  \node[dot] (00) at (-35: 0.7) {};
\draw (1) -- node[midway, above right, pos=1] {\tiny $(r - 1) l_b$} (11);
\draw (11) -- node[midway, above right, pos=0.8] {\tiny $l_a$} (111);
\draw (1) -- node {}  (3);
\draw (1) -- node {}  (4);
\draw (1) -- node {}  (5);
\draw (1) -- node {}  (6);
\end{tikzpicture} \; ; \; {\rm T}_{ab}^{m+p-l_0} (r)
\right) \nonumber
\\
& \otimes \; \left(
\begin{tikzpicture}[baseline={(0, 0cm-\MathAxis pt)}, thick,
main/.style = {draw, circle, fill=black, minimum size=6pt},
dot/.style={inner sep=0pt,fill=black, circle, minimum size=3pt},
dots/.style={inner sep=0pt,fill=black, circle, minimum size=1pt}] 
  \node[dot] (1) at (0, 0) {};
  \node[dot] (11) at (-1.5, 0) {};
  \node[main] (111) at (-2.5, 0) {};
  \node[main] (3) at (40:1) {};
  \node[main] (4) at (15:1) {};
  \node[main] (5) at (-15:1) {};
  \node[main] (6) at (-40:1) {};
  \node[dots] (0) at (-25: 0.7) {};
  \node[dots] (00) at (-35: 0.7) {};
  \node[dots] (01) at (-2, 0) {};
  \node[dots] (02) at (-2, -0.3) {};
\draw (1) -- node[midway, above right, pos=1] {\tiny $(s-1) f_l$} (11);
\draw (11) [dashed] -- node[below=2mm] {\tiny ${\bar{\gamma}^{n_0} (-(s-1) f_l - {\rm S}_{ij} (q) - {\rm T}_{ab} (r))\over ( (s-1)f_l + {\rm S}_{ij} (q) + {\rm T}_{ab} (r))^{2n_0}}$} (111);
\draw (1) -- node {}  (3);
\draw (1) -- node {}  (4);
\draw (1) -- node {}  (5);
\draw (1) -- node {}  (6);
\draw (02)[->] -- node {} (01);
\end{tikzpicture} \;  + \; 
\begin{tikzpicture}[baseline={(0, 0cm-\MathAxis pt)}, thick,
main/.style = {draw, circle, fill=black, minimum size=6pt},
dot/.style={inner sep=0pt,fill=black, circle, minimum size=3pt},
dots/.style={inner sep=0pt,fill=black, circle, minimum size=1pt}] 
  \node[dot] (1) at (0, 0) {};
  \node[dot] (11) at (-1.5, 0) {};
  \node[main] (111) at (-3, 0) {};
  \node[dot] (112) at (-1.5, 1) {};
  \node[main] (1121) at (-2, 1.5) {};
  \node[main] (1122) at (-1, 1.5) {};
  \node[main] (3) at (40:1) {};
  \node[main] (4) at (15:1) {};
  \node[main] (5) at (-15:1) {};
  \node[main] (6) at (-40:1) {};
  \node[dot] (0) at (-25: 0.7) {};
  \node[dot] (00) at (-35: 0.7) {};
  \node[dots] (01) at (-2, 0) {};
  \node[dots] (02) at (-2, -0.3) {};
\draw (1) -- node[midway, above right, pos=1] {\tiny $(s - 3) f_g$} (11);
\draw (11) [dashed] -- node[below=2mm] {\tiny ${\bar{\gamma}^{n_0} (-2f_h -(s-3) f_g - {\rm S}_{ij} (q) - {\rm T}_{ab} (r))\over (2f_h + (s-3)f_g + {\rm S}_{ij} (q) + {\rm T}_{ab} (r))^{2n_0}}$} (111);
\draw (11) -- node[midway, left, pos=0.8] {\tiny $2f_h$} (112);
\draw (112) -- node[midway, left, pos=0.8] {} (1121);
\draw (112) -- node[midway, left, pos=0.8] {} (1122);
\draw (1) -- node {}  (3);
\draw (1) -- node {}  (4);
\draw (1) -- node {}  (5);
\draw (1) -- node {}  (6);
\draw (02)[->] -- node {} (01);
\end{tikzpicture} \;  + ... \right. \nonumber
\\
& \left. \; ... + \; 
\begin{tikzpicture}[baseline={(0, 0cm-\MathAxis pt)}, thick,
main/.style = {draw, circle, fill=black, minimum size=6pt},
dot/.style={inner sep=0pt,fill=black, circle, minimum size=3pt},
dots/.style={inner sep=0pt,fill=black, circle, minimum size=1pt}] 
  \node[dot] (1) at (0, 0) {};
  \node[main] (11) at (-2.5, 0) {} ;
  \node[dot] (13) at (0, 0.8) {};
  \node[dot] (14) at (0, 1.5) {};
  \node[dot] (15) at (0, -1.2) {};
  \node[main] (121) at (1, 0) {};
  \node[main] (131) at (1, 0.8) {};
  \node[main] (141) at (1, 1.5) {};
  \node[main] (151) at (1, -1.2) {};
  \node[dots] (01) at (0, -0.3) {};
  \node[dots] (01) at (0, -0.7) {};
  \node[dots] (01) at (0, -1) {};
  \node[dots] (02) at (-1.5, 0) {};
  \node[dots] (03) at (-1.5, -0.3) {};
  \node[dots] (04) at (-2.4, 0) {};
\draw (1) [dashed] -- node {} (04);
\draw (04) [dashed] -- node [below=3mm] {\tiny$\frac{\bar{\gamma}^{n_0} \left(-\sum\limits_{p=1}^{s-1} f_{h_p} - {\rm S}_{ij} (q) - {\rm T}_{ab} (r)\right)}{\left(-\sum\limits_{p=1}^{s-1} f_{h_p} + {\rm S}_{ij} (q) + {\rm T}_{ab} (r)\right)^{2n_0}}$} (11);
\draw (1) -- node[midway, above right, pos=0.2] {\tiny $f_{h_3}$} (121);
\draw (13) -- node[midway, above right, pos=0.2] {\tiny $f_{h_2}$} (131);
\draw (14) -- node[midway, above right, pos=0.2] {\tiny $f_{h_1}$} (141);
\draw (15) -- node[midway, below right, pos=0] {\tiny $f_{h_{s-1}}$} (151);
\draw (1) -- node {} (13);
\draw (13) -- node {} (14);
\draw (03)[->] -- node {} (02);
\end{tikzpicture} 
\right)
\nd}
where ${\rm S}_{ij}(q) \equiv 2k_j + (q-2)k_i$ and ${\rm T}_{ab}(r) \equiv l_a + (r-1)l_b$. Note that all the nodal diagrams from $\varphi_3$ sector have different values for $\overline\gamma/f^2_{w_s}$ that are sensitive to ${\rm S}_{ij}(q), {\rm T}_{ab}(r)$ as well as the distribution of momenta in the $\varphi_3$ nodal diagrams. This fact can be easily verified by keeping \eqref{luna22} fixed but changing \eqref{luna7} to \eqref{luna10}. The diagrams from the $\varphi_3$ sector contributing now may be represented by:
\bg\label{luna33}
& \sum\limits_{l_0}
\begin{pmatrix}
p \\
l_0
\end{pmatrix}
\left(
\begin{tikzpicture}[baseline={(0, 0cm-\MathAxis pt)}, thick,
main/.style = {draw, circle, fill=black, minimum size=6pt},
dot/.style={inner sep=0pt,fill=black, circle, minimum size=3pt},
ball/.style={ball  color=white, circle,  minimum size=15pt},
dots/.style={inner sep=0pt,fill=black, circle, minimum size=1pt}] 
  \node[dot] (1) at (0, 0) {};
  \node[ball] (11) at (-1.5, 0) {i};
  \node[dot] (111) at (-2.4, 0) {};
  \node[main] (1111) at (-2.8, 0.5) {};
  \node[main] (1112) at (-2.8, -0.5) {};
  \node[main] (3) at (40:1) {};
  \node[main] (4) at (15:1) {};
  \node[main] (5) at (-15:1) {};
  \node[main] (6) at (-40:1) {};
  \node[dots] (0) at (-25: 0.7) {};
  \node[dots] (00) at (-35: 0.7) {};
\draw (1) -- node[midway, above right, pos=1] {\tiny{$(q-2) k_i$}} (11);
\draw (11) -- node[midway, above right, pos=1] {\tiny{$2 k_j$}} (111);
\draw (111) -- node {} (1111);
\draw (111) -- node {} (1112);
\draw (1) -- node {}  (3);
\draw (1) -- node {}  (4);
\draw (1) -- node {}  (5);
\draw (1) -- node {}  (6);
\end{tikzpicture}  \; ; \; {\rm S}_{ij}^{n+l_0} (q)
\right) \; \otimes \nonumber
\\
& \otimes \;  
\left( 
\frac{1}{V^2} \sum\limits_{a, b} \; 
\begin{tikzpicture}[baseline={(0, 0cm-\MathAxis pt)}, thick,
main/.style = {draw, circle, fill=black, minimum size=6pt},
dot/.style={inner sep=0pt,fill=black, circle, minimum size=3pt},
ball/.style={ball  color=white, circle,  minimum size=15pt},
dots/.style={inner sep=0pt,fill=black, circle, minimum size=1pt}] 
  \node[dot] (1) at (0, 0) {};
  \node[dot] (11) at (-1.5, 0) {};
  \node[dot] (111) at (-2.4, 0) {};
  \node[main] (1111) at (-2.8, 0.5) {};
  \node[main] (1112) at (-2.8, -0.5) {};
  \node[main] (3) at (40:1) {};
  \node[main] (4) at (15:1) {};
  \node[main] (5) at (-15:1) {};
  \node[main] (6) at (-40:1) {};
  \node[dots] (0) at (-25: 0.7) {};
  \node[dots] (00) at (-35: 0.7) {};
\draw (1) -- node[midway, above right, pos=1] {\tiny $(r-2) l_b$} (11);
\draw (11) -- node[midway, above right, pos=1] {\tiny $2 l_a$} (111);
\draw (111) -- node {} (1111);
\draw (111) -- node {} (1112);
\draw (1) -- node {}  (3);
\draw (1) -- node {}  (4);
\draw (1) -- node {}  (5);
\draw (1) -- node {}  (6);
\end{tikzpicture}  \; ; \; {\rm U}_{ab}^{m+p-l_0} (r)
\right) \nonumber
\\
& \otimes \; \left(
\begin{tikzpicture}[baseline={(0, 0cm-\MathAxis pt)}, thick,
main/.style = {draw, circle, fill=black, minimum size=6pt},
dot/.style={inner sep=0pt,fill=black, circle, minimum size=3pt},
dots/.style={inner sep=0pt,fill=black, circle, minimum size=1pt}] 
  \node[dot] (1) at (0, 0) {};
  \node[dot] (11) at (-1.5, 0) {};
  \node[main] (111) at (-2.5, 0) {};
  \node[main] (3) at (40:1) {};
  \node[main] (4) at (15:1) {};
  \node[main] (5) at (-15:1) {};
  \node[main] (6) at (-40:1) {};
  \node[dots] (0) at (-25: 0.7) {};
  \node[dots] (00) at (-35: 0.7) {};
  \node[dots] (01) at (-2, 0) {};
  \node[dots] (02) at (-2, -0.3) {};
\draw (1) -- node[midway, above right, pos=1] {\tiny $(s-1) f_l$} (11);
\draw (11) [dashed] -- node[below=2mm] {\tiny ${\bar{\gamma}^{n_0} (-(s-1) f_l - {\rm S}_{ij} (q) - {\rm U}_{ab} (r))\over ( (s-1)f_l + {\rm S}_{ij} (q) + {\rm U}_{ab} (r))^{2n_0}}$} (111);
\draw (1) -- node {}  (3);
\draw (1) -- node {}  (4);
\draw (1) -- node {}  (5);
\draw (1) -- node {}  (6);
\draw (02)[->] -- node {} (01);
\end{tikzpicture} \;  + \; 
\begin{tikzpicture}[baseline={(0, 0cm-\MathAxis pt)}, thick,
main/.style = {draw, circle, fill=black, minimum size=6pt},
dot/.style={inner sep=0pt,fill=black, circle, minimum size=3pt},
dots/.style={inner sep=0pt,fill=black, circle, minimum size=1pt}] 
  \node[dot] (1) at (0, 0) {};
  \node[dot] (11) at (-1.5, 0) {};
  \node[main] (111) at (-3, 0) {};
  \node[dot] (112) at (-1.5, 1) {};
  \node[main] (1121) at (-2, 1.5) {};
  \node[main] (1122) at (-1, 1.5) {};
  \node[main] (3) at (40:1) {};
  \node[main] (4) at (15:1) {};
  \node[main] (5) at (-15:1) {};
  \node[main] (6) at (-40:1) {};
  \node[dot] (0) at (-25: 0.7) {};
  \node[dot] (00) at (-35: 0.7) {};
  \node[dots] (01) at (-2, 0) {};
  \node[dots] (02) at (-2, -0.3) {};
\draw (1) -- node[midway, above right, pos=1] {\tiny $(s - 3) f_g$} (11);
\draw (11) [dashed] -- node[below=2mm] {\tiny ${\bar{\gamma}^{n_0} (-2f_h -(s-3) f_g - {\rm S}_{ij} (q) - {\rm U}_{ab} (r))\over (2f_h + (s-3)f_g + {\rm S}_{ij} (q) + {\rm U}_{ab} (r))^{2n_0}}$} (111);
\draw (11) -- node[midway, left, pos=0.8] {\tiny $2f_h$} (112);
\draw (112) -- node[midway, left, pos=0.8] {} (1121);
\draw (112) -- node[midway, left, pos=0.8] {} (1122);
\draw (1) -- node {}  (3);
\draw (1) -- node {}  (4);
\draw (1) -- node {}  (5);
\draw (1) -- node {}  (6);
\draw (02)[->] -- node {} (01);
\end{tikzpicture} \;  + ... \right. \nonumber
\\
&\left. \; ... + \; 
\begin{tikzpicture}[baseline={(0, 0cm-\MathAxis pt)}, thick,
main/.style = {draw, circle, fill=black, minimum size=6pt},
dot/.style={inner sep=0pt,fill=black, circle, minimum size=3pt},
dots/.style={inner sep=0pt,fill=black, circle, minimum size=1pt}] 
  \node[dot] (1) at (0, 0) {};
  \node[main] (11) at (-2.5, 0) {} ;
  \node[dot] (13) at (0, 0.8) {};
  \node[dot] (14) at (0, 1.5) {};
  \node[dot] (15) at (0, -1.2) {};
  \node[main] (121) at (1, 0) {};
  \node[main] (131) at (1, 0.8) {};
  \node[main] (141) at (1, 1.5) {};
  \node[main] (151) at (1, -1.2) {};
  \node[dots] (01) at (0, -0.3) {};
  \node[dots] (01) at (0, -0.7) {};
  \node[dots] (01) at (0, -1) {};
  \node[dots] (02) at (-1.5, 0) {};
  \node[dots] (03) at (-1.5, -0.3) {};
  \node[dots] (04) at (-2.4, 0) {};
\draw (1) [dashed] -- node {} (04);
\draw (04) [dashed] -- node [below=3mm] {\tiny$\frac{\bar{\gamma}^{n_0} \left(-\sum\limits_{p=1}^{s-1} f_{h_p} - {\rm S}_{ij} (q) - {\rm U}_{ab} (r)\right)}{\left(\sum\limits_{p=1}^{s-1} f_{h_p} + {\rm S}_{ij} (q) + {\rm U}_{ab} (r)\right)^{2n_0}}$} (11);
\draw (1) -- node[midway, above right, pos=0.2] {\tiny $f_{h_3}$} (121);
\draw (13) -- node[midway, above right, pos=0.2] {\tiny $f_{h_2}$} (131);
\draw (14) -- node[midway, above right, pos=0.2] {\tiny $f_{h_1}$} (141);
\draw (15) -- node[midway, below right, pos=0] {\tiny $f_{h_{s-1}}$} (151);
\draw (1) -- node {} (13);
\draw (13) -- node {} (14);
\draw (03)[->] -- node {} (02);
\end{tikzpicture} 
\right)
\nd
where ${\rm U}_{ab}(r) \equiv 2l_a + (r-2)l_b$. It is now easy to compare the two set of diagrams from \eqref{luna29} and \eqref{luna30}, and we will only consider the $\varphi_3$ nodal diagrams with all unequal momenta. The key difference comes from the $\overline\gamma/f^2_{w_s}$ factors that take the form:

{\footnotesize
\bg\label{milenj}
{\overline\gamma^{n_o}\Big(-\sum\limits_{p=1}^{s-1}f_{h_p} - {\rm S}_{ij}(q) - {\rm T}_{ab}(r)\Big) \over 
~~~~~\Big(\sum\limits_{p=1}^{s-1}f_{h_p} + {\rm S}_{ij}(q) + {\rm T}_{ab}(r)\Big)^{2n_o}} ~~~~{\rm and} ~~~~{\overline\gamma^{n_o}\Big(-\sum\limits_{p=1}^{s-1}f_{h_p} - {\rm S}_{ij}(q) - {\rm U}_{ab}(r)\Big) \over 
~~~~~\Big(\sum\limits_{p=1}^{s-1}f_{h_p} + {\rm S}_{ij}(q) + {\rm U}_{ab}(r)\Big)^{2n_o}}, \nd}
respectively, where ${\rm S}_{ij}(q), {\rm T}_{ab}(r)$ and ${\rm U}_{ab}(r)$ are as defined earlier. The deviation from the product structure appears precisely from the different functional forms for ${\rm T}_{ab}(r)$ and ${\rm U}_{ab}(r)$. In fact, if we take a different nodal diagram from the $\varphi_1$ sector, say for example \eqref{luna25}, even the function 
${\rm S}_{ij}(q)$ needs to be changed to ${\rm S}_{mnl}(q) \equiv k_m + k_n + (q - 2)k_l$. Therefore we see that only in the limit \eqref{perfume}, if we can Taylor expand $\overline\gamma$ and $f^2_{w_s}$, then the product structure of \eqref{2summers} ensues.

Interestingly, whether or not we demand a product structure {\it a la} 
\eqref{2summers}, we see that to the first order in the coupling constant
$c_{nmpqrs}$ for every nodal diagrams from the $\varphi_1$ sector there are {at most} $\sum\limits_{j = 1}^r j!\sum\limits_{t = 1}^{s-1}t!$ contributions from the $\varphi_2$ and $\varphi_3$ sectors. The highest 
contribution of $r!(s-1)!$ comes from the nodal diagrams with completely unequal momenta from both the sectors respectively\footnote{The $r!$ and $(s-1)!$ growth of the number of terms in the $\varphi_2$ and $\varphi_3$ sectors respectively should not be confused with the spurious combinatorial factors of $r!$ and $(s-1)!$. These have already been removed by the redefinition of the coupling constant as $c_{nmpqrs} \to {c_{nmpqrs}\over q!r!(s-1)!}$ earlier.}. These are of course the diagrams that would respectively dominate over all other diagrams in the individual sectors. 

What happens to ${\cal O}({\rm N})$? We basically want to see how the amplitudes of the nodal diagrams grow when we are in the regime 
\eqref{macharrsn}. Since infinite possible coupling constants of the form 
$c_{nmpqrs}$ for integer values of $(n, ..., s)$ can participate now, and the fact that the nodal diagrams for a given choice of the coupling constant do not allow product structures (or only allow approximate product structures), the situation at hand would appear to be technically challenging. We can simplify this for the time being by resorting to one set of coupling constant (by fixing the values of $(n, ..., s)$). Even for this simplified scenario there would be various subtleties, as we shall see below. 

\subsection{How {\it not} to Borel resum the path integral \eqref{2summers} \label{tagcorfu}}

In the following we will start by discussing how Borel resummation may be done for the path integral \eqref{2summers}. However as we will soon see, the scenario is much more subtle and a naive implementation of the resurgence idea will lead to certain contradictions. It is however instructive to follow this route to point out some subtleties with the nodal diagrams 
that we would have inadvertently missed otherwise. 

The growth of the nodal diagrams to ${\cal O}({\rm N})$, for the simpler case when we allow only one set of coupling constant $c_{nmpqrs}$ with 
fixed choices of $(n, m, ..., s)$ and an approximate product structure 
as in \eqref{2summers}, is controlled by the terms in the expansion 
\eqref{ivabst2} that have completely unequal distribution of the 
momenta\footnote{This means there would be no sum over $(n, .., s)$. We will stick to this choice for illustrative purposes here and elaborate on the generic case in section \ref{sec4.4}.}. Alternatively, this means that the growth of the terms will be controlled by the combinatorial factors from the usual ${\rm N}^{\rm th}$ order expansion of \eqref{ivabst2} $-$ including similar ${\rm N}^{\rm th}$ order expansions from the $\varphi_2$ and $\varphi_3$ sectors $-$ as well as the actual growths of the amplitudes of the corresponding nodal diagrams. Combining these together, we expect the number of terms ${\cal N}({\rm N}; q, r, s)$ contributing at ${\rm N}^{\rm th}$ order from the three sectors to the numerator of the path integral \eqref{2summers} can be expressed 
in the following way:

{\footnotesize
\bg\label{meypach}
{\cal N}({\rm N}; q, r, s) \equiv {1\over {\rm N}!} \cdot  \sum_{a_1 = 0}^{{\rm N}q} c^{(1)}_{a_1}\left({\rm N}q + 1 - a_1\right)!\sum_{a_2 = 0}^{{\rm N}r - 1}c^{(2)}_{a_2}\left({\rm N}r - a_2\right)! \sum_{a_3 = 1}^{{\rm N}(s - 1) -1}c^{(3)}_{a_3}\left({\rm N}(s-1) - a_3\right)!, \nd}
where the first term is from the exponential factor in the path integral 
\eqref{lincoln}, the coefficients $c^{(i)}_{a_i}$ for $i = 1, 2, 3$ are the {standard} combinatorial factor that we expect when we expand, for example \eqref{ivabst2} from $\varphi_1$ sector and similar ones from $(\varphi_2, \varphi_3)$ sectors, to ${\rm N}^{\rm th}$ order. The remain three set of terms are the {\it presumed} 
growths of the amplitudes of the nodal diagrams. The combinatorial factors $c^{(i)}_{a_i}$ should also have volume suppressions (see discussions above). 
It is then clear that the dominant contributions would come from $c^{(i)}_{0} = {\rm N}!$, and we get:
\bg\label{fakhost}
{\rm dom}\Big[{\cal N}({\rm N}; q, r, s)\Big] \equiv {1\over {\rm N}!} \cdot \left({\rm N}!\right)^3 \left({\rm N}q + 1\right)!\left({\rm N}r\right)! \left({\rm N}(s-1)\right)!, \nd
which is again motivated by the approximate product structure of the nodal amplitudes in \eqref{2summers}. Either of these growths, even if we ignore the usual combinatorial factors $c^{(i)}_{a_i}$, cannot lead to a convergent series in the path integral \eqref{2summers} for both the numerator and the denominator. Incidentally, the denominator of the path integral \eqref{2summers} will also grow as \eqref{meypach} with the only difference being the replacement ${\rm N} q + 1 ~ \to ~ {\rm N}q$. Such asymptotic growth of the amplitudes of the nodal diagrams means that we have use the technique of Borel resummation of both the numerator and the denominator of the path integral \eqref{2summers}. 

Few re-definitions will ease the computational process. As we noticed above, there are three sectors whose nodal diagrams are captured by 
$\mathcal{N}^{(1)}_{nmp}(k; q)$ from \eqref{luna29} for $\varphi_1$, $\mathcal{N}^{(2)}_{nmp}(\mu; r)$ from \eqref{jul13crisis} for $\varphi_2$, and 
$\mathcal{N}^{(3)}_{nmp}(\mu; s)$ from \eqref{luna31} for $\varphi_3$. The overall coupling constant is $c_{nmpqrs}$, which may be distributed over the three sectors as:
\bg\label{kelirely}
c_{nmpqrs} \equiv c^{(1)}_{nq} c^{(2)}_{mr} c^{(3)}_{ps} = g_1 g_2 g_3, \nd
where $g_i$, or equivalently $c^{(i)}_{..}$, is related to the sector with field $\varphi_i$. This means, to ${\rm N}^{\rm th}$ order, we are basically looking at interactions with coupling $g_1^{\rm N}g_2^{\rm N} g_3^{\rm N}$ in the path integral \eqref{2summers}. It is also easy to see that if we extract out the dominant growth \eqref{fakhost} from \eqref{meypach}, we are in principle looking at the series that takes the form:

{\footnotesize
\bg\label{pachgond}
{\cal N}({\rm N}; q, r, s) & = & {\left({\rm N}r\right)! \over {\rm N}!}
\left[c^{(2)}_0 + {c^{(2)}_1\over {\rm N}r} + 
{c^{(2)}_2\over {\rm N}r\left({\rm N}r - 1\right)} + 
{c^{(2)}_3\over {\rm N}r \left({\rm N}r - 1\right)\left({\rm N}r - 2\right)} + ..... + {c^{(2)}_{{\rm N}r -1}\over \left({\rm N}r\right)!}\right]\\
& \times & {\left({\rm N}(s-1)\right)!}
\left[c^{(3)}_0 + {c^{(3)}_1\over {\rm N}(s-1)} + 
{c^{(3)}_2\over {\rm N}(s-1)\left({\rm N}(s-1) - 1\right)} + 
 ..... + {c^{(3)}_{{\rm N}(s-1) -1}\over \left({\rm N}(s-1)\right)!}\right]
\nonumber\\
& \times & \left({\rm N}q + 1\right)!
\left[c^{(1)}_0 + {c^{(1)}_1\over {\rm N}q + 1} + 
{c^{(1)}_2\over {\rm N} q\left({\rm N}q + 1\right)} + 
{c^{(1)}_3\over {\rm N} q \left({\rm N}q + 1\right)\left({\rm N}q - 1\right)} + ..... + {c^{(1)}_{{\rm N}q}\over \left({\rm N}q + 1\right)!}\right],\nonumber \nd}
with $c_0^{(i)} = {\rm N}!$ with no volume suppression, but the other 
$c^{(i)}_k$ with $k > 0$ and $k \in \mathbb{Z}_+$ all have both the combinatorial factors and volume suppression. In principle we should take 
${\rm V} \to \infty$ to covert all the summations to integrals. But due to the underlying UV/IR mixing we have to impose an IR cut-off (the UV is well defined from our choice of the Wilsonian effective action at the scale $\mu << {\rm M}_p$ or $\mu << 1$). This IR cut-off translates to an overall upper limit to the volume ${\rm V}$. Clearly for our previous computations to make sense, we 
use ${\rm V} >> 1$, but {\it not} ${\rm V} \to \infty$. Under this limit, the summations would still be converted to their respective integrals as before, but now we no longer have to worry about the zero momentum issues.
We will come back to this in section \ref{sec4.4} when we connect the amplitudes of the nodal diagrams with the results from the Schwinger-Dyson's equations.

To see the resurgence structure of the path integral \eqref{2summers}, let us start with the numerator. However we now make a small change in the form of the path integral: we will instead take the expectation value as 
$\langle g_1^{1/q}\varphi_1\rangle$ where $g_1 \equiv c^{(1)}_{nq}$ from 
\eqref{kelirely}. The numerator of the path integral may now be expressed as:

{\footnotesize
\bg\label{toronto}
{\rm Num}\langle g_1^{1/q}\varphi_1\rangle & = & \sum_{\rm N} {g_1^{\rm N} g_2^{\rm N} g_3^{\rm N} {\cal N}_1({\rm N}) {\cal N}_2({\rm N})
{\cal N}_3({\rm N}) g_1^{1/q} \over 
\left({\rm N}q + 1\right)! \left({\rm N} r\right)! \left({\rm N}(s-1)\right)!} \prod_{j = 1}^5\int d{\cal S}_j~{\rm exp}\left(-\sum_{i = 1}^5 {\cal S}_i\right) 
{\cal S}^{{\rm N}q + 1}_1 {\cal S}^{{\rm N}r}_2 
{\cal S}^{{\rm N}(s-1)}_3 {\cal S}^{{\rm N}}_4 {\cal S}^{{\rm N}}_5 \nonumber\\
& = & \prod_{j = 1}^5\int d{\cal S}_j~{\rm exp}\left(-{\cal S}_j\right) \sum_{\rm N} {\left({\cal S}_1 g_1^{1/q}\right)^{{\rm N}q + 1}
\left({\cal S}_2 g_2^{1/r}\right)^{{\rm N}r} \left({\cal S}_1 g_3^{1/(s-1)}\right)^{{\rm N}(s-1)}   \over \left({\rm N}q + 1\right)! \left({\rm N} r\right)! \left({\rm N}(s-1)\right)!}
~\left({\cal S}_4{\cal S}_5\right)^{\rm N} {\cal N}_{123}({\rm N}), \nonumber\\ \nd}
where ${\cal N}_i({\rm N})$ for $i = 1, 2, 3$ are the values of the nodal diagrams \eqref{luna29}, \eqref{jul13crisis} and \eqref{luna31} respectively at ${\rm N}^{\rm th}$ order, and ${\cal N}_{123}({\rm N}) \equiv {\cal N}_1({\rm N}) {\cal N}_2({\rm N})
{\cal N}_3({\rm N})$. We have also converted $\left({\rm N}!\right)^2$ as an exponential integral using $({\cal S}_4, {\cal S}_5)$. One of the important assumption we took going from the first line to the second in
\eqref{toronto} is possibility of an exchange between the summation over 
${\rm N}$ and the integral over ${\rm S}_i$. This is already alarming, but not fatal enough for us to discontinue this line of thought. Thus, for the time being we will assume this to be true, and study the consequences emanating from this. To this effect, let us now redefine some of the variables in the following way:
\bg\label{3host}
{\cal S}_1 g_1^{1/q} \equiv {\rm S}_1,~~~~
{\cal S}_2 g_2^{1/r} \equiv {\rm S}_2, ~~~~
{\cal S}_3 g_3^{1/(s-1)} \equiv {\rm S}_3, ~~~~{\cal S}_4 \equiv {\rm S}_4,
~~~~{\cal S}_5 \equiv {\rm S}_5, \nd
and insert in the path integral \eqref{toronto}. Before moving ahead we should point out that, had we not used an extra $g_1^{1/q}$ factor in the 
path integral, we would not have been able to make such clean replacements as \eqref{3host}. Plugging \eqref{3host} in \eqref{toronto}, we get:

{\footnotesize
\bg\label{linetag}
{\rm Num}\langle g_1^{1/q}\varphi_1\rangle = {1\over g}\int \prod_{i = 1}^5 d{\rm S}_i ~{\rm exp}\left(-{{\rm S}_1\over g_1^{1/q}}\right) {\rm exp}\left(-{{\rm S}_2\over g_2^{1/r}}\right) {\rm exp}\left(-{{\rm S}_3\over g_3^{1/(s-1)}}\right)
{\rm exp}\left(-{\rm S}_4 - {\rm S}_5\right)~{\cal B}{\rm Z}\left({\rm S}_1,..., {\rm S}_5\right), \nonumber\\ \nd}
where $g \equiv g_1^{1/q} g_2^{1/r} g_3^{1/(s-1)}$. Note that, expectedly
\eqref{linetag} is the Laplace transformation of the Borel transform
${\cal B}{\rm Z}\left({\rm S}_1,..., {\rm S}_5\right)$. But this is also a {\it non-perturbative} form of the numerator of the path integral 
\eqref{toronto} but with a few key differences and a few new facts. The new facts are the appearance of non-perturbative effects of the form 
${\rm exp}\left(-{{\rm S}_1\over g_1^{1/q}}\right),  {\rm exp}\left(-{{\rm S}_2\over g_2^{1/r}}\right)$ and  ${\rm exp}\left(-{{\rm S}_3\over g_3^{1/(s-1)}}\right)$. Since we expect $q \ge 3, r\ge 3$ and $s \ge 3$, the lowest non-perturbative effects are of the form
${\rm exp}\left(-{{\rm S}_1\over g_1^{1/3}}\right),  {\rm exp}\left(-{{\rm S}_2\over g_2^{1/3}}\right)$, ${\rm exp}\left(-{{\rm S}_3\over g_3^{1/2}}\right)$ etc. Recall that in \cite{desitter2, coherbeta, coherbeta2}, the analysis of the Schwinger-Dyson's equations led us to predict the existence of non-perturbative effects of the form 
${\rm exp}\left(-{1\over \left(g_s/{\rm HH}_o\right)^{1/3}}\right)$. Although the $g_s$ used therein is not directly related to $g_i$ used here, nevertheless, the Borel resummation of the path integral gives a hint that such non-perturbative effects are indeed feasible with our construction. On the other hand, the key differences are the appearance of 
$({\rm S}_4, {\rm S}_5)$ in \eqref{linetag}. They are not accompanied by any inverse couplings, so they are not really non-perturbative per se. In fact they are more like perturbative corrections, and their appearance 
seem to indicate a spurious nature of the combinatorial coefficients $\left({\rm N}!\right)^2$. Finally, the Borel transform is given by:
\bg\label{toronmey}
{\cal B}{\rm Z}\left({\rm S}_1,..., {\rm S}_5\right) \equiv 
\sum_{\rm N}~{{\cal N}_1({\rm N}) {\cal N}_2({\rm N}) {\cal N}_3({\rm N})\over \left({\rm N}q + 1\right)! \left({\rm N} r\right)! \left({\rm N}(s-1)\right)!}~{\rm S}_1^{{\rm N}q+1}
{\rm S}_2^{{\rm N}r} {\rm S}_3^{{\rm N}(s-1)} \left({\rm S}_4{\rm S}_5\right)^{\rm N}, \nd
where ${\rm S}_i$ needs to be further complexified so that we are dealing with five-dimensional complex surfaces, and ${\cal N}_i({\rm N})$ are the values of the nodal diagrams \eqref{luna29}, \eqref{jul13crisis} and 
\eqref{luna31} for $i = 1, 2, 3$ respectively. If we allow no IR cut-off then ${\rm V} \to \infty$ and the dominant diagrams to any order in the coupling constant $g_1 g_2 g_3$ are the ones with momenta all different. In fact these are the ones classified by \eqref{fakhost}. There are now two ways to proceed: one, determine the functional forms for 
${\cal N}_i({\rm N})$, and two, determine the singularity structures in the five-dimensional Borel planes. 

Unfortunately this straight-forward way of implementing resurgence and Borel resummability to the series in the numerator and the denominator of the path integral \eqref{2summers} is plagued with numerous subtleties that we have ignored. In the following let us point them out.

\vskip.1in

\noindent $\bullet$ The non-perturbative corrections appear to take the form ${\rm exp}\Big(-{{\rm S}_1\over g_1^{1/q}}\Big),  {\rm exp}\Big(-{{\rm S}_2\over g_2^{1/r}}\Big)$ and  ${\rm exp}\Big(-{{\rm S}_3\over g_3^{1/(s-1)}}\Big)$ which are clearly sensitive to the momentum conservation choice. As we pointed out earlier, no physical parameters should depend on how we implement momentum conservation in the nodal diagrams, therefore the non-perturbative effects should be independent of this.

\vskip.1in

\noindent $\bullet$ The non-perturbative effects of the form 
${\rm exp}\Big(-{{\rm S}_i\over g_i^{1/p}}\Big)$ with $p \in \mathbb{Z}$
is already problematic because for $g_i << 1$, the effects of 
${\rm exp}\Big(-{{\rm S}_i\over g_i^{1/p}}\Big)$ {\it grow} as 
$p$ grows. This means the non-perturbative effects cannot be controlled 
here, and therefore the trans-series would fail to make sense.

\vskip.1in

\noindent $\bullet$ The splitting of the three couplings $g_1, g_2$ and $g_3$ to construct non-perturbative structure of the form 
\eqref{linetag} appears completely arbitrary. We could have divided the three couplings over the five ${\cal S}_i$ in any fractional ways to allow for non-perturbative effects of the form 
${\rm exp}\Big(-{{\rm S}_k\over g_l^{1/f(q, r, s)}}\Big)$ where 
$k = 1, .., 5$ and $l = 1, 2, 3$ with $f(q, r, s)$ being any function of $(q, r, s)$. Such a distribution would lead to ambiguities, so cannot be the right picture.

\vskip.1in

\noindent $\bullet$ Our analysis leads to a five-dimensional Borel plane, whereas we only have three degrees of freedom (parameterized by $\varphi_i$). This itself is a bit puzzling. The singularities are now classified by co-dimensional one surfaces so that some aspect of the residue theorem may be applied. However since the non-perturbative structure is ambiguous (see discussion above), the appearance of co-dimension one singularities would also become ambiguous. This points 
towards some serious flaws in the picture. 

\vskip.1in

\noindent What went wrong? Clearly we missed out on some subtle facts while preforming the Borel resummation. These facts have to do with the actual values of the nodal diagrams which were represented by 
${\cal N}_i({\rm N})$. Contrary to the usual case with the Feynman diagrams, introducing the {\it amplitudes} of the nodal diagrams will change the outcome of the summation process. Plus there are a few more subtleties that we kept under the rug. These and other details will be elaborated in the following section.

\subsection{How to correctly perform the Borel resummation \label{sec4.2}}

To perform the Borel resummation correctly, we will have to look at the amplitudes of the nodal diagrams from all the three sectors carefully. One would now worry because the amplitudes of the nodal diagrams were all expressed in terms of nested integrals involving $\overline{\alpha}(k), 
\overline{\beta}(k)$ and $\overline{\gamma}(k)$ (including their corresponding propagators $a(k), b(k)$ and $c(k)$ respectively), whose functional forms are yet to be determined. However as we shall see below, this will not be an issue. 

To proceed we will not introduce any IR cut-offs in the system right now and only insert them later. This would mean that the contributing nodal diagrams all have unequal momenta in their respective sectors. (This will lead to some issues along the way, but let us push on. We will rectify them in section \ref{sec4.4}.) We will also assume that the eleven-dimensional momenta may be split into a radial part and a temporal part, and we will ignore the subtleties associated with the internal eight-manifold. Thus, with some abuse of notation, we denote the dimensionless momentum\footnote{See footnote \ref{yaya} and section \ref{sec1.3} for the conventions used here.} $k$ as 
$k = (k_0, {\bf k}) \equiv (k_0, \vert {\bf k}\vert, \{k_\Omega\})$ with
$\{k_\Omega\}$ being the set of angular components. Similar distributions can be arranged for 
$(l_0, {\bf l})$ and $(f_0, {\bf f})$ for $\varphi_1$, $\varphi_2$ and $\varphi_3$ sectors respectively.
More involved choices may be taken without losing much of the physics from the simpler choice, and therefore we will keep the generic picture for later study. We now, using \eqref{nehaprit}, make the following choices for the functional forms for $\alpha(k_0, {\bf k}), \beta(l_0, {\bf l})$ and 
$\gamma(f_0, {\bf f})$ (or $\overline\alpha(k_0, {\bf k}), \overline\beta(l_0, {\bf l})$ and 
$\overline\gamma(f_0, {\bf f})$):

{\scriptsize
\bg\label{greek}
&& \beta(l_0, {\bf l}) = {\rm V} \overline\beta(l_0, {\bf l}) \equiv \sum_{n, m} b_{nm} {\bf H}_n(\mathtt{l}) {\bf H}_m(l_0) = \sum_{n > 0} b_{n0} {\bf H}_n(\mathtt{l}) + \sum_{m > 0} b_{0m} 
{\bf H}_m(l_0) + \sum_{(n, m)> 0} b_{nm} {\bf H}_n(\mathtt{l}) {\bf H}_m(l_0)\nonumber\\
&& \alpha(k_0, {\bf k}) = {\rm V}\overline\alpha(k_0, {\bf k}) \equiv \sum_{n, m} c_{nm} 
{\bf H}_n(\mathtt{k}) {\bf H}_m(k_0) = \sum_{n > 0} c_{n0} {\bf H}_n(\mathtt{k}) + \sum_{m>0} c_{0m} 
{\bf H}_m(k_0) + \sum_{(n, m)> 0} c_{nm} {\bf H}_n(\mathtt{k}) {\bf H}_m(k_0) \nonumber\\
&& \gamma(f_0, {\bf f}) = {\rm V}\overline\gamma(f_0, {\bf f}) \equiv \sum_{n, m} d_{nm} {\bf H}_n(\mathtt{f}) {\bf H}_m(f_0) = \sum_{n > 0} d_{n0} {\bf H}_n(\mathtt{f}) + \sum_{m>0} d_{0m} 
{\bf H}_m(f_0) + \sum_{(n, m)> 0} d_{nm} {\bf H}_n(\mathtt{f}) {\bf H}_m(f_0), \nonumber\\
\nd}
where ${\bf H}_n(x)$ are the Hermite polynomials and, again with some abuse of notations, we have defined 
$\mathtt{k} \equiv \vert {\bf k}\vert$ (with similar definitions for $\mathtt{l}\equiv \vert{\bf l}\vert$ and $\mathtt{f}\equiv \vert{\bf f}\vert$). As discussed after \eqref{lincoln}, we have taken the momentum factors to be dimensionless so all the Hermite polynomials are defined using dimensionless parameters. The coefficients $c_{nm}, b_{nm}$ and $d_{nm}$ are functions of the angular variables and arise from our aforementioned decomposition of the momenta into radial and angular components. Standard useful identities, like:

{\footnotesize
\bg\label{besbest}
&&{\bf H}_n(k^2) = {\bf H}_n(-k_0^2 + \vert{\bf k}\vert^2) = 
2^{-n/2}\sum_{p = 0}^n \left(\begin{matrix} n \\ p\\ \end{matrix}\right)
{\bf H}_{n-p}(-\sqrt{2} k_0^2) {\bf H}_p(\sqrt{2}\mathtt{k}^2)\nonumber\\
&& \partial_{\mathtt{k}}^m {\bf H}_n(\mathtt{k}) \equiv {\bf H}^{(m)}_n(\mathtt{k}) = 2^m m! \left(\begin{matrix} n \\ m\\ \end{matrix}\right){\bf H}_{n-m}(\mathtt{k}), ~~~\mathtt{k}^n \equiv {n!\over 2^n} \sum_{m = 0}^{\lfloor{n\over 2}\rfloor} 
{{\bf H}_{n - 2m}(\mathtt{k})\over m!(n - 2m)!}, \nd}
will be helpful in replacing derivatives over and powers of the arguments of the Hermite polynomials in terms of series of Hermite polynomials themselves. Interestingly ${\bf H}_n^{(n)}(\mathtt{k}) = 2^n n!$, which will be useful soon. We have defined $\mathtt{k} = \vert{\bf k}\vert$ as before, and the first identity in \eqref{besbest} is a statement of Lorentz invariance. In the decomposition \eqref{greek} we do not relate $k_0$ with 
${\bf k}$ so $\overline\alpha(k)$ should actually be interpreted as 
$\overline\alpha(k_0, {\bf k})$ (similar argument goes for $\overline\beta(l)$ and $\overline\gamma(f)$). We could have also made a finer division by using Hermite polynomials for the angular directions 
$\{k_\Omega\}$ but we won't do it here. This means, henceforth, we shall directly assume $({\bf k, l, f})$ to be the radial coordinate and ignore the angular directions (thus assume a slice), to avoid unnecessary clutter.

With these we are ready to compute the amplitudes corresponding to the nodal diagrams. We will start by the nodal diagrams in the $\varphi_2$ sector with momenta $l_{v_j}$. To ${\rm N}^{\rm th}$ order in the coupling, {\it i.e.} to order $g_2^{\rm N}$, the amplitude of the {\it dominant} nodal diagram will take the following form:

{\footnotesize
\bg\label{luna34}
&&  \begin{tikzpicture}[baseline={(0, 0cm-\MathAxis pt)}, thick,
main/.style = {draw, circle, fill=black, minimum size=3pt},
dot/.style={inner sep=0pt,fill=black, circle, minimum size=3pt},
dot1/.style={inner sep=0pt,fill=black, circle, minimum size=2pt},
ball/.style={ball  color=white, circle,  minimum size=15pt},
dots/.style={inner sep=0pt,fill=black, circle, minimum size=1pt}] 
  
  \node[dot] (0) at (0, 0) {};
  \node[dot] (1) at (1, 0) {};
  \node[dot] (2) at (-1, 0) {};
  
  \node[dot] (11) at (1, 2) {};
  \node[dot] (12) at (1, -0.3) {};
  \node[dot] (13) at (1, -2.5) {};
  
  \node[dot] (100) at (4, 2) {};
  \node[dot1] (101) at (4, 2.8) {};
  \node[dot1] (102) at (4, 2.2) {};
  \node[dot1] (103) at (4, 1.6) {};
  \node[main] (111) at (5, 2.8) {};
  \node[main] (112) at (5, 2.2) {};
  \node[main] (113) at (5, 1.6) {};
  \node[dots] (1dd) at (4.5, 2) {};
  \node[dots] (1dd) at (4.5, 1.85) {};
  \node[dots] (1dd) at (4.5, 1.7) {};
  
  \node[dot] (120) at (4, -0.3) {};
  \node[dot1] (121) at (4, 0.5) {};
  \node[dot1] (122) at (4, -0.1) {};
  \node[dot1] (123) at (4, -0.8) {};
  \node[main] (1211) at (5, 0.5) {};
  \node[main] (1222) at (5, -0.1) {};
  \node[main] (1233) at (5, -0.8) {};
  \node[dots] (1dd) at (4.5, -0.3) {};
  \node[dots] (1dd) at (4.5, -0.4) {};
  \node[dots] (1dd) at (4.5, -0.5) {};
  \node[dots] (1dd) at (4.5, -0.6) {};
  
  \node[dot] (130) at (4, -2.5) {};
  \node[dot1] (131) at (4, -1.7) {};
  \node[dot1] (132) at (4, -2.3) {};
  \node[dot1] (133) at (4, -3.1) {};
  \node[main] (1311) at (5, -1.7) {};
  \node[main] (1322) at (5, -2.3) {};
  \node[main] (1333) at (5, -3.1) {};
  \node[dots] (1dd) at (4.5, -2.5) {};
  \node[dots] (1dd) at (4.5, -2.7) {};
  \node[dots] (1dd) at (4.5, -2.8) {};
  \node[dots] (1dd) at (4.5, -2.9) {};

  \node[dot] (21) at (-1, 2) {};
  \node[dot] (22) at (-1, -0.3) {};
  \node[dot] (23) at (-1, -2.5) {};
  
  \node[dot] (200) at (-4, 2) {};
  \node[dot1] (201) at (-4, 2.8) {};
  \node[dot1] (202) at (-4, 2.2) {};
  \node[dot1] (203) at (-4, 1.6) {};
  \node[main] (211) at (-5, 2.8) {};
  \node[main] (212) at (-5, 2.2) {};
  \node[main] (213) at (-5, 1.6) {};
  \node[dots] (2dd) at (-4.5, 2) {};
  \node[dots] (2dd) at (-4.5, 1.85) {};
  \node[dots] (2dd) at (-4.5, 1.7) {};
  
  \node[dot] (220) at (-4, -0.3) {};
  \node[dot1] (221) at (-4, 0.5) {};
  \node[dot1] (222) at (-4, -0.1) {};
  \node[dot1] (223) at (-4, -0.8) {};
  \node[main] (2211) at (-5, 0.5) {};
  \node[main] (2222) at (-5, -0.1) {};
  \node[main] (2233) at (-5, -0.8) {};
  \node[dots] (2dd) at (-4.5, -0.3) {};
  \node[dots] (2dd) at (-4.5, -0.4) {};
  \node[dots] (2dd) at (-4.5, -0.5) {};
  \node[dots] (2dd) at (-4.5, -0.6) {};
  
  \node[dot] (230) at (-4, -2.5) {};
  \node[dot1] (231) at (-4, -1.7) {};
  \node[dot1] (232) at (-4, -2.3) {};
  \node[dot1] (233) at (-4, -3.1) {};
  \node[main] (2311) at (-5, -1.7) {};
  \node[main] (2322) at (-5, -2.3) {};
  \node[main] (2333) at (-5, -3.1) {};
  \node[dots] (2dd) at (-4.5, -2.5) {};
  \node[dots] (2dd) at (-4.5, -2.7) {};
  \node[dots] (2dd) at (-4.5, -2.8) {};
  \node[dots] (2dd) at (-4.5, -2.9) {};
\draw (0)  -- node[midway, above right, pos=0.8] {} (1);
\draw (0)  -- node[midway, above right, pos=0.8] {} (2);

\draw (1) -- node {} (11);
\draw (1) -- node {} (12);
\draw (12) [dashed]--  node {} (13);
\draw (121) -- node {} (1211);
\draw (122) -- node {} (1222);
\draw (123) -- node {} (1233);
\draw (121) -- node{} (123);

\draw (13) -- node[midway, above right, pos=0] {\tiny$\zeta_{\rm N} = \sum\limits_{j =({\rm N}-1)r}^{{\rm N}r} l_{v_j}$} (130);
\draw (131) -- node{} (133);
\draw (131) -- node {} (1311);
\draw (132) -- node {} (1322);
\draw (133) -- node {} (1333);

\draw (11) -- node[midway, above right, pos=0] {\tiny$\zeta_{p + 1} = \sum\limits_{j = p+r+1}^{p + 2r} l_{v_j}$} (100);
\draw (101) -- node{} (103);
\draw (101) -- node{} (111);
\draw (102) -- node{} (112);
\draw (103) -- node{} (113);

\draw (12) -- node[midway, above right, pos=0] {\tiny$\zeta_{p + 2} = \sum\limits_{j = p+2r+1}^{p + 3r} l_{v_j}$} (120);

\draw (2) -- node {} (21);
\draw (21) -- node {} (22);
\draw (22) [dashed]-- node {} (23);

\draw (201) -- node {} (211);
\draw (202) -- node {} (212);
\draw (203) -- node {} (213);
\draw (201) -- node{} (203);
\draw (21) -- node[midway, above right, pos=0.8] {\tiny$\zeta_1 = \sum\limits_{j = 1}^r l_{v_j}$} (200);

\draw (221) -- node {} (2211);
\draw (222) -- node {} (2222);
\draw (223) -- node {} (2233);
\draw (221) -- node{} (223);
\draw (22) -- node[midway, above right, pos=0.8] {\tiny$\zeta_2 = \sum\limits_{j = r + 1}^{2r} l_{v_j}$} (220);

\draw (231) -- node {} (2311);
\draw (232) -- node {} (2322);
\draw (233) -- node {} (2333);
\draw (231) -- node{} (233);
\draw (23) -- node[midway, above right, pos=0.8] {\tiny$\zeta_p = \sum\limits_{j = p}^{p + r} l_{v_j}$} (230);
\end{tikzpicture} 
\nonumber\\
&& = g_2^{\rm N} \prod_{s = 1}^{{\rm N}r}\left({\pi_s\over b_s}\right)^{1/2} \sum_{\{m_i\}}\mathbb{C}\left(m_1, m_2, m_3, ..., m_{{\rm N}r-1}, m_{{\rm N}r}\right)\nonumber\\
&&\times \int_{k_{\rm IR}}^\mu d^{11}l_1~{\overline\beta(l_1) l_1^{m_1}\over b(l_1)}
\int_{l_1}^\mu d^{11}l_2~{\overline\beta(l_2) l_2^{m_2}\over b(l_2)}....
\int_{l_{r-1}}^\mu d^{11}l_r~{\overline\beta(l_r) l_r^{m_r}\over b(l_r)}....
\int_{l_{2r-1}}^\mu d^{11}l_{2r}~{\overline\beta(l_{2r}) l_{2r}^{m_{2r}}\over b(l_{2r})}....\nonumber\\
&& \times \int_{l_{{\rm N}r-2}}^\mu d^{11}l_{{\rm N}r - 1}~{\overline\beta(l_{{\rm N}r - 1}) l_{{\rm N}r - 1}^{m_{{\rm N}r - 1}}\over b(l_{{\rm N}r - 1})}
\int_{l_{{\rm N}r-1}}^\mu d^{11}l_{{\rm N}r}~{\overline\beta(l_{{\rm N}r}) l_{{\rm N}r}^{m_{{\rm N}r}}\over b(l_{{\rm N}r})} + ~~\left(({\rm N}r)! - 1\right)~{\rm terms},
\nd}
where $\mathbb{C}\left(m_1, m_2, m_3, ..., m_{{\rm N}r-1}, m_{{\rm N}r}\right)$ is the combinatoric coefficient that may be derived from 
expanding $\prod\limits_{i = 1}^{\rm N}\zeta^r_i$ in the nodal diagram with $\zeta_i$ as in the nodal diagram of \eqref{luna34}. According to footnote \ref{loraseri}, powers of momenta $l_i^j$ accompanying $\overline\beta(l_i)$ may even include generalizations of the interaction \eqref{angelslune} once the combinatoric factor is appropriately defined. For the present analysis we will stick with the simpler case of \eqref{angelslune} and integrate over the nested integral
by substituting the form of $\overline\beta(l_0, {\bf l})$ from 
\eqref{greek}. The integrations may be explicitly performed since we know the form of the propagator $b(l) \equiv l^2$ and the form of the Hermite polynomials from \eqref{besbest}:
\bg\label{nightangle}
{\bf H}_n({\bf l}) \equiv {n!} \sum_{m = 0}^{\lfloor{n\over 2}\rfloor} 
{(-1)^m  \over m!(n - 2m)!}~(2{\bf l})^{n -2m}, \nd
for even or odd $n$. A similar formula holds for ${\bf H}_m(l_0)$. Combining the $({\rm N}r)!$ terms under one roof, as elaborated earlier, we see that a particular piece of the chain integrates to:

{\footnotesize
\bg\label{ww84}
\int_{k_{\rm IR}}^\mu d^{11}l_r~{\overline\beta(l_r) l_r^{m_r}\over b(l_r)} &=&
\sum_{\cal C} {(-1)^{p+q} 2^{n + m - 2p -2q}n! m!\over 
p! q!(n - 2p)! (m - 2q)!}\cdot{{\vert\mu\vert}^{n - 2p + m_r +8}\mu_0^{m - 2q + 1}\over (m - 2q+1)(n - 2p + m_r + 8)}\left(\int b_{nm} d\Omega\right) + ...\nonumber\\
& = & \sum_{n = 0}^\infty \sum_{m = 0}^{\lfloor{n\over 2}\rfloor}
{2^{n - 2m}(-1)^m n!  \over m!(n - 2m)!(n - 2m + m_r + 8)}\left(\int b_{n0} d\Omega\right)~\mu_0 \vert\mu\vert^{n - 2m + m_r + 8} + ....., \nonumber\\ \nd}
where $b(l_r) = -l_r^2 \equiv l^2_{r, 0} - {\bf l}^2_r = -{\bf l}^2_r\left(1 - {l^2_{r, 0}\over {\bf l}^2_r}\right)$; and ${\cal C}$ involves sum over $(n, m, p, q)$ such that $n \in [0, \infty], m \in [0, \infty], p \in \left[0, \lfloor{n\over 2}\rfloor\right]$ and $q \in \left[0, \lfloor{m\over 2}\rfloor\right]$. 
In the second line of 
\eqref{ww84} we have isolated one term in the series expansion over 
${\cal C}$ to show the behavior over the energy scale $\mu \equiv (\mu_0, \vert\mu\vert)$. Interestingly, looking at the individual piece in the chain of terms from \eqref{luna34} we see that the amplitude in \eqref{luna34} {\it does not} grow as $({\rm N}r)!$. This means, although there are $({\rm N}r)!$ terms contributing to the nodal diagram, each piece of the nodal diagrams is suppressed by at least ${1\over ({\rm N}r)!}$ such that the system does not show exponential growth\footnote{Since the coefficients $b_{nm}$ from \eqref{greek} are still unknown, there is no clear-cut integral growth of the term \eqref{ww84} that one may combine with other equivalent terms to predict a $({\rm N}r)!$ growth. The combinatorial factor ${\cal C}(m_1, m_2, ..., m_{{\rm N}r})$ doesn't help either because it would at most show a growth of $(r!)^{\rm N}$. It is easy to see that for large ${\rm N}$, $(r!)^{\rm N} << ({\rm N}r)!$ as $({\rm N}r)! \approx (r!)^{\rm N} {\rm N}^{{\rm N}r}$ for large ${\rm N}$ and large $r$.\label{deva}}.

Similar story will unfold for the $\varphi_3$ sector by using the Hermite polynomial expansion from \eqref{greek}. The analysis therein will involve {\it derivatives} of $\overline\gamma(f)$, as seen from \eqref{luna15} till \eqref{luna18}. Derivatives of the Hermite polynomials can be replaced by other Hermite polynomials along the lines of \eqref{besbest}, so the analysis is similar to what we did for the $\varphi_2$ sector. Again, a careful study reveals that the amplitudes of the nodal diagrams do not grow as $({\rm N}(s-1))!$
way, despite the increase in the number of diagrams being $({\rm N}(s-1))!$. This subtlety at least partially resolves the conundrum we encountered before regarding the placement of the momentum conservation 
condition. Of course we haven't checked the $\varphi_1$ sector, so it will all depend how the amplitudes in that sector behave.

What happens in the $\varphi_1$ sector? This sector is different from the other two sectors because of the presence of the source field. This introduces a new ingredient $-$ namely the wave-function of the source field $-$ in the amplitudes of the nodal diagrams as could be seen from \eqref{luna20} till \eqref{luna25}. The amplitude of the most dominant nodal diagram to ${\cal O}({\rm N})$ then takes the following form:

{\footnotesize
\bg\label{luna35}
&& 
\begin{tikzpicture}[baseline={(0, 0cm-\MathAxis pt)}, thick,
main/.style = {draw, circle, fill=black, minimum size=3pt},
dot/.style={inner sep=0pt,fill=black, circle, minimum size=3pt},
dot1/.style={inner sep=0pt,fill=black, circle, minimum size=2pt},
ball/.style={ball  color=white, circle,  minimum size=15pt},
dots/.style={inner sep=0pt,fill=black, circle, minimum size=1pt}] 
  
  \node[ball] (0) at (0, 0) {i};
  \node[dot] (1) at (1, 0) {};
  \node[dot] (2) at (-1, 0) {};
  
  \node[dot] (11) at (1, 2) {};
  \node[dot] (12) at (1, -0.3) {};
  \node[dot] (13) at (1, -2.5) {};
  
  \node[dot] (100) at (4, 2) {};
  \node[dot1] (101) at (4, 2.8) {};
  \node[dot1] (102) at (4, 2.2) {};
  \node[dot1] (103) at (4, 1.6) {};
  \node[main] (111) at (5, 2.8) {};
  \node[main] (112) at (5, 2.2) {};
  \node[main] (113) at (5, 1.6) {};
  \node[dots] (1dd) at (4.5, 2) {};
  \node[dots] (1dd) at (4.5, 1.85) {};
  \node[dots] (1dd) at (4.5, 1.7) {};
  
  \node[dot] (120) at (4, -0.3) {};
  \node[dot1] (121) at (4, 0.5) {};
  \node[dot1] (122) at (4, -0.1) {};
  \node[dot1] (123) at (4, -0.8) {};
  \node[main] (1211) at (5, 0.5) {};
  \node[main] (1222) at (5, -0.1) {};
  \node[main] (1233) at (5, -0.8) {};
  \node[dots] (1dd) at (4.5, -0.3) {};
  \node[dots] (1dd) at (4.5, -0.4) {};
  \node[dots] (1dd) at (4.5, -0.5) {};
  \node[dots] (1dd) at (4.5, -0.6) {};
  
  \node[dot] (130) at (4, -2.5) {};
  \node[dot1] (131) at (4, -1.7) {};
  \node[dot1] (132) at (4, -2.3) {};
  \node[dot1] (133) at (4, -3.1) {};
  \node[main] (1311) at (5, -1.7) {};
  \node[main] (1322) at (5, -2.3) {};
  \node[main] (1333) at (5, -3.1) {};
  \node[dots] (1dd) at (4.5, -2.5) {};
  \node[dots] (1dd) at (4.5, -2.7) {};
  \node[dots] (1dd) at (4.5, -2.8) {};
  \node[dots] (1dd) at (4.5, -2.9) {};

  \node[dot] (21) at (-1, 2) {};
  \node[dot] (22) at (-1, -0.3) {};
  \node[dot] (23) at (-1, -2.5) {};
  
  \node[dot] (200) at (-4, 2) {};
  \node[dot1] (201) at (-4, 2.8) {};
  \node[dot1] (202) at (-4, 2.2) {};
  \node[dot1] (203) at (-4, 1.6) {};
  \node[main] (211) at (-5, 2.8) {};
  \node[main] (212) at (-5, 2.2) {};
  \node[main] (213) at (-5, 1.6) {};
  \node[dots] (2dd) at (-4.5, 2) {};
  \node[dots] (2dd) at (-4.5, 1.85) {};
  \node[dots] (2dd) at (-4.5, 1.7) {};
  
  \node[dot] (220) at (-4, -0.3) {};
  \node[dot1] (221) at (-4, 0.5) {};
  \node[dot1] (222) at (-4, -0.1) {};
  \node[dot1] (223) at (-4, -0.8) {};
  \node[main] (2211) at (-5, 0.5) {};
  \node[main] (2222) at (-5, -0.1) {};
  \node[main] (2233) at (-5, -0.8) {};
  \node[dots] (2dd) at (-4.5, -0.3) {};
  \node[dots] (2dd) at (-4.5, -0.4) {};
  \node[dots] (2dd) at (-4.5, -0.5) {};
  \node[dots] (2dd) at (-4.5, -0.6) {};
  
  \node[dot] (230) at (-4, -2.5) {};
  \node[dot1] (231) at (-4, -1.7) {};
  \node[dot1] (232) at (-4, -2.3) {};
  \node[dot1] (233) at (-4, -3.1) {};
  \node[main] (2311) at (-5, -1.7) {};
  \node[main] (2322) at (-5, -2.3) {};
  \node[main] (2333) at (-5, -3.1) {};
  \node[dots] (2dd) at (-4.5, -2.5) {};
  \node[dots] (2dd) at (-4.5, -2.7) {};
  \node[dots] (2dd) at (-4.5, -2.8) {};
  \node[dots] (2dd) at (-4.5, -2.9) {};
\draw (0)  -- node[midway, above right, pos=0.8] {} (1);
\draw (0)  -- node[midway, above right, pos=0.8] {} (2);

\draw (1) -- node {} (11);
\draw (1) -- node {} (12);
\draw (12) [dashed]--  node {} (13);
\draw (121) -- node {} (1211);
\draw (122) -- node {} (1222);
\draw (123) -- node {} (1233);
\draw (121) -- node{} (123);

\draw (13) -- node[midway, above right, pos=0] {\tiny$\xi_{\rm N} = \sum\limits_{j =({\rm N}-1)q}^{{\rm N}q} k_{u_j}$} (130);
\draw (131) -- node{} (133);
\draw (131) -- node {} (1311);
\draw (132) -- node {} (1322);
\draw (133) -- node {} (1333);

\draw (11) -- node[midway, above right, pos=0] {\tiny$\xi_{p + 1} = \sum\limits_{j = p+q+1}^{p + 2q} k_{u_j}$} (100);
\draw (101) -- node{} (103);
\draw (101) -- node{} (111);
\draw (102) -- node{} (112);
\draw (103) -- node{} (113);

\draw (12) -- node[midway, above right, pos=0] {\tiny$\xi_{p + 2} = \sum\limits_{j = p+2q+1}^{p + 3q} k_{u_j}$} (120);

\draw (2) -- node {} (21);
\draw (21) -- node {} (22);
\draw (22) [dashed]-- node {} (23);

\draw (201) -- node {} (211);
\draw (202) -- node {} (212);
\draw (203) -- node {} (213);
\draw (201) -- node{} (203);
\draw (21) -- node[midway, above right, pos=0.8] {\tiny$\xi_1 = \sum\limits_{j = 1}^q k_{u_j}$} (200);

\draw (221) -- node {} (2211);
\draw (222) -- node {} (2222);
\draw (223) -- node {} (2233);
\draw (221) -- node{} (223);
\draw (22) -- node[midway, above right, pos=0.8] {\tiny$\xi_2 = \sum\limits_{j = q + 1}^{2q} k_{u_j}$} (220);

\draw (231) -- node {} (2311);
\draw (232) -- node {} (2322);
\draw (233) -- node {} (2333);
\draw (231) -- node{} (233);
\draw (23) -- node[midway, above right, pos=0.8] {\tiny$\xi_p = \sum\limits_{j = p}^{p + q} k_{u_j}$} (230);
\end{tikzpicture} 
\nonumber\\
&& = \Gamma({\bf k}_1, k_{1, 0}) \equiv g_1^{\rm N} \prod_{s = 1}^{{\rm N}q + 1}\left({\pi_s\over b_s}\right)^{1/2} \sum_{\{m_i\}}\mathbb{D}\left(m_1, m_2, m_3, ..., m_{{\rm N}q-1}, m_{{\rm N}q}\right)\nonumber\\
&&\times ~{\overline\alpha(k_1) k_1^{m_1}\over a(k_1)}
\int_{k_1}^\mu d^{11}k_2~{\overline\alpha(k_2) k_2^{m_2}\over a(k_2)}....
\int_{k_{q-1}}^\mu d^{11}k_q~{\overline\alpha(k_q) k_q^{m_q}\over a(k_q)}....
\int_{k_{2q-1}}^\mu d^{11}k_{2q}~{\overline\alpha(k_{2q}) k_{2q}^{m_{2q}}\over a(k_{2q})}....\nonumber\\
&& \times \int_{k_{{\rm N}q-2}}^\mu d^{11}k_{{\rm N}q - 1}~{\overline\alpha(k_{{\rm N}q - 1}) k_{{\rm N}q - 1}^{m_{{\rm N}q - 1}}\over a(k_{{\rm N}q - 1})}
\int_{k_{{\rm N}q-1}}^\mu d^{11}k_{{\rm N}q}~{\overline\alpha(k_{{\rm N}q}) k_{{\rm N}q}^{m_{{\rm N}q}}\over a(k_{{\rm N}q})}
\int_{k_{{\rm N}q}}^\mu d^{11}k' \psi_{{\bf k}'}({\rm X})e^{-ik'_0t} {\overline\alpha(k')\over 
a(k')} \nonumber\\
&& + ~~\left(({\rm N}q + 1)! - 1\right)~{\rm terms},
\nd}
where $k' = ({\bf k}', k'_0) \equiv k_{{\rm N}q+1}$ and ${\rm X} = ({\bf x}, y, z)$. Note that we did not integrate the outer-most term. This is in accordance with what we discussed earlier, and also in \cite{coherbeta, coherbeta2}, namely, that the amplitudes from the $\varphi_1$ sector may be used to directly match with the Fourier transform of the kind of background that we want (here it is of course the uplift of the four-dimensional de Sitter background in type IIB to M-theory). In this language all the $({\rm N}q + 1)!$ nested integrals are {\it different} and cannot be absorbed into one set of integrals with $({\rm N}q + 1)$-terms. One might then naively think that the number of terms grow as $({\rm N}q + 1)!$, thus leading to the conclusion that we had earlier, at least for this sector.  

This naive way of thinking again misses the actual value of the amplitude of the nodal diagram in \eqref{luna35}. However now we do have a wave-function accompanying one of the term, so the question is whether this could change the outcome of the amplitude. To see this we will study the 
{\it integrated} version of \eqref{luna35}, namely:

{\footnotesize
\bg\label{lorenp}
&&\int_{k_{\rm IR}}^\mu d^{11} k_1 \Gamma({\bf k}_1, k_{1, 0}) \equiv g_1^{\rm N} \prod_{s = 1}^{{\rm N}q + 1}\left({\pi_s\over b_s}\right)^{1/2} \sum_{\{m_i\}}\mathbb{D}\left(m_1, m_2, m_3, ..., m_{{\rm N}q-1}, m_{{\rm N}q}\right)\nonumber\\
&&\times \int_{k_{\rm IR}}^\mu d^{11}k_1 {\overline\alpha(k_1) k_1^{m_1}\over a(k_1)}
\int_{k_{\rm IR}}^\mu d^{11}k_2~{\overline\alpha(k_2) k_2^{m_2}\over a(k_2)}....
\int_{k_{\rm IR}}^\mu d^{11}k_q~{\overline\alpha(k_q) k_q^{m_q}\over a(k_q)}....
\int_{k_{\rm IR}}^\mu d^{11}k_{2q}~{\overline\alpha(k_{2q}) k_{2q}^{m_{2q}}\over a(k_{2q})}....\nonumber\\
&& \times \int_{k_{\rm IR}}^\mu d^{11}k_{{\rm N}q - 1}~{\overline\alpha(k_{{\rm N}q - 1}) k_{{\rm N}q - 1}^{m_{{\rm N}q - 1}}\over a(k_{{\rm N}q - 1})}
\int_{k_{\rm IR}}^\mu d^{11}k_{{\rm N}q}~{\overline\alpha(k_{{\rm N}q}) k_{{\rm N}q}^{m_{{\rm N}q}}\over a(k_{{\rm N}q})}
\int_0^\mu d^{11}k' \psi_{{\bf k}'}({\rm X})e^{-ik'_0t} {\overline\alpha(k')\over 
a(k')}, \nd}
which expectedly combines all the $({\rm N}q + 1)!$ terms under one roof. We now see that the integrals in \eqref{lorenp} splits into two sets: one set contains all the integrals that do not involve the wave-function of the source, and the other set contains the source wave-function. The former set may be easily integrated following \eqref{ww84}, and for the latter we can use the Riemann-Lebesgue lemma:
\bg\label{aug19visa}
\int_a^b d\mathtt{k}~e^{i\mathtt{kx}} f(\mathtt{k}) = \sum_{n = 0}^{{\rm N} - 1}~ (-1)^n ~ e^{i\mathtt{kx}} ~{f^{(n)}(\mathtt{k})\over (i\mathtt{x})^{n+1}}\Bigg\vert_a^b~+ ~ {\cal O}\left({1\over \mathtt{x}^{{\rm N} + 1}}\right), \nd
where $f^{(n)}(\mathtt{k}) = {d^n f(\mathtt{k})\over d\mathtt{k}^n}$ 
and $f^{(0)}(\mathtt{k}) = f(\mathtt{k})$. A sufficient, but not necessary, condition is that $f(\mathtt{k})$ is continuously differentiable for 
$a \le \mathtt{k} \le b$ and $\int_a^b \vert f(\mathtt{k})\vert ~d\mathtt{k} < \infty$. In the limit of $\mathtt{x} \to \infty$, the factor $e^{i\mathtt{kx}}$ oscillates faster and faster such
that $e^{i\mathtt{kx}} f(\mathtt{k})$ averages out to zero over any finite region of $\mathtt{k}$ inside the interval.

Our ans\"atze for $\overline\alpha$ (as well for $\overline\beta$ and 
$\overline\gamma$) remains the same as in \eqref{greek}, but now we do need some specific form for the wave-function $\psi_{\bf k}({\rm X})$. Without loss of generalities, we will replace $\psi_{\bf k}({\rm X})$ by 
$\psi_{k}({\rm X})$, and define the total wave-function in the following way:
\bg\label{creolet}
\psi_{k}({\rm X}) e^{-ik_0t} \equiv {k^2\over \pi \vert\omega\vert {\bf k}^9}
~{\rm exp}\left(-{({\bf k} - \overline{\bf k}_{\rm IR})^2\over \omega^2} - ik_0t\right)~ 
{\cal F}\left({\bf k}, {\rm X}\right), \nd
where ${\cal F}\left({\bf k}, {\rm X}\right)$ solves a Schr\"odinger type equation with a potential fixed by the solitonic background (see 
eq. (1.5) in the second reference of \cite{coherbeta} for details on the Schr\"odinger equation\footnote{The small fluctuations over the solitonic background, which is a supersymmetric Minkowski background that we took here, typically leads to a wave-function of the form:
\bg\label{pugh}
\chi_k({\rm X}, t) = {\cal F}({\bf k}, {\rm X})~e^{-ik_0t}, \nd
such that for ${\rm X} \to \infty$, the function ${\cal F}({\bf k}, {\rm X}) \to e^{i({\bf k}-\overline{\bf k}_{\rm IR}). {\rm X}}$, where ${\rm X} \equiv ({\bf x}, y, z)$ as shown in eq. (1.5) in the second reference of \cite{coherbeta}. For compact internal eight-manifold, parametrized by $(y, z)$ the above limit of large distances is more subtle to implement, but we will not worry too much on this for the analysis presented here as the subtleties are inconsequential.
The wave-function that we took in \eqref{creolet} is the modulated version of the above wave-function. \label{cheer}}.
The choice of $k \equiv (k_0, {\bf k}) = (0, \overline{\bf k}_{\rm IR})$ is made such that the wave-function for non-zero values of $\omega$ approaches a constant. In particular this means ${\cal F}\left({\bf k} = \overline{\bf k}_{\rm IR}, {\rm X}\right)=$ constant and, in the limit $\omega \to 0$, the coefficient of ${k^2\over \pi \vert\omega\vert \mathtt{k}^9}$ is chosen in a way that we get only temporal dependence. Note that, due to the IR-UV mixing we expect $\overline{\bf k}^2_{\rm IR} \ge k^2_{\rm IR}$. (As such $k^2$ in \eqref{creolet} appears to be unnecessary because the delta function makes $k^2 = -k^2_0 + \overline{\bf k}^2_{\rm IR}$. In fact we could have modified \eqref{greek} itself to insert such a coefficient and kept the wave-function simple (see section \ref{sec4.4}). In either case no loss of generalities occur as alluded to above. We will elaborate further on this a little later.) Plugging \eqref{nehaprit}, \eqref{creolet} and \eqref{greek} in the 
wave-function part of \eqref{lorenp}, we get:

{\footnotesize
\bg\label{milady}
&&\lim_{\omega \to 0}\int_{k_{\rm IR}}^\mu {d^{11}k}~{\bf Re}\Big(\psi_k({\rm X}) e^{-ik_0 t}\Big) 
~{\overline\alpha({\bf k}, k_0)\over a(k)} =  
\lim_{\omega \to 0}\int_{k_{\rm IR}}^\mu {d^{11}k\over k^2}~{\bf Re}\Big(\psi_k({\rm X}) e^{-ik_0 t}\Big)~\sum\limits_{n, m} c_{nm} ~{\bf H}_n(\mathtt{k}){\bf H}_m(k_0)\nonumber\\
&& = \sum_{n, m} \left(\int d\Omega~c_{2n,m}~{\bf H}_{2n}(\overline{\bf k}_{\rm IR})\right) 
\left[e^{ik_0 t}\sum_{p = 0}^\m {(-1)^p 2^p m!\over (m - p)!} ~
{{\bf H}_{m - p}(k_0)\over (it)^{p + 1}} + 
e^{-ik_0 t}\sum_{p = 0}^\m {(-1)^p 2^p m!\over (m - p)!} ~
{{\bf H}_{m - p}(k_0)\over (-it)^{p + 1}}
\right]_{\kappa_{\rm IR}}^{\mu_0}, \nonumber\\ \nd}
where only ${\bf k} = \overline{\bf k}_{\rm IR}$ is realized via delta function constraint in the limit $\omega \to 0$, which also restricts one set of Hermite functions (associated with ${\bf k}$) to be even. Any other constants are absorbed in the definition of the coefficients $c_{mn}$. The terms inside the square bracket in \eqref{milady} are universal for any order in ${\rm N}$, as we shall also clarify below. The upper limits for the sum over $p$ is $m$, so as we go to higher values of $m$, the number of terms would increase proportionately. 

The other ${\rm N}q$ number of integrals in \eqref{lorenp} may be easily integrated out following \eqref{ww84} as mentioned earlier. We can combine these together with the coefficients in the round brackets in \eqref{milady} to define a new coefficient that reflects the {\it order} of the expansion (here it being ${\rm N}$):

{\footnotesize
\bg\label{altomaa}
\mathbb{C}^{({\rm N})}_m &\equiv & 
\prod_{t = 1}^{{\rm N}q + 1}\left({\pi_{t}\over b_{t}}\right)^{1/2} \sum_{\{m_i\}}\mathbb{D}\left(m_1, m_2, m_3, ..., m_{{\rm N}q-1}, m_{{\rm N}q}\right)\mathbb{D}_{\varphi_2}({\rm N}r) \mathbb{D}_{\varphi_3}({\rm N}(s-1))\\
&\times & \int_{k_{\rm IR}}^\mu d^{11}k_1 {\overline\alpha(k_1) k_1^{m_1}\over a(k_1)}
\int_{k_{\rm IR}}^\mu d^{11}k_2~{\overline\alpha(k_2) k_2^{m_2}\over a(k_2)}....
\int_{k_{\rm IR}}^\mu d^{11}k_q~{\overline\alpha(k_q) k_q^{m_q}\over a(k_q)}....
\int_{k_{\rm IR}}^\mu d^{11}k_{2q}~{\overline\alpha(k_{2q}) k_{2q}^{m_{2q}}\over a(k_{2q})}....\nonumber\\
& \times & \int_{k_{\rm IR}}^\mu d^{11}k_{{\rm N}q - 1}~{\overline\alpha(k_{{\rm N}q - 1}) k_{{\rm N}q - 1}^{m_{{\rm N}q - 1}}\over a(k_{{\rm N}q - 1})}
\int_{k_{\rm IR}}^\mu d^{11}k_{{\rm N}q}~{\overline\alpha(k_{{\rm N}q}) k_{{\rm N}q}^{m_{{\rm N}q}}\over a(k_{{\rm N}q})}
\sum_{n = 1}^\infty \left(\int d\Omega~c_{2n,m}~{\bf H}_{2n}(\overline{\bf k}_{\rm IR})\right), \nonumber
\nd}
where $\mathbb{D}\left(m_1, m_2, m_3, ..., m_{{\rm N}q-1}, m_{{\rm N}q}\right)$ is a combinatorial factor that has been defined earlier. However there are also the set of ${\rm N}r$ and 
${\rm N}(s-1)$ number of integrals from the $\varphi_2$ and $\varphi_3$ sectors respectively. These have been accommodated in $\mathbb{D}_{\varphi_2}({\rm N}r) \mathbb{D}_{\varphi_3}({\rm N}(s-1))$.
Since both these have combinatoric factors of ${\rm N}!$, $\mathbb{C}^{\rm N}_m$ grows\footnote{As alluded to in footnote \ref{deva}, the terms in 
\eqref{altomaa} are still controlled by as yet unknown coefficients 
$b_{mn}, c_{mn}$ and $d_{nm}$ from \eqref{greek}. However once we fix them at the lowest order in the coupling constant, the growth at higher orders in the coupling constant are determined by two factors. One, the expected power law growth at each order, and two, the {\it additional} $({\rm N}!)^2$ growth from the combinatoric factors in $\mathbb{D}_{\varphi_2}({\rm N}r)$
and $\mathbb{D}_{\varphi_3}({\rm N}(s-1))$. The latter is the genuine growth of the coefficients $\mathbb{C}^{({\rm N})}_m$ at ${\cal O}({\rm N})$ (and not {\it spurious} as we had thought in section \ref{tagcorfu}) implying that the series constructed from $\mathbb{C}^{({\rm N})}_m$ cannot be a convergent series.} 
at least as $({\rm N}!)^2$.
Note also that since we are only integrating over polynomial powers of momenta from the set of Hermite polynomials in \eqref{greek}, the final result should not depend on which sector we impose the momentum conservation condition. This means, by definition, \eqref{altomaa} has two free parameters: ${\rm N}$ from the ${\rm N} q, {\rm N}r$ and ${\rm N}(s-1)$ number of integrals (we suppress $q, r$ and $s$ in the subsequent discussions as they are kept fixed here but would eventually be summed over in a more generic setting), and $m$. The Hermite polynomials in the square bracket of
\eqref{milady} can then be expanded over $m \in \mathbb{Z}$ at every order 
${\rm N}$. 

With these we are ready to study the amplitude of the nodal diagrams to 
${\cal O}({\rm N})$. This would imply combining the nodal diagrams 
\eqref{luna34}, \eqref{luna35} and the one corresponding to the 
$\varphi_3$ sector. We will also combine $g_1^{\rm N}, g_2^{\rm N}$ and
$g_3^{\rm N}$ and call it ${g}^{\rm N}$. This way Borel resummation may be performed over $g$, and not over $g_1, g_2$ and $g_3$, thus rectifying another short-comings from our earlier consideration.
The amplitude of the nodal diagrams to ${\cal O}({\rm N})$ can then be evaluated in the following way:

{\footnotesize
\bg\label{luna36}
 & \mathcal{S}^{({\rm N})}(t) = g^{\rm N} 
\left( \begin{tikzpicture}[baseline={(0, 0cm-\MathAxis pt)}, thick,
main/.style = {draw, circle, fill=black, minimum size=2pt},
dot/.style={inner sep=0pt,fill=black, circle, minimum size=3pt},
dot1/.style={inner sep=0pt,fill=black, circle, minimum size=2pt},
ball/.style={ball  color=white, circle,  minimum size=15pt},
dots/.style={inner sep=0pt,fill=black, circle, minimum size=1pt}] 
  
  \node[ball] (0) at (0, 0) {i};
  \node[dot] (1) at (1, 0) {};
  \node[dot] (2) at (-1, 0) {};
  
  \node[dot] (11) at (1, 1.5) {};
  \node[dot] (12) at (1, -0.3) {};
  \node[dot] (13) at (1, -1.7) {};
  
  \node[dot] (100) at (2, 1.5) {};
  \node[dot1] (101) at (2, 2) {};
  \node[dot1] (102) at (2, 1.6) {};
  \node[dot1] (103) at (2, 1) {};
  \node[main] (111) at (2.5, 2) {};
  \node[main] (112) at (2.5, 1.6) {};
  \node[main] (113) at (2.5, 1) {};
  \node[dots] (1dd) at (2.3, 1.4) {};
  \node[dots] (1dd) at (2.3, 1.3) {};
  \node[dots] (1dd) at (2.3, 1.2) {};
  
  \node[dot] (120) at (2, -0.3) {};
  \node[dot1] (121) at (2, 0.4) {};
  \node[dot1] (122) at (2, -0.2) {};
  \node[dot1] (123) at (2, -0.7) {};
  \node[main] (1211) at (2.5, 0.4) {};
  \node[main] (1222) at (2.5, -0.2) {};
  \node[main] (1233) at (2.5, -0.7) {};
  \node[dots] (1dd) at (2.3, -0.3) {};
  \node[dots] (1dd) at (2.3, -0.4) {};
  \node[dots] (1dd) at (2.3, -0.5) {};
  \node[dots] (1dd) at (2.3, -0.6) {};
  
  \node[dot] (130) at (2, -1.7) {};
  \node[dot1] (131) at (2, -1.2) {};
  \node[dot1] (132) at (2, -1.8) {};
  \node[dot1] (133) at (2, -2.3) {};
  \node[main] (1311) at (2.5, -1.2) {};
  \node[main] (1322) at (2.5, -1.8) {};
  \node[main] (1333) at (2.5, -2.3) {};
  \node[dots] (1dd) at (2.3, -1.9) {};
  \node[dots] (1dd) at (2.3, -2) {};
  \node[dots] (1dd) at (2.3, -2.1) {};
  \node[dots] (1dd) at (2.3, -2.2) {};

  \node[dot] (21) at (-1, 1.5) {};
  \node[dot] (22) at (-1, -0.3) {};
  \node[dot] (23) at (-1, -1.7) {};
  
  \node[dot] (200) at (-2, 1.5) {};
  \node[dot1] (201) at (-2, 2) {};
  \node[dot1] (202) at (-2, 1.6) {};
  \node[dot1] (203) at (-2, 1) {};
  \node[main] (211) at (-2.5, 2) {};
  \node[main] (212) at (-2.5, 1.6) {};
  \node[main] (213) at (-2.5, 1) {};
  \node[dots] (2dd) at (-2.3, 1.5) {};
  \node[dots] (2dd) at (-2.3, 1.4) {};
  \node[dots] (2dd) at (-2.3, 1.2) {};
  
  \node[dot] (220) at (-2, -0.3) {};
  \node[dot1] (221) at (-2, 0.4) {};
  \node[dot1] (222) at (-2, -0.2) {};
  \node[dot1] (223) at (-2, -0.7) {};
  \node[main] (2211) at (-2.5, 0.4) {};
  \node[main] (2222) at (-2.5, -0.2) {};
  \node[main] (2233) at (-2.5, -0.7) {};
  \node[dots] (2dd) at (-2.3, -0.3) {};
  \node[dots] (2dd) at (-2.3, -0.4) {};
  \node[dots] (2dd) at (-2.3, -0.5) {};
  \node[dots] (2dd) at (-2.3, -0.6) {};
  
  \node[dot] (230) at (-2, -1.7) {};
  \node[dot1] (231) at (-2, -1.2) {};
  \node[dot1] (232) at (-2, -1.8) {};
  \node[dot1] (233) at (-2, -2.2) {};
  \node[main] (2311) at (-2.5, -1.2) {};
  \node[main] (2322) at (-2.5, -1.8) {};
  \node[main] (2333) at (-2.5, -2.2) {};
  \node[dots] (2dd) at (-2.3, -1.9) {};
  \node[dots] (2dd) at (-2.3, -2) {};
  \node[dots] (2dd) at (-2.3, -2.1) {};
  
  \node[dots] (22d) at (-1, -0.5) {};
  \node[dots] (22d) at (-1, -0.7) {};
  \node[dots] (22d) at (-1, -0.9) {};
  \node[dots] (22d) at (-1, -1.1) {};
  \node[dots] (22d) at (-1, -1.3) {};
  \node[dots] (22d) at (-1, -1.5) {};
  
  \node[dots] (22d) at (1, -0.5) {};
  \node[dots] (22d) at (1, -0.7) {};
  \node[dots] (22d) at (1, -0.9) {};
  \node[dots] (22d) at (1, -1.1) {};
  \node[dots] (22d) at (1, -1.3) {};
  \node[dots] (22d) at (1, -1.5) {};

\draw (0)  -- node[midway, above right, pos=0.8] {} (1);
\draw (0)  -- node[midway, above right, pos=0.8] {} (2);

\draw (1) -- node {} (11);
\draw (1) -- node {} (12);

\draw (121) -- node {} (1211);
\draw (122) -- node {} (1222);
\draw (123) -- node {} (1233);
\draw (121) -- node{} (123);

\draw (13) -- node[midway, above right, pos=0] {\tiny$\xi_{\rm N} $} (130);
\draw (131) -- node{} (133);
\draw (131) -- node {} (1311);
\draw (132) -- node {} (1322);
\draw (133) -- node {} (1333);

\draw (11) -- node[midway, above right, pos=0] {\tiny$\xi_{p + 1} $} (100);
\draw (101) -- node{} (103);
\draw (101) -- node{} (111);
\draw (102) -- node{} (112);
\draw (103) -- node{} (113);

\draw (12) -- node[midway, above right, pos=0] {\tiny$\xi_{p + 2} $} (120);

\draw (2) -- node {} (21);
\draw (21) -- node {} (22);

\draw (201) -- node {} (211);
\draw (202) -- node {} (212);
\draw (203) -- node {} (213);
\draw (201) -- node{} (203);
\draw (21) -- node[midway, above right, pos=0.8] {\tiny$\xi_1 $} (200);

\draw (221) -- node {} (2211);
\draw (222) -- node {} (2222);
\draw (223) -- node {} (2233);
\draw (221) -- node{} (223);
\draw (22) -- node[midway, above right, pos=0.8] {\tiny$\xi_2 $} (220);

\draw (231) -- node {} (2311);
\draw (232) -- node {} (2322);
\draw (233) -- node {} (2333);
\draw (231) -- node{} (233);
\draw (23) -- node[midway, above right, pos=0.8] {\tiny$\xi_p $} (230);
\end{tikzpicture} 
\right) \quad \otimes \quad
\left( 
\begin{tikzpicture}[baseline={(0, 0cm-\MathAxis pt)}, thick,
main/.style = {draw, circle, fill=black, minimum size=2pt},
dot/.style={inner sep=0pt,fill=black, circle, minimum size=3pt},
dot1/.style={inner sep=0pt,fill=black, circle, minimum size=2pt},
ball/.style={ball  color=white, circle,  minimum size=15pt},
dots/.style={inner sep=0pt,fill=black, circle, minimum size=1pt}] 
  
  \node[dot] (0) at (0, 0) {};
  \node[dot] (1) at (1, 0) {};
  \node[dot] (2) at (-1, 0) {};
  
  \node[dot] (11) at (1, 1.5) {};
  \node[dot] (12) at (1, -0.3) {};
  \node[dot] (13) at (1, -1.7) {};
  
  \node[dot] (100) at (2, 1.5) {};
  \node[dot1] (101) at (2, 2) {};
  \node[dot1] (102) at (2, 1.6) {};
  \node[dot1] (103) at (2, 1) {};
  \node[main] (111) at (2.5, 2) {};
  \node[main] (112) at (2.5, 1.6) {};
  \node[main] (113) at (2.5, 1) {};
  \node[dots] (1dd) at (2.3, 1.4) {};
  \node[dots] (1dd) at (2.3, 1.3) {};
  \node[dots] (1dd) at (2.3, 1.2) {};
  
  \node[dot] (120) at (2, -0.3) {};
  \node[dot1] (121) at (2, 0.4) {};
  \node[dot1] (122) at (2, -0.2) {};
  \node[dot1] (123) at (2, -0.7) {};
  \node[main] (1211) at (2.5, 0.4) {};
  \node[main] (1222) at (2.5, -0.2) {};
  \node[main] (1233) at (2.5, -0.7) {};
  \node[dots] (1dd) at (2.3, -0.3) {};
  \node[dots] (1dd) at (2.3, -0.4) {};
  \node[dots] (1dd) at (2.3, -0.5) {};
  \node[dots] (1dd) at (2.3, -0.6) {};
  
  \node[dot] (130) at (2, -1.7) {};
  \node[dot1] (131) at (2, -1.2) {};
  \node[dot1] (132) at (2, -1.8) {};
  \node[dot1] (133) at (2, -2.3) {};
  \node[main] (1311) at (2.5, -1.2) {};
  \node[main] (1322) at (2.5, -1.8) {};
  \node[main] (1333) at (2.5, -2.3) {};
  \node[dots] (1dd) at (2.3, -1.9) {};
  \node[dots] (1dd) at (2.3, -2) {};
  \node[dots] (1dd) at (2.3, -2.1) {};
  \node[dots] (1dd) at (2.3, -2.2) {};

  \node[dot] (21) at (-1, 1.5) {};
  \node[dot] (22) at (-1, -0.3) {};
  \node[dot] (23) at (-1, -1.7) {};
  
  \node[dot] (200) at (-2, 1.5) {};
  \node[dot1] (201) at (-2, 2) {};
  \node[dot1] (202) at (-2, 1.6) {};
  \node[dot1] (203) at (-2, 1) {};
  \node[main] (211) at (-2.5, 2) {};
  \node[main] (212) at (-2.5, 1.6) {};
  \node[main] (213) at (-2.5, 1) {};
  \node[dots] (2dd) at (-2.3, 1.5) {};
  \node[dots] (2dd) at (-2.3, 1.4) {};
  \node[dots] (2dd) at (-2.3, 1.2) {};
  
  \node[dot] (220) at (-2, -0.3) {};
  \node[dot1] (221) at (-2, 0.4) {};
  \node[dot1] (222) at (-2, -0.2) {};
  \node[dot1] (223) at (-2, -0.7) {};
  \node[main] (2211) at (-2.5, 0.4) {};
  \node[main] (2222) at (-2.5, -0.2) {};
  \node[main] (2233) at (-2.5, -0.7) {};
  \node[dots] (2dd) at (-2.3, -0.3) {};
  \node[dots] (2dd) at (-2.3, -0.4) {};
  \node[dots] (2dd) at (-2.3, -0.5) {};
  \node[dots] (2dd) at (-2.3, -0.6) {};
  
  \node[dot] (230) at (-2, -1.7) {};
  \node[dot1] (231) at (-2, -1.2) {};
  \node[dot1] (232) at (-2, -1.8) {};
  \node[dot1] (233) at (-2, -2.2) {};
  \node[main] (2311) at (-2.5, -1.2) {};
  \node[main] (2322) at (-2.5, -1.8) {};
  \node[main] (2333) at (-2.5, -2.2) {};
  \node[dots] (2dd) at (-2.3, -1.9) {};
  \node[dots] (2dd) at (-2.3, -2) {};
  \node[dots] (2dd) at (-2.3, -2.1) {};
  
  \node[dots] (22d) at (-1, -0.5) {};
  \node[dots] (22d) at (-1, -0.7) {};
  \node[dots] (22d) at (-1, -0.9) {};
  \node[dots] (22d) at (-1, -1.1) {};
  \node[dots] (22d) at (-1, -1.3) {};
  \node[dots] (22d) at (-1, -1.5) {};
  
  \node[dots] (22d) at (1, -0.5) {};
  \node[dots] (22d) at (1, -0.7) {};
  \node[dots] (22d) at (1, -0.9) {};
  \node[dots] (22d) at (1, -1.1) {};
  \node[dots] (22d) at (1, -1.3) {};
  \node[dots] (22d) at (1, -1.5) {};

\draw (0)  -- node[midway, above right, pos=0.8] {} (1);
\draw (0)  -- node[midway, above right, pos=0.8] {} (2);

\draw (1) -- node {} (11);
\draw (1) -- node {} (12);

\draw (121) -- node {} (1211);
\draw (122) -- node {} (1222);
\draw (123) -- node {} (1233);
\draw (121) -- node{} (123);

\draw (13) -- node[midway, above right, pos=0] {\tiny$\chi_{\rm N} $} (130);
\draw (131) -- node{} (133);
\draw (131) -- node {} (1311);
\draw (132) -- node {} (1322);
\draw (133) -- node {} (1333);

\draw (11) -- node[midway, above right, pos=0] {\tiny$\chi_{p + 1} $} (100);
\draw (101) -- node{} (103);
\draw (101) -- node{} (111);
\draw (102) -- node{} (112);
\draw (103) -- node{} (113);

\draw (12) -- node[midway, above right, pos=0] {\tiny$\chi_{p + 2} $} (120);

\draw (2) -- node {} (21);
\draw (21) -- node {} (22);

\draw (201) -- node {} (211);
\draw (202) -- node {} (212);
\draw (203) -- node {} (213);
\draw (201) -- node{} (203);
\draw (21) -- node[midway, above right, pos=0.8] {\tiny$\chi_1 $} (200);

\draw (221) -- node {} (2211);
\draw (222) -- node {} (2222);
\draw (223) -- node {} (2233);
\draw (221) -- node{} (223);
\draw (22) -- node[midway, above right, pos=0.8] {\tiny$\chi_2 $} (220);

\draw (231) -- node {} (2311);
\draw (232) -- node {} (2322);
\draw (233) -- node {} (2333);
\draw (231) -- node{} (233);
\draw (23) -- node[midway, above right, pos=0.8] {\tiny$\chi_p $} (230);
\end{tikzpicture} 
\right)
\nonumber 
\\
& \otimes \quad \quad 
\left(
\begin{tikzpicture}[baseline={(0, 0cm-\MathAxis pt)}, thick,
main/.style = {draw, circle, fill=black, minimum size=2pt},
dot/.style={inner sep=0pt,fill=black, circle, minimum size=3pt},
dot1/.style={inner sep=0pt,fill=black, circle, minimum size=2pt},
ball/.style={ball  color=white, circle,  minimum size=15pt},
dots/.style={inner sep=0pt,fill=black, circle, minimum size=1pt}] 
  
  \node[dot] (0) at (0, 0) {};
  \node[dot] (1) at (1, 0) {};
  \node[dot] (2) at (-1, 0) {};
  
  \node[dot] (11) at (1, 1.5) {};
  \node[dot] (12) at (1, -0.3) {};
  \node[dot] (13) at (1, -1.7) {};
  
  \node[dot] (100) at (2, 1.5) {};
  \node[dot1] (101) at (2, 2) {};
  \node[dot1] (102) at (2, 1.6) {};
  \node[dot1] (103) at (2, 1) {};
  \node[main] (111) at (2.5, 2) {};
  \node[main] (112) at (2.5, 1.6) {};
  \node[main] (113) at (2.5, 1) {};
  \node[dots] (1dd) at (2.3, 1.4) {};
  \node[dots] (1dd) at (2.3, 1.3) {};
  \node[dots] (1dd) at (2.3, 1.2) {};
  \node[dot1] (10d) at (2, 1.3 ) {};
  \node[main] (11d) at (3, 1.3 ) {};
  
  \node[dot] (120) at (2, -0.3) {};
  \node[dot1] (121) at (2, 0.4) {};
  \node[dot1] (122) at (2, -0.2) {};
  \node[dot1] (123) at (2, -0.7) {};
  \node[main] (1211) at (2.5, 0.4) {};
  \node[main] (1222) at (2.5, -0.2) {};
  \node[main] (1233) at (2.5, -0.7) {};
  \node[dots] (1dd) at (2.3, -0.3) {};
  \node[dots] (1dd) at (2.3, -0.4) {};
  \node[dots] (1dd) at (2.3, -0.5) {};
  \node[dots] (1dd) at (2.3, -0.6) {};
  \node[dot1] (12d) at (2, 0.1 ) {};
  \node[main] (122d) at (3, 0.1 ) {};
  
  \node[dot] (130) at (2, -1.7) {};
  \node[dot1] (131) at (2, -1.2) {};
  \node[dot1] (132) at (2, -1.8) {};
  \node[dot1] (133) at (2, -2.3) {};
  \node[main] (1311) at (2.5, -1.2) {};
  \node[main] (1322) at (2.5, -1.8) {};
  \node[main] (1333) at (2.5, -2.3) {};
  \node[dots] (1dd) at (2.3, -1.9) {};
  \node[dots] (1dd) at (2.3, -2) {};
  \node[dots] (1dd) at (2.3, -2.1) {};
  \node[dots] (1dd) at (2.3, -2.2) {};
  \node[dot1] (13d) at (2, -1.5 ) {};
  \node[main] (133d) at (3, -1.5 ) {};

  \node[dot] (21) at (-1, 1.5) {};
  \node[dot] (22) at (-1, -0.3) {};
  \node[dot] (23) at (-1, -1.7) {};
  
  \node[dot] (200) at (-2, 1.5) {};
  \node[dot1] (201) at (-2, 2) {};
  \node[dot1] (202) at (-2, 1.6) {};
  \node[dot1] (203) at (-2, 1) {};
  \node[main] (211) at (-2.5, 2) {};
  \node[main] (212) at (-2.5, 1.6) {};
  \node[main] (213) at (-2.5, 1) {};
  \node[dots] (2dd) at (-2.3, 1.5) {};
  \node[dots] (2dd) at (-2.3, 1.4) {};
  \node[dots] (2dd) at (-2.3, 1.2) {};
  \node[dot1] (20d) at (-2, 1.3 ) {};
  \node[main] (21d) at (-3, 1.3 ) {};
  
  \node[dot] (220) at (-2, -0.3) {};
  \node[dot1] (221) at (-2, 0.4) {};
  \node[dot1] (222) at (-2, -0.2) {};
  \node[dot1] (223) at (-2, -0.7) {};
  \node[main] (2211) at (-2.5, 0.4) {};
  \node[main] (2222) at (-2.5, -0.2) {};
  \node[main] (2233) at (-2.5, -0.7) {};
  \node[dots] (2dd) at (-2.3, -0.3) {};
  \node[dots] (2dd) at (-2.3, -0.4) {};
  \node[dots] (2dd) at (-2.3, -0.5) {};
  \node[dots] (2dd) at (-2.3, -0.6) {};
  \node[dot1] (22dot) at (-2, 0.1 ) {};
  \node[main] (222dot) at (-3, 0.1 ) {};
  
  \node[dot] (230) at (-2, -1.7) {};
  \node[dot1] (231) at (-2, -1.2) {};
  \node[dot1] (232) at (-2, -1.8) {};
  \node[dot1] (233) at (-2, -2.2) {};
  \node[main] (2311) at (-2.5, -1.2) {};
  \node[main] (2322) at (-2.5, -1.8) {};
  \node[main] (2333) at (-2.5, -2.2) {};
  \node[dots] (2dd) at (-2.3, -1.9) {};
  \node[dots] (2dd) at (-2.3, -2) {};
  \node[dots] (2dd) at (-2.3, -2.1) {};
  \node[dot1] (23d) at (-2, -1.5 ) {};
  \node[main] (233d) at (-3, -1.5 ) {};
  
  \node[dots] (22d) at (-1, -0.5) {};
  \node[dots] (22d) at (-1, -0.7) {};
  \node[dots] (22d) at (-1, -0.9) {};
  \node[dots] (22d) at (-1, -1.1) {};
  \node[dots] (22d) at (-1, -1.3) {};
  \node[dots] (22d) at (-1, -1.5) {};
  
  \node[dots] (22d) at (1, -0.5) {};
  \node[dots] (22d) at (1, -0.7) {};
  \node[dots] (22d) at (1, -0.9) {};
  \node[dots] (22d) at (1, -1.1) {};
  \node[dots] (22d) at (1, -1.3) {};
  \node[dots] (22d) at (1, -1.5) {};

\draw (0)  -- node[midway, above right, pos=0.8] {} (1);
\draw (0)  -- node[midway, above right, pos=0.8] {} (2);

\draw (1) -- node {} (11);
\draw (1) -- node {} (12);

\draw (121) -- node {} (1211);
\draw (122) -- node {} (1222);
\draw (123) -- node {} (1233);
\draw (121) -- node{} (123);

\draw (13) -- node[midway, above right, pos=0] {\tiny$\eta_{\rm N} $} (130);
\draw (131) -- node{} (133);
\draw (131) -- node {} (1311);
\draw (132) -- node {} (1322);
\draw (133) -- node {} (1333);

\draw (11) -- node[midway, above right, pos=0] {\tiny$\eta_{p + 1} $} (100);
\draw (101) -- node{} (103);
\draw (101) -- node{} (111);
\draw (102) -- node{} (112);
\draw (103) -- node{} (113);

\draw (12) -- node[midway, above right, pos=0] {\tiny$\eta_{p + 2} $} (120);

\draw (2) -- node {} (21);
\draw (21) -- node {} (22);

\draw (201) -- node {} (211);
\draw (202) -- node {} (212);
\draw (203) -- node {} (213);
\draw (201) -- node{} (203);
\draw (21) -- node[midway, above right, pos=0.8] {\tiny$\eta_1 $} (200);

\draw (221) -- node {} (2211);
\draw (222) -- node {} (2222);
\draw (223) -- node {} (2233);
\draw (221) -- node{} (223);
\draw (22) -- node[midway, above right, pos=0.8] {\tiny$\eta_2 $} (220);

\draw (231) -- node {} (2311);
\draw (232) -- node {} (2322);
\draw (233) -- node {} (2333);
\draw (231) -- node{} (233);

\draw (23) -- node[midway, above right, pos=0.8] {\tiny$\eta_p $} (230);

\draw (10d) [dashed] -- node {} (11d);
\draw (12d) [dashed] -- node {} (122d);
\draw (13d) [dashed] -- node {} (133d);
\draw (20d) [dashed] -- node {} (21d);
\draw (22dot) [dashed] -- node {} (222dot);
\draw (23d) [dashed] -- node {} (233d);
\end{tikzpicture}
\right)
\nonumber
\\
& = g^{\rm N}\Bigg[\sum\limits_{p = 0}^1{(-1)^p 2^p\mathbb{C}^{({\rm N})}_1
{\bf H}_{1-p}(k_0)\over (1-p)! (it)^{p+1}} + 
\sum\limits_{p = 0}^2{(-1)^p 2^p 2!\mathbb{C}^{({\rm N})}_2
{\bf H}_{2-p}(k_0)\over (2-p)! (it)^{p+1}} + 
\sum\limits_{p = 0}^3{(-1)^p 2^p 3!\mathbb{C}^{({\rm N})}_3
{\bf H}_{3-p}(k_0)\over (3-p)! (it)^{p+1}} \\
& ~+ \sum\limits_{p = 0}^4{(-1)^p 2^p 4!\mathbb{C}^{({\rm N})}_4
{\bf H}_{4-p}(k_0)\over (4-p)! (it)^{p+1}} + 
\sum\limits_{p = 0}^5{(-1)^p 2^p 5!\mathbb{C}^{({\rm N})}_5
{\bf H}_{5-p}(k_0)\over (5-p)! (it)^{p+1}} + ... 
+ \sum\limits_{p = 0}^{\rm N}{(-1)^p 2^p {\rm N}!\mathbb{C}^{({\rm N})}_{\rm N}
{\bf H}_{{\rm N}-p}(k_0)\over ({\rm N}-p)! (it)^{p+1}} \nonumber\\
&~~~~~~~~~~~ + \sum\limits_{p = 0}^{{\rm N} + 1}{(-1)^p 2^p ({\rm N} + 1)!
\mathbb{C}^{({\rm N})}_{{\rm N} + 1}
{\bf H}_{{\rm N}+1 -p}(k_0)\over ({\rm N} + 1-p)! (it)^{p+1}} + 
\sum\limits_{s = 2}^\infty \sum\limits_{p = 0}^{{\rm N} + s}{(-1)^p 2^p
({\rm N} + s)!\mathbb{C}^{({\rm N})}_{{\rm N} + s}
{\bf H}_{{\rm N}+s -p}(k_0)\over ({\rm N} + s -p)! (it)^{p+1}}\Bigg] {\rm exp}\left(ik_0t\right)\Bigg\vert_{k_{\rm IR}}^\mu 
+ \mathtt{c.c}\Bigg\vert_{k_{\rm IR}}^\mu, \nonumber \nd}
where note two things: {\Su one}, the series goes to infinity in the sense that all the Hermite polynomials and powers of $(it)$ contribute; and {\Su two}, we have isolated, somewhat arbitrarily, the ${\rm N}$-th piece so that the series could also be interpreted as though it is centered on ${\rm N}$ with decreasing and increasing values on both sides. The decreasing side is bounded from below, but the increasing side has no bounds as it can grow to infinity. We can also rearrange the above series in the following suggestive way:

{\footnotesize
\bg\label{alilara}
{\cal S}^{({\rm N})}(t) =&&\Bigg[\sum_{a = 0}^1 {\mathbb{C}_a^{({\rm N})} {\bf H}_a(k_0) \over it} 
- {2\cdot 1! \mathbb{C}_1^{({\rm N})} {\bf H}_0(k_0) \over (it)^2}\nonumber\\
&& + {\mathbb{C}_2^{({\rm N})} {\bf H}_2(k_0) \over it} - 
{2\cdot 2! \mathbb{C}_2^{({\rm N})} {\bf H}_1(k_0) \over (it)^2} + 
{2^2\cdot 2!\mathbb{C}_2^{({\rm N})} {\bf H}_0(k_0) \over (it)^3}\nonumber\\
&& + {\mathbb{C}_3^{({\rm N})} {\bf H}_3(k_0) \over it} - 
{2\cdot 3!\mathbb{C}_3^{({\rm N})} {\bf H}_2(k_0) \over 2!(it)^2} + 
{2^2\cdot 3!\mathbb{C}_3^{({\rm N})} {\bf H}_1(k_0) \over (it)^3} 
- {2^3\cdot 3!\mathbb{C}_3^{({\rm N})} {\bf H}_0(k_0) \over (it)^4} \nonumber\\
&& + {\mathbb{C}_4^{({\rm N})} {\bf H}_4(k_0) \over it} - 
{2\cdot 4!\mathbb{C}_4^{({\rm N})} {\bf H}_3(k_0) \over 3!(it)^2} + 
{2^2\cdot 4!\mathbb{C}_4^{({\rm N})} {\bf H}_2(k_0) \over 2!(it)^3}
- {2^3\cdot 4!\mathbb{C}_4^{({\rm N})} {\bf H}_1(k_0) \over (it)^4}
+ {2^4\cdot 4!\mathbb{C}_4^{({\rm N})} {\bf H}_0(k_0) \over (it)^5}\nonumber\\
&&+ ....... \nonumber\\
&& + {\mathbb{C}_{\rm N}^{({\rm N})} {\bf H}_{\rm N}(k_0) \over it} 
- {2\cdot {\rm N}!\mathbb{C}_{\rm N}^{({\rm N})} {\bf H}_{{\rm N}-1}(k_0) \over ({\rm N} - 1)!(it)^2} + 
{2^2\cdot {\rm N}!\mathbb{C}_{\rm N}^{({\rm N})} {\bf H}_{{\rm N}-2}(k_0) \over ({\rm N} - 2)!(it)^3} + .... + 
{(-1)^{\rm N} 2^{\rm N}\cdot {\rm N}!\mathbb{C}_{\rm N}^{({\rm N})} {\bf H}_0(k_0) \over (it)^{{\rm N}+ 1}} \nonumber\\
&& + \sum_{s = 1}^\infty \left({\mathbb{C}_{{\rm N} + s}^{({\rm N})} {\bf H}_{{\rm N} + s}(k_0) \over it} - {2\cdot ({\rm N} + s)!\mathbb{C}_{{\rm N}+ s}^{({\rm N})} {\bf H}_{{\rm N}+s - 1}(k_0) \over ({\rm N} + s - 1)!(it)^2} + ..+ {(-2)^{{\rm N}+s}\mathbb{C}^{({\rm N})}_{{\rm N} + s}({\rm N} + s)!{\bf H}_0(k_0) \over 
(it)^{{\rm N} + s + 1}}\right)\Bigg]\nonumber\\
&& \times~ {\rm exp}\left(ik_0 t\right)\Bigg\vert_{k_{\rm IR}}^\mu g^{\rm N}
+ {\rm exp}\left(-ik_0t\right){\bf \Gamma}^{({\rm N})}\left(k_0; it \leftrightarrow -it\right)\Bigg\vert_{k_{\rm IR}}^\mu g^{\rm N},\nd}
where ${\bf \Gamma}^{({\rm N})}\left(k_0; it\right)$ is the term inside the square bracket of \eqref{alilara}; and $\mathbb{C}^{({\rm N})}_m$ is defined in \eqref{altomaa}. From the form of the Hermite polynomials
we see that the first terms of each line dominates over the other terms in that line (as ${\bf H}_m(\mu) > {\bf H}_{m - r}(\mu)$), implying that the dominant contribution will come from the coefficient of 
${{\rm sin}(\mu t)\over t}$ piece for finite values of $t$ (we will soon make $t$ dimensionless using an appropriate factor). Thus if we extract the {\it dominant} contributions at every ${\cal O}({\rm N})$ for $\mu > 1$ and finite $t$, we get:

{\scriptsize
\bg\label{fajarjee2}
{\cal S}^{({\rm N})}_{\rm dom} & = &  g^{\rm N} \sum_{r = -\infty}^{{\rm N} - 1} {\mathbb{C}^{({\rm N})}_{{\rm N} - r} \over \left[({\rm N} - r)!\right]^\alpha}~\left[({\rm N} - r)!\right]^\alpha{\bf H}_{{\rm N} - r}(k_0)\Bigg\vert_{k_{\rm IR}}^\mu\\
& = & g^{\rm N}\left[\sum_{r = 1}^{{\rm N} - 1} {\mathbb{C}^{({\rm N})}_{{\rm N} - r} \over \left[({\rm N} - r)!\right]^\alpha}\left[({\rm N} - r)!\right]^\alpha{\bf H}_{{\rm N} - r}(k_0) + 
{\mathbb{C}^{({\rm N})}_{{\rm N}} \over \left[{\rm N}!\right]^\alpha}\left[{\rm N}!\right]^\alpha{\bf H}_{\rm N} (k_0) + 
\sum_{s = 1}^\infty {\mathbb{C}^{({\rm N})}_{{\rm N} + s} \over \left[({\rm N} + s)!\right]^\alpha}\left[({\rm N} + s)!\right]^\alpha{\bf H}_{{\rm N} + s}(k_0)\right]_{k_{\rm IR}}^\mu, \nonumber \nd}
where ${\cal S}^{({\rm N})}_{\rm dom} = {\rm coeff}\left[{\sin(\mu t)\over t}\right]$. Note also the placement of the factorials which are arranged to facilitate the Borel resummation. However the series we get in \eqref{fajarjee2} differs from the usual asymptotic series in at least 
three ways: {\Su one}, by the presence of decreasing and increasing values of factorials over ${\rm N}$ at every order in ${\rm N}$; {\Su two}, by the presence of {\it all} the infinite non-zero contributions from ${\rm coeff}\left[{\sin(\mu t)\over t^{2w+1}}\right]$ and ${\rm coeff}\left[{\cos(\mu t)\over t^{2w}}\right]$ for all $w \in \mathbb{Z}_+$ even at ${\cal O}({\rm N})$; and {\Su three}, by the asymptotic growth being captured by a Gevrey-$\alpha$ series \cite{gevreyorig} where $\alpha > 1$ with $\alpha \in \mathbb{Z}$ (here $\alpha \ge 2$). 

The Gevrey order $\alpha$ growth clearly implies the necessity of resummation \cite{gevrey} because the series on both the numerator and the denominator of the path integral \eqref{2summers} are asymptotic. One may also easily check that for large ${\rm N}$ we expect the following simplification by using the Stirling's approximation:
\bg\label{corfu1}
g^{\rm N} \left({\rm N}!\right)^\alpha \approx {g^{\rm N}\left(\alpha{\rm N}\right)!\over 
\alpha^{\alpha{\rm N}}} = {g^{\rm N}\left(\alpha{\rm N}\right)!\over 
1 + \sum\limits_{k = 1}^\infty {(\alpha{\rm N})^k {\rm log}^k \alpha\over 
k!}} ~ =  ~ \widetilde{g}^{\rm N} \left(\alpha {\rm N}\right)!, \nd
for $\widetilde{g} \equiv {g\over \alpha^\alpha} << 1$ and $\alpha << {\rm N}$. Clearly both the conditions {\it can} be maintained for our case because the total number of field components can only grow to 256 in M-theory, whereas ${\rm N}$ can be arbitrarily large; and\footnote{We will not distinguish between $\widetilde{g}$ and $g$ henceforth unless mentioned otherwise. \label{kaludi}} $g << 1$. Notice that 
while $\alpha$ has an upper bound (controlled by the supersymmetric solitonic background), the parameters of the interaction in say \eqref{lincoln}, {\it i.e.} $(q, r, s)$, can be arbitrarily large. Fortunately they do not control the order of the Gevrey series, thus 
releasing us of the conundrum that we faced in section \ref{tagcorfu} (more on this below) and converting \eqref{fajarjee2} to the following:

{\scriptsize
\bg\label{fajarjee3}
&&{\cal S}^{({\rm N})}_{\rm dom} =  g^{\rm N} \sum_{r = -\infty}^{{\rm N} - 1} {\mathbb{C}^{({\rm N})}_{{\rm N} - r} \over \big(\alpha({\rm N} - r)\big)!}~\big(\alpha({\rm N} - r)\big)!~{\bf H}_{{\rm N} - r}(k_0)\Bigg\vert_{k_{\rm IR}}^\mu \\
&& = g^{\rm N}\left[\sum_{r = 1}^{{\rm N} - 1} {\mathbb{C}^{({\rm N})}_{{\rm N} - r} \over \left(\alpha({\rm N} - r)\right)!}\left(\alpha({\rm N} - r)\right)!~{\bf H}_{{\rm N} - r}(k_0) + 
{\mathbb{C}^{({\rm N})}_{{\rm N}} \over \left(\alpha{\rm N}\right)!}\left(\alpha{\rm N}\right)!~{\bf H}_{\rm N} (k_0) + 
\sum_{s = 1}^\infty {\mathbb{C}^{({\rm N})}_{{\rm N} + s} \over \left(\alpha({\rm N} + s)\right)!}\left(\alpha({\rm N} + s)\right)!~{\bf H}_{{\rm N} + s}(k_0)\right]_{k_{\rm IR}}^\mu, \nonumber \nd}
which should now be Borel resummed. On the other hand, if we do not restrict ourselves to the most dominant contributions then the resulting picture would be more involved. To ${\cal O}({\rm N})$ the numerator of the path integral \eqref{2summers} may be represented in the following compact form:

{\footnotesize
\bg\label{harmonik}
g^{\rm N} {\bf \Pi}^{({\rm N})}(t) \equiv 
\sum_{s = 1 - {\rm N}}^{\infty} \sum_{p = 0}^{{\rm N} + s} 
{(-1)^p 2^p ({\rm N} + s)!\over ({\rm N} + s - p)!}\left[{{\rm exp}(ik_0 t)\over (it)^{p + 1}} + {{\rm exp}(-ik_0t)\over (-it)^{p + 1}}\right] 
\mathbb{C}^{({\rm N})}_{{\rm N} + s}~ {\bf H}_{{\rm N} + s - p}(k_0)
\Bigg\vert_{k_{\rm IR}}^\mu, \nd}
where one may easily relate ${\bf \Gamma}^{({\rm N})}(k_0; it)$ from 
\eqref{alilara} with ${\bf \Pi}^{({\rm N})}(t)$ above using complex conjugations. Note that in \eqref{harmonik} keeping $p \in \mathbb{Z}_+$ fixed and varying
$s \in \mathbb{Z}$ implies a vertical motion in \eqref{alilara}, whereas 
keeping $s$ fixed and varying $p$ implies a horizontal motion. For a fixed $p$, the series we get by summing over all ${\rm N} \in \mathbb{Z}_+$ and $s \in \mathbb{Z}$ is asymptotic. Since $p \in \mathbb{Z}_+$, there is in fact an infinite class of asymptotic series possible at ${\cal O}({\rm N})$. Most are sub-dominant, as we saw above, but are nevertheless necessary for the consistency of the background. 

For the denominator of the path integral \eqref{2summers} we can compute all the diagrams in ${\cal N}^{(1)'}_{nmp}(k; q)$ as discussed in 
section \ref{3.2.6}. Alternatively, we can simplify the process by rewriting \eqref{lucliu} in the following way\footnote{We denote the operators using the hatted symbols following the conventions laid out in section \ref{sec1.3} and the discussion after \eqref{nehaprit}.}:
\bg\label{lucliu2}
\widehat{\mathbb{D}}_0(\overline\sigma)\vert\Omega\rangle = \lim_{{\rm T} \to \infty(1-i\epsilon)} \widehat{\mathbb{D}}_0(\overline\sigma)~{\rm exp}\left(-i{\rm M}_p\int_{-{\rm T}}^t 
dt~\widehat{\bf H}_{\rm int}\right)\vert 0\rangle \approx 
\widehat{\mathbb{D}}_0(\overline\sigma)\vert 0 \rangle, \nd
where $\vert 0 \rangle$ is the collection of free vacua for all momenta $k$. Plugging this in the denominator of 
the path integral formula \eqref{wilsonrata} implies that there is a {\it convergent} perturbative series in the absence of the source term as in \eqref{shaystone}. We should however caution the reader that this is in general {\it not} true and therefore the dominant diagrams should {\it not} contribute. Nevertheless the condition \eqref{lucliu2} introduces a controlled laboratory to use the simplest nodal diagrams (which are the dominant ones) to analyze the consequences of our computations. A more refined technique, which will be bit more involved, will be discussed in section \ref{sec4.4}. Such a choice then imposes the following 
restrictions on the coefficients $\mathbb{C}^{({\rm N})}_m$:
\bg\label{turkish}
\sum_{{\rm N} = 0}^\infty g^{\rm N}\sum_{m = 0}^\infty {(-1)^p 2^p ~m!~\mathbb{C}^{({\rm N})}_m ~{\bf H}_{m-p}(k_0)\over (m - p)!}\Bigg\vert_{k_{\rm IR}}^\mu = f(p), \nd
for all $p \in \mathbb{Z}_+$ starting with $p = 0$ and $f(p)$ will be determined below. We have denoted the tree-level coefficients\footnote{The tree-level result may be easily derived from the amplitude of the nodal diagram \eqref{luna3} by using the wave-function ans\"atze from \eqref{creolet} and then applying the Riemann-Lebesgue lemma to restrict the integral to lie in $k_0 \in [k_{\rm IR}, \mu]$ where $k_{\rm IR}$ is an IR cutoff.} as $\mathbb{C}^{(0)}_m$ $\forall m \in \mathbb{Z}_+$, and combined the series from \eqref{alilara}, \eqref{fajarjee2} and \eqref{fajarjee3} under one roof by summing\footnote{One might wonder why we do not keep the series as in \eqref{fajarjee2} or \eqref{fajarjee3}. The reason will become apparent from \eqref{hicksmel} and \eqref{hicksmel2}: the Gevrey-$\alpha$ growth of the coefficients $\mathbb{C}^{({\rm N})}_m$ are controlled by the superscript ${\rm N}$ and do not depend on the subscript $m$. Thus $\mathbb{C}^{({\rm N})}_{{\rm N} \pm s}$ for any $s$ will grow in the same way, namely as $({\rm N}!)^\alpha$. As such \eqref{turkish} would make more sense here.} over $m$ for a given choice of $p$. The summation over $m$ in \eqref{turkish} however requires some care: since the Hermite polynomials cannot have negative subscripts, $m$ is bounded from below by $p$, {\it i.e} for a given choice of $p$, $m \ge p$. We have also inserted an IR cutoff to make sense of the integrals. The function $f(p)$ can now be determined from first taking the Fourier cosine transformation\footnote{The Fourier transform we use here ranges from 0 to $\infty$ to comply with the reality of the functions as alluded to in footnote \ref{redskix3}. As such the Fourier transforms, using the modes \eqref{pugh}, lead to:
\bg\label{FT} 
&&{\bf FT}\left[\left({g_s\over {\rm HH}_o}\right)^{-a}\right]_{\sin} = 
\sqrt{2\over \pi\Lambda^{8/3}}~\cos\left({a\pi\over 2}\right)~\Gamma(1-a) ~k_0^{a-1}\nonumber\\
&&{\bf FT}\left[\left({g_s\over {\rm HH}_o}\right)^{-a}\right]_{\cos} = 
\sqrt{2\over \pi\Lambda^{8/3}}~\cos\left({(1-a)\pi\over 2}\right)~\Gamma(1-a) ~k_0^{a-1},\nonumber \nd
where $\Lambda$ is the cosmological constant. The subscript denotes whether it is a cosine or a sine Fourier transform. For our purpose, we will stick with the cosine formula.}
of ${\bf g}_{\mu\nu} \equiv 
\left({g_s\over {\rm HH}_o}\right)^{-8/3}$, and then using the Riemann-Lebesgue lemma \eqref{aug19visa} to write this back in the finite range $k_{\rm IR} \le k_o \le \mu$. Such a procedure will exactly produce a series in ${\sin(\mu t)\over t^{2w+1}}$ and ${\cos(\mu t)\over t^{2w}}$ for all $w\in \mathbb{Z}_+$ which we can then compare with the equivalent series we got above. (We will assume that the delta function integral over ${\bf k}$ provides a constant factor even in the finite range.) After the dust settles, the answer we get for $f(p)$ is:
\bg\label{incal}
f(p) = {(-1)^{p+1}\Gamma\left(-{5\over 3}\right)\over \Lambda^{4/3} \sqrt{2\pi/3}}~{\left({5\over 3}\right)! \over \left({5\over 3}-p\right)!}
~k_0^{{5\over 3} - p}\Bigg\vert_{k_{\rm IR}}^\mu, \nd
where $\left({5\over 3}\right)! \equiv {5\over 3}\left({5\over 3} -1\right)
\left({5\over 3} -2\right)\left({5\over 3} -3\right)....$ and the series does not end, implying that it is the series in the denominator of \eqref{incal} which is responsible in terminating the expression at the $p$-th order. Note that, since $p \in \mathbb{Z}_+$ can be arbitrarily large, an IR cutoff is necessary to make sense of the expression in \eqref{incal}. On the other hand, the LHS of \eqref{turkish} the scenario is more subtle. For any given value of $p \in \mathbb{Z}_+$, the LHS is a series in ${\rm N}$ as well as in $m \ge p$. The series in ${\rm N}$ is asymptotic as the growth of the coefficients $\mathbb{C}_m^{({\rm N})}$ is of Gevrey-$\alpha$ kind. This would mean that Borel resummation would be necessary to make sense of the LHS of \eqref{turkish}. Again, for a given value of ${\rm N}$ in \eqref{turkish} we are also explicitly summing over $m$. The growth there is interesting: for a fixed value of $p$ the coefficients grow as ${m!\over (m - p)!}$, implying that for large values of $m$ the coefficients do become large. For small values of $p$, but large values of $m$, the RHS of \eqref{turkish} can be small, but the coefficients on the LHS would be large implying that $\mathbb{C}^{({\rm N})}_m$ will either have to be very small or could carry different signs to make the sum very small for \eqref{turkish} to make sense.

What does this imply for the $\mathbb{C}^{({\rm N})}_m$ coefficients? From the aforementioned discussions, it appears that we can restrict ${\rm N}$ to some upper limit up to where the non-perturbative effects set in, or simply Borel resum the whole series over ${\rm N}$. On the other hand, for large values of $m$, there doesn't appear any strong reason to allow for the coefficients $\mathbb{C}^{({\rm N})}_m$ to be arbitrarily small. Although for practical purpose one might be able to arrange the series with some cutoff for large values of $m$, we will not put any restrictions at this stage. For the former however a more detailed study is required to show how the non-perturbative effects set in for Gevrey-$\alpha$ series. This is what we turn to next.

\subsection{Borel resummation of the Gevrey-$\alpha$ series and 
resurgence \label{sec4.3}}

In the above section we argued how the nodal diagrams show non-trivial factorial growths of Gevrey kind. Since this subject may be new to many physicists, we will first elaborate
on the basic mathematical structure of the Gevrey-$\alpha$ series before going into more detailed discussion on the Borel resummation of such a series.

\subsubsection{Mathematical structure of the Gevrey-$\alpha$ series \label{sec4.3.1}}

The Gevrey series \cite{gevreyorig, gevrey} which generalizes the usual asymptotic series $-$ where the coefficients show factorial growths $-$ appears in the mathematics literature when the truncation procedure of a power series is no longer restricted to some fixed parameter (usually denoted by ${\rm N}$). As an example let us consider a formal power series of the form:
\bg\label{stasuk}
\widetilde{\rm F}(g) = \sum_{k = 0}^\infty c_k g^k, \nd
with power-one factorial divergence, and let us assume ${\rm F}(g)$ to be a function asymptotic to it. Here $g$ is a parameter that could in general be complex and we will consider the limit when $g \to 0$. Few definitions then helps us understand some of the aforementioned statements. An asymptotic expansion of a function ${\rm F}(g)$ at a given point $g_o$ may be expressed as a formal series of simpler functions, much like what we have in \eqref{stasuk}, in which each successive term is much smaller than its predecessors. However this definition provides estimates of the value of ${\rm F}(g)$ for small $g$, within ${\cal O}(g^{\rm N})$, ${\rm N} \in \mathbb{N}$, which are insufficient to determine a unique ${\rm F}$ associated to $\widetilde{\rm F}$. As an example function like 
${\rm F}(g) \equiv {\rm exp}\left(-{1\over g^{1/\alpha}}\right)$ is beyond all orders of $\widetilde{\rm F}(g)$ in a sector of angle almost $\alpha\pi$. To deal with functions like that leads us to the notion of Gevrey asymptotics where a formal series like \eqref{stasuk} is by definition Gevrey of order $\alpha$, or Gevrey-$\alpha$ if:
\bg\label{mikasta}
\vert c_k \vert ~\le ~ {\rm C}_1 {\rm C}_2^k \left(k!\right)^\alpha, \nd
for some ${\rm C}_i$. The above bound on $c_k$ tells us that $c_k$ can have at most polynomial growth as ${\rm C}_2^k$ and factorial growth as $(k!)^\alpha$. Looking at $\mathbb{C}^{({\rm N})}_m$ in \eqref{altomaa} we see that there is indeed a polynomial growth of ${\rm N}$ and {\it at least} a factorial growth of $({\rm N}!)^2$. (See \eqref{hicksmel} and \eqref{hicksmel2} for the precise growth.) Thus the series constructed from 
$\mathbb{C}^{({\rm N})}_m$ would indeed be classified as a Gevrey series.
It is also easy to see that, if we define $g \equiv \lambda^\alpha$, then 
$\widetilde{\rm G}(\lambda) \equiv \widetilde{\rm F}(g)$ is in fact a Gevrey-1 series. Thus every Gevrey-$\alpha$ series may be converted to Gevrey-1. This is also in accordance to \eqref{corfu1} derived using Sterling approximation. Such a consideration leads us to the following observation:
\vskip.1in

\noindent {\it The Gevrey order of the series 
$\widetilde{\rm F}(g) =\sum\limits_{k = 0}^\infty (k!)^\alpha g^k$ where 
$\alpha > 0$ and $g \to 0$, is the same as that of the series $\widetilde{\rm G}(g) = \sum\limits_{k = 0}^\infty (\alpha k)! g^k$, thus reducing everything to Gevrey-1.}

\vskip.1in

\noindent One can make the above observation even more precise using the so-called {\it acceleration operators} of \'Ecalle \cite{ecalle}. This means that any finite order Gevrey series is Borel resummable after a repeated application of the acceleration operators (thus reducing to Gevrey-1). If this can be done, then the system has the property of multi-summability according to 
\'Ecalle \cite{ecalle}. Once we reduce to Gevrey-1 then a function ${\rm F}(g)$ is Gevrey-1 asymptotic to $\widetilde{\rm F}(g)$ as $g \to 0$ if:
\bg\label{align}
\Delta(g; {\rm N}) \equiv \Big\vert {\rm F}(g) - \sum_{k = 0}^{\rm N} c_k g^k\Big\vert ~ \le ~ 
{\rm C}_1 {\rm C}_2^{{\rm N} + 1} \vert g\vert^{{\rm N} + 1} \left({\rm N} + 1\right)!, \nd
{\it i.e.} if the error $\Delta(g; {\rm N})$ is of the same size as the first omitted terms in $\widetilde{\rm F}(g)$ (up to possible powers of constants). The question now is how to perform Borel sum and Borel transform of the function $\widetilde{\rm F}(g)$. 

Mathematicians distinguish between Borel transformation and {\it formal} Borel transformation, the latter of which is generically considered as Borel transformation in the physics literature. The Borel transformation in the mathematics literature is typically denoted by ${\bf B}_d$, the Borel transform in the direction $d$, via:
\bg\label{borolgum}
{\bf B}_d f(\zeta) = {\bf f}(\zeta) = {1\over 2\pi i} \int_{\gamma_d} 
{dx} ~{f(x)\over x^2} ~{\rm exp}\left({\zeta\over x}\right), \nd
with $\gamma_d$ $-$ where the subscript $d$ means that we have chosen an appropriate $\zeta$ along direction $d$ for convergence $-$ is a contour (with the corresponding deformation) shown in the figure.

\begin{equation*}
% Contour original 
\begin{tikzpicture}
  \def\ul{0.6}
  \def\Ra{1.4}
  \def\Rb{0.8}
  \coordinate (A)  at (0,0);
  \coordinate (Ma) at (-1,1.5);
  \coordinate (Mb) at (3,1);
  \coordinate (Mc) at (-0.5,-1);
  \coordinate (B)  at (0,0);
  
  % PATHS
  \draw[xcol,thick, midarr=0.6]
    (A) to[out=140,in=-110] (Ma)
        to[out=70,in=90] (Mb)to[out=-90, in=-70] (Mc)  node[above right=-2] {$\gamma_d$}   to[out=140, in=80] (B);
        
   \draw [->] [black,thick] (0, 0) -- node [above right=-1,  near end] {$d$} (4,1.2) ;
  
  \fill[black] (A)  circle (0.08) node[above right=-1] {$O$}; 
  
% Deformation 
  \coordinate (a)  at (4,0);
  \coordinate (b) at (6,0);
  
  \draw [->] [black, thick] (a) -- node[above] {contour} node[below] {deformation} (b);
  
% Contour deformed 
  \def\ul{0.6}
  \def\Ra{1.4}
  \def\Rb{0.8}
  \coordinate (A1)  at (8,0);
  \coordinate (Ma1) at (7,1.5);
  \coordinate (Mb1) at (11,1);
  \coordinate (Mc1) at (7.5,-1);
  \coordinate (B1)  at (8,0);
  
  % PATHS
  \draw[xcol,thick, midarr=0.6]
    (Ma1) to[out=70,in=90] (Mb1) node[above right=-2] {$\gamma_d$}   to[out=-90, in=-70] (Mc1);
    
  \draw[red,dashed]  (A1) to[out=140,in=-110] (Ma1);
  \draw[red,dashed]  (Mc1) to[out=140, in=80] (B1);
  \draw[xcol, thick, midarr=0.6] (Mc1) to[out=115, in=-100] (Ma1);
      
  \fill[black] (A1)  circle (0.08) node[above right=-1] {$O$};
\end{tikzpicture}
\end{equation*}

\noindent The above form is also denoted by ${\bf B} f(\zeta)$ if the Borel transformation is independent of the direction $d$. Additionally, 
the above form makes sense once a reasonably good hypothesis of $f(x)$ is proposed. One such choice for $f(x)$
is $f(x) = x^n$ with $n \in {\bf N}^\ast$, using which the Borel transform takes the form:
\bg\label{borlgum2}
{\bf B}f(\zeta) = {\zeta^{n - 1}\over \Gamma(n)} = {\zeta^{n - 1}\over (n-1)!}. \nd
Note the appearance of the Gamma function $\Gamma(n)$, which is in fact a more natural way (than the factorial) to present the Borel transformation.
On the other hand, a {\it formal} Borel transformation $\hat{\bf B}$ for a function 
$\hat{f} \in {\bf C}$, $\hat{f}(x) = \sum\limits_{n = 1}^\infty c_n x^n$ with $n \in \Lambda^\ast = \Lambda - \{0\}$, may then be expressed as:
\bg\label{borlgum3}
\hat{\bf B}\hat{f}(\zeta) = \sum_{n \in \Lambda^\ast} 
{a_n \zeta^{n - 1}\over \Gamma(n)}. \nd
For physicists this would be an ideal way to express the Borel transformation and we will take this as our definition (and remove the hats over $\hat{\bf B}$ and $\hat{f}$). The Borel resummation could then be performed by taking a Laplace transform of \eqref{borlgum3}.

Generalization of the above story exists that leads us to the Gevrey series. One way to achieve this would be via the so-called ramification operators $\rho_\alpha$ ($\alpha > 0$) that acts on the function $f(x)$ in \eqref{borolgum} as $\rho_\alpha f(x) = f(x^{1/\alpha})$ (which would imply that 
$\rho_{1/\alpha} = \rho^{-1}_\alpha$). If $d^\alpha$ corresponds to $d$ by the ramification $\rho_\alpha$, the $\alpha$-Borel transform may be defined as:
\bg\label{melhicks}
{\bf B}_{\alpha; d} = \rho^{-1}_\alpha {\bf B}_{d^\alpha} \rho_\alpha, \nd
which would be a natural way to implement the Borel transformation using the ramification operators. In general the above procedure leads to the following definition of the $\alpha$-Borel transformation:
\bg\label{bessmel}
{\bf B}_{\alpha; d}f(\zeta_\alpha) = {\bf f}_\alpha(\zeta_\alpha) = {1\over 2\pi i}\int_{\gamma_\alpha} dx~{\alpha f(x)\over x^{\alpha+1}}~{\rm exp}\left({\zeta_\alpha^\alpha\over x^\alpha}\right), \nd
which could be compared to our earlier expression \eqref{borolgum}. A Laplace transformation of ${\bf f}_\alpha(\zeta_\alpha)$ should lead us back to the function that we want. This immediately gives us the following expression for the Laplace transformation:
\bg\label{bessador}
{\bf L}_{\alpha; d}{\bf f}_\alpha(\zeta_\alpha) = f(x) = \int_l 
d\zeta_\alpha~\alpha {\bf f}_\alpha(\zeta_\alpha)~\zeta_\alpha^{\alpha - 1} ~{\rm exp}\left(-{\zeta_\alpha^\alpha\over x^\alpha}\right). \nd
The above should more or less suffice, but we would like to extend this a bit further to incorporate the {\it formal} Borel transformation which would be useful for the physicists. To this end, we can go back to the standard case and define an operator ${\bf S}$ which relates ${\bf f}(\zeta)$ with $\hat{\bf f}(\zeta)$ via: ${\bf f} = {\bf S}\hat{\bf f}$. Clearly if $\hat{f} \in {\bf C}\{x\}$, ${\bf S}\hat{\bf B} = {\bf B}$ and 
$f_d(x) \equiv {\bf S}\hat{f}(x)$, the ${\bf S}_d \equiv {\bf L}_d {\bf S}\hat{\bf B}$ extends the operator ${\bf S}$ in $d$ direction. In terms of Borel transformation, this means:
\bg\label{henaot}
f_d(x) = {\bf L}_d {\bf S} \hat{\bf f} = {\bf L}_d {\bf S} \hat{\bf B} \hat{f}, \nd
as a way to implement the {\it formal} Borel transformation from the usual one and thus easily reproduces \eqref{borolgum} from the aforementioned considerations. 

For the $\alpha$-Borel case, the ${\bf S}$ operator is again useful and is implemented here via ${\bf f}_\alpha = {\bf S} \hat{\bf f}_\alpha$. The {\it formal} Borel transformation then is simply $\hat{\bf f}_\alpha = \hat{\bf B}_\alpha \hat{f} \in {\bf C}\{\zeta_\alpha\}$. Then by definition the ``$\alpha$-sum" of $\hat{f}$ in the direction $d$, denoted by $f_{\alpha; d}$, may be implemented via:
\bg\label{coribess}
f_{\alpha;d}(x) = {\bf L}_{\alpha;d} {\bf S} \hat{\bf f}_\alpha = {\bf L}_{\alpha; d} {\bf S}
\hat{\bf B}_\alpha \hat{f} \equiv {\bf S}_{\alpha; d} \hat{f}, \nd
implying that the operator ${\bf S}_{\alpha; d}$ is an injective morphism of differential algebras much like ${\bf S}_d$ above (the difference is that the former is defined over differential algebra of $\alpha$-summable series wheras the latter is defined over differential algebra of Borel summable series). In general the usefulness of ${\bf S}_{\alpha; d}$ lies in the fact that ${\bf S}_{\alpha; d}\hat{f} = {\bf S}_{\alpha'; d}\hat{f}$ for $\alpha' > \alpha, (\alpha, \alpha') > 0$ and $\hat{f} \in {\bf C}[[x]]$ is $\alpha$-summable and $\alpha'$-summable in direction $d$. On the other hand, for $(\alpha, \alpha') > 0$ with $\alpha'>\alpha$ and 
$\hat{f} \in {\bf C}[[x]]_{1/\alpha'}$, if $\hat{f}$ is $\alpha$-summable then it is also a convergent series. This leads us to conclude that if
$f(x) = \sum\limits_{n = 0}^\infty c_n x^n$ is a power series, then the {\it formal} $\alpha$-Borel transform, which we shall henceforth denote as $\alpha$-Borel transform, may be expressed in the following suggestive way:
\bg\label{gabloz}
{\bf B}_\alpha f(\zeta) = \zeta^{-\alpha} \sum_{n = 0}^\infty {c_n \zeta^n\over 
\Gamma\left({n + \alpha\over \alpha}\right)}, \nd
where we have removed the hat from $f$ and assumed that ${\bf B}_{\alpha; d} = {\bf B}_\alpha$. Note the expected appearance of the Gamma-function instead of the factorial. On the other hand if $g(\zeta)$ is a function of the complex variable $\zeta$ and $d$
is a ray in the $\zeta$-plane, we can define the $\alpha$-Laplace transform in the
direction $d$ by:
\bg\label{coricha}
{\bf L}_{\alpha; d}~ g \equiv \alpha\int_{\gamma_d} d\zeta~\zeta^{\alpha - 1} g(\zeta) 
~{\rm exp}\left(-{\zeta^\alpha\over x^\alpha}\right), \nd
provided it converges. Putting everything together, we can see that the $\alpha$-Borel sum of $f$ in the direction $d$ is given by the following procedure:

\vskip.1in

\noindent $\bullet$ 
Compute ${\bf B}_\alpha f$ as above; assume it converges in a neighbourhood of $\zeta = 0$ to some function $g_0$.

\vskip.1in

\noindent $\bullet$ 
Analytically continue $g_0$ to a function $g$ defined in a neighbourhood
of the ray $d$.

\vskip.1in

\noindent $\bullet$ 
Compute the Laplace transform ${\bf L}_{\alpha; d}~ g$.

\vskip.1in

\noindent Note that the convergence in first case is equivalent to the Gevrey-$\alpha$ condition.
  This however does not always guarantee the existence of an analytic continuation as
in second case, and even if the latter exists, there is no guarantee that the
integral defining ${\bf L}_{\alpha; d} ~g$ converges. These are 
of course all things that are needed to be justified in a given example, and the
proof usually requires some additional structure, such as a differential
equation satisfied by $f$. For our case, as we shall see below this is not going to be an issue because the nodal diagrams will generically produce ``nice" functions whose $\alpha$-Borel transform will be well-defined.

\begin{figure}[h]
\centering
\begin{tabular}{c}
\includegraphics[width=3in]{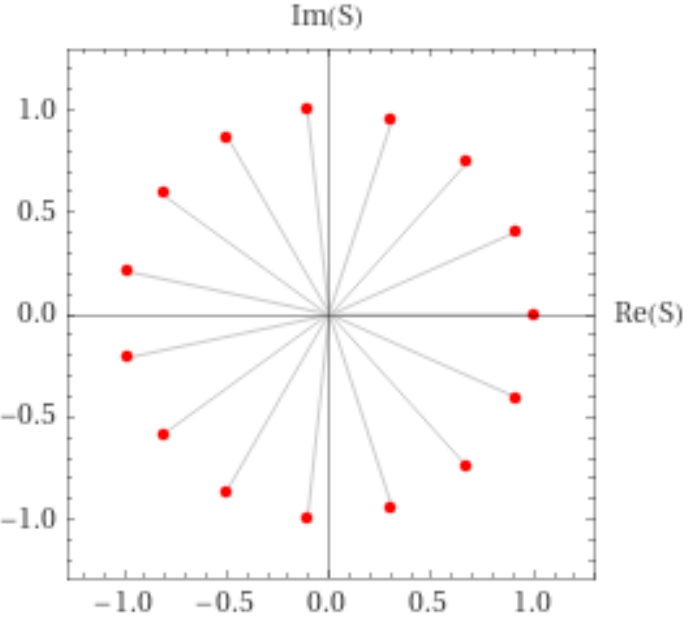}
\end{tabular}
\caption[]{The number of poles in the Borel plane when $\alpha$ is odd in 
\eqref{macdonald}. Note that there is only one positive pole on the Borel axis, rest are all complex.}
\label{roots1}
\end{figure} 

\subsubsection{Borel resummation of the Gevrey-$\alpha$ series \label{sec4.3.2}}

\begin{figure}[h]
\centering
\begin{tabular}{c}
\includegraphics[width=3in]{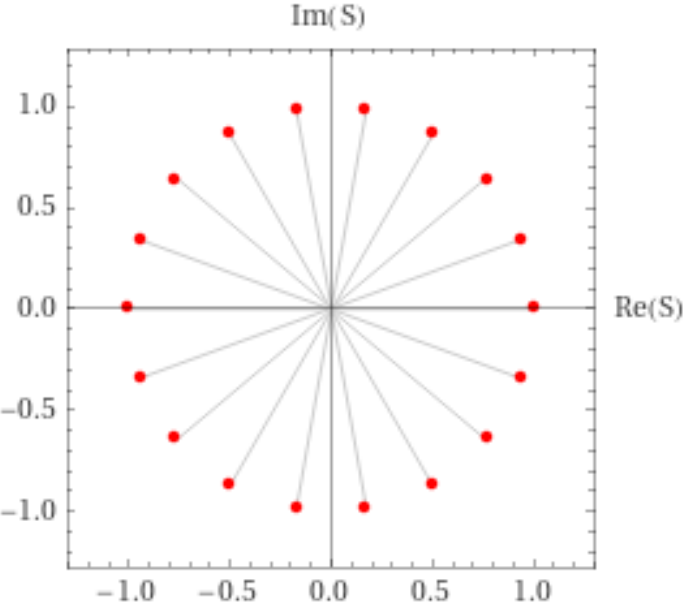}
\end{tabular}
\caption[]{The number of poles in the Borel plane when $\alpha$ is even in 
\eqref{macdonald}. As in figure \ref{roots1}, there is only one positive pole on the Borel axis, rest are all complex except one that is negative real. This appears to be generic implying that irrespective of the choice of $\alpha \in \mathbb{Z}_+$, there is only one contribution from the positive pole to \eqref{fajarjee6} or \eqref{fajarjee5}.}
\label{roots2}
\end{figure}

Nothing what we said so far in the above sub-section is new or original, and in fact has been studied extensively in the maths literature. In physics most examples studied so far (see for example \cite{unsal} and the review article in \cite{dorigoni}) have been for the usual Borel case with very little literature on Gevrey-$\alpha$ for $\alpha > 1$. 
Let us then see how we can apply the aforementioned ideas to the explicit case of \eqref{turkish} and \eqref{incal} associated with Gevrey-$\alpha$ with $\alpha \ge 2$. We will start with $p = 0$, which makes it $f(0)$ from \eqref{incal} and converts \eqref{turkish} to \eqref{fajarjee3} with $\alpha \ge 2$. The coefficients $\mathbb{C}^{({\rm N})}_m$ from \eqref{altomaa} show a typical growth as:
\bg\label{hicksmel}
\mathbb{C}^{({\rm N})}_m ~ = ~ {\cal A}^{\otimes{\rm N}} \sum_{n = 1}^\infty \left(\int d\Omega c_{2n, m}~{\bf H}_{2n}(\overline{\bf k}_{\rm IR})\right) \left({\rm N}!\right)^\alpha, \nd
which is somewhat similar to the growth proposed in \eqref{mikasta} with a few minor differences. 
Note that we have not specified the precise value for $\alpha$. This is in accordance to our plan of generalizing the analysis to any Gevrey-$\alpha$ case implying, in turn, that the Borel resummability will be performed keeping $\alpha$ arbitrary. The coefficient ${\cal A}$ controlling the polynomial growth takes the form:

{\scriptsize
\bg\label{coldjuly}
{\cal A} &\equiv& {\rm A}(\{m_i\}, \{n_i\}, \{f_i\})\\
& = & \mathbb{B}_1(m_1, .., m_q)\mathbb{B}_2(n_1,..,n_r)\mathbb{B}_3(f_1,..,f_{s-1})
\prod_{i = 1}^q \int_{k_{\rm IR}}^\mu d^{11} k_i{\overline{\alpha}(k_i)k_i^{m_i}\over a(k_i)}\prod_{j = 1}^r \int_{k_{\rm IR}}^\mu d^{11} l_j{\overline{\beta}(l_j)l_j^{n_j}\over b(l_j)}\prod_{t = 1}^{s-1} 
\int_{k_{\rm IR}}^\mu d^{11} f_t{\overline{\gamma}(f_t)f_t^{p_t}\over c(f_t)}, \nonumber \nd}
where all the parameters have been defined earlier in \eqref{altomaa}. We can use \eqref{coldjuly} to define any powers of ${\cal A}$ by multiplying equivalent polynomials together and summing over sets like $\{{\rm U}\} = \left(\{m_i\}, \{n_i\}, \{f_i\}\right)$ which would also explain how 
${\rm N}$ copies of $\mathbb{B}_i$ from \eqref{coldjuly} may be related to $\mathbb{D}, \mathbb{D}_{\varphi_2}$ and 
$\mathbb{D}_{\varphi_3}$ in \eqref{altomaa}. For example 
${\cal A}^{\otimes 2} \equiv \sum\limits_{{\rm U}, {\rm U}'} 
{\rm A}(\{{\rm U}\}){\rm A}(\{{\rm U}'\})$. It is easy to see that for simpler cases we consider ${\cal A}^{\otimes {\rm N}} \approx {\cal A}^{\rm N}$ which could then be identified with ${\rm C}_2$ in \eqref{mikasta}. This will convert the growth of the coefficients $\mathbb{C}^{({\rm N})}_m$ to take the form:
\bg\label{hicksmel2}
\mathbb{C}^{({\rm N})}_m ~ = ~ {\cal A}^{\rm N} \left({\rm N}!\right)^\alpha {\bf a}_m, \nd
where ${\bf a}_m = \sum\limits_{n = 1}^\infty\left(\int d\Omega c_{2n, m}~{\bf H}_{2n}(\overline{\bf k}_{\rm IR})\right)$. The coefficients ${\cal A}$ and ${\bf a}_m$ are not unrelated, as both involve $c_{nm}$ from \eqref{greek}, but we will not worry too much about this and keep them separate. 
Putting everything together the dominant piece from \eqref{turkish} for 
$p = 0$ may now be Borel resummed to take the following form:

{\footnotesize
\bg\label{fajarjee5}
\sum_{{\rm N} = 0}^\infty g^{\rm N}\left[\sum_{m = 0}^\infty\mathbb{C}^{({\rm N})}_{m} {\bf H}_m(k_0) \right]^\mu_{k_{\rm IR}} &= & {1\over g^{1/\alpha}} \int_0^\infty d{\rm S} ~{\rm exp}\left(-{{\rm S}\over g^{1/\alpha}}\right) \sum_{{\rm N} = 0}^\infty\left[\sum_{m = 0}^\infty {\mathbb{C}_{m}^{({\rm N})} \over (\alpha{\rm N}!)} \Big({\bf H}_{m}(\mu) - {\bf H}_{m}(k_{\rm IR})\Big) {\rm S}^{\alpha {\rm N}}\right] \nonumber\\
& = & {1\over g^{1/\alpha}} \int_0^\infty d{\rm S} ~{\rm exp}\left(-{{\rm S}\over g^{1/\alpha}}\right)\left[{\sum\limits_{m = 0}^\infty {\bf a}_m \Big({\bf H}_{m}(\mu) - {\bf H}_{m}(k_{\rm IR})\Big)\over 1 - {\cal A} {\rm S}^\alpha}\right], \nd}
where note that in the second line we have exchanged the two summations and then Borel resummed over ${\rm N}$ following our above discussions\footnote{A more careful usage of \eqref{corfu1} in the first line of \eqref{fajarjee5} would suggest that ${\cal A} \to {\cal A}' \equiv {{\cal A}\over \alpha^\alpha}$. Since this is just a constant rescaling of ${\cal A}$, we will not distinguish between ${\cal A}$ and 
${\cal A}'$ and simply rename ${\cal A}'$ as ${\cal A}$ in the second line of \eqref{fajarjee5} henceforth.}. The ${\rm S}$ appearing above may be related to the Laplace transformation \eqref{coricha} as ${\rm S} = \zeta^\alpha$. One may then make the appropriate contour deformation to bring \eqref{coricha} in the above form. Interestingly the Borel resummation leads to non-perturbative effects of the form ${\rm exp}\left(-{1\over g^{1/\alpha}}\right)$, which are fixed in a given theory, and {\it not} of the form ${\rm exp}\Big(-{{\rm S}_1\over g_1^{1/q}}\Big),  {\rm exp}\Big(-{{\rm S}_2\over g_2^{1/r}}\Big)$ and  ${\rm exp}\Big(-{{\rm S}_3\over g_3^{1/(s-1)}}\Big)$ that we entertained earlier. This relieves us from yet another conundrum that we faced in 
the sub-section \ref{tagcorfu}. Finally, 
the poles in the Borel plane may be ascertained from the roots of the following polynomial function:
\bg\label{macdonald}
{\rm S}^\alpha - {1\over {\cal A}} \equiv \prod_{i = 1}^\alpha\left({\rm S} - a_i\right), \nd
with ${\cal A}$ as in \eqref{coldjuly} and $\sum\limits_{i = 1}^\alpha a_i = 0$. For $\alpha = 2$ there are two roots: $a_1 = {1\over \sqrt{\cal A}}, a_2 = -{1\over \sqrt{\cal A}}$ in the Borel axis. The above pole structure remains unchanged even if we take different values of $p \in \mathbb{Z}_+$ in \eqref{turkish}. This may be easily seen by Borel resumming over ${\rm N}$ in \eqref{turkish}, in the same vein as \eqref{fajarjee5}, to get:

{\scriptsize
\bg\label{fajarjee6}
\sum_{{\rm N} = 0}^\infty g^{\rm N}\sum_{m = 0}^\infty {(-1)^p 2^p ~m!~\mathbb{C}^{({\rm N})}_m ~{\bf H}_{m-p}(k_0)\over (m - p)!}\Bigg\vert_{k_{\rm IR}}^\mu = 
{1\over g^{1/\alpha}} \int_0^\infty d{\rm S} ~{\rm exp}\left(-{{\rm S}\over g^{1/\alpha}}\right) \sum_{m = 0}^\infty {(-1)^p 2^p m!\over (m - p)!} 
\left[{{\bf a}_m {\bf H}_{m - p}(k_0)\over 1 - {\cal A}{\rm S}^\alpha}\right]_{k_{\rm IR}}^\mu, \nonumber\\ \nd}
for any values of $p \in \mathbb{Z}_+$. For $p = 0$ we reproduce 
\eqref{fajarjee5}. The poles in the Borel plane is controlled by the roots of \eqref{macdonald} $\forall p \in \mathbb{Z}_+$. For the present case there is only {\it one} positive real root that contributes irrespective of the choice of $\alpha$ as shown in {\Su Figures} \ref{roots1} and \ref{roots2}.

\subsection{New contributions and additional subtleties on the Borel 
plane \label{sec4.4}}

There are a few more subtleties\footnote{And one cheat: we have assumed that the denominator of the path integral may be expressed as a convergent series.} that we kept under the rug so far, and it is now time to elaborate on them. In the following we will study them in details and point out under what conditions certain assumptions that we made earlier remain true.

\subsubsection{Off-shell contributions to the Glauber-Sudarshan states \label{sec4.4.1}}

The first one is on the presence of 
off-shell contributions to the Glauber-Sudarshan states first described in \cite{coherbeta}. These off-shell contributions appear from $k_0 = 0$ modes and we should be able to incorporate them via the wave-function \eqref{creolet}. Note that \eqref{creolet} was used earlier to allow for 
${\bf k} = \overline{\bf k}_{\rm IR}$ states to the nodal diagrams which, in turn, allowed us to incorporate purely the temporal dependence expected of the $2+1$ dimensional part of the metric \eqref{carrie} (for the simpler case when we take ${\rm H}(y) = 1$. This can be easily generalized, but we will not do so here). Question now is whether we can incorporate both $k_0 = 0$ and ${\bf k} = \overline{\bf k}_{\rm IR}$, again keeping ${\rm H}(y) = 1$, from the $2+1$ dimensional part of the solitonic configuration \eqref{sissys}, and $\overline{\bf k}^2_{\rm IR} \ge k^2_{\rm IR}$ due to UV/IR mixing. The answer is in the affirmative if we incorporate the following change to the wave-function \eqref{creolet}:

{\footnotesize
\bg\label{creolet2}
\psi_{k}({\rm X}) e^{-ik_0t} \equiv {k^2\over \pi \vert\omega\vert {\bf k}^9}
~{\rm exp}\left(-{({\bf k} - \overline{\bf k}_{\rm IR})^2\over \omega^2} - ik_0t\right)\left[1 - 
{1\over \pi \vert\omega\vert} {\rm exp}\left(-{k_0^2\over \omega^2}\right)\right]~ 
{\cal F}\left({\bf k}, {\rm X}\right), \nd}
where ${\cal F}\left({\bf k}, {\rm X}\right)$ continues to solve the Schr\"odinger equation constructed over the solitonic configuration 
\eqref{sissys} and satisfies ${\cal F}\left({\bf k} = \overline{\bf k}_{\rm IR}, {\rm X}\right) = 1$ in the limit $\omega \to 0$. The additional contributions come with an overall minus sign to remove the background solitonic configuration at least from its most probable amplitude. This means we now have to solve an equation of the form:
\bg\label{turkish2}
\sum_{{\rm N} = 0}^\infty g^{\rm N} \sum_{m = 0}^\infty \mathbb{C}_m^{({\rm N)}} {\bf H}_{2m}(0) = {1\over \Lambda^{4/3}}, \nd
which should be compared to \eqref{turkish}. There are a few major differences: {\Su one}, we only have one equation instead of an infinite set classified by $p \in \mathbb{Z}_+$, and {\Su two}, because of the delta function $\delta(k_0)$ from \eqref{creolet2}, there is no need of implementing the Riemann-Lebesgue integral from \eqref{aug19visa} and the Hermite polynomial takes the simple form of ${\bf H}_{2m}(0)$. On the other hand, the coefficients $\mathbb{C}_m^{({\rm N})}$ from these additional contributions become:

{\footnotesize
\bg\label{altomaa3}
\mathbb{C}^{({\rm N})}_m &\equiv & 
\prod_{t = 1}^{{\rm N}q + 1}\left({\pi_t\over b_t}\right)^{1/2} \sum_{\{m_i\}}\mathbb{D}\left(m_1, m_2, m_3, ..., m_{{\rm N}q-1}, m_{{\rm N}q}\right)\mathbb{D}_{\varphi_2}({\rm N}r) \mathbb{D}_{\varphi_3}({\rm N}(s-1))\\
&\times & \int_{k_{\rm IR}}^\mu d^{11}k_1 {\overline\alpha(k_1) k_1^{m_1}\over a(k_1)}
\int_{k_{\rm IR}}^\mu d^{11}k_2~{\overline\alpha(k_2) k_2^{m_2}\over a(k_2)}....
\int_{k_{\rm IR}}^\mu d^{11}k_q~{\overline\alpha(k_q) k_q^{m_q}\over a(k_q)}....
\int_{k_{\rm IR}}^\mu d^{11}k_{2q}~{\overline\alpha(k_{2q}) k_{2q}^{m_{2q}}\over a(k_{2q})}....\nonumber\\
& \times & \int_{k_{\rm IR}}^\mu d^{11}k_{{\rm N}q - 1}~{\overline\alpha(k_{{\rm N}q - 1}) k_{{\rm N}q - 1}^{m_{{\rm N}q - 1}}\over a(k_{{\rm N}q - 1})}
\int_{k_{\rm IR}}^\mu d^{11}k_{{\rm N}q}~{\overline\alpha(k_{{\rm N}q}) k_{{\rm N}q}^{m_{{\rm N}q}}\over a(k_{{\rm N}q})}
\sum_{n = 1}^\infty \left(\int d\Omega~c_{2n,2m}~{\bf H}_{2n}(\overline{\bf k}_{\rm IR})\right), \nonumber
\nd}
which is almost what we had in \eqref{altomaa} except with one minor difference: the coefficients are all even of the form $c_{2n, 2m}$ instead of $c_{2n, m}$ before. The growth of these coefficients then remain the same as in \eqref{hicksmel} or \eqref{hicksmel2}, implying that for the off-shell case we should again Borel resum the LHS of \eqref{turkish2}
to get an equation of the form:
\bg\label{henaois}
{1\over g^{1/\alpha}} \int_0^\infty d{\rm S} ~{\rm exp}\left(-{{\rm S}\over g^{1/\alpha}}\right) \sum_{m = 0}^\infty \left({{\bf a}_m {\bf H}_{2m}(0)\over 1 - {\cal A}{\rm S}^\alpha}\right) = {1\over \Lambda^{4/3}}, \nd
where ${\bf a}_m = \sum\limits_{n = 1}^\infty\left(\int d\Omega c_{2n, 2m}~{\bf H}_{2n}(\overline{\bf k}_{\rm IR})\right)$. We notice that the pole structure has remained unchanged from what we had earlier and therefore from {\Su Figures} \ref{roots1} and \ref{roots2} we know that only the positive real root of \eqref{macdonald} will contribute.

\subsubsection{Alternative on-shell and off-shell computations \label{sec4.4.2}}

The story has developed very cleanly so far, although one might wonder if there is an alternative way to implement the delta function constraints instead of implementing them via the wave-functions \eqref{turkish} and 
\eqref{turkish2}. If this is possible then we can keep the wave-functions simple, as in \eqref{pugh}, and implement the necessary requirements via the ans\"atze \eqref{greek} appropriately. In the following we shall see that this is indeed possible, and we will demonstrate this using the mode $\overline{\alpha}(k_0, {\bf k})$ only. Again, this could be extended to the other two set of modes but we will not do so here. Our starting point would be to split the coefficients $c_{mn}$, which are in general functions of the angular coordinates, in the following simple way:
\bg\label{chrishnell}
c_{nm} = c_n b_m + \bar{c}_n {\bar b}_m + \tilde{c}_n \tilde{b}_m, \nd
where $\bar{c}$ and $\bar{b}$ do not imply complex conjugations as $c_{nm}$ is taken to be real here. The above splitting is only for convenience and we could have proceeded with our computations without invoking \eqref{chrishnell}, as will become clearer soon. Using \eqref{chrishnell} we can rewrite the 
ans\"atze for ${\alpha}(k_0, {\bf k})$ from \eqref{greek} in the following way:

{\tiny
\bg\label{bullock}
{\alpha}(k_0, {\bf k}) = {\rm V}\overline{\alpha}(k_0, {\bf k}) = \sum_{n = 0}^\infty c_n {\bf H}_n({\bf k}) \sum_{m = 0}^\infty b_m{\bf H}_m(k_0) + 
\sum_{n = 0}^\infty \bar{c}_n {\bf H}_n({\bf k}) \sum_{m = 0}^\infty \bar{b}_m{\bf H}_m(k_0) + 
\sum_{n = 0}^\infty \tilde{c}_n {\bf H}_n({\bf k}) \sum_{m = 0}^\infty \tilde{b}_m{\bf H}_m(k_0), \nonumber\\ \nd}
and ask what values of $(c_n, \bar{c}_n, \tilde{c}_n)$ and 
$(b_n, \bar{b}_n, \tilde{b}_n)$ would reproduce the same results that we got from the wave-function ans\"atze \eqref{creolet2} or \eqref{creolet}.
One possibility would be to allow for the following choices:

{\scriptsize
\bg\label{rocreyna}
&& \sum_{n = 0}^\infty c_n {\bf H}_n({\bf k}) ~ = ~
\sum_{n = 0}^\infty \bar{c}_n {\bf H}_n({\bf k}) ~ = ~ {1\over \pi \vert\omega\vert {\bf k}^9} {\rm exp}\left[-{({\bf k} - \overline{\bf k}_{\rm IR})^2\over \omega^2}\right]\\
&& \sum_{m = 0}^\infty \bar{b}_m {\bf H}_m(k_0) = {k_0^2\over \pi \vert\omega\vert} {\rm exp}\left[-{(k_0 -\overline\kappa_{\rm IR})^2\over \omega^2}\right] = 
\sum_{m = 0}^\infty \tilde{b}_m {\bf H}_m(k_0) k_0^2, ~~~~
\sum_{n = 0}^\infty \tilde{c}_n {\bf H}_n({\bf k}) = -{1\over \pi \vert\omega\vert {\bf k}^7} {\rm exp}\left[-{({\bf k} - \overline{\bf k}_{\rm IR})^2\over \omega^2}\right], \nonumber \nd}
where the choice $(k_0, {\bf k}) = (\overline\kappa_{\rm IR}, \overline{\bf k}_{\rm IR})$ would keep the wave-function constant. Needless to say, we require $-\overline{\kappa}^2_{\rm IR} + \overline{\bf k}^2_{\rm IR} \ge k^2_{\rm IR}$. (Earlier we had taken $\overline{\kappa}_{\rm IR} = 0$. Here we generalize this.) Interestingly the ${\bf k}^{-9}$ and ${\bf k}^{-7}$ is not necessary in the presence of the IR cut-off (although we take them here) and one may show that all the coefficients 
$(c_n, \bar{c}_n, \bar{b}_m, \tilde{c}_n, \tilde{b}_m)$ may be easily determined from \eqref{rocreyna}. In the limit $\omega \to 0$ these coefficients take the following form:
\bg\label{lou}
&&\bar{b}_m = {\overline\kappa^2_{\rm IR} {\bf H}_m(\overline\kappa_{\rm IR}) 
~{\rm exp}(-{\overline\kappa}^2_{\rm IR}) \over 2^m m! \sqrt{\pi}}, ~~~~~
\tilde{b}_m = {{\bf H}_m(\overline\kappa_{\rm IR}) 
~{\rm exp}(-{\overline\kappa}^2_{\rm IR}) \over 2^m m! \sqrt{\pi}} \nonumber\\
&& c_n ~= ~ \bar{c}_n = {{\bf H}_n(\overline{\bf k}_{\rm IR}) ~{\rm exp}(-\overline{\bf k}^2_{\rm IR}) \over 2^n n! \sqrt{\pi}~ \overline{\bf k}^9_{\rm IR}}, ~~~~
\tilde{c}_n = -{{\bf H}_n(\overline{\bf k}_{\rm IR}) ~{\rm exp}(-\overline{\bf k}^2_{\rm IR}) \over 2^n n! \sqrt{\pi}~ \overline{\bf k}^7_{\rm IR}}, \nd
which is the limit that we will be interested in here.
The only remaining unknown quantities are $b_m$ whose values will be determined soon. Since we choose non-zero $\overline{\kappa}_{\rm IR}$, we may allow 
a wave-function ans\"atze of the form (compare \eqref{pugh}):
\bg\label{mabel}
\psi_{k}({\rm X}, t) \equiv
{\cal F}({\bf k}, {\rm X}) ~e^{-i(k_0 - \overline\kappa_{\rm IR})t}, \nd
in \eqref{lincoln} and in \eqref{fontini} which is now simple with the constraint ${\cal F}(\overline{\bf k}_{\rm IR}, {\rm X}) = 1$ so as to not generate any 
spatial or temporal dependencies when ${\bf k} = \overline{\bf k}_{\rm IR}$ and 
$k_0 = \overline\kappa_{\rm IR}$. Plugging \eqref{rocreyna} and \eqref{lou} in \eqref{bullock}, we get the following ans\"atze for $\overline{\alpha}(k_0, {\bf k})$:

{\scriptsize  
\bg\label{rocketr}
\overline{\alpha}(k_0, {\bf k}) = 
{1\over \pi{\rm V}\vert\omega\vert {\bf k}^9} {\rm exp}\left[-{({\bf k} - \overline{\bf k}_{\rm IR})^2\over \omega^2}\right] \sum_{m = 0}^\infty b_m {\bf H}_m(k_0) - {k^2\over \pi^2{\rm V}\vert\omega\vert^2 {\bf k}^9} {\rm exp}\left[-{({\bf k} - \overline{\bf k}_{\rm IR})^2 + (k_0 - \overline\kappa_{\rm IR})^2\over \omega^2}\right], \nonumber\\ \nd}
expressed in terms of the as-yet-unknown parameters $b_m$. 
One may also invoke similar ans\"atze for $\overline{\beta}(l_0, {\bf l})$
and $\overline{\gamma}(f_0, {\bf f})$, but for that we will have to solve the corresponding Schwinger-Dyson's equations to figure out how the fluxes and the Rarita-Schwinger fermions behave. They impose more technical challenges so we will avoid elaborating them any further here. Instead we will ask how the integrated amplitude \eqref{lorenp} behaves with the choice \eqref{rocketr} in the limit $\omega \to 0$. There are two key pieces in \eqref{lorenp} which give us:

{\tiny
\bg\label{leah146}
&& \lim_{\omega \to 0} \int_{k_{\rm IR}}^\mu d^{11}k~{\overline{\alpha}(k)\over a(k)}~k^{m_1} \equiv {\cal B} = \int d\Omega 
\left[\int_{\kappa_{\rm IR}}^{\mu_0} dk_0 \sum_{m = 0}^\infty b_m {\bf H}_m(k_0) (-k_0^2 + \overline{\bf k}^2_{\rm IR})^{{m_1\over 2} - 1} - \overline{k}_{\rm IR}^{m_1}\right] \\
&& \lim_{\omega \to 0} \int_{k_{\rm IR}}^\mu d^{11}k~{\cal F}({\bf k}, {\rm X}) {\rm exp}(-i(k_0-\overline\kappa_{\rm IR})t)~{\overline{\alpha}(k)\over a(k)} = 
 \int d\Omega 
\left[\int_{\kappa_{\rm IR} - \overline{\kappa}_{\rm IR}}^{\mu_0 - \overline\kappa_{\rm IR}} {dk_0\over \overline{\bf k}^2_{\rm IR} - (k_0 + \overline\kappa_{\rm IR})^2} \sum_{m = 0}^\infty b_m {\bf H}_m(k_0+\overline\kappa_{\rm IR})~{\rm exp}(-ik_0 t) - 1\right], \nonumber \nd}
where we have defined $\overline{k}^2_{\rm IR} \equiv -\overline\kappa^2_{\rm IR} + \overline{\bf k}^2_{\rm IR} \ge k^2_{\rm IR}$, which will help us to control the second integral in \eqref{leah146}, and absorbed any constants in $b_m$. The presence of $k^2$ in \eqref{rocketr} keeps the second terms in both the above equations simple. The Lorentz invariance is broken because $\overline{\alpha}(k_0, {\bf k})$ is arranged to choose a specific set of momenta in \eqref{rocketr} which are bounded from below by the IR cut-off. Thus taking these IR cut-offs for both temporal and spatial momenta, the real part of the second integral in \eqref{leah146} takes the following form:

{\footnotesize
\bg\label{taglaurent}
&& \lim_{\omega \to 0}\int_{k_{\rm IR}}^\mu d^{11}k~{\bf Re}\Big(\psi_{\bf k}({\rm X}) {\rm exp}(-i(k_0-\overline\kappa_{\rm IR})t)\Big){\overline{\alpha}(k)\over a(k)} \nonumber\\
& = & 
\sum_{m = 0}^\infty \int d\Omega~b_m\left[\sum_{p = 0}^\infty (-1)^p e^{ik_0 t} {\partial^p\over \partial k_0^p} \left({{\bf H}_m(k_0+\overline\kappa_{\rm IR})\over 
{\bf k}^2_{\rm IR} - (k_0+\overline\kappa_{\rm IR})^2}\right) {1\over (it)^{p+1}} 
+ {\bf c.c}\right]_{\kappa_{\rm IR} - \overline{\kappa}_{\rm IR}}^{\mu_0 - \overline{\kappa}_{\rm IR}} - \int d\Omega, \nd}
where we have used the Riemann-Lebesgue lemma from \eqref{aug19visa}, and 
$\psi_{\bf k}({\rm X}) = {\cal F}({\bf k}, {\rm X})$. Additionally, on the LHS of \eqref{taglaurent} we have imposed the limit $\omega \to 0$ and on the RHS the two limiting values for $k_0$ are $\kappa_{\rm IR} - \overline{\kappa}_{\rm IR}$ and $\mu_0 - \overline{\kappa}_{\rm IR}$ for both \eqref{leah146} and \eqref{taglaurent} instead of $\kappa_{\rm IR}$ and $\mu_0$. If we take $\overline{\kappa}_{\rm IR} = 0$ we recover the usual limiting values. Plugging \eqref{taglaurent} and \eqref{leah146} back in 
\eqref{lorenp} and then adding back the contributions from the $\varphi_2$ and $\varphi_3$ sectors, we can easily see that $\mathbb{C}_m^{({\rm N})}$
coefficients (which we will call as $\mathbb{C}_{m; 1}^{({\rm N})}$ and $\mathbb{C}_{m; 2}^{({\rm N})}$ to distinguish between the ${\bf k} = \overline{\bf k}_{\rm IR}$ and the off-shell $k_0 = \overline{\kappa}_{\rm IR}, {\bf k} = \overline{\bf k}_{\rm IR}$ cases) take the following form:

{\footnotesize
\bg\label{altomaa8}
\mathbb{C}^{({\rm N})}_{m; 1} &=& \mathbb{C}_{m; 2}^{({\rm N})} \int d\Omega~b_m \equiv
\prod_{t = 1}^{{\rm N}q + 1}\left({\pi_t\over b_t}\right)^{1/2} \sum_{\{m_i\}}\mathbb{D}\left(m_1, m_2, m_3, ..., m_{{\rm N}q-1}, m_{{\rm N}q}\right)\mathbb{D}_{\varphi_2}({\rm N}r) \mathbb{D}_{\varphi_3}({\rm N}(s-1))\nonumber\\
&\times & \int_{k_{\rm IR}}^\mu d^{11}k_1 {\overline\alpha(k_1) k_1^{m_1}\over a(k_1)}
\int_{k_{\rm IR}}^\mu d^{11}k_2~{\overline\alpha(k_2) k_2^{m_2}\over a(k_2)}....
\int_{k_{\rm IR}}^\mu d^{11}k_q~{\overline\alpha(k_q) k_q^{m_q}\over a(k_q)}....
\int_{k_{\rm IR}}^\mu d^{11}k_{2q}~{\overline\alpha(k_{2q}) k_{2q}^{m_{2q}}\over a(k_{2q})}....\nonumber\\
& \times & \int_{k_{\rm IR}}^\mu d^{11}k_{{\rm N}q - 1}~{\overline\alpha(k_{{\rm N}q - 1}) k_{{\rm N}q - 1}^{m_{{\rm N}q - 1}}\over a(k_{{\rm N}q - 1})}
\int_{k_{\rm IR}}^\mu d^{11}k_{{\rm N}q}~{\overline\alpha(k_{{\rm N}q}) k_{{\rm N}q}^{m_{{\rm N}q}}\over a(k_{{\rm N}q})}\int d\Omega~b_m,
\nd}
where $b_m$ is the coefficient appearing in \eqref{bullock} and the $\overline{\alpha}(k)$ integrals follow the ones in \eqref{leah146}, thus involve the $b_m$ series therein. The above series should be compared to 
\eqref{altomaa} and \eqref{altomaa3} and we can notice few key differences: {\Su One}, is of course the form of the $\overline{\alpha}(k)$ integrals as discussed above. Similar changes are expected for the $\overline{\beta}(l)$ and $\overline{\gamma}(f)$ integrals. And {\Su two}, the $m$ dependence of 
$\mathbb{C}^{({\rm N})}_m$ appears solely from $b_m$ instead of $c_{2n, m}$ or $c_{2n, 2m}$ coefficients earlier. The $b_n$ dependence also appears in the $\overline{\alpha}(k)$ integrals, so there is a mixture. The growth of the $\mathbb{C}^{({\rm N})}_{m; i}$ coefficients follow a similar pattern as in 
\eqref{hicksmel} or \eqref{hicksmel2}:
\bg\label{hicksmel4}
\mathbb{C}^{({\rm N})}_{m; 1} = {\cal A}^{\rm N} \left({\rm N}!\right)^\alpha b_m,~~~~~ \mathbb{C}^{({\rm N})}_{m; 2} = {\cal A}^{\rm N} \left({\rm N}!\right)^\alpha \nd
with the difference being that $b_m$ from \eqref{bullock} replaces 
${\bf a}_m$ therein, but ${\cal A}$ takes exactly the same form as in \eqref{coldjuly} (with the integrals now defined from \eqref{taglaurent}).
The growth of the off-shell case is also a bit different from what we had earlier in \eqref{altomaa3}. After the dust settles, the path integral with the nodal diagrams leads to the following two equations:
\bg\label{camilahawk}
&& {1\over g^{1/\alpha}} \int_0^\infty d{\rm S} ~{\rm exp}\left(-{{\rm S}\over g^{1/\alpha}}\right) {1\over 1 - {\cal A}{\rm S}^\alpha} = 
{1\over \Lambda^{4/3}}\\
&& {1\over g^{1/\alpha}} \int_0^\infty d{\rm S} ~{\rm exp}\left(-{{\rm S}\over g^{1/\alpha}}\right) {(-1)^p\over 1 - {\cal A}{\rm S}^\alpha}
\sum_{m = 0}^\infty b_m {\partial\over \partial k_0^p} \left({{\bf H}_m(k_0+\overline\kappa_{\rm IR})\over 
\overline{\bf k}^2_{\rm IR} - (k_0 +\overline\kappa_{\rm IR})^2}\right)\Bigg\vert_{\kappa_{\rm IR} - \overline{\kappa}_{\rm IR}}^{\mu_0 - \overline\kappa_{\rm IR}} = f(p), \nonumber \nd
where $f(p)$ as in \eqref{incal}. The above two set of equations 
may now be compared with \eqref{fajarjee6} and \eqref{henaois}: the off-shell case using the wave-function modification \eqref{creolet2} gave us \eqref{henaois} which still had some imprints of ${\bf a}_m$. Now, looking at the first equation in \eqref{camilahawk} we see that we can make a clean separation between the non-perturbative (in $g$) and perturbative (in $m$) series, giving us the following relations between the coefficients $b_m$:
\bg\label{maycamil}
\sum_{m = 0}^\infty b_m {\partial^p\over \partial k_0^p} \left({{\bf H}_m(k_0+\overline\kappa_{\rm IR})\over 
(k_0+\overline\kappa_{\rm IR})^2 - \overline{\bf k}^2_{\rm IR}}\right)\Bigg\vert_{\kappa_{\rm IR} - \overline{\kappa}_{\rm IR}}^{\mu_0 - \overline\kappa_{\rm IR}} = {\Gamma\left(-{5\over 3}\right)\over \sqrt{2\pi/3}}~{\left({5\over 3}\right)! \over \left({5\over 3}-p\right)!}
~k_0^{{5\over 3} - p}\Bigg\vert_{\kappa_{\rm IR} - \overline{\kappa}_{\rm IR}}^{\mu_0 - \overline{\kappa}_{\rm IR}}, \nd
for all values of $p \in \mathbb{Z}_+$, with ${\cal A}$ satisfying the first equation in \eqref{camilahawk}. We have also regulated the RHS of \eqref{maycamil} using $\overline\kappa_{\rm IR}$. (As mentioned earlier, we can keep $\overline{\kappa}_{\rm IR} = 0$, so there is no reason to presume $\kappa_{\rm IR} = \overline{\kappa}_{\rm IR}$ ever. It is easy to see therein how the RHS of \eqref{maycamil} does not blow up.) There are literally an infinite possible $b_m$'s here, but there are also an infinite number of equations determined from an infinite number of values for $p \in \mathbb{Z}_+$, so in-principle the system will have solutions. We can restrict ourselves to a finite (but large) set of $b_m$'s and solve \eqref{maycamil} to express $b_m$ in terms of $(\mu_0, \kappa_{\rm IR})$. Since ${\cal A}$ depends on the values of $b_m$, we can use the first equation in \eqref{camilahawk} to relate $\Lambda, \mu_0$ and $\kappa_{\rm IR}$.

\subsubsection{The denominators of the path integrals \eqref{wilsonrata} and \eqref{2summers} \label{sec4.4.3}}

The above conclusion relies on the fact that the denominator of the path integral in \eqref{2summers} does not change any of the equations in \eqref{camilahawk}. In fact we have used \eqref{lucliu2}, or \eqref{shaystone}, to keep the denominator of the path integral in \eqref{wilsonrata} simple by assuming that under special cases it may be expressed in terms of a convergent series. Unfortunately the scenario is more complicated 
because of the form of the interactions in \eqref{angelslune}. The interactions include {\it sum} 
over 
$(n, m, .., s)$ which means even at the first order in the coupling constants there is a sum of three sets of interactions. 
The {\Su first} set 
involves one kind of fields with various denominations depending on what values of 
$(q, r, s)$ we take. They all appear with different coupling constants. The {\Su second} set consists of the sum of two kinds of fields again with various denominations which we may denote as:
\bg\label{katedidi}
\sum_{i, j} c_{ij} ~\widetilde\varphi_i^{q_i} \otimes \widetilde\varphi_j^{q_j}, \nd
where $i \ne j = (1, 2, 3)$, $q_i \in \mathbb{Z}_+$ and $c_{ij}$ are the coupling constants. The {\Su third} set consists of the sum of all the three field interactions. This is the part we have been dealing with so far for fixed values of $(n, m, ..., s)$. To ${\rm N}$-th order the first set shows a growth of at least $({\rm N}!)^0$, the second set shows a growth of at least ${\rm N}!$, and the third set shows a growth of at least 
$({\rm N}!)^2$. The second and the third set may all be Borel resummed and added to the first set to get a mixed term of the form:
\bg\label{charlieht}
\sum_{{\rm N} = 0}^\infty g^{\rm N} ~{\cal B}^{\rm N} + \sum_{\alpha \ge 1} {1\over g^{1/\alpha}} \int_0^\infty d{\rm S} ~{\rm exp}\left(-{{\rm S}\over g^{1/\alpha}}\right) {1\over 1 - {\cal A}_\alpha{\rm S}^\alpha}, \nd
where ${\cal B}$ could be more general than the one defined in \eqref{leah146}, ${\cal A}_1$ appears from the $\varphi_1$ sector coupled to $(\varphi_2, \varphi_3)$ sectors and ${\cal A}_2$ is ${\cal A}$ defined earlier. There would also be additional contributions to \eqref{charlieht} from the $(\varphi_2, \varphi_3)$ sectors which we do not show here.

The above equation \eqref{charlieht} shows that there is a possibility that we can get some convergent series because of the $({\rm N}!)^0$ growth of some of the interactions. Does this then justify the usage of \eqref{lucliu2} in the denominator of the path integral \eqref{wilsonrata} which, in turn, allowed us to keep only the {\it dominant} nodal diagrams?

To see this we will have to develop some machinery. We will call $\mathbb{T}$ as the value of the tree-level diagram from \eqref{luna3} with the condition that \eqref{luna2} is normalized to unity. Our aim is to compute the path integral \eqref{wilsonrata} and 
\eqref{2summers} with due care for the denominator. We can denote the nodal diagrams from \eqref{luna30} using a parameter $r$, $r \in \mathbb{Z}_+$, as:
\bg\label{luna37}
\left[{\rm dom}\right]_{\varphi_1}^{(-r)} ~ \equiv ~ \frac{1}{{\rm V}^{q-r}} 
\sum_{\rm U} \quad
\footnotesize {
 \begin{tikzpicture}[baseline={(0, 0.3cm-\MathAxis pt)}, thick,
main/.style = {draw, circle, fill=black, minimum size=3pt},
dot/.style={inner sep=0pt,fill=black, circle, minimum size=3pt},
dots/.style={inner sep=0pt,fill=black, circle, minimum size=1pt}] 
  \node[dot] (1) at (0, 0.3) {};
  
  \node[dot] (11) at (-0.5, 0.3) {} ;
  \node[dot] (111) at (-1, 0.3) {};
  
  \node[dot] (12) at (0, 0.8) {};
  \node[dot] (13) at (0, 1.8) {};
  \node[dot] (14) at (0, -0.2) {};
  \node[dot] (15) at (0, -1.2) {};
  \node[main] (121) at (1, 0.8) {};
  \node[main] (131) at (1, 1.8) {};
  \node[main] (141) at (1, -0.2) {};
  \node[main] (151) at (1, -1.2) {};
  
  \node[dot] (112) at (-1, 0.8) {};
  \node[dot] (113) at (-1, 1.8) {};
  \node[dot] (114) at (-1, -0.2) {};
  \node[dot] (115) at (-1, -1.2) {};
  \node[main] (1121) at (-2, 0.8) {};
  \node[dot] (1131) at (-2, 1.8) {};
  \node[main] (1141) at (-2, -0.2) {};
  \node[main] (1151) at (-2, -1.2) {};
  
  \node[main] (11311) at (-2.9, 2.3) {};
  \node[main] (11312) at (-2.9, 1.9) {};
  \node[main] (11313) at (-2.9, 1.4) {};
  \node[dots] (01) at (-2.7, 1.8) {};
  \node[dots] (01) at (-2.7, 1.7) {};
  \node[dots] (01) at (-2.7, 1.6) {};

\draw (1)  -- node {} (11);
\draw (11) -- node {} (111);

\draw (12) -- node[midway, above right, pos=0.2] {} (121);
\draw (13) -- node[midway, above right, pos=0.2] {} (131);
\draw (14) -- node[midway, above right, pos=0.2] {} (141);
\draw (15) -- node[midway, below right, pos=0] {\tiny $k_{u_{q-r-1}}$} (151);

\draw (112) -- node[midway, above right, pos=0.2] {} (1121);
\draw (113) -- node[midway, above right, pos=1] {\tiny $(r+1) k_j$} (1131);
\draw (114) -- node[midway, above right, pos=0.2] {} (1141);
\draw (115) -- node[midway, below right, pos=0] {} (1151);

\draw (1131) -- node {} (11311);
\draw (1131) -- node {} (11312);
\draw (1131) -- node {} (11313);

\draw (13) -- node {} (14);
\draw (14) [dashed] -- node {} (15);
\draw (113) -- node {} (114);
\draw (114) [dashed] -- node {} (115);

\end{tikzpicture} }
\nd
where $r = 0$ denotes the most dominant diagram, $r > 0$ are all the nodal diagrams suppressed by ${\rm V}^{-r}$, and ${\rm U} = (j, {u_1}, ..., 
{u_{q - r-1}})$. The subscript $\varphi_1$ implies the diagrams from the $\varphi_1$ fields, and therefore there would be equivalent set of diagrams from the $(\varphi_2, \varphi_3)$ sectors (which we don't show here). As mentioned in \eqref{luna30}, these diagrams would contribute to the denominator of \eqref{2summers}. For the numerator, we will need to diagrammatically represent how the source interacts with 
\eqref{luna37}. There are two possible ways. The simplest ones are denoted by:
\bg\label{luna38}
\left[{\rm dom}\right]_{\varphi_1}^{(-r)} ~\cup~\mathbb{T} ~\equiv ~ \frac{1}{{\rm V}^{q-r + 1}} \sum_{\rm U} 
\footnotesize {
\begin{tikzpicture}[baseline={(0, 0.3cm-\MathAxis pt)}, thick,
main/.style = {draw, circle, fill=black, minimum size=3pt},
dot/.style={inner sep=0pt,fill=black, circle, minimum size=2pt},
dots/.style={inner sep=0pt,fill=black, circle, minimum size=1pt}, 
ball/.style={ball  color=white, circle,  minimum size=15pt}] 
  \node[dot] (1) at (0, 0.3) {};
  
  \node[ball] (11) at (-0.5, 0.3) {i} ;
  \node[dot] (111) at (-1, 0.3) {};
  
  \node[dot] (12) at (0, 0.8) {};
  \node[dot] (13) at (0, 1.8) {};
  \node[dot] (14) at (0, -0.2) {};
  \node[dot] (15) at (0, -1.2) {};
  \node[main] (121) at (1, 0.8) {};
  \node[main] (131) at (1, 1.8) {};
  \node[main] (141) at (1, -0.2) {};
  \node[main] (151) at (1, -1.2) {};
  
  \node[dot] (112) at (-1, 0.8) {};
  \node[dot] (113) at (-1, 1.8) {};
  \node[dot] (114) at (-1, -0.2) {};
  \node[dot] (115) at (-1, -1.2) {};
  \node[main] (1121) at (-2, 0.8) {};
  \node[dot] (1131) at (-2, 1.8) {};
  \node[main] (1141) at (-2, -0.2) {};
  \node[main] (1151) at (-2, -1.2) {};
  
  \node[main] (11311) at (-2.9, 2.3) {};
  \node[main] (11312) at (-2.9, 1.9) {};
  \node[main] (11313) at (-2.9, 1.4) {};
  \node[dots] (01) at (-2.7, 1.8) {};
  \node[dots] (01) at (-2.7, 1.7) {};
  \node[dots] (01) at (-2.7, 1.6) {};

\draw (1)  -- node {} (11);
\draw (11) -- node {} (111);

\draw (12) -- node[midway, above right, pos=0.2] {} (121);
\draw (13) -- node[midway, above right, pos=0.2] {} (131);
\draw (14) -- node[midway, above right, pos=0.2] {} (141);
\draw (15) -- node[midway, below right, pos=0] {\tiny $k_{u_{q-r-1}}$} (151);

\draw (112) -- node[midway, above right, pos=0.2] {} (1121);
\draw (113) -- node[midway, above right, pos=1] {\tiny $(r+1) k_j$} (1131);
\draw (114) -- node[midway, above right, pos=0.2] {} (1141);
\draw (115) -- node[midway, below right, pos=0] {} (1151);

\draw (1131) -- node {} (11311);
\draw (1131) -- node {} (11312);
\draw (1131) -- node {} (11313);

\draw (13) -- node {} (14);
\draw (14) [dashed] -- node {} (15);
\draw (113) -- node {} (114);
\draw (114) [dashed] -- node {} (115);

\end{tikzpicture} } 
\nd
where ${\rm U} = (i, j, {u_1}, ..., {u_{q - r-1}})$ and $i \ne j \ne u_k$. 
(Henceforth all elements of ${\rm U}$ will be unequal unless mentioned otherwise.)
We note that the source momenta do not equal the momenta from the internal legs. The $r = 0$ nodal diagram was used earlier because this is the dominant one. Again, these diagrams for $r > 0$ are suppressed by 
${\rm V}^{-r}$. The more non-trivial diagrams are the ones in which the source momenta equal the momenta of the internal legs of the nodal diagrams. They may be represented by:

{\footnotesize{ 
\bg\label{luna39}
\left[{\rm dom}\right]_{\varphi_1}^{(-r)} 
~\cup~\overset{\curvearrowleft}{\mathbb{T}} =\frac{1}{{\rm V}^{q-r}} 
\sum_{\rm U} 
 \begin{tikzpicture}[baseline={(0, 0.3cm-\MathAxis pt)}, thick,
main/.style = {draw, circle, fill=black, minimum size=4pt},
dot/.style={inner sep=0pt,fill=black, circle, minimum size=3pt},
dots/.style={inner sep=0pt,fill=black, circle, minimum size=1pt}, 
ball/.style={ball  color=white, circle,  minimum size=15pt}] 
  \node[dot] (1) at (0, 0.3) {};
  
  \node[ball] (11) at (-0.5, 0.3) {\red j} ;
  \node[dot] (111) at (-1, 0.3) {};
  
  \node[dot] (12) at (0, 0.8) {};
  \node[dot] (13) at (0, 1.8) {};
  \node[dot] (14) at (0, -0.2) {};
  \node[dot] (15) at (0, -1.2) {};
  \node[main] (121) at (1, 0.8) {};
  \node[main] (131) at (1, 1.8) {};
  \node[main] (141) at (1, -0.2) {};
  \node[main] (151) at (1, -1.2) {};
  
  \node[dot] (112) at (-1, 0.8) {};
  \node[dot] (113) at (-1, 1.8) {};
  \node[dot] (114) at (-1, -0.2) {};
  \node[dot] (115) at (-1, -1.2) {};
  \node[main] (1121) at (-2, 0.8) {};
  \node[dot] (1131) at (-2, 1.8) {};
  \node[main] (1141) at (-2, -0.2) {};
  \node[main] (1151) at (-2, -1.2) {};
  
  \node[main] (11311) at (-2.9, 2.3) {};
  \node[main] (11312) at (-2.9, 1.9) {};
  \node[main] (11313) at (-2.9, 1.4) {};
  \node[dots] (01) at (-2.7, 1.8) {};
  \node[dots] (01) at (-2.7, 1.7) {};
  \node[dots] (01) at (-2.7, 1.6) {};

\draw (1)  -- node {} (11);
\draw (11) -- node {} (111);

\draw (12) -- node[midway, above right, pos=0.2] {} (121);
\draw (13) -- node[midway, above right, pos=0.2] {} (131);
\draw (14) -- node[midway, above right, pos=0.2] {} (141);
\draw (15) -- node[midway, below right, pos=0] {\tiny $k_{u_{q-r-1}}$} (151);

\draw (112) -- node[midway, above right, pos=0.2] {} (1121);
\draw (113) -- node[midway, above right, pos=1] {\tiny ${\red (r+1) k_j}$} (1131);
\draw (114) -- node[midway, above right, pos=0.2] {} (1141);
\draw (115) -- node[midway, below right, pos=0] {} (1151);

\draw (1131) -- node {} (11311);
\draw (1131) -- node {} (11312);
\draw (1131) -- node {} (11313);

\draw (13) -- node {} (14);
\draw (14) [dashed] -- node {} (15);
\draw (113) -- node {} (114);
\draw (114) [dashed] -- node {} (115);

\end{tikzpicture} 
 \; + \; \frac{1}{{\rm V}^{q-r}} 
\sum_{\rm U} 
 \begin{tikzpicture}[baseline={(0, 0.3cm-\MathAxis pt)}, thick,
main/.style = {draw, circle, fill=black, minimum size=4pt},
dot/.style={inner sep=0pt,fill=black, circle, minimum size=3pt},
dots/.style={inner sep=0pt,fill=black, circle, minimum size=1pt}, 
ball/.style={ball  color=white, circle,  minimum size=15pt}] 
  \node[dot] (1) at (0, 0.3) {};
  
  \node[ball] (11) at (-0.5, 0.3) {\red i} ;
  \node[dot] (111) at (-1, 0.3) {};
  
  \node[dot] (12) at (0, 0.8) {};
  \node[dot] (13) at (0, 1.8) {};
  \node[dot] (14) at (0, -0.2) {};
  \node[dot] (15) at (0, -1.2) {};
  \node[main] (121) at (1, 0.8) {};
  \node[main] (131) at (1, 1.8) {};
  \node[main] (141) at (1, -0.2) {};
  \node[main] (151) at (1, -1.2) {};
  
  \node[dot] (112) at (-1, 0.8) {};
  \node[dot] (113) at (-1, 1.8) {};
  \node[dot] (114) at (-1, -0.2) {};
  \node[dot] (115) at (-1, -1.2) {};
  \node[main] (1121) at (-2, 0.8) {};
  \node[dot] (1131) at (-2, 1.8) {};
  \node[main] (1141) at (-2, -0.2) {};
  \node[main] (1151) at (-2, -1.2) {};
  
  \node[main] (11311) at (-2.9, 2.3) {};
  \node[main] (11312) at (-2.9, 1.9) {};
  \node[main] (11313) at (-2.9, 1.4) {};
  \node[dots] (01) at (-2.7, 1.8) {};
  \node[dots] (01) at (-2.7, 1.7) {};
  \node[dots] (01) at (-2.7, 1.6) {};

\draw (1)  -- node {} (11);
\draw (11) -- node {} (111);

\draw (12) -- node[midway, above right, pos=0.2] {} (121);
\draw (13) -- node[midway, above right, pos=0.2] {} (131);
\draw (14) -- node[midway, above right, pos=0.2] {} (141);
\draw (15) -- node[midway, below right, pos=0] {\tiny $k_{u_{q-r-1}}$} (151);

\draw (112) -- node[midway, above right, pos=0.8] {\tiny ${\red k_i}$} (1121);
\draw (113) -- node[midway, above right, pos=1] {\tiny $(r+1) k_j$} (1131);
\draw (114) -- node[midway, above right, pos=0.2] {} (1141);
\draw (115) -- node[midway, below right, pos=0] {} (1151);

\draw (1131) -- node {} (11311);
\draw (1131) -- node {} (11312);
\draw (1131) -- node {} (11313);

\draw (13) -- node {} (14);
\draw (14) [dashed] -- node {} (15);
\draw (113) -- node {} (114);
\draw (114) [dashed] -- node {} (115);

\end{tikzpicture} \nonumber\\
\nd}}
where we see that there are two possible choices: {\Su one}, corresponds to the source momenta equal to the momenta $k_j$ of the internal multiple legs with 
${\rm U} = (j, {u_1}, ..., {u_{q - r - 1}})$. And {\Su two}, corresponds to the source momenta equal to the momenta $k_i$ of the internal single legs of the nodal diagrams with 
${\rm U} = (j, i, {u_1}, ..., {u_{q - r - 2}})$. Both these set of diagrams are suppressed by ${\rm V}^{-r}$. (The matching momenta are shown in {\red red}.)

The above are clearly not the only set of diagrams we can draw. We can allow nodal diagrams with different set of multiple legs. These we can denote with superscripts $(r_1, r_2, r_3,..)$  etc. As an example the denominator of \eqref{2summers} can have nodal diagrams of the following form:
\bg\label{luna40}
\left[{\rm dom}\right]_{\varphi_s}^{(-r_1, -r_2, -r_3)} = 
 \frac{1}{{\rm V}^{q-r_1-r_2-r_3}} \sum_{\rm U}~~~ 
 \footnotesize {
 \begin{tikzpicture}[baseline={(0, 0.3cm-\MathAxis pt)}, thick,
main/.style = {draw, circle, fill=black, minimum size=4pt},
dot/.style={inner sep=0pt,fill=black, circle, minimum size=3pt},
dots/.style={inner sep=0pt,fill=black, circle, minimum size=1pt}, 
ball/.style={ball  color=white, circle,  minimum size=15pt}] 
  \node[dot] (1) at (0, 0.3) {};
  
  \node[dot] (11) at (-0.5, 0.3) {} ;
  \node[dot] (111) at (-1, 0.3) {};
  
  \node[dot] (12) at (0, 0.8) {};
  \node[dot] (13) at (0, 1.8) {};
  \node[dot] (14) at (0, -0.2) {};
  \node[dot] (15) at (0, -1.2) {};
  \node[main] (121) at (1, 0.8) {};
  \node[main] (131) at (1, 1.8) {};
  \node[main] (141) at (1, -0.2) {};
  \node[main] (151) at (1, -1.2) {};
  
  \node[dot] (112) at (-1, 0.8) {};
  \node[dot] (113) at (-1, 1.8) {};
  \node[dot] (114) at (-1, -0.2) {};
  \node[dot] (115) at (-1, -1.2) {};
  \node[dot] (1121) at (-2.5, 0.8) {};
  \node[dot] (1131) at (-2, 1.8) {};
  \node[dot] (1141) at (-2, -0.2) {};
  \node[main] (1151) at (-2, -1.2) {};
  
  \node[main] (11211) at (-3.5, 1.3) {};
  \node[main] (11212) at (-3.5, 0.9) {};
  \node[main] (11213) at (-3.5, 0.4) {};
  \node[dots] (01) at (-3.1, 0.8) {};
  \node[dots] (01) at (-3.1, 0.7) {};
  \node[dots] (01) at (-3.1, 0.6) {};
  
  \node[main] (11311) at (-2.9, 2.3) {};
  \node[main] (11312) at (-2.9, 1.9) {};
  \node[main] (11313) at (-2.9, 1.4) {};
  \node[dots] (01) at (-2.7, 1.8) {};
  \node[dots] (01) at (-2.7, 1.7) {};
  \node[dots] (01) at (-2.7, 1.6) {};

  \node[main] (11411) at (-2.9, 0.3) {};
  \node[main] (11412) at (-2.9, -0.1) {};
  \node[main] (11413) at (-2.9, -0.7) {};
  \node[dots] (01) at (-2.7, -0.5) {};
  \node[dots] (01) at (-2.7, -0.3) {};
  \node[dots] (01) at (-2.7, -0.4) {};

\draw (1)  -- node {} (11);
\draw (11) -- node {} (111);

\draw (12) -- node[midway, above right, pos=0.2] {} (121);
\draw (13) -- node[midway, above right, pos=0.2] {} (131);
\draw (14) -- node[midway, above right, pos=0.2] {} (141);
\draw (15) -- node[midway, below right, pos=0] {\tiny $k_{u_{q-r_1-r_2-r_3 -3}}$} (151);

\draw (112) -- node[midway, above right, pos=1] {\tiny $(r_2+1) k_i$} (1121);
\draw (113) -- node[midway, above right, pos=1] {\tiny $(r_1+1) k_j$} (1131);
\draw (114) -- node[midway, below right, pos=1] {\tiny $(r_3+1) k_l$} (1141);
\draw (115) -- node[midway, below right, pos=0] {} (1151);

\draw (1131) -- node {} (11311);
\draw (1131) -- node {} (11312);
\draw (1131) -- node {} (11313);

\draw (1121) -- node {} (11211);
\draw (1121) -- node {} (11212);
\draw (1121) -- node {} (11213);

\draw (1141) -- node {} (11411);
\draw (1141) -- node {} (11412);
\draw (1141) -- node {} (11413);

\draw (13) -- node {} (14);
\draw (14) [dashed] -- node {} (15);
\draw (113) -- node {} (114);
\draw (114) [dashed] -- node {} (115);

\end{tikzpicture} }
\nd
where $s = 1, 2, 3$ corresponding to the three fields $(\varphi_1, \varphi_2, \varphi_3)$; $r_i$ for $i = 1, 2, 3$ denotes the multiplicity of three legs, with 
$r_i \ge 0$; and ${\rm U} = (i, j, l, {u_1}, ..., {u_{q-r_1-r_2-r_3-3}})$. The dominant diagram from \eqref{luna37} may be denoted alternatively as $\left[{\rm dom}\right]_{\varphi_1}^{(0, 0, 0, ..)}$, and the other diagrams are suppressed by ${\rm V}^{-r_1-r_2-r_3}$. Clearly the number of superscripts cannot exceed $q$, {\it i.e} 
$\sum\limits_{i = 1}^{\rm R} r_i = q$, with $q$ being the power of field $\varphi_1$ in \eqref{lincoln} and $r_{\rm R}$ being the upper limit. It is also easy to see that nodal diagrams corresponding to 
$\left[{\rm dom}\right]_{\varphi_1}^{(-r_1, -r_2, .., -r_{\rm R})} \cup 
\mathbb{T}$ would correspond to adding a source term whose momenta do not match with any of the momenta of the internal legs. A more non-trivial set of diagrams would be when the source momenta match with any one set of the internal legs. For the case with $(r_1, .., r_3)$ we can represent the nodal diagrams by:
\bg\label{luna41}
& \left[{\rm dom}\right]_{\varphi_1}^{(-r_1, -r_2, -r_3)}
~\cup~\overset{\curvearrowleft}{\mathbb{T}} = 
 \frac{1}{{\rm V}^{q-r_1-r_2-r_3}} \sum\limits_{\rm U} \quad 
 \footnotesize {
 \begin{tikzpicture}[baseline={(0, 0.3cm-\MathAxis pt)}, thick,
main/.style = {draw, circle, fill=black, minimum size=4pt},
dot/.style={inner sep=0pt,fill=black, circle, minimum size=3pt},
dots/.style={inner sep=0pt,fill=black, circle, minimum size=1pt}, 
ball/.style={ball  color=white, circle,  minimum size=15pt}] 
  \node[dot] (1) at (0, 0.3) {};
  
  \node[ball] (11) at (-0.5, 0.3) {\red i} ;
  \node[dot] (111) at (-1, 0.3) {};
  
  \node[dot] (12) at (0, 0.8) {};
  \node[dot] (13) at (0, 1.8) {};
  \node[dot] (14) at (0, -0.2) {};
  \node[dot] (15) at (0, -1.2) {};
  \node[main] (121) at (1, 0.8) {};
  \node[main] (131) at (1, 1.8) {};
  \node[main] (141) at (1, -0.2) {};
  \node[main] (151) at (1, -1.2) {};
  
  \node[dot] (112) at (-1, 0.8) {};
  \node[dot] (113) at (-1, 1.8) {};
  \node[dot] (114) at (-1, -0.2) {};
  \node[dot] (115) at (-1, -1.2) {};
  \node[dot] (1121) at (-2.5, 0.8) {};
  \node[dot] (1131) at (-2, 1.8) {};
  \node[dot] (1141) at (-2, -0.2) {};
  \node[main] (1151) at (-2, -1.2) {};
  
  \node[main] (11211) at (-3.5, 1.3) {};
  \node[main] (11212) at (-3.5, 0.9) {};
  \node[main] (11213) at (-3.5, 0.4) {};
  \node[dots] (01) at (-3.1, 0.8) {};
  \node[dots] (01) at (-3.1, 0.7) {};
  \node[dots] (01) at (-3.1, 0.6) {};
  
  \node[main] (11311) at (-2.9, 2.3) {};
  \node[main] (11312) at (-2.9, 1.9) {};
  \node[main] (11313) at (-2.9, 1.4) {};
  \node[dots] (01) at (-2.7, 1.8) {};
  \node[dots] (01) at (-2.7, 1.7) {};
  \node[dots] (01) at (-2.7, 1.6) {};

  \node[main] (11411) at (-2.9, 0.3) {};
  \node[main] (11412) at (-2.9, -0.1) {};
  \node[main] (11413) at (-2.9, -0.7) {};
  \node[dots] (01) at (-2.7, -0.5) {};
  \node[dots] (01) at (-2.7, -0.3) {};
  \node[dots] (01) at (-2.7, -0.4) {};

\draw (1)  -- node {} (11);
\draw (11) -- node {} (111);

\draw (12) -- node[midway, above right, pos=0.2] {\tiny $k_m$} (121);
\draw (13) -- node[midway, above right, pos=0.2] {} (131);
\draw (14) -- node[midway, above right, pos=0.2] {} (141);
\draw (15) -- node[midway, below right, pos=0] {\tiny $k_{u_{q-r_1-r_2-r_3-3}}$} (151);

\draw (112) -- node[midway, above right, pos=1] {\tiny ${\red (r_2+1) k_i}$} (1121);
\draw (113) -- node[midway, above right, pos=1] {\tiny $(r_1+1) k_j$} (1131);
\draw (114) -- node[midway, below right, pos=1] {\tiny $(r_3+1) k_l$} (1141);
\draw (115) -- node[midway, below right, pos=0] {} (1151);

\draw (1131) -- node {} (11311);
\draw (1131) -- node {} (11312);
\draw (1131) -- node {} (11313);

\draw (1121) -- node {} (11211);
\draw (1121) -- node {} (11212);
\draw (1121) -- node {} (11213);

\draw (1141) -- node {} (11411);
\draw (1141) -- node {} (11412);
\draw (1141) -- node {} (11413);

\draw (13) -- node {} (14);
\draw (14) [dashed] -- node {} (15);
\draw (113) -- node {} (114);
\draw (114) [dashed] -- node {} (115);

\end{tikzpicture} } 
\nonumber\\
& ~~~~~~~~~~~~~~~~~~~~~~~~ \quad + \quad \footnotesize{\frac{1}{{\rm V}^{q-r_1-r_2-r_3}} \sum\limits_{\rm U} \quad
 \begin{tikzpicture}[baseline={(0, 0.3cm-\MathAxis pt)}, thick,
main/.style = {draw, circle, fill=black, minimum size=4pt},
dot/.style={inner sep=0pt,fill=black, circle, minimum size=3pt},
dots/.style={inner sep=0pt,fill=black, circle, minimum size=1pt}, 
ball/.style={ball  color=white, circle,  minimum size=15pt}] 
  \node[dot] (1) at (0, 0.3) {};
  
  \node[ball] (11) at (-0.5, 0.3) {\red j} ;
  \node[dot] (111) at (-1, 0.3) {};
  
  \node[dot] (12) at (0, 0.8) {};
  \node[dot] (13) at (0, 1.8) {};
  \node[dot] (14) at (0, -0.2) {};
  \node[dot] (15) at (0, -1.2) {};
  \node[main] (121) at (1, 0.8) {};
  \node[main] (131) at (1, 1.8) {};
  \node[main] (141) at (1, -0.2) {};
  \node[main] (151) at (1, -1.2) {};
  
  \node[dot] (112) at (-1, 0.8) {};
  \node[dot] (113) at (-1, 1.8) {};
  \node[dot] (114) at (-1, -0.2) {};
  \node[dot] (115) at (-1, -1.2) {};
  \node[dot] (1121) at (-2.5, 0.8) {};
  \node[dot] (1131) at (-2, 1.8) {};
  \node[dot] (1141) at (-2, -0.2) {};
  \node[main] (1151) at (-2, -1.2) {};
  
  \node[main] (11211) at (-3.5, 1.3) {};
  \node[main] (11212) at (-3.5, 0.9) {};
  \node[main] (11213) at (-3.5, 0.4) {};
  \node[dots] (01) at (-3.1, 0.8) {};
  \node[dots] (01) at (-3.1, 0.7) {};
  \node[dots] (01) at (-3.1, 0.6) {};
  
  \node[main] (11311) at (-2.9, 2.3) {};
  \node[main] (11312) at (-2.9, 1.9) {};
  \node[main] (11313) at (-2.9, 1.4) {};
  \node[dots] (01) at (-2.7, 1.8) {};
  \node[dots] (01) at (-2.7, 1.7) {};
  \node[dots] (01) at (-2.7, 1.6) {};

  \node[main] (11411) at (-2.9, 0.3) {};
  \node[main] (11412) at (-2.9, -0.1) {};
  \node[main] (11413) at (-2.9, -0.7) {};
  \node[dots] (01) at (-2.7, -0.5) {};
  \node[dots] (01) at (-2.7, -0.3) {};
  \node[dots] (01) at (-2.7, -0.4) {};

\draw (1)  -- node {} (11);
\draw (11) -- node {} (111);

\draw (12) -- node[midway, above right, pos=0.2] {\tiny $k_m$} (121);
\draw (13) -- node[midway, above right, pos=0.2] {} (131);
\draw (14) -- node[midway, above right, pos=0.2] {} (141);
\draw (15) -- node[midway, below right, pos=0] {\tiny $k_{u_{q-r_1-r_2-r_3-3}}$} (151);

\draw (112) -- node[midway, above right, pos=1] {\tiny $(r_2+1) k_i$} (1121);
\draw (113) -- node[midway, above right, pos=1] {\tiny ${\red (r_1+1) k_j}$} (1131);
\draw (114) -- node[midway, below right, pos=1] {\tiny $(r_3+1) k_l$} (1141);
\draw (115) -- node[midway, below right, pos=0] {} (1151);

\draw (1131) -- node {} (11311);
\draw (1131) -- node {} (11312);
\draw (1131) -- node {} (11313);

\draw (1121) -- node {} (11211);
\draw (1121) -- node {} (11212);
\draw (1121) -- node {} (11213);

\draw (1141) -- node {} (11411);
\draw (1141) -- node {} (11412);
\draw (1141) -- node {} (11413);

\draw (13) -- node {} (14);
\draw (14) [dashed] -- node {} (15);
\draw (113) -- node {} (114);
\draw (114) [dashed] -- node {} (115);

\end{tikzpicture} } \nonumber
\\
& ~~~~~~~~~~~~~~~~~~~~~~~~\quad + \quad \footnotesize { 
\frac{1}{{\rm V}^{q-r_1-r_2-r_3}} \sum\limits_{\rm U} \quad
 \begin{tikzpicture}[baseline={(0, 0.3cm-\MathAxis pt)}, thick,
main/.style = {draw, circle, fill=black, minimum size=4pt},
dot/.style={inner sep=0pt,fill=black, circle, minimum size=3pt},
dots/.style={inner sep=0pt,fill=black, circle, minimum size=1pt}, 
ball/.style={ball  color=white, circle,  minimum size=15pt}] 
  \node[dot] (1) at (0, 0.3) {};
  
  \node[ball] (11) at (-0.5, 0.3) {\red l} ;
  \node[dot] (111) at (-1, 0.3) {};
  
  \node[dot] (12) at (0, 0.8) {};
  \node[dot] (13) at (0, 1.8) {};
  \node[dot] (14) at (0, -0.2) {};
  \node[dot] (15) at (0, -1.2) {};
  \node[main] (121) at (1, 0.8) {};
  \node[main] (131) at (1, 1.8) {};
  \node[main] (141) at (1, -0.2) {};
  \node[main] (151) at (1, -1.2) {};
  
  \node[dot] (112) at (-1, 0.8) {};
  \node[dot] (113) at (-1, 1.8) {};
  \node[dot] (114) at (-1, -0.2) {};
  \node[dot] (115) at (-1, -1.2) {};
  \node[dot] (1121) at (-2.5, 0.8) {};
  \node[dot] (1131) at (-2, 1.8) {};
  \node[dot] (1141) at (-2, -0.2) {};
  \node[main] (1151) at (-2, -1.2) {};
  
  \node[main] (11211) at (-3.5, 1.3) {};
  \node[main] (11212) at (-3.5, 0.9) {};
  \node[main] (11213) at (-3.5, 0.4) {};
  \node[dots] (01) at (-3.1, 0.8) {};
  \node[dots] (01) at (-3.1, 0.7) {};
  \node[dots] (01) at (-3.1, 0.6) {};
  
  \node[main] (11311) at (-2.9, 2.3) {};
  \node[main] (11312) at (-2.9, 1.9) {};
  \node[main] (11313) at (-2.9, 1.4) {};
  \node[dots] (01) at (-2.7, 1.8) {};
  \node[dots] (01) at (-2.7, 1.7) {};
  \node[dots] (01) at (-2.7, 1.6) {};

  \node[main] (11411) at (-2.9, 0.3) {};
  \node[main] (11412) at (-2.9, -0.1) {};
  \node[main] (11413) at (-2.9, -0.7) {};
  \node[dots] (01) at (-2.7, -0.5) {};
  \node[dots] (01) at (-2.7, -0.3) {};
  \node[dots] (01) at (-2.7, -0.4) {};

\draw (1)  -- node {} (11);
\draw (11) -- node {} (111);

\draw (12) -- node[midway, above right, pos=0.2] {\tiny $k_m$} (121);
\draw (13) -- node[midway, above right, pos=0.2] {} (131);
\draw (14) -- node[midway, above right, pos=0.2] {} (141);
\draw (15) -- node[midway, below right, pos=0] {\tiny $k_{u_{q-r_1-r_2-r_3-3}}$} (151);

\draw (112) -- node[midway, above right, pos=1] {\tiny $(r_2+1) k_i$} (1121);
\draw (113) -- node[midway, above right, pos=1] {\tiny $(r_1+1) k_j$} (1131);
\draw (114) -- node[midway, below right, pos=1] {\tiny ${\red (r_3+1) k_l}$} (1141);
\draw (115) -- node[midway, below right, pos=0] {} (1151);

\draw (1131) -- node {} (11311);
\draw (1131) -- node {} (11312);
\draw (1131) -- node {} (11313);

\draw (1121) -- node {} (11211);
\draw (1121) -- node {} (11212);
\draw (1121) -- node {} (11213);

\draw (1141) -- node {} (11411);
\draw (1141) -- node {} (11412);
\draw (1141) -- node {} (11413);

\draw (13) -- node {} (14);
\draw (14) [dashed] -- node {} (15);
\draw (113) -- node {} (114);
\draw (114) [dashed] -- node {} (115);

\end{tikzpicture} } \nonumber
\\
& ~~~~~~~~~~~~~~~~~~~~~~~~\quad + \quad
 \footnotesize { 
\frac{1}{{\rm V}^{q-r_1-r_2-r_3}} \sum\limits_{\rm U} \quad
 \begin{tikzpicture}[baseline={(0, 0.3cm-\MathAxis pt)}, thick,
main/.style = {draw, circle, fill=black, minimum size=4pt},
dot/.style={inner sep=0pt,fill=black, circle, minimum size=3pt},
dots/.style={inner sep=0pt,fill=black, circle, minimum size=1pt}, 
ball/.style={ball  color=white, circle,  minimum size=15pt}] 
  \node[dot] (1) at (0, 0.3) {};
  
  \node[ball] (11) at (-0.5, 0.3) {\red m} ;
  \node[dot] (111) at (-1, 0.3) {};
  
  \node[dot] (12) at (0, 0.8) {};
  \node[dot] (13) at (0, 1.8) {};
  \node[dot] (14) at (0, -0.2) {};
  \node[dot] (15) at (0, -1.2) {};
  \node[main] (121) at (1, 0.8) {};
  \node[main] (131) at (1, 1.8) {};
  \node[main] (141) at (1, -0.2) {};
  \node[main] (151) at (1, -1.2) {};
  
  \node[dot] (112) at (-1, 0.8) {};
  \node[dot] (113) at (-1, 1.8) {};
  \node[dot] (114) at (-1, -0.2) {};
  \node[dot] (115) at (-1, -1.2) {};
  \node[dot] (1121) at (-2.5, 0.8) {};
  \node[dot] (1131) at (-2, 1.8) {};
  \node[dot] (1141) at (-2, -0.2) {};
  \node[main] (1151) at (-2, -1.2) {};
  
  \node[main] (11211) at (-3.5, 1.3) {};
  \node[main] (11212) at (-3.5, 0.9) {};
  \node[main] (11213) at (-3.5, 0.4) {};
  \node[dots] (01) at (-3.1, 0.8) {};
  \node[dots] (01) at (-3.1, 0.7) {};
  \node[dots] (01) at (-3.1, 0.6) {};
  
  \node[main] (11311) at (-2.9, 2.3) {};
  \node[main] (11312) at (-2.9, 1.9) {};
  \node[main] (11313) at (-2.9, 1.4) {};
  \node[dots] (01) at (-2.7, 1.8) {};
  \node[dots] (01) at (-2.7, 1.7) {};
  \node[dots] (01) at (-2.7, 1.6) {};

  \node[main] (11411) at (-2.9, 0.3) {};
  \node[main] (11412) at (-2.9, -0.1) {};
  \node[main] (11413) at (-2.9, -0.7) {};
  \node[dots] (01) at (-2.7, -0.5) {};
  \node[dots] (01) at (-2.7, -0.3) {};
  \node[dots] (01) at (-2.7, -0.4) {};

\draw (1)  -- node {} (11);
\draw (11) -- node {} (111);

\draw (12) -- node[midway, above right, pos=0.2] {\tiny ${\red k_m}$} (121);
\draw (13) -- node[midway, above right, pos=0.2] {} (131);
\draw (14) -- node[midway, above right, pos=0.2] {} (141);
\draw (15) -- node[midway, below right, pos=0] {\tiny $k_{u_{q-r_1-r_2-r_3-3}}$} (151);

\draw (112) -- node[midway, above right, pos=1] {\tiny $(r_2+1) k_i$} (1121);
\draw (113) -- node[midway, above right, pos=1] {\tiny $(r_1+1) k_j$} (1131);
\draw (114) -- node[midway, below right, pos=1] {\tiny $(r_3+1) k_l$} (1141);
\draw (115) -- node[midway, below right, pos=0] {} (1151);

\draw (1131) -- node {} (11311);
\draw (1131) -- node {} (11312);
\draw (1131) -- node {} (11313);

\draw (1121) -- node {} (11211);
\draw (1121) -- node {} (11212);
\draw (1121) -- node {} (11213);

\draw (1141) -- node {} (11411);
\draw (1141) -- node {} (11412);
\draw (1141) -- node {} (11413);

\draw (13) -- node {} (14);
\draw (14) [dashed] -- node {} (15);
\draw (113) -- node {} (114);
\draw (114) [dashed] -- node {} (115);

\end{tikzpicture} }
\nd
where ${\rm U} = (i, j, l, m, {u_1}, ..., {u_{q-r_1-r_2-r_3-3}})$ as above with each set of diagrams suppressed by ${\rm V}^{-r_1-r_2-r_3-1}$. (The matching momenta with the external legs are shown in {\red red}.) As mentioned earlier, we can draw similar sets of diagrams for the $(\varphi_2, \varphi_3)$ sectors. Note that we have avoided using any approximations so far, and therefore the aforementioned computations are exact. It is not too hard to see the following commutative algebra for the $\overset{\curvearrowleft}{\mathbb{T}}$ operator:
\bg\label{readyapril}
&&\left(\left[{\rm dom}\right]_{\varphi_2}^{(-r_1, -r_2, -r_3, ..)}\right)^{\otimes {\rm N}_1} \cup \left(\left[{\rm dom}\right]_{\varphi_3}^{(-r_1, -r_2, -r_3, ..)}\right)^{\otimes 
{\rm N}_2} 
~\cup~\overset{\curvearrowleft}{\mathbb{T}} \nonumber\\
= &&
\overset{\curvearrowleft}{\mathbb{T}}~\cup~\left(\left[{\rm dom}\right]_{\varphi_2}^{(-r_1, -r_2, -r_3, ..)}
\right)^{\otimes {\rm N}_1} \cup \left(\left[{\rm dom}\right]_{\varphi_3}^{(-r_1, -r_2, -r_3, ..)}\right)^{\otimes 
{\rm N}_2}, 
\nd
where ${\rm N}_i \in \mathbb{Z}_+$, and the $\otimes {\rm N}_i$ signify ${\rm N}_i$ copies of the internal interactions but with due care to the combinatoric factors that accompany the components of momenta from \eqref{lincoln}. Because of these combinatoric factors, the amplitude doesn't generically follow a simple power law (as we also saw in
\eqref{hicksmel}).

With these machinery, we are ready to express the numerator and the denominator of the path integral \eqref{wilsonrata} or \eqref{2summers} 
in a precise way. We will follow the latter, {\it i.e.} \eqref{2summers} with fixed $c_{nmpqrs}$. The expectation value of $\varphi_1$ over the Glauber-Sudarshan state may now be represented exactly as:

{\tiny
\bg\label{arose}
\langle\varphi_1\rangle_{\overline\sigma} & = & 
{\mathbb{T} + \sum\limits_{{\rm N} = 1}^\infty {g^{\rm N}\over {\rm N}!}
\left(\sum\limits_{q_i = 0}^{\rm Q}\left[{\rm dom}\right]_{\varphi_1}^{(-q_1, -q_2, ...)}\right)^{\otimes {\rm N}} \cup 
\left(\sum\limits_{r_i = 0}^{\rm R}\left[{\rm dom}\right]_{\varphi_2}^{(-r_1, -r_2, ...)}\right)^{\otimes {\rm N}} \cup
\left(\sum\limits_{s_i = 0}^{\rm S}\left[{\rm dom}\right]_{\varphi_3}^{(-s_1, -s_2, ...)}\right)^{\otimes {\rm N}} \cup
\left(\mathbb{T} + \overset{\curvearrowleft}{\mathbb{T}}\right) \over 
\mathbb{I} + \sum\limits_{{\rm N} = 1}^\infty {g^{\rm N}\over {\rm N}!}
\left(\sum\limits_{q_i = 0}^{\rm Q}\left[{\rm dom}\right]_{\varphi_1}^{(-q_1, -q_2, ...)}\right)^{\otimes {\rm N}} \cup 
\left(\sum\limits_{r_i = 0}^{\rm R}\left[{\rm dom}\right]_{\varphi_2}^{(-r_1, -r_2, ...)}\right)^{\otimes {\rm N}} \cup
\left(\sum\limits_{s_i = 0}^{\rm S}\left[{\rm dom}\right]_{\varphi_3}^{(-s_1, -s_2, ...)}\right)^{\otimes {\rm N}}} \nonumber\\ 
&= & \mathbb{T} + 
{\sum\limits_{{\rm N} = 1}^\infty {g^{\rm N}\over {\rm N}!}
\left(\sum\limits_{q_i = 0}^{\rm Q}\left[{\rm dom}\right]_{\varphi_1}^{(-q_1, -q_2, ...)}\right)^{\otimes {\rm N}} \cup 
\left(\sum\limits_{r_i = 0}^{\rm R}\left[{\rm dom}\right]_{\varphi_2}^{(-r_1, -r_2, ...)}\right)^{\otimes {\rm N}} \cup
\left(\sum\limits_{s_i = 0}^{\rm S}\left[{\rm dom}\right]_{\varphi_3}^{(-s_1, -s_2, ...)}\right)^{\otimes {\rm N}} \cup
 \overset{\curvearrowleft}{\mathbb{T}} \over 
\mathbb{I} + \sum\limits_{{\rm N} = 1}^\infty {g^{\rm N}\over {\rm N}!}
\left(\sum\limits_{q_i = 0}^{\rm Q}\left[{\rm dom}\right]_{\varphi_1}^{(-q_1, -q_2, ...)}\right)^{\otimes {\rm N}} \cup 
\left(\sum\limits_{r_i = 0}^{\rm R}\left[{\rm dom}\right]_{\varphi_2}^{(-r_1, -r_2, ...)}\right)^{\otimes {\rm N}} \cup
\left(\sum\limits_{s_i = 0}^{\rm S}\left[{\rm dom}\right]_{\varphi_3}^{(-s_1, -s_2, ...)}\right)^{\otimes {\rm N}}},
\nonumber\\
\nd}
which removes the dominant nodal diagrams and keeps all the sub-dominant ones. Once we go beyond the fixed choice of $c_{nmpqrs}$, then $g$ in \eqref{arose} may be replaced by $g_{\rm QRS}$ (for fixed choices of derivatives in \eqref{angelslune} and \eqref{lincoln}), and we can sum over all choices of $({\rm Q, R, S})$. In that case, for choices like $({\rm Q, 0, 0}), ({\rm 0, R, 0})$ and $({\rm 0, 0, S})$, the series would grow as 
$({\rm N}!)^0$ as pointed out in \eqref{charlieht}. For such a case we can approximate the denominator by $\mathbb{I}$ but it would appear that this still removes the dominant nodal diagrams in the $\varphi_1$ sector keeping only the sub-dominant ones. This means the only way the dominant nodal diagrams would contribute if we express the expectation value as in 
\eqref{shaystone} which is related to \eqref{wilsonrata} when $\langle\overline\sigma\vert\overline\sigma\rangle = \langle\Omega\vert \Omega\rangle\left(1 + {\cal O}(g_{\rm QRS})\right)$. The above is a special case of a more general picture that we discuss here, but the usefulness of \eqref{shaystone} is that it makes the subsequent analysis very simple as we saw earlier. However the full story is a bit more involved, although the final answer does not change significantly from what we had earlier. In the following we will illustrate this. Using the coupling $g_{\rm QRS}$, the second line in 
\eqref{arose} changes to:

{\tiny
\bg\label{aroseready}
\langle\varphi_1\rangle_{\overline\sigma} = 
\mathbb{T} + 
{\sum\limits_{{\rm N} = 1}^\infty \sum\limits_{\rm U}{g_{\rm QRS}^{\rm N}\over {\rm N}!}
\left(\sum\limits_{q_i = 0}^{\rm Q}\left[{\rm dom}\right]_{\varphi_1}^{(-q_1, -q_2, ...)}\right)^{\otimes {\rm N}} \cup 
\left(\sum\limits_{r_i = 0}^{\rm R}\left[{\rm dom}\right]_{\varphi_2}^{(-r_1, -r_2, ...)}\right)^{\otimes {\rm N}} \cup
\left(\sum\limits_{s_i = 0}^{\rm S}\left[{\rm dom}\right]_{\varphi_3}^{(-s_1, -s_2, ...)}\right)^{\otimes {\rm N}} \cup
 \overset{\curvearrowleft}{\mathbb{T}} \over 
\mathbb{I} + \sum\limits_{{\rm N} = 1}^\infty \sum\limits_{\rm U}{g_{\rm QRS}^{\rm N}\over {\rm N}!}
\left(\sum\limits_{q_i = 0}^{\rm Q}\left[{\rm dom}\right]_{\varphi_1}^{(-q_1, -q_2, ...)}\right)^{\otimes {\rm N}} \cup 
\left(\sum\limits_{r_i = 0}^{\rm R}\left[{\rm dom}\right]_{\varphi_2}^{(-r_1, -r_2, ...)}\right)^{\otimes {\rm N}} \cup
\left(\sum\limits_{s_i = 0}^{\rm S}\left[{\rm dom}\right]_{\varphi_3}^{(-s_1, -s_2, ...)}\right)^{\otimes {\rm N}}}, \nonumber\\ \nd}
where ${\rm U} = (\{{\rm Q}\}, \{{\rm R}\}, \{{\rm S}\})$ and we are summing over all possible choices for $({\rm Q, R, S})$. Since now the denominator can have perturbative expansions as of the form \eqref{charlieht}, we may expand over the denominator and keep the numerator of \eqref{aroseready} alongwith the tree-level contribution $\mathbb{T}$. Since there are now multiple couplings involved, for $({\rm Q, R, S}) \in \mathbb{Z}_+$, we can concentrate on one particular set and define $g \equiv g_{\rm Q_oR_oS_o}$. Using the commutative algebra \eqref{readyapril}, we can transfer the action of $\overset{\curvearrowleft}{\mathbb{T}}$ only on the first term to express \eqref{aroseready} in the following suggestive way:

{\tiny
\bg\label{arosecame}
\langle\varphi_1\rangle_{\overline\sigma} = 
\mathbb{T} + 
\sum\limits_{{\rm N} = 1}^\infty {g^{\rm N}\over {\rm N}!}
\left[\left(\sum\limits_{q_i = 0}^{\rm Q}\left[{\rm dom}\right]_{\varphi_1}^{(-q_1, -q_2, ...)}\right)^{\otimes {\rm N}}
\cup
 \overset{\curvearrowleft}{\mathbb{T}}\right]
\cup 
\left(\sum\limits_{r_i = 0}^{\rm R}\left[{\rm dom}\right]_{\varphi_2}^{(-r_1, -r_2, ...)}\right)^{\otimes {\rm N}} \cup
\left(\sum\limits_{s_i = 0}^{\rm S}\left[{\rm dom}\right]_{\varphi_3}^{(-s_1, -s_2, ...)}\right)^{\otimes {\rm N}}, 
\nonumber\\ \nd}
with additional contributions coming from various choices of $({\rm Q, R, S})$, including the expansions of the denominator, that we don't show here.
We believe our diligent readers will be able to work them out for themselves. The terms in the square bracket in \eqref{arosecame} are captured by the nodal diagrams in \eqref{luna41} whereas the remaining two terms are captured by the nodal diagrams in \eqref{luna40} with $s = 2, 3$. Interestingly, due to the commutative algebra \eqref{readyapril}, we are allowed to keep only the dominant nodal diagrams from sectors $(\varphi_2, \varphi_3)$, which is what we had earlier, but the $\overset{\curvearrowleft}{\mathbb{T}}$ operation on the dominant nodal diagrams in sector $\varphi_1$ will keep the average of the two diagrams in \eqref{luna39}. Since both these diagrams are identical for $r_i = 0$, the dominant contribution to the path integral will be exactly given by the nodal diagrams in \eqref{luna36}, with two minor differences: {\Su one}, there will be a volume suppression of the form 
${\rm V}^{-1}$ and {\Su two}, the external momenta $k_i$ has to match with the momenta of any one leg in $\zeta_i$ for $i = 1, ..., {\rm N}$. However to implement the second condition we have to revisit the path integral computation from section \ref{sec3.2.5}.

Before moving ahead, one comment is in order. In the square bracket of
\eqref{arosecame}, as we increase the values of $q_i$ in the superscript, the $\overset{\curvearrowleft}{\mathbb{T}}$ operation will lead to more complicated amplitudes. On the other hand, the denominator from \eqref{aroseready} cannot be very complicated, so the ratio in 
\eqref{aroseready} will be non-trivial and cannot be simplified further, implying that \eqref{arosecame} will also be equally non-trivial. The dominant contribution from \eqref{arosecame} then takes the following form:
\bg\label{sisi}
&& {g^{\rm N}\over {\rm N}!}
\left[\left(\left[{\rm dom}\right]_{\varphi_1}^{(0, 0, ...)}\right)^{\otimes {\rm N}}
\cup
 \overset{\curvearrowleft}{\mathbb{T}}\right]
\cup 
\left(\left[{\rm dom}\right]_{\varphi_2}^{(0, 0, ...)}\right)^{\otimes {\rm N}} \cup
\left(\left[{\rm dom}\right]_{\varphi_3}^{(0, 0, ...)}\right)^{\otimes {\rm N}}\\
&= & {g^{\rm N}\over {\rm V}}
\prod_{t = 1}^{{\rm N}q + 1}\left({\pi_t\over b_t}\right)^{1/2} \sum_{\{m_i\}, m_l}\mathbb{D}\left(m_1, m_2, m_3, [m_l],..., m_{{\rm N}q-1}, m_{{\rm N}q}\right)\mathbb{D}_{\varphi_2}({\rm N}r) \mathbb{D}_{\varphi_3}({\rm N}(s-1))\nonumber\\
&\times & \prod_{j = 1}^{{\rm N}q-1}\ast \left(\int_{k_{\rm IR}}^\mu d^{11}k_j {\overline\alpha(k_j) k_j^{m_j}\over a(k_j)}\right) 
\int_{k_{\rm IR}}^\mu d^{11}k \left(
{\overline{\alpha}^2(k)\over a^2(k)} + {1\over 2a(k)}\right) k^{m_l}~ \psi_k({\rm X})~{\rm exp}\left(-ik'_0 t\right), \nonumber \nd
where $\ast$ denotes product over all terms except $m_l$, 
$k_{\rm IR} \equiv (\kappa_{\rm IR}, {\bf k}_{\rm IR})$ and 
$k'_0 = k_0 - \overline\kappa_{\rm IR}$. It is easy to see that $m_l$ can hop between any of the $\zeta_i$ sets in the nodal diagram
\eqref{luna36}. The other combinatorial factors, namely $\mathbb{D}(\{m_i\}, m_l), \mathbb{D}_{\varphi_2}$ and $\mathbb{D}_{\varphi_3}$, remain almost the same as in say \eqref{luna36}. We can also express $\overline{\alpha}^2(k_0, {\bf k}) \equiv \sum\limits_{m, n} f_{mn} {\bf H}_m({\bf k}) {\bf H}_n(k_0)$, where \cite{hermite}:

{\scriptsize
\bg\label{thenest}
f_{mn} = \sum_{u_1} c_{m'n'} c_{p'q'} {m'!n'!p'!q'!\over (m'+p')!(n'+q')!}\sum_{r = 0}^{u_2} \sum_{s = 0}^{u_3} 
\left(\begin{matrix} m' \\ r \\ \end{matrix}\right) 
\left(\begin{matrix} p' \\ r \\ \end{matrix}\right)
\left(\begin{matrix} n' \\ s \\ \end{matrix}\right)
\left(\begin{matrix} q' \\ s \\ \end{matrix}\right) r! s! 2^{r+s} 
\delta_{m, m'+p'-2r} \delta_{n, n' + q' -2s}, \nonumber\\ \nd}
and where $u_1 \equiv (\{m'\},\{n'\},\{p'\},\{q'\}), u_2 \equiv {\rm min}(m',p')$, $u_3 \equiv {\rm min}(n',q')$ and 
$c_{m'n'}$ are the coefficients from \eqref{greek}. The above representation of the Hermite polynomials suggests that we can use similar techniques as before, namely, introduce the on-shell and off-shell constraints either through the wave-function or through a proper representation of  $\overline{\alpha}(k_0, {\bf k})$. The latter however is more non-trivial now, because the powers of $\overline{\alpha}(k_0, {\bf k})$ would change as $q_i$ increases in $[{\rm dom}]^{(-q_1, -q_2, ...)}_{\varphi_1}$, so we will start with the former. For the dominant case we consider here, where $q_i = 0$, the {\it modulated} wave-function takes the following form:

{\tiny
\bg\label{creolet4}
\psi_{\bf k}({\rm X}) e^{-i(k_0 - \overline\kappa_{\rm IR})t} \equiv {1\over \pi \vert\omega\vert {\bf k}^9}
~{\rm exp}\left(-{({\bf k} -\overline{\bf k}_{\rm IR})^2\over \omega^2} - i(k_0 - \overline\kappa_{\rm IR})t\right)\left[1 - 
{1\over \pi \vert\omega\vert} {\rm exp}\left(-{(k_0 - \overline\kappa_{\rm IR})^2\over \omega^2}\right)\right]~ 
{\cal F}\left({\bf k}, {\rm X}\right), \nd}
where ${\rm X} = ({\bf x}, y, z)$ and we will take $\omega \to 0$ as before. We have also assumed that ${\cal F}(\overline{\bf k}_{\rm IR}, {\rm X}) = 1$ to avoid generating spatial or temporal dependences when $k_0 = \overline\kappa_{\rm IR}$ and ${\bf k} = \overline{\bf k}_{\rm IR}$ for the off-shell situation, as mentioned earlier\footnote{In other words, the choice of $(\overline{\kappa}_{\rm IR}, \overline{\bf k}_{\rm IR})$ is done to keep the wave-function {\it constant} when $k_0 = \overline{\kappa}_{\rm IR}$ and ${\bf k} = \overline{\bf k}_{\rm IR}$ with the constraint that 
$\overline{k}^2_{\rm IR} \equiv -\overline{\kappa}^2_{\rm IR} + \overline{\bf k}^2_{\rm IR} \ge k^2_{\rm IR}$.}. For further computational efficiency, we will define:
\bg\label{lynskey}
2\overline\alpha^2(k) + a(k) = 2\sum_{m, n} f_{mn} {\bf H}_m({\bf k}) {\bf H}_n(k_0) + a(k) \equiv \sum_{m, n} \overline{f}_{mn} {\bf H}_m({\bf k}) {\bf H}_n(k_0), \nd
which tells us that we can determine $\overline{f}_{mn}$ in terms of $f_{mn}$ which, in turn, may be determined in terms of the coefficients $c_{mn}$ from \eqref{greek}. However one should note that, due to the fact that the $f_{mn}$ coefficients from \eqref{lynskey} are non-trivially connected to the $c_{mn}$ coefficients from \eqref{thenest}, the process of extracting $c_{mn}$ from $f_{mn}$ can become increasing complicated as we go to higher orders in $q_i$. Of course since the nodal diagrams with higher values of $q_i$ are also suppressed by higher powers of the volume factors, they are naturally highly sub-leading. Now plugging \eqref{lynskey} and \eqref{creolet4} in the relevant part of \eqref{sisi}, we get:

{\footnotesize
\bg\label{owilde}
&&\int_{k_{\rm IR}}^\mu d^{11}k \left(
{\overline{\alpha}^2(k)\over a^2(k)} + {1\over 2a(k)}\right) k^{m_l}~ \psi_k({\rm X})~{\rm exp}\left(-ik'_0 t\right)\\
&=& 
{1\over 2}\int d\Omega\left[\sum_{m = 0}^\infty b_m 
\int_{\kappa_{\rm IR} - \overline{\kappa}_{\rm IR}}^{\mu_0 - \overline\kappa_{\rm IR}} dk_0\left(\overline{\bf k}^2_{\rm IR} - (k_0 + \overline\kappa_{\rm IR})^2\right)^{{m_l\over 2}-2} {\bf H}_m(k_0 + \overline\kappa_{\rm IR})
e^{-ik_0 t} - a_a\overline{k}^{m_l-4}_{\rm IR}\right], \nonumber \nd}
which, up-to different powers of $\overline{\bf k}^2_{\rm IR} - (k_0 + \overline\kappa_{\rm IR})^2$, is very similar to what we had section \ref{sec4.4.2}; and 
$k'_0 = k_0 - \overline\kappa_{\rm IR}$. The limits of the integral is shifted by $\overline\kappa_{\rm IR}$, but as mentioned earlier 
$\kappa_{\rm IR} > \overline\kappa_{\rm IR}$ so in principle we can keep 
$\overline\kappa_{\rm IR} = 0$. We have also defined:
\bg\label{lyonne}
b_m \equiv \sum\limits_n \overline{f}_{mn} {\bf H}_n(\overline{\bf k}_{\rm IR}), ~~~~~
a_a \equiv  \sum_{m, n} \overline{f}_{mn} {\bf H}_m(\overline{\bf k}_{\rm IR}) {\bf H}_n(\overline\kappa_{\rm IR}), \nd
where the second relation suggests that once we know all the values of 
$b_m$, we will know $a_a$. (From \eqref{nehaprit} and \eqref{lynskey} both $b_m$ and $a_a$ are proportional to ${\rm V}$.) Further
defining 
${\cal A}$ as in \eqref{coldjuly}, and taking a median value\footnote{The median value is defined here as the integer powers in a given expansion that provides the dominant contribution. (A more refined definition will be given soon.) Such a choice is not essential, and is taken here to simplify the computations. One could take all possible values of $m_l$, allowed by the binomial expansion, and study the series.} for $m_l \equiv {m}^\ast$, we see that the coefficients $\mathbb{C}^{({\rm N})}_m$ (which we shall divide as $\mathbb{C}^{({\rm N})}_{m; 1}$ and 
$\mathbb{C}^{({\rm N})}_{m; 2}$) are related in exactly the same way as in 
\eqref{altomaa8}. The growths of these coefficients then become:
\bg\label{hicksmel8}
&&\mathbb{C}^{({\rm N})}_{m; 2} = {\cal A}^{\rm N}\left(1 + \sum_{r = 1}^\infty {(-1)^r{\rm log}^r{\cal A}\over r! q^r}\right) \left({\rm N}!\right)^\alpha \le  {\cal A}^{\rm N} \left({\rm N}!\right)^\alpha\nonumber\\
&& \mathbb{C}^{({\rm N})}_{m; 1} = {\cal A}^{\rm N}\left(1 + \sum_{r = 1}^\infty {(-1)^r{\rm log}^r{\cal A}\over r! q^r}\right) \left({\rm N}!\right)^\alpha b_m \le  {\cal A}^{\rm N} \left({\rm N}!\right)^\alpha b_m,  \nd
for ${\cal A} > 1$ which are in-turn controlled by the growths of the most dominant nodal diagrams 
from \eqref{hicksmel4} as $q \ge 3$. There is however one subtlety that requires clarification. The ${\cal A}$ defined in \eqref{hicksmel8} should in principle depend on $m_l$, the binomial factor appearing as the power of $k$ in \eqref{sisi} and \eqref{owilde}. Since we are summing over all possible values of $m_l$ allowed by the binomial expansion, the correct way to interpret ${\cal A}$ in \eqref{hicksmel8} is then the following. First define $m_l$ as $m_l \equiv m_l({\rm N})$ to signify the fact that $m_l$ can hop between any of the $\zeta_i$, with $i = 1, ..., {\rm N}$, sets in the nodal diagram
\eqref{luna36}. We can now use the following upper bound:
\bg\label{yvesrocherme}
\sum\limits_{m_l({\rm N})}{\cal A}^{\rm N}\left[m_l({\rm N})\right] k^{m_l({\rm N})} \le {\cal A}_{\rm eff}^{\rm N} k^{m^\ast}, ~~~~~ \forall {\rm N} \in \mathbb{Z}_+ \nd 
where ${\cal A}_{\rm eff}$ is now independent of $m_l({\rm N})$ in \eqref{hicksmel8} and $m^\ast$ is fixed for any ${\rm N}$. (The above is possible because both ${\cal A}$ and $k$ are dimensionless with $k < 1$ from footnote \ref{yaya} and for any ${\cal A}$ positive or negative, although we will soon see that a constant value of $m^\ast$ doesn't suffice.) In other words, using some abuse of notation to avoid clutter, we define ${\cal A}_{\rm eff} \equiv {\cal A}$ in \eqref{yvesrocherme} and it is this factor that appears on the RHS of \eqref{hicksmel8} and henceforth in all other equations that follow from here. The equation \eqref{yvesrocherme} should also serve as the definition of the median value ${m}^\ast$.
If we now fix the coupling $g << 1$
such that ${\rm V} g^{1/\alpha} = h =$ constant, and extract the real values, then the coefficients 
$b_m$ that are used to define ${\rm G}(k_0)$ in \eqref{violmill} satisfy the following infinite sets of equations $\forall p \in \mathbb{Z}_+$:

{\footnotesize
\bg\label{maycamil3}
\sum_{m = 0}^\infty b_m {\partial^p\over \partial k_0^p} 
\left({{\bf H}_m(k_0 + \overline\kappa_{\rm IR})\over 
\left(\overline{\bf k}^2_{\rm IR} - (k_0 + \overline\kappa_{\rm IR})^2\right)^{2-m^\ast/2}}\right)
\Bigg\vert_{\kappa_{\rm IR}- \overline{\kappa}_{\rm IR}}^{\mu_0 - \overline\kappa_{\rm IR}} = -{\Gamma\left(-{5\over 3}\right)\over \sqrt{2\pi/3}}~{\left({5\over 3}\right)! \over \left({5\over 3}-p\right)!}
~b_a k_0^{{5\over 3} - p}\Bigg\vert_{\kappa_{\rm IR}- \overline{\kappa}_{\rm IR}}^{\mu_0 - \overline{\kappa}_{\rm IR}}, \nd}
resembling \eqref{maycamil} up-to a factor of $m^\ast$, with the RHS regulated using $\overline\kappa_{\rm IR}$ as before and 
$b_a \equiv a_a \overline{k}^{m^\ast - 4}_{\rm IR}$ where $a_a$ as in \eqref{lyonne}. (There is an 
extra factor of $\left(\overline{\bf k}^2_{\rm IR} -(k_0 + \overline\kappa_{\rm IR})^2\right)^{2 - m^\ast/2}$
that may be absorbed in the definition of the wave-function \eqref{creolet4} itself.)
Additionally, looking at the bounds on the growths of the $\mathbb{C}^{({\rm N})}_{m; i}$ coefficients 
\eqref{hicksmel8} for $i = 1, 2$ and comparing the on-shell and off-shell contributions, it is easy to see that the Borel resummation of the Gevrey-$\alpha$ series leads to the following equation:
\bg\label{akkim}
\int_0^\infty d{\rm S} ~{\rm exp}\left(-{{\rm S}\over g^{1/\alpha}}\right) {1\over 1 - {\cal A}{\rm S}^\alpha} = 
{1 \over \Lambda^{4/3}}\left({2h \over b_a}\right), \nd
where $h = {\rm V} g^{1/\alpha}$ and $b_a$ is defined earlier in terms of $a_a$ from \eqref{lyonne}. (We can take $b_a \equiv {{\rm V}\sqrt{2\pi}\over \Gamma\left(-{5/3}\right)}$ for the time being and we can restore it back later\footnote{This is possible because we have a tunable parameter $a_o$ in the definition of the metric \eqref{sissys}. We can adjust $a_o$ to keep $b_a = {{\rm V}\sqrt{2\pi}\over \Gamma\left(-{5/3}\right)}$ without over-constraining the system. \label{killie}}.) The above result matches exactly with what we had in \eqref{camilahawk} if we absorb all the extra constant factors in the definition of $h$. This is because we expect the dynamics to be controlled by the dominant nodal diagrams despite the fact that they are eliminated when taking the ratio. Note also that the RHS is a positive definite quantity because we have taken $\overline{k}^2_{\rm IR} \equiv -\overline\kappa^2_{\rm IR} + \overline{\bf k}^2_{\rm IR} > k^2_{\rm IR} > 0$ and $\overline{\bf k}_{\rm IR} > \overline\kappa_{\rm IR} > 0$; as well as the cosmological constant $\Lambda > 0$. ($a_a$ from \eqref{lyonne} is automatically positive definite because the combination in \eqref{lynskey} can be kept positive definite.) The question then is whether the LHS is a positive definite quantity for all values of ${\cal A}$ $-$ both positive and negative $-$ and for all values of $\alpha \ge 1$. This will be the subject of section \ref{sec4.5}, but before moving ahead let us clarify a couple more subtleties. 

\subsubsection{Revisiting the tree level contribution and more subtleties \label{sec4.4.4}}

Our analysis in the previous section has resulted in two set of equations, namely \eqref{maycamil3} and \eqref{akkim}. Unfortunately there are still a couple more subtleties that we kept under the rug. First, has to do with the tree-level contribution from \eqref{luna3}, and the second, with the form of $\overline\alpha(k)$ itself. Let us begin with the first case, {\it i.e.} with the tree-level contribution. The tree level term from \eqref{luna3} is {\it not} similar to what we have in \eqref{owilde}. In particular it is linear in $\overline\alpha(k)$, instead of being quadratic as in \eqref{owilde} plus there is no propagator correction nor is there a volume suppression. The latter however is crucial, and originates from the fact that the dominating nodal diagrams contributing to the expectation value $\langle\varphi_1\rangle_{\overline\sigma}$ all have volume suppression whereas the tree-level diagram does not. If $\mathbb{T}_1$ is the tree-level diagram contributing to \eqref{maycamil3}, then $\mathbb{T}$ in \eqref{arosecame} is related to $\mathbb{T}_1$ by $\mathbb{T} \equiv 
\mathbb{T}_1 + \delta\mathbb{T}$, with $\delta\mathbb{T}$ given by:
\bg\label{600ween}
\delta\mathbb{T} = \int d^{11} k\left[{\overline\alpha(k)\over a(k)} - 
{k^{m^\ast}\over {\rm V}}\left({\overline{\alpha}^2(k)\over a^2(k)} + {1\over 2a(k)}\right) + {\cal O}\left({1\over {\rm V}^2}\right)\right]\psi_k({\rm X})e^{-i(k_0 - \overline\kappa_{\rm IR})t}, \nd
which may be derived from \eqref{luna3}, \eqref{maycamil3} and \eqref{yvesrocherme}. The ${\cal O}\left({1\over {\rm V}^2}\right)$ term appears from the higher order nodal diagrams. We have also used $\overline{\alpha}(k) \to -\overline{\alpha}(k)$ in \eqref{luna3} to allow for a positive metric configuration. The integrand in \eqref{600ween} vanishes for:
\bg\label{casintagra}
\overline\alpha(k) = {a(k)\over 2k^{m^\ast}}\left({\rm V} \pm {\rm V} \sqrt{1 - {2k^{2m^\ast}\over a(k){\rm V}^2}}\right), \nd
where, from our choice of conventions {\it et cetera} in section \ref{sec1.3} and elsewhere, all the terms appearing in \eqref{casintagra} are dimensionless\footnote{One might however worry that $\overline\alpha(k)$ in \eqref{casintagra} may not be an even function of $k$ for $m^\ast \in 2\mathbb{Z}_+ + 1$. This is actually not the case because we should define $k^{m^\ast}$ or $k^{\nu(k)}$, which is a more generalized version that we shall use later, as $k^{\nu(k)} = \left(-k_0^2 + {\bf k}^2\right)^{\nu(k)/2}$ or alternatively as $\vert k \vert^{\nu(k)}$. Thus $\overline\alpha(k) \equiv \overline\alpha(-k)$ (similar definitions exist for $\overline\beta(k)$ and $\overline\gamma(k)$).}.
Of course we could have also demanded the integral in \eqref{600ween} to vanish using the wave-function ans\"atze \eqref{creolet4}, instead of just the integrand. This would have led to more complicated analysis with $k_0$ integral which may nevertheless be done but we shall avoid it here. The reason is that the vanishing of the integrand in \eqref{600ween} doesn't appear to over-constrain the system as we shall see in the following.
The result \eqref{casintagra} is without the ${\cal O}\left({1\over {\rm V}^2}\right)$ contribution. In general, the two solutions for \eqref{casintagra} are (see \eqref{nehaprit}):
\bg\label{mariec}
\overline{\alpha}(k) = {\alpha(k)\over {\rm V}} = \begin{cases} ~ {k^{m^\ast}\over 2{\rm V}} + 
{\cal O}\left({1\over {\rm V}^2}\right)\\
~{a(k) {\rm V}\over k^{m^\ast}} - {k^{m^\ast}\over 2{\rm V}} + 
{\cal O}\left({1\over {\rm V}^2}\right)\\
\end{cases}
\nd
implying that $\overline{\alpha}(k)$ can either take very large or very small values. The results in \eqref{mariec} somehow suggest that the system might have exact solutions, but there are a few subtleties that prohibit us to allow that conclusion. First, for large values of 
$\overline\alpha(k)$ we cannot terminate the nodal diagrams to any order, and since the powers of the fields appearing in \eqref{lincoln} could be arbitrary, the system may not have any solutions (unless we find some hierarchy of scales). Secondly, for small values of $\overline{\alpha}(k)$, while the system happily has a controlled behavior and we can terminate the nodal diagrams at any order, it is not clear whether $\overline{\alpha}(k) = {k^{m^\ast}\over 2{\rm V}} + 
{\cal O}\left({1\over {\rm V}^2}\right)$ would also solve \eqref{lynskey} and \eqref{maycamil3} consistently, the latter being an infinite set of equations (we will have more to say on this soon). Therefore conservatively, and by comparing with \eqref{maycamil3} and \eqref{akkim}, we can take the following ans\"atze for $\overline\alpha(k)$:
\bg\label{evaBcaku}
\overline{\alpha}(k) = {\alpha(k)\over {\rm V}} = {1\over 2{\rm V}}\left(k^{11/3} + \chi(k)\right) + {\cal O}\left({1\over {\rm V}^2}\right), \nd 
implying a scaling of $m^\ast = {11\over 3}$
which would still keep it very small and thus easily make\footnote{The meaning of ${\cal O}\left({1\over {\rm V}^2}\right)$ in \eqref{evaBcaku} is the following. As we go to higher orders in the nodal diagrams, $\overline\alpha(k)$ gets corrected accordingly that consistently keeps 
$\delta\mathbb{T} \to 0$.}
$\delta\mathbb{T} \to 0$. The latter may be guaranteed in the following way. The functional form for $\chi(k)$ can be determined directly or by solving for $b_m$ in \eqref{maycamil3}, which in-turn will determine the values of $\overline{f}_{mn}$ in \eqref{lynskey}, thus reproducing the values of $f_{mn}$. Because of the volume suppression of $\overline\alpha(k)$, it is sub-dominant in \eqref{lynskey}, but we can still determine the form of $\chi(k)$ from the aforementioned procedure. For $(\mu_0, \kappa_{\rm IR}) << 1$ with $\mu_0 > \kappa_{\rm IR}$ (recall that they are dimensionless quantities), the RHS of \eqref{maycamil3} can be small thus rendering $b_m$'s small. This would at least keep $\delta\mathbb{T} \to 0$. Combining \eqref{arosecame} and \eqref{600ween}, we get:

{\tiny
\bg\label{arosecame3}
\langle\varphi_1\rangle_{\overline\sigma} & = & 
{\bf Re}~\mathbb{T} + 
\sum\limits_{{\rm N} = 1}^\infty {g^{\rm N}\over {\rm N}!}
\left[\left(\sum\limits_{q_i = 0}^{\rm Q}\left[{\rm dom}\right]_{\varphi_1}^{(-q_1, -q_2, ...)}\right)^{\otimes {\rm N}}
\cup
 \overset{\curvearrowleft}{\mathbb{T}}\right]
\cup 
\left(\sum\limits_{r_i = 0}^{\rm R}\left[{\rm dom}\right]_{\varphi_2}^{(-r_1, -r_2, ...)}\right)^{\otimes {\rm N}} \cup
\left(\sum\limits_{s_i = 0}^{\rm S}\left[{\rm dom}\right]_{\varphi_3}^{(-s_1, -s_2, ...)}\right)^{\otimes {\rm N}}, 
\nonumber\\ 
& = & \int d^{11} k\left[{\overline\alpha(k)\over a(k)} - 
{k^{m^\ast}\over {\rm V}}\left({\overline{\alpha}^2(k)\over a^2(k)} + {1\over 2a(k)}\right)\right]{\bf Re}\left(\psi_k({\rm X})e^{-i(k_0 - \overline\kappa_{\rm IR})t}\right) \nonumber\\
&+ & {1\over {\rm V}}\sum_{{\rm N} = 0}^\infty g^{\rm N} ({\rm N}!)^\alpha \sum_{m_l} {\cal A}^{\rm N}(m_l) \int_{k_{\rm IR}}^\mu d^{11}k \left(
{\overline{\alpha}^2(k)\over a^2(k)} + {1\over 2a(k)}\right) k^{m_l}~ {\bf Re}\left(\psi_k({\rm X})~e^{-i(k_0 - \overline\kappa_{\rm IR})t}\right)\nonumber\\
& = & {\bf Re}~\delta\mathbb{T} + {1\over {\rm V}g^{1/\alpha}}
\int_0^\infty d{\rm S} ~{\rm exp}\left(-{{\rm S}\over g^{1/\alpha}}\right) {1\over 1 - {\cal A}{\rm S}^\alpha} 
\int_{k_{\rm IR}}^\mu d^{11}k \left(
{\overline{\alpha}^2(k)\over a^2(k)} + {1\over 2a(k)}\right) k^{m^\ast}~ {\bf Re}\left(\psi_k({\rm X})~e^{-i(k_0 - \overline\kappa_{\rm IR})t}\right)\nonumber\\
& = & {\bf Re}~\delta\mathbb{T} + {1\over g^{1/\alpha}}
\int_0^\infty d{\rm S} ~{\rm exp}\left(-{{\rm S}\over g^{1/\alpha}}\right) {1\over 1 - {\cal A}{\rm S}^\alpha} 
\int_{k_{\rm IR}}^\mu d^{11}k ~{\overline\alpha(k)\over a(k)}~
{\bf Re}\left(\psi_k({\rm X})~e^{-i(k_0 - \overline\kappa_{\rm IR})t}\right), \nd}
where going from first to the second and third lines we have restricted ourselves to $q_i = 0$, third to fourth line we have used \eqref{yvesrocherme} and from fourth to fifth line we used the vanishing of the integrand in \eqref{600ween}. The form of the final result is very similar to what we had earlier 
with the dominant diagram before it was eliminated by the denominator, but the differences from \eqref{camilahawk} are what we had in 
\eqref{maycamil3} and \eqref{akkim}. Thus, despite similarities with the dominant nodal diagrams in sections \ref{sec4.4.1} and \ref{sec4.4.2} in the form of the final answer, these differences are crucial. Using \eqref{evaBcaku}, $\delta \mathbb{T}$ takes the form:
\bg\label{sally}
\delta\mathbb{T} = \int d^{11}k ~{\chi(k)\over k^2} ~
{\bf Re}\left(\psi_k({\rm X})~e^{-i(k_0 - \overline\kappa_{\rm IR})t}\right) + {\cal O}\left({1\over {\rm V}^2}\right), \nd
which is highly sub-dominant compared to the second term in the last line of \eqref{arosecame3} for $\Lambda << 1$ justifying the ans\"atze 
\eqref{evaBcaku}. If we had used the dominant nodal diagrams, there would not have been a simple way to formulate an ans\"atze for $\overline\alpha(k)$ although the cosmological constant $\Lambda$ would have been independent of ${\rm V}$ (see \eqref{camilahawk}).

The ans\"atze \eqref{evaBcaku} has few subtleties as alluded to earlier. {\Su One}, because of the volume suppression, one might assume that the shift of the interacting vacua becomes arbitrarily small within the energy scale $k \le \mu$. However because of an additional volume suppression in the definition of $\overline\alpha(k)$ in terms of $\alpha(k)$ from \eqref{nehaprit}, this is not the case. Nevertheless the Glauber-Sudarshan state will almost be like a {\it vacuum} configuration but differ slightly from it by powers of $k$ with $k \le \mu$. This difference is crucial and is solely responsible for all the nice properties that we demand for an effective field theory with de Sitter isometries in four space-time dimensions from string theory. And {\Su two}, the functional form for 
$\overline\alpha(k)$ gives us a hint how we can improve on the bound 
\eqref{yvesrocherme} which will provide a more precise determination of the constant factor ${\cal A}$. Thus instead of using $m^\ast$ to specify the upper bound, we can say that ${\cal A}_{\rm eff}$ $-$ which we shall call ${\cal A}$ 
henceforth $-$ is bounded from below by the following:
\bg\label{beatudi}
{\cal A}\ge \left(\sum\limits_{m_l({\rm N})}{\cal A}^{\rm N}\left[m_l({\rm N})\right] k^{m_l({\rm N}) - \nu(k)}\right)^{1\over {\rm N}}, \nd
where ${\cal A}\left[m_l({\rm N})\right]$ may be read-off from \eqref{sisi}, and the functional form for $\nu(k)$ will be determined below. Note that, since $k_{\rm IR} \le k \le \mu$ the lower bound in \eqref{beatudi} will be fixed by the value of $k \in [k_{\rm IR}, \mu]$ that provides the largest contribution on the RHS of \eqref{beatudi}. The functional form for $\nu(k)$ may be given by:
\bg\label{mylenecake}
\nu(k) = {11\over 3} + {\rm log}_k\left(1 + {\chi(k)\over k^{11/3}}\right), \nd
where $\chi(k)$ is the same function that appeared in \eqref{evaBcaku}. Plugging the contribution from \eqref{mylenecake} to \eqref{sally}, we see that the $\chi(k)$ contribution is removed making:
\bg\label{duibone}
\delta\mathbb{T} = 0 + {\cal O}\left({1\over {\rm V}^2}\right), \nd
thus bringing \eqref{maycamil3} and \eqref{akkim} to a solid footing. Note also that,  
since both $(k_{\rm IR}, \mu) << 1$ and ${\rm V} >> 1$ but 
${\overline\alpha(k)\over a(k)}$ independent of ${\rm V}$ (similarly for the $(\overline\beta, \overline\gamma)$ sectors), ${\cal A}$ can be small and finite but not arbitrarily small. We can also work out the higher order nodal diagrams to see the changes to the scaling of $k$ in \eqref{yvesrocherme}
that will in turn effect both \eqref{beatudi} and \eqref{mylenecake} cancelling the higher order corrections to $\delta\mathbb{T}$.

Our analysis above was based on the observation that the wave-function always appears linearly at any order of the nodal diagrams and therefore it is easier to implement the on-shell and off-shell constraints by expressing the wave-function as in \eqref{creolet4}. Looking at \eqref{arosecame3}, wherein now $\overline\alpha(k)$ also appears linearly, one might wonder if we can keep the wave-function simple as in \eqref{mabel} but change the ans\"atze for $\overline\alpha(k)$ instead, much like what we did in section \ref{sec4.4.2}. There is however one immediate difference: $\overline\alpha(k)$ also appears in the definition of ${\cal A}$ as seen from \eqref{coldjuly}. This fortunately doesn't affect much because it is not conjugated with the wave-function and therefore, although it changes the value of ${\cal A}$, it cannot affect the on-shell and off-shell constraints. The above conclusion relies on the fact that making $\delta \mathbb{T} = 0$ replaces the volume-suppressed quadratic piece in $\overline\alpha(k)$ by a linear piece (see the transformation from line four to line five in \eqref{arosecame3}). In a more generic setting with higher order interactions this is not {\it a-priori} guaranteed, so it will be worthwhile to see what happens if we cannot always go from line four to line five in \eqref{arosecame3}. Unfortunately the story now gets a little harder to tackle analytically because $\overline\alpha(k)$ appears with different powers at different orders of the nodal diagrams. Nevertheless, if we use the Hermite polynomial expansion for $\overline\alpha(k)$, we can express any powers of $\overline\alpha(k)$ in terms of Hermite polynomials using \cite{hermite}
and use appropriate constraints. Furthermore because of the volume suppressions of the higher order nodal diagrams we do not have to consider beyond quadratic pieces in $\overline\alpha(k)$ for terms conjugated with the wave-function. To see this more explicitly let us consider the following example. Taking $\overline\alpha^2(k)$, 
the constraint on $f_{mn}$
from \eqref{thenest} takes the following form:

{\scriptsize
\bg\label{violmill}
\overline\alpha^2(k) \equiv \sum_{m, n} f_{mn} {\bf H}_m({\bf k}) {\bf H}_n(k_0) = {a^2(k)\over \pi\vert\hat\omega\vert}\left[\widetilde{\rm G}(k_0) - {1\over \pi\vert\hat\omega\vert}{\rm exp}\left(-{(k_0 - \overline\kappa_{\rm IR})^2\over \hat\omega^2}\right)\right]
{\rm exp}\left(-{({\bf k} - \overline{\bf k}_{\rm IR})^2\over \hat\omega^2}\right),\nd}
where $a(k)$ is the propagator and  $\widetilde{\rm G}(k_0)$ is any arbitrary function of $k_0$ which in turn may be represented by a linear combination of the Hermite polynomials (much like how we represented a similar function using the $\widetilde{b}_m$ coefficients in \eqref{bullock}).  
If we follow a similar procedure here by expressing $\widetilde{\rm G}(k_0) = \sum\limits_m \widetilde{b}_m {\bf H}_m(k_0)$, then the following set of subtleties appear. {\Su One}, we need to make sure that $\delta\mathbb{T}$ from 
\eqref{600ween} is still sub-dominant, and {\Su two}, the second-last line of 
\eqref{arosecame3} remains consistent with the Fourier transform relating it to the on-shell and off-shell pictures discussed earlier. For the second case, the discussion is very similar to what we encountered in section \ref{sec4.4.2}, wherein the wave-function was kept simple as in \eqref{mabel}, but $\overline\alpha(k)$ was non-trivial and took the form \eqref{rocketr}. We can rewrite this as:

{\scriptsize
\bg\label{violmill7}
\overline\alpha(k) \equiv \sum_{m, n} c_{mn} {\bf H}_m({\bf k}) {\bf H}_n(k_0) = {a(k)\over \pi{\rm V}\vert\omega\vert}\left[{\rm G}(k_0) - {1\over \pi\vert\omega\vert}{\rm exp}\left(-{(k_0 - \overline\kappa_{\rm IR})^2\over \omega^2}\right)\right]
{\rm exp}\left(-{({\bf k} - \overline{\bf k}_{\rm IR})^2\over \omega^2}\right), \nd}
where ${\rm G}(k_0) = \sum\limits_m \hat{b}_m {\bf H}_m(k_0)$, and \eqref{violmill7} differs slightly from \eqref{rocketr}. Note that in both \eqref{violmill} and \eqref{violmill7}, we use $(\omega, \hat\omega) << 1$, so that the exponential pieces therein form distributions, but not sharply peaked delta functions.
Expectedly, \eqref{violmill7} doesn't exactly match with 
\eqref{evaBcaku}, plus it has an additional ${\rm V}^{-1}$ dependence. The question then is: how can we take care of the first case discussed above? 

To deal with the first case, namely the behavior of $\delta\mathbb{T}$, we first note another difference from the analysis in section \ref{sec4.4.2}, namely that the expression for the cosmological constant $\Lambda$ in \eqref{camilahawk} doesn't have a volume ${\rm V}$ dependence. Multiplying and dividing the second term in the last line of \eqref{arosecame3} by ${\rm V}$, the volume dependence from \eqref{violmill7} cancels out, and we can fix the coefficients $\hat{b}_m$ 
from the Fourier transforms of the on-shell and the off-shell backgrounds.
On the other hand, squaring the expression in \eqref{violmill7} we get:

{\tiny
\bg\label{denise}
\overline\alpha^2(k) = {a^2(k)\over \pi^2 {\rm V}^2 \vert\omega\vert^2}
\left[{\rm G}^2(k_0) + {1\over \pi^2\vert\omega\vert^2}{\rm exp}\left(-{2(k_0 - \overline\kappa_{\rm IR})^2\over \omega^2}\right)
- {2{\rm G}(k_0)\over \pi\vert\omega\vert}{\rm exp}\left(-{(k_0 - \overline\kappa_{\rm IR})^2\over \omega^2}\right)\right]
{\rm exp}\left(-{2({\bf k} - \overline{\bf k}_{\rm IR})^2\over \omega^2}\right), \nd}
which doesn't appear to match with \eqref{violmill}. However, while
\eqref{denise} does have an additional ${\rm V}^{-2}$ dependence, \eqref{violmill} doesn't have any extra volume factor in the expression. This gives us a hint that if we fix ${\rm V} \vert\omega\vert^c = {\rm constant} \equiv 1$ with $c > 0$, then \eqref{denise} becomes:

{\tiny
\bg\label{denise2}
\overline\alpha^2(k) = {a^2(k)\over \pi^2 \vert\omega\vert}
\left[{{\rm G}^2(k_0)\over \vert\omega\vert^{1-2c}} + {1\over \pi^2\vert\omega\vert^{3-2c}}{\rm exp}\left(-{2(k_0 - \overline\kappa_{\rm IR})^2\over \omega^2}\right)
- {2{\rm G}(k_0)\over \pi\vert\omega\vert^{2-2c}}{\rm exp}\left(-{(k_0 - \overline\kappa_{\rm IR})^2\over \omega^2}\right)\right]
{\rm exp}\left(-{2({\bf k} - \overline{\bf k}_{\rm IR})^2\over \omega^2}\right), \nd}
where we see that the third term inside the square bracket vanishes for $c = 1$ and in the limit $\omega << 1$. Unfortunately, the relative sign {\it does not} match, showing that the simplified ans\"atze \eqref{violmill7} needs to be modified further to satisfy all the constraints in the system.
The relative minus sign survives
for $c \ne 1$. For example, with $c = {3\over 2}$ we have:

{\tiny
\bg\label{barbarian}
\overline\alpha^2(k) & = & {a^2(k)\over \pi^2 \vert\omega\vert}
\left[{{\rm G}^2(k_0)\over {\rm V}^{4/3}} + {1\over \pi^2} {\rm exp}\left(-{2(k_0 - \overline\kappa_{\rm IR})^2\over \omega^2}\right)
- {2{\rm G}(k_0)\over \pi {\rm V}^{4/3}\vert\omega\vert}{\rm exp}\left(-{(k_0 - \overline\kappa_{\rm IR})^2\over \omega^2}\right)\right]
{\rm exp}\left(-{2({\bf k} - \overline{\bf k}_{\rm IR})^2\over \omega^2}\right)\nonumber\\
& \to & {a^2(k)\over \sqrt{2}\pi}\left[{{\rm G}^2(k_0)\over {\rm V}^{4/3}} + {1\over \pi^2} {\rm exp}\left(-{2(k_0 - \overline\kappa_{\rm IR})^2\over \omega^2}\right) - {2{\rm G}(\overline\kappa_{\rm IR})\over {\rm V}^{4/3}}
\delta(k_0 - \overline\kappa_{\rm IR})\right] \delta({\bf k} - \overline{\bf k}_{\rm IR}), \nd}
which is similar to \eqref{violmill} once $\hat\omega \to 0$. (We can restrict ourselves to distributions where $(\omega, \hat\omega) << 1$, as mentioned earlier.) This is encouraging but not enough: we still need to show that $\delta\mathbb{T}$ from \eqref{600ween} is sub-leading. Since $\overline\alpha^2(k) \propto {1\over {\rm V}^{2}}$, the dominant terms contributing to the integrand of $\delta\mathbb{T}$ in \eqref{600ween} will be ${1\over a(k)}\left(\overline\alpha(k) 
- {k^{\nu(k)}\over 2{\rm V}}\right)$ with $\nu(k)$ as in \eqref{mylenecake}. Both these terms are suppressed by the volume factor ${\rm V}$, so $\delta\mathbb{T}$ is again sub-dominant (recall that the cosmological constant terms has a $h \equiv {\rm V} g^{1/\alpha}$ factor, and the ${\rm V}$ dependence of $\overline\alpha(k)$ term cancels out). One may also confirm, from the consistencies of the picture, that $m^\ast = {11\over 3}$ thus resembling what we had earlier.

From the discussion above it appears therefore that a similar story could be developed with non-trivial $\overline\alpha(k)$ but a simple choice for the wave-function. As we increase powers of $\overline\alpha(k)$, the volume suppression factor also increases, thus helping us to ignore the higher order nodal diagrams. One could make other consistency checks, but
we will not pursue this any further here and leave it for our diligent readers to complete it. All in all it appears that a consistent solution may be easily found by either taking the modulated wave-function \eqref{creolet4} and a Glauber-Sudarshan state \eqref{evaBcaku}, or taking the Glauber-Sudarshan state \eqref{violmill7} and a wave-function \eqref{mabel}, leading to a four-dimensional de Sitter state with a {\it positive} cosmological constant $\Lambda$ given by \eqref{akkim}. 

One last thing $-$ namely the positivity of the cosmological constant $-$ and a few other subtleties still remain. 
In the following sections we will start by demonstrating why $\Lambda$ from \eqref{akkim} is always positive definite, after which we will turn towards clarifying the remaining subtleties.

\subsection{The positivities of the Borel sum and the cosmological constant \label{sec4.5}}

From the various computational procedures we followed above, for example
\eqref{akkim}, \eqref{camilahawk} and \eqref{henaois}, one specific result stood out from Borel resummation of the Gevrey-$\alpha$ series, namely that the cosmological constant may be given by a non-perturbative series in 
the coupling constant $g$ in the following way:
\bg\label{myjohn}
\int_0^\infty d{\rm S} ~{\rm exp}\left(-{{\rm S}\over g^{1/\alpha}}\right) {1\over 1 - {\cal A}{\rm S}^\alpha} = 
{1 \over \Lambda^{4/3}}, \nd
where the dimensions are taken care of because ${\rm S}$ has appropriate dimension, although for computational purpose we will take ${\rm S}, g, {\cal A}$ and $\Lambda$ to be dimensionless. (Note that we have absorbed 
${2h \over b_a}$ from \eqref{akkim}, which is a dimensionless positive definitive quantity, in the definition of the cosmological constant.) For our analysis in the earlier sections to make sense we will require the LHS of \eqref{myjohn} to be a positive definite quantity for all values of $\alpha \in \mathbb{Z}_+$, and for all positive and negative values of ${\cal A}$. Such a condition will justify the positivity of the cosmological constant in our set-up. 

One might wonder if, because of the positivity of the quantity 
$\Lambda^{4/3}$, any signs for $\Lambda$ should work. The answer however lies in the specific ans\"atze for a positive cosmological constant background that we took in the IIB side, namely \eqref{rihan}, and its M-theory 
uplift in \eqref{carrie}. The $\Lambda^{4/3}$ factor came from a Fourier transform of such a background. Had we taken a negative cosmological constant background, the result would have been very different (although its not clear if an AdS space-time can be realized as a coherent state over a Minkowski background). Thus the form of $\Lambda$ in \eqref{myjohn} works only for {\it positive} $\Lambda$. 

There is another satisfying feature of the relation \eqref{myjohn}, namely the non-perturbative nature of the LHS. If we succeed in demonstrating the positivity of the LHS, then we can justify the oft-quoted fact that the cosmological constant can only be realized non-perturbatively. There are no perturbative limits of the above expression, showing that no perturbative computations can ever give us the form of the cosmological constant! Such a conclusion matches rather well with the computations performed in \cite{desitter2}\footnote{See eq. (4.233) in the first reference of \cite{desitter2} where a formal expression for the cosmological constant is given in terms of the quantum terms. The positivity therein comes only from the quantum effects which are in turn only realized non-perturbatively. Perturbative corrections can never contribute to the positivity of the cosmological constant and as such they are simply red herrings in the problem.}. 

\begin{figure}[h]
\centering
\begin{tabular}{c}
\includegraphics[width=3in]{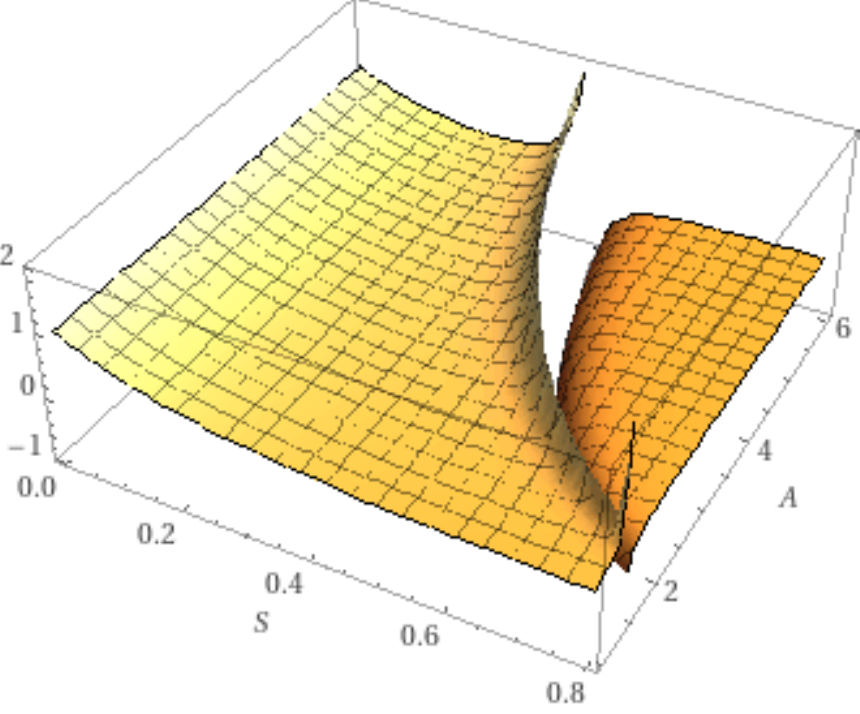}
\end{tabular}
\caption[]{The plot of the LHS of \eqref{myjohn} for $g = {1\over 25}$ and 
$\alpha = 2$. We have taken the range where $0 \le {\rm S} \le 0.8$ and 
$1 \le {\cal A} \le 6$. Clearly as ${\cal A}$ increases, the positive pole in the Borel axis appears for small values of ${\rm S}$.}
\label{gevrey1}
\end{figure} 

\begin{figure}[h]
\centering
\begin{tabular}{c}
\includegraphics[width=3in]{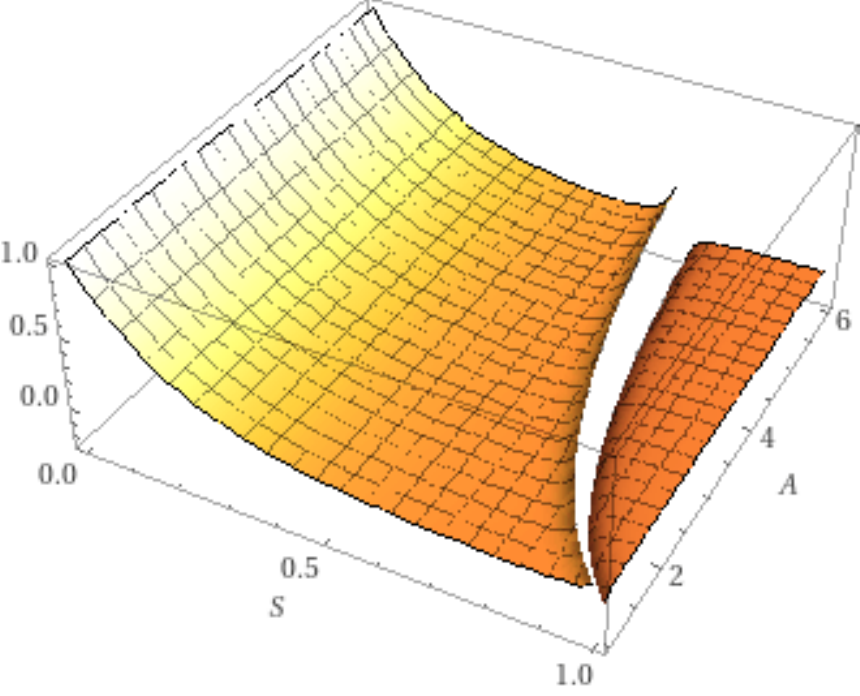}
\end{tabular}
\caption[]{Similar plot of the LHS of \eqref{myjohn} as in figure \ref{gevrey1} for $g = {1\over 25}$ 
but now $\alpha = 5$. We have taken the range where $0 \le {\rm S} \le 1$ and $1 \le {\cal A} \le 6$. We also observe expected behavior of the poles on the Borel axis for larger values of ${\cal A}$.}
\label{gevrey2}
\end{figure}

When ${\cal A} < 0$ the integral on the LHS of \eqref{myjohn} is trivially positive for any values of $\alpha$. The interesting question is what happens when ${\cal A} > 0$ (by construction $g$ and ${\rm S}$ are always positive here). For positive ${\cal A}$, poles appear on the Borel axis, and in {\Su Figures} \ref{gevrey1} and \ref{gevrey2} we show the behavior of the poles for various positive values of ${\cal A}$ with $\alpha = 2$ and 
$\alpha = 5$ respectively. Because of these poles the analysis becomes slightly non-trivial, but since only {\it one} positive pole contributes for any values of $\alpha$ one may at least perform an indefinite integral of the LHS of \eqref{myjohn}. The answer for $\alpha = 2$ becomes:

{\tiny
\bg\label{mytaglen}
\int d{\rm S} ~{\rm exp}\left(-{{\rm S}\over \sqrt{g}}\right) {1\over 1 - {\cal A}{\rm S}^2} = 
{1\over 2\sqrt{\cal A}}\left[{\rm exp}\left({1\over \sqrt{{\cal A}g}}\right) {\bf Ei}\left(-{{\rm S}\over \sqrt{g}} - {1\over \sqrt{{\cal A}g}}\right) - {\rm exp}\left(-{1\over \sqrt{{\cal A}g}}\right)
{\bf Ei}\left({1\over \sqrt{{\cal A}g}} - {{\rm S}\over \sqrt{g}}\right)\right] + {\rm constant}, \nd}
where ${\bf Ei}(x) \equiv -\int\limits_{-x}^\infty {dt\over t}~{\rm exp}(-t)$ is the Exponential integral. Expectedly the value of the above integral is undefined at the two roots of ${\rm S}$ because of 
${\bf Ei}(0)$. In general, for any $\alpha \in \mathbb{Z}_+$ and 
$({\cal A}, g) > 0$, the value of the integral in \eqref{myjohn} gives us:

{\footnotesize
\bg\label{milinmey}
\int d{\rm S} ~{\rm exp}\left(-{{\rm S}\over g^{1/\alpha}}\right) {1\over 1 - {\cal A}{\rm S}^\alpha} = -\sum_{j = 1}^\alpha {1\over \alpha {\cal A} a_j^{\alpha - 1}}~{\rm exp}\left(-{a_j\over g^{1/\alpha}}\right) 
{\bf Ei}\left({a_j - {\rm S}\over g^{1/\alpha}}\right) + {\rm constant}, \nd}
where $a_j$ are the roots of the polynomial in \eqref{macdonald}, implying that the value of the integral may be expressed in terms of the sum over the roots of \eqref{macdonald}. For $\alpha > 2$ most of the roots are in the complex Borel plane (see {\Su Figures} \ref{roots1} and \ref{roots2}), but since the range for ${\rm S}$ is ${\rm S} \in [0, \infty]$, only one positive real root contributes as we also saw earlier. Unfortunately however even for this root, we have ${\bf Ei}(0)$ as the value of the integral. This means we have to take the {\it principle value} of the integral 
\eqref{milinmey} from $0$ to $\infty$ to extract a real answer. Question 
is: what is the reason for choosing the principal value, and what happens to the contribution from the contour that passes through the complex plane (in other words, what happens to the complex residue at the pole)? 

Before answering this, a quick comment on the ${\cal A} < 0$ case. The result of the integral with ${\cal A} < 0$ is exactly the same as in \eqref{milinmey} with the exception that the roots satisfy \eqref{macdonald} with a relative positive sign on the LHS. A plot of the poles in the Borel plane shows that either all the poles are in the complex plane or there is one pole in the negative real axis with all others in the complex plane for even or odd values of $\alpha$ respectively. The integrand itself is a decaying function for all values of ${\cal A} < 0$ and $\alpha \in \mathbb{Z}_+$. This justifies the positivity of the LHS of \eqref{myjohn} for negative values of ${\cal A}$.

Returning to the case ${\cal A} > 0$, we must deal with the pole of the integrand at ${\rm S} = {\cal A}^{-1/\alpha}$.  The theory of resurgence gives a definite recipe in situations such as this, where the Borel transform of a series has singularities on the path of integration used to define the Borel sum.  A naive approach would be to slightly deform the integration contour, to make a detour around any singularity.  However, different choices of such contours will give different results, so this produces ambiguities; moreover since such contours depart from the real axis, the resulting integrals will, in general not be real numbers, even if the original power series has real coefficients.   This means that such a naive approach cannot produce a physically meaningful answer.  However, these mathematical issues are resolved by \'Ecalle's procedure of ``median summation''~\cite{ecalle}, which is a sort of weighted average of all the integrals over all possible contours, which respects basic algebraic operations (such as multiplication of series) and returns real-valued sums for series with real coefficients.  The exact recipe for the median summation is subtle, involving a careful analysis of the Stokes factors that express the differences between the Borel sums on different contours (see e.g.~\cite[Section 7]{dorigoni} for a summary).   However, in the case at hand, since the Borel transform is single-valued and has only a single simple pole in the path of integration (at ${\rm S} = {\cal A}^{-1/\alpha}$), the median summation reduces to the ordinary average of the integrals over two contours: one passing just above the real axis, and the other just below, and this average can be expressed as the classical Cauchy principal value integral
\bg
\PV \int_0^\infty d{\rm S} {e^{-{\rm S}/g^{1/\alpha}} \over 1 - {\cal A}{\rm S^\alpha}} &\equiv & \lim_{\epsilon \to 0} \left({\int_0^{{\cal A}^{-1/\alpha} - \epsilon} + \int^\infty_{{\cal A}^{-1/\alpha} + \epsilon}}\right){e^{-{\rm S}/g^{1/\alpha}}d{\rm S}\over 1 - {\cal A}{\rm S^\alpha}}. \label{pv-integral}
\nd
We claim that this principal value integral evaluates to a positive number.  To see this, we first make the change of variables  ${\rm S} = u/{\cal A}^{1/\alpha}$ and combine the constants $g$ and ${\cal A}$ into the constant $c = (g{\cal A})^{1/\alpha} > 0$, which reduces the problem to the following result.

\begin{proposition}\label{prop:positive}
For any $c > 0$ and any integer $\alpha \ge 2$, we have
\bg\label{borel3}
\PV\int_0^\infty\frac{e^{-u/c}\,du}{1-u^\alpha} > 0
\nd
\end{proposition}

\noindent To prove this proposition, note that
\bg\label{borel4}
\frac{e^{-u/c}}{1-u^\alpha} = \frac{e^{-u/c}-e^{-1/c}}{1-u^\alpha} + \frac{e^{-1/c}}{1-u^\alpha},
\nd
so by linearity of principal value integrals (when they exist), \autoref{prop:positive} follows immediately from \autoref{lem:first-integral} and \autoref{lem:second-integral} below.
\begin{lemma}\label{lem:first-integral}
The function
\bg\label{borel5}
f(u) = \frac{e^{-u/c}-e^{-1/c}}{1-u^\alpha}
\nd
is analytic, strictly positive and absolutely integrable on the interval $[0,\infty)$.  Hence
\bg\label{borel6}
\PV \int_0^\infty f(u) \,du = \int_0^\infty f(u) \,du > 0
\nd
\end{lemma}
{\it Proof}:
$f$ is the ratio of two entire functions of $u$ and is therefore analytic away from the zeros of the denominator, which are the $\alpha$th roots of unity.  In particular, the only point on the positive real axis where the function could fail to be analytic is at $u=1$, where the denominator vanishes to order one.  However the numerator also vanishes to order one there, and hence there is no pole, and $f$ extends analytically over $[0,\infty)$.  

To see that $f$ is strictly positive, note that the only possible zeros of $f$ are where the numerator vanishes, i.e.~$e^{-u/c}=e^{-1/c}$, or equivalently $u = 1 + 2\pi i k c$ for some integer $k$. Hence the only possible zero on the interval $[0,\infty)$ would be at $u=1$, but since the denominator vanishes there also, we have $f(1)\neq 0$.  Hence $f$ is nonvanishing on $[0,\infty)$, and since $c>0$ we have $f(0) = 1 - e^{-1/c} > 0$, so that $f$ is positive on all of $[0,\infty)$ by the intermediate value theorem, as desired.

Finally to see that $f$ is integrable, it suffices to note that $f \sim \tfrac{1}{u^\alpha}$ as $u \to \infty$. \hfill$\Box$

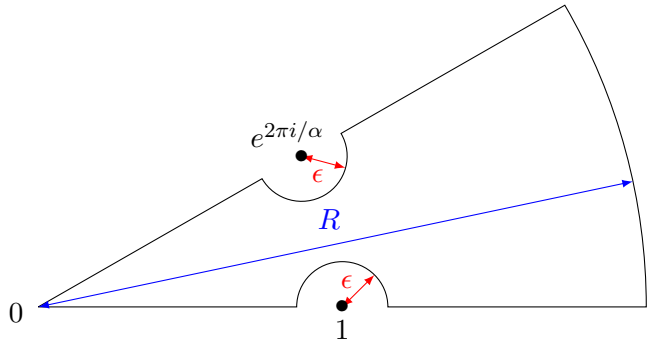
\begin{figure}[b]
\begin{center}
\begin{tikzpicture}
\def\eps{0.6}
\def\bigR{8}
\def\smallR{4}
\def\sectorSize{30}
\draw (0,0) -- ({\smallR-\eps},0) arc (180:0:\eps) -- (\bigR,0) arc (0:{\sectorSize}:\bigR) -- (\sectorSize:{\smallR+\eps}) arc (\sectorSize:{\sectorSize-180}:\eps) -- (0,0);
\draw[red,<->] (0:\smallR) -- +(45:\eps);
\draw[red] (0:\smallR)+(80:{\eps/1.8}) node {${\epsilon}$};
\draw[red,<->] (\sectorSize:\smallR) -- +({\sectorSize-45}:\eps);
\draw[red] (\sectorSize:\smallR)+({\sectorSize-80}:{\eps/1.8}) node {${\epsilon}$};
\draw (0:\smallR) node {$\bullet$};
\draw (\sectorSize:\smallR) node {$\bullet$};
\draw[blue,<->] (0,0) -- ({\sectorSize/2.5}:\bigR);
\draw[blue] ({\sectorSize/2.5+5}:{\bigR/2}) node {$R$};
\draw (\smallR,-0.3) node {$1$};
\draw ({180+\sectorSize/2}:0.3) node {$0$};
\draw ({\sectorSize+5}:\smallR) node {$e^{2\pi i / \alpha}$};
\end{tikzpicture}
\end{center}
\caption{The contour used in the proof of \autoref{lem:second-integral}.}\label{contour}
\end{figure}

\begin{lemma}\label{lem:second-integral}
We have
\bg\label{borel7}
\PV \int_0^\infty \frac{e^{-1/c}}{1-u^\alpha}\,du = \frac{\pi\cot\left(\frac{\pi}{\alpha}\right)}{\alpha e^{1/c}} \ge 0
\nd
for every  integer $\alpha \ge 2$.
\end{lemma}

\noindent {\it Proof}:
Let $h(u) = \frac{1}{1-u^\alpha}$ so the integal is $e^{-1/c}\PV\int_0^\infty h(u)\,du$.
Consider the contour $\Gamma$ given by a sector of radius $R > 0$ and opening angle $2\pi/\alpha$, indented with small semi-circles of radius $\epsilon$ around the point $u =1$ and $u = \exp(2\pi i/\alpha)$ where $h$ has a first order-pole, as shown in \autoref{contour}.  Since $h$ is analytic inside $\Gamma$ we have $\int_\Gamma h(u)\,du = 0$.

In the limit as $R \to \infty$, the integral over the outer arc of the sector tends to zero, because the integrand decays like $\frac{1}{R^\alpha}$ and $\alpha \ge 2$.  Meanwhile, in the limit as $\epsilon \to 0$, the integral over each small semi-circle tends to $-\mathrm{i}\pi$ times the corresponding residue.  Finally, as $R \to \infty$ and $\epsilon \to 0$, the integrals over the straight segments tend to the sum of the Cauchy principal value integrals along the rays from $0$ to $+\infty$ and from $e^{2 \pi i/\alpha}\infty$ to $0$.  We therefore have
\begin{align*}
0 &= \lim_{R\to \infty,\epsilon\to 0} \int_\Gamma h(u)\,du \\
&= \PV \int_0^\infty h(u)\,du - i \pi \Res_{u=1}h(u)du -  \PV \int_0^{e^{2\pi i/\alpha}\infty} h(u)\,du - i\pi \Res_{u=e^{2\pi i /\alpha}}h(u)\,du\\
&= (1- e^{2\pi i/\alpha})\PV \int_0^\infty h(u)\,du + i \pi\left(\frac{-1}{\alpha} + \frac{1}{-\alpha e^{2\pi i /\alpha}}\right)
\end{align*}
The formula for the integral follows by rearranging.  It remains to observe that since $\alpha \ge 2$, we have $0 < \pi/\alpha \le \pi/2$, which implies that $\cot(\pi/\alpha) \ge 0$, with equality if and only if $\alpha = 2$. \hfill$\Box$

Finally, the following result will be used in sub-section \ref{sec4.6.2} to determine the cosmological constant in our model, in the small coupling limit.

\begin{proposition}
We have
\bg\label{brent111}
\lim_{~~~c \to 0^+} \frac{1}{c} \pv \int_0^\infty\frac{e^{-u/c}\,du}{1-u^\alpha} = 0.
\nd
for all $\alpha \ge 2$.
\end{proposition}

\noindent {\it Proof}: Let $\Gamma$ be the contour given by the dented horizontal ray in {\Su Figure \ref{contour}}.  Then we have
\begin{align*}
\frac{1}{c} \PV\int_0^\infty\frac{e^{-u/c}\,du}{1-u^\alpha} &=  \int_\gamma \frac{e^{-u/c}\,du}{c(1-u^\alpha)} + i \pi \Res_{u=1}\frac{e^{-u/c}\,du}{c(1-u^\alpha)} \\
&=  \int_\Gamma \frac{e^{-u/c}\,du}{c(1-u^\alpha)}  - \frac{i \pi e^{-1/c}}{c\alpha} \\ 
&= \left.-\frac{e^{-u/c}}{1-u^\alpha}\right|^{u=+\infty}_{u=0} + \int_\Gamma \frac{\alpha e^{-u/c}u^{\alpha -1}}{(1-u^\alpha)^2} + \frac{i \pi e^{-1/c}}{c\alpha} \\
&= 1 +  \int_\Gamma \frac{\alpha e^{-u/c}u^{\alpha -1}}{(1-u^\alpha)^2}- \frac{i \pi e^{-1/c}}{c\alpha}
\end{align*}
where for the third equality we used integration by parts.  Since $\lim\limits_{~~~c\to 0^+} \tfrac{e^{-1/c}}{c} = 0$, it remains to prove that
\begin{align*}
\lim_{~~~c \to 0^+}\int_\Gamma \frac{\alpha e^{-u/c}u^{\alpha -1}}{(1-u^\alpha)^2} &=0.
\end{align*}
To see this, note that the integrand has the following properties:
\begin{itemize}
\item $\lim\limits_{~~~c \to 0^+} \frac{\alpha e^{-u/c}u^{\alpha -1}}{(1-u^\alpha)^2} = 0$ for all points $u \neq 0$ lying on $\Gamma$.
\item $\left|\tfrac{\alpha e^{-u/c}u^{\alpha -1}}{(1-u^\alpha)^2}\right|$ is bounded above by the function $\left|\frac{\alpha e^{-u/c}u^{\alpha -1}}{(1-u^\alpha)^2}\right|$ on $\Gamma$, and the latter is integrable on $\Gamma$ since it is continuous, it decays like $\frac{1}{u^{\alpha+1}}$ as $u\to \infty$, and $\alpha \ge 2$.
\end{itemize}
Hence the result follows by application of Lebesgue's dominated convergence theorem. \hfill$\Box$

\subsection{Solving equations \eqref{maycamil3} and \eqref{akkim} for a toy model of de Sitter \label{sec4.6}}

Our derivation in the previous sub-section has demonstrated the positivity of the integral on the LHS of \eqref{akkim} for both negative and positive values of ${\cal A}$. However there are still a couple of subtleties that need clarifications so, before moving forward, let us elaborate on them. After which we will study the two set of equations \eqref{maycamil3} and \eqref{akkim}.

\subsubsection{Subtleties with ${\cal A}$ and the Schwinger-Dyson's equations \label{sec4.6.1}}

The {\red first} is the tree-level result studied in section \ref{sec3.2.1}. We have used the Riemann-Lebesgue lemma from 
\eqref{aug19visa} to express it as a series in inverse powers of $(it)^p$ with 
$p \in \mathbb{Z}_+$, but now, instead of adding higher order contributions from $g$, we could have compared it directly with the series we get from the Fourier transform of $\left({g_s\over {\rm HH}_o}\right)^{-{8/3}}$. Question is what is the necessity of adding higher powers of 
$g$? Shouldn't the tree-level result suffice? The puzzle deepens once we note that non-perturbative effects are absolutely {\it necessary} from the point of view of the Schwinger-Dyson's equation because, as we showed in \cite{coherbeta, coherbeta2}, we are actually solving a background EOM of the form:
\bg\label{toula}
\langle\widehat{\bf R}_{\mu\nu}\rangle_{\overline\sigma} - {1\over 2} 
\langle\widehat{\bf g}_{\mu\nu}\rangle_{\overline\sigma} 
\langle\widehat{\bf R}\rangle_{\overline\sigma} = 
\langle\widehat{\bf T}_{\mu\nu}\rangle_{\overline\sigma}, \nd
to get the bulk metric ${\bf g}_{\mu\nu} \equiv 
\langle\widehat{\bf g}_{\mu\nu}\rangle_{\overline\sigma} = \left({g_s\over {\rm HH}_o}\right)^{-{8/3}}$. Here ${\bf T}_{\mu\nu} \equiv \langle\widehat{\bf T}_{\mu\nu}\rangle_{\overline\sigma}$ is the quantum energy-momentum tensor
that includes all the non-perturbative corrections (see details in 
\cite{desitter2, coherbeta, coherbeta2}). We are ignoring subtleties with Faddeev-Popov ghosts, but as discussed in \cite{coherbeta}, they decouple at least from an equation like \eqref{toula}. We have also used the conventions denoted in section \ref{sec1.3}.

Now to rephrase the aforementioned question: couldn't we have directly computed 
$\langle\widehat{\bf g}_{\mu\nu}\rangle_{\overline\sigma}$ $-$ here of course it is 
$\langle \varphi_1\rangle_{\overline\sigma}$ $-$ from the tree-level analysis of the path-integral? Why go to higher orders in $g$? The answer, in retrospect, lies in the Gevrey-$\alpha$ growth of the expectation value as we go to higher orders in $g$. This means, once the series goes to the ${\rm N}$-th order with:
\bg\label{nomwatts}
{\rm N} = {\rm exp}\left(1 - {\log ~g\over \alpha}\right), \nd
the results from higher nodal diagrams start to dominate over the tree-level result, rendering the whole process meaningless unless we Borel resum the Gevrey series. The latter then incorporates the non-perturbative
effects directly at the level of the expectation value 
$\langle \varphi_1\rangle_{\overline\sigma}$ itself. This means, while the Schwinger-Dyson's equations incorporate non-perturbative effects from the {\it bulk} $-$ or from the {\it global} $-$ point of view, the Borel resummation incorporates non-perturbative effects directly from the path-integral $-$ or from the {\it local} $-$ point of view.

This correlation between the {\it global} and the {\it local} points of view is crucial and important. Consider the situation in which the path-integral for the expectation value of a certain field is computed by Borel resumming a Gevrey series. We could have also determined the dynamics of the field by solving a Schwinger-Dyson equation of the form \eqref{toula}. (This is if the field whose expectation value we seek is the metric, otherwise the Schwinger-Dyson equation will look 
different from 
\eqref{toula}, and might even include Faddeev-Popov ghost contributions.) In the Schwinger-Dyson case we expect the non-perturbative effects to play an equally important role as it played for the case where we Borel resummed the Gevrey series. The reason is simple: the interacting Lagrangian is the same for both cases! This shows that there should be a deeper connection between the expectation values of fields over a Glauber-Sudarshan state and the corresponding Schwinger-Dyson's equations.

There is still a subtlety related to the aforementioned connection once we look at the operator $\widehat{\bf S}_{\rm int} \equiv -\widehat{\bf H}_{\rm int}$, which is more general than the ones studied in section \ref{sec2.2}. (Note the slight change of notation from section \ref{sec2.2} as mentioned in section \ref{sec1.3}.) There are two possibilities of defining the expectation values of the energy-momentum tensors, namely:
\bg\label{watcher}
\langle\widehat{\bf T}^{(1)}_{\mu\nu}\rangle_{\overline\sigma} \equiv \left\langle 
{\delta \widehat{\bf S}_{\rm int}\over \delta\widehat{\bf g}^{\mu\nu}}\right\rangle_{\overline\sigma},
~~~~\langle\widehat{\bf T}^{(2)}_{\mu\nu}\rangle_{\overline\sigma} \equiv {\delta \langle 
\widehat{\bf S}_{\rm int} \rangle_{\overline\sigma} \over \delta\langle \widehat{\bf g}^{\mu\nu}\rangle_{\overline\sigma}}, \nd
where the former can be related to the expectation value of the energy-momentum tensor used in \eqref{toula}. For this case, if we express 
$\langle\widehat{\bf T}^{(1)}_{\mu\nu}\rangle_{\overline\sigma}$ in a path-integral form, much like how we studied $\langle\varphi_1\rangle_{\overline\sigma}$ here, then the non-perturbative corrections should automatically be inserted in because of (a) the Gevrey nature of the series, and (b) the subsequent Borel resummation. On the other hand, if we use $\langle\widehat{\bf T}^{(2)}_{\mu\nu}\rangle_{\overline\sigma}$ to represent the RHS of \eqref{toula}, then the scenario is a bit more involved because the non-perturbative contributions are not automatic as in the previous choice
of the energy-momentum tensor\footnote{There is another subtlety here that needs some elaboration. Both the expectation values, namely $\langle 
\widehat{\bf S}_{\rm int} \rangle_{\overline\sigma}$ and $\langle \widehat{\bf g}^{\mu\nu}\rangle_{\overline\sigma}$, will individually make sense after Borel resummation because of their Gevrey natures. However, as discussed in \cite{desitter2, coherbeta} we will also need non-local counter-terms as well as additional non-perturbative contributions to compare the ${g_s\over {\rm HH}_o}$ dependences on both sides of \eqref{toula}. In this sense the second choice of the energy-momentum tensor needs additional ingredients beyond the contributions from the Borel resummed perturbation theory for completeness. We will expand more on this in a future work. \label{tasty}}. For this case one has to add the non-perturbative effects by hand. Despite this, the analysis with the latter choice of the energy-momentum tensor, {\it i.e.} with $\langle\widehat{\bf T}^{(2)}_{\mu\nu}\rangle_{\overline\sigma}$, in \eqref{toula} turns out to be {\it easier} than the one with the choice 
$\langle\widehat{\bf T}^{(1)}_{\mu\nu}\rangle_{\overline\sigma}$. As an example, using $\langle\widehat{\bf T}^{(2)}_{\mu\nu}\rangle_{\overline\sigma}$ and incorporating the non-perturbative and non-local corrections carefully there, one can get a formally exact expression for the cosmological constant $\Lambda$ in the following way \cite{desitter2}:

{\footnotesize
\bg\label{hathway16}
\Lambda &=& 
{1\over 12 \mathbb{V}} \left(\left\langle [\mathbb{C}^i_{i}]^{(0, 0)} \right\rangle -
{1\over 2{\rm H}_o^4} \left\langle\left[\mathbb{C}_a^a \right]^{(3, 0)} \right\rangle
-{1\over 4{\rm H}_o^4} \left\langle\left[\mathbb{C}_m^m \right]^{(0, 0)} \right\rangle
-{1\over 4 {\rm H}_o^4} \left\langle\left[\mathbb{C}_\alpha^\alpha\right]^{(0, 0)}\right\rangle\right) \\ 
&-& {\kappa^2 {\rm T}_2 n_b \over 6 \mathbb{V} {\rm H}_o^8}
-{5\over 384 \mathbb{V} {\rm H}_o^8}\left(\left\langle {\bf  G}^{(3/2)}_{mnab}{\bf G}^{(3/2)mnab} \right\rangle 
+ \left\langle {\bf G}^{(3/2)}_{m\alpha ab}{\bf G}^{(3/2)m\alpha ab} \right\rangle +
\left\langle {\bf G}^{(3/2)}_{\alpha\beta ab}{\bf G}^{(3/2)\alpha\beta ab} \right\rangle\right), \nonumber
 \nd}
where $[\mathbb{C}^{\rm A}_{\rm A}]^{(a, b)}$ are the traces of the quantum terms  with the superscripts denoting the level (see details in \cite{desitter2}). $\mathbb{V}$ denotes the un-warped volume of the internal six-manifold (not to be confused with ${\rm V}$ used earlier), $(\kappa, {\rm T}_2, n_b)$ denote parameters of the dynamical integer and fractional two-branes, and ${\rm H}_o$ is the constant warp-factor. The modes of the G-fluxes are denoted by superscripts (again all details are in \cite{desitter2}). The take-away points from \eqref{hathway16} are the following. {\Su One}, due to the volume suppression, the cosmological constant can be small. {\Su Two}, all the perturbative corrections cannot provide a positive cosmological constant and are therefore red-herrings in the problem. {\Su Three}, the traces of the quantum terms are {\it non-perturbative} in ${\rm M}_p$, thus only the non-perturbative terms can conspire to give a positive cosmological constant. In this sense \eqref{hathway16} matches somewhat with the expression for $\Lambda$ we have from \eqref{akkim}.
Unfortunately however a detailed comparison between the two results is beyond the scope of this paper and will therefore be dealt in a future work.

\begin{figure}[h]
\centering
\begin{tabular}{c}
\includegraphics[width=5in]{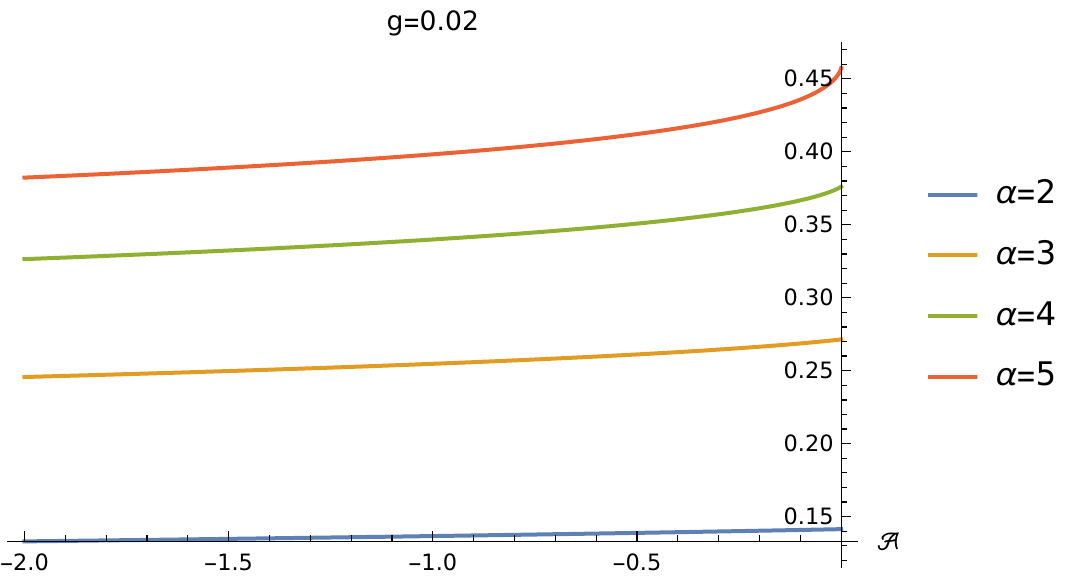}
\end{tabular}
\caption[]{The behavior of the integral on the LHS of \eqref{akkim3} for negative values of ${\cal A}$ and for $\alpha = 2$ to $\alpha = 5$. We have taken $g = 0.02$. The principal values are always positive.}
\label{negativeA}
\end{figure}

The {\red second} issue is in the form of ${\cal A}$ given in \eqref{coldjuly}. This involves integrals over $(\overline\alpha(k), \overline\beta(l), \overline\gamma(f))$, which in turn would depend on the Hermite polynomial expansions from \eqref{greek}. Thus the full knowledge of ${\cal A}$ is possible once we know the coefficients $(b_{mn}, c_{mn}, d_{mn})$ precisely. Our above analysis has given us a way to fix the $c_{nm}$ coefficients, but to fix the other two set of coefficients, namely $b_{nm}$ and $d_{nm}$, we need to compute $\langle\varphi_2\rangle_{\overline\sigma}$ and 
$\langle\varphi_3\rangle_{\overline\sigma}$ and compare the results with the ones from the corresponding Schwinger-Dyson's equations. This is an elaborate process, but fortunately, whether or not ${\cal A}$ is positive, the integral \eqref{akkim} is always positive definite. This is also clear from the lower bound of ${\cal A}$ given in \eqref{beatudi}. Because of this, we are no longer needed to compute the precise value of ${\cal A}$, and in the following we will consider a range of values for both positive and negative choices of ${\cal A}$.

\subsubsection{Solutions to \eqref{akkim} and the cosmological constant \label{sec4.6.2}}

With these two points taken into considerations, we are now ready to analyze the two equations \eqref{maycamil3} and \eqref{akkim} for a toy model of de Sitter space. We will start with \eqref{akkim}, that relates the cosmological constant with the non-perturbative corrections.
What remains now is to see how specific values of the integral, for given choices of ${\cal A}$ and $\alpha$, fit with the values of the RHS of \eqref{akkim}. To see this we can rewrite \eqref{akkim} as:
\bg\label{akkim3}
\PV \int_0^\infty d{\rm S} ~{\rm exp}\left(-{{\rm S}\over g^{1/\alpha}}\right) {1\over 1 - {\cal A}{\rm S}^\alpha}  + {\cal O}\left({1\over {\rm V}}\right) = 
{1 \over \Lambda^{4/3}}\left({2{\rm V}g^{1/\alpha} \over b_a}\right), \nd
where $b_a = {\rm V}\overline{k}_{\rm IR}^{\nu(k_{\rm IR}) - 4}$, and $\nu(k)$ is given by \eqref{mylenecake}. The volume ${\rm V}$ appears in \eqref{akkim3} $-$ which is eventually removed by ${\rm V}$ appearing in the definition of $b_a$ $-$ because the denominator of the path-integral eliminates all nodal diagrams in which the source does not couple with the interactions. As emphasized earlier, such elimination is an expected property of any path-integral, whether we are dealing with the Feynman diagrams or the nodal diagrams. The net result is that $\Lambda$ remains independent of ${\rm V}$.

For negative values of ${\cal A}$ the behavior of the principal value of the integral on the LHS of \eqref{akkim3}, by ignoring the sub-dominant corrections, is always positive and is plotted for a range of ${\cal A}$ and $\alpha$ as shown in {\Su Figure} \ref{negativeA}. It appears that for a given value of $g$ $-$ here we take $g = 0.02$ $-$ the principal value 
increases for small values of $\vert{\cal A}\vert$ and large values of $\alpha$. However $\alpha$ is bounded by the number of field components in the theory, so the principal value cannot be arbitrarily large for negative ${\cal A}$.

\begin{figure}[h]
\centering
\begin{tabular}{c}
\includegraphics[width=5in]{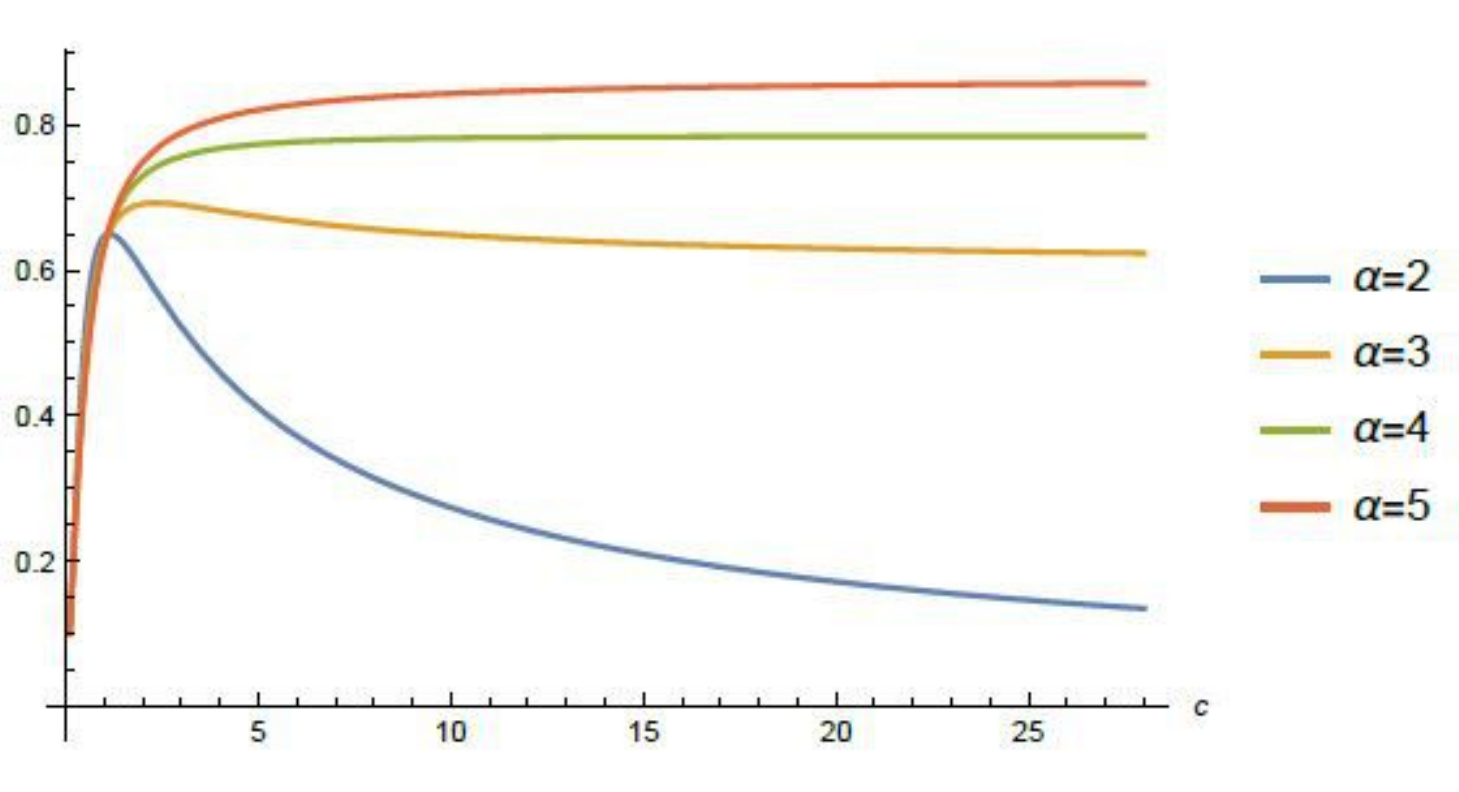}
\end{tabular}
\caption[]{The behavior of the integral for positive values of ${\cal A}$ and for $2 \le \alpha \le 5$ with $\alpha \in \mathbb{Z}_+$. We take $g = 0.02$ as before but now use $c \equiv {(g{\cal A})^{{1\over \alpha}}}$ to parameterize the LHS of \eqref{akkim3}. The principal values are again all positives.}
\label{mmfkeshav1}
\end{figure} 

\begin{figure}[h]
\centering
\begin{tabular}{c}
\includegraphics[width=5in]{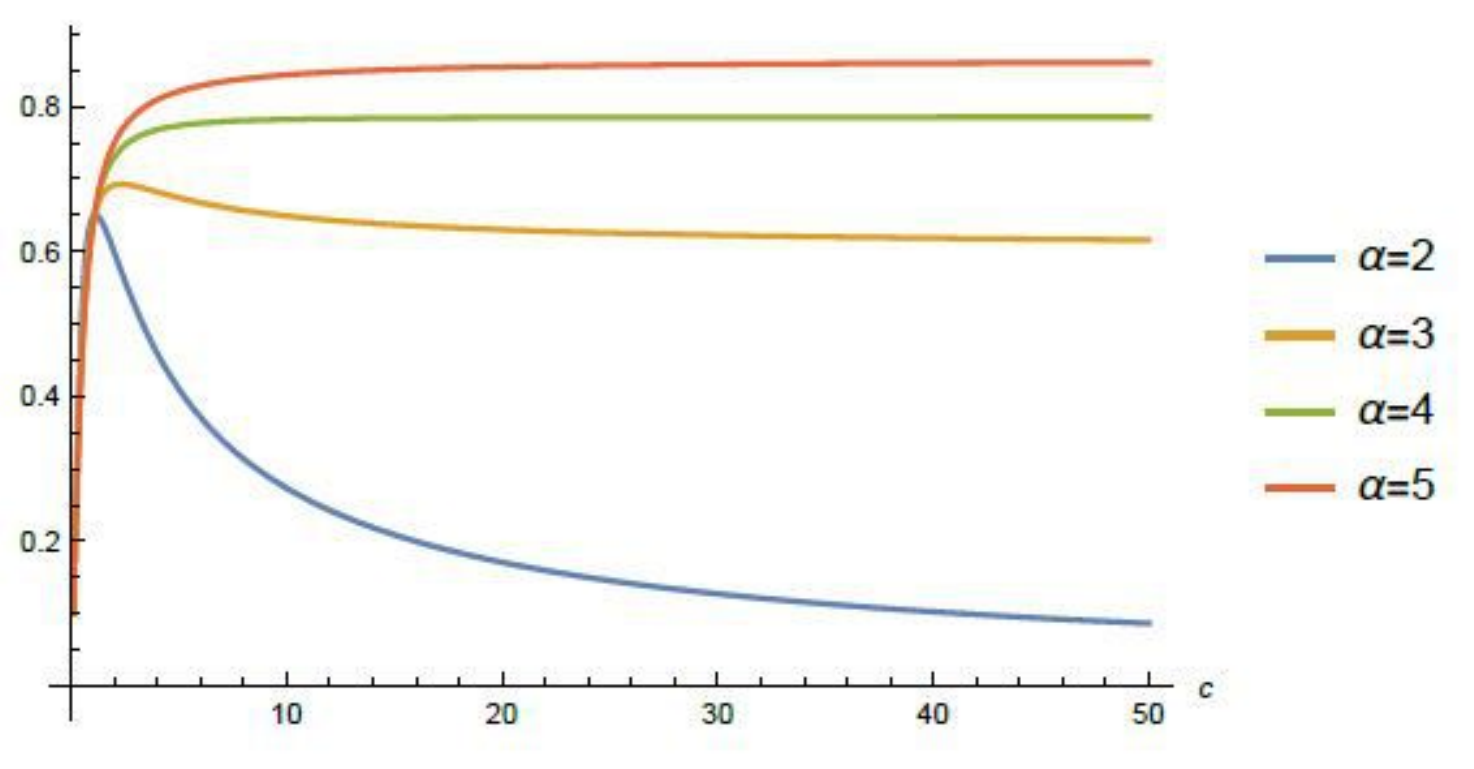}
\end{tabular}
\caption[]{Changing the range of $c$ does not change the positivity of the principal value of the integral in \eqref{borel3} or \eqref{akkim3} above for positive values of ${\cal A}$. The other parameters take the same values as before.}
\label{mmfkeshav2}
\end{figure}

We can also go to positive values of ${\cal A}$ and parameterize the LHS of \eqref{akkim3} as in \eqref{borel3}. The principal values are again positive as shown in {\Su Figure} \ref{mmfkeshav1}. In fact even if we change the range of $c \equiv (g{\cal A})^{{1\over\alpha}}$, the positivity of the principal value does not change as shown in {\Su Figure} \ref{mmfkeshav2}. This of course consistent with the proof that we gave in section \ref{sec4.5} for all values of $({\cal A}, \alpha) > 0$.

\begin{figure}[h]
\centering
\begin{tabular}{c}
\includegraphics[width=5in]{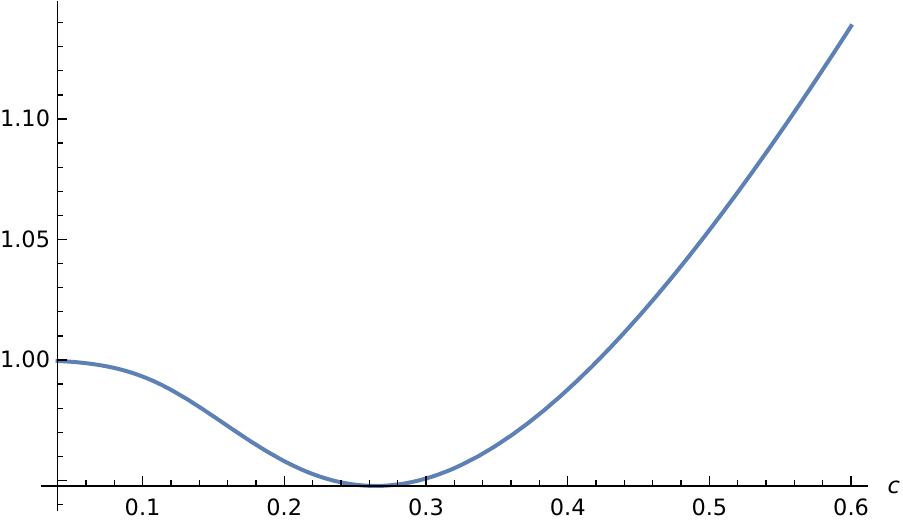}
\end{tabular}
\caption[]{Plot of $\left({c\over \mathbb{I}_{c, 3}}\right)^{3/4}$ versus $c \equiv (g{\cal A})^{1/3}$ where $\mathbb{I}_{c, 3}$ is the principal value integral given in \eqref{akkim00} for $\alpha = 3$. Again, observe that for $c \to 0$, ${c\over \mathbb{I}_{c, 3}} \to 1$ and the cosmological constant $\Lambda$ from \eqref{akkim00} takes the specific value of \eqref{akkim000}.}
\label{lambda2}
\end{figure}

\begin{figure}[h]
\centering
\begin{tabular}{c}
\includegraphics[width=5in]{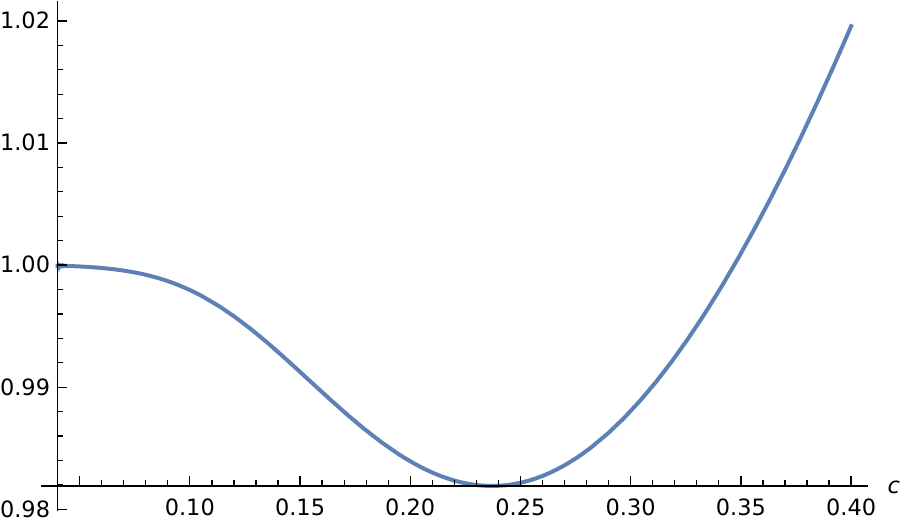}
\end{tabular}
\caption[]{Similar plot as above but now we plot $\left({c\over \mathbb{I}_{c, 4}}\right)^{3/4}$ versus $c \equiv (g{\cal A})^{1/4}$ where $\mathbb{I}_{c, 4}$ is the principal value integral given in \eqref{akkim00} for $\alpha = 4$. For $c \to 0$, ${c\over \mathbb{I}_{c, 4}} \to 1$ and the cosmological constant $\Lambda$ from \eqref{akkim00} takes the specific value of \eqref{akkim000}.}
\label{lambdap}
\end{figure}

Let us clarify a few other details related to the above plots. As mentioned above, the principal value cannot be arbitrarily large, and in fact for 
$g = 0.02$ and small values of $\vert{\cal A}\vert$ and $\alpha$ it appears to be smaller than 1. On the other hand, the RHS of \eqref{akkim3} has parameters that can take large and small values. To see how the 
equality in 
\eqref{akkim3} may work out, let us consider some limiting values of the parameters on the RHS. For simplicity we will take the temporal part of 
$\overline{k}^2_{\rm IR} \equiv -\overline{\kappa}^2_{\rm IR} + \overline{\bf k}^2_{\rm IR}$ to be zero, {\it i.e.} 
$\overline{\kappa}_{\rm IR} = 0$. We then parameterize the {\it small} values of the parameters by powers of $\epsilon$ ($\epsilon << 1$) and {\it large} values of the parameters by powers of $\epsilon^{-1}$. In this language, we have:
\bg\label{watts}
{\rm V} ~ \to ~ \epsilon^{-a}, ~~~ g ~ \to ~ \epsilon^b, ~~~ 
\overline{\bf k}_{\rm IR} ~ \to ~ \epsilon^{\bar{c}}, ~~~ \Lambda ~ \to ~ \epsilon^{d}, \nd
where $(a, b, \bar{c}, d) \in \mathbb{Z}_+$. The above representation simply tells us that the volume ${\rm V}$ can be large, but not infinite, due to the UV/IR mixing and the IR cut-off discussed in section \ref{sec3.2.3}.
The IR spatial momentum $\overline{\bf k}_{\rm IR}$, that keeps the wave-function constant when $k_0 = \overline{\kappa}_{\rm IR} = 0$ and 
${\bf k} = \overline{\bf k}_{\rm IR}$, is bounded from below by the IR cut-off $k_{\rm IR}$ because $k_{\rm IR} \le \overline{\bf k}_{\rm IR} \le \mu$. This means, in principle it can be made small, but not smaller than the IR cut-off implying that $\bar{c}$ can be taken to be zero. The remaining 
parameters can be small, but not zero. (The {\it dimensions} have already been taken care of from the beginning $-$ recall that we kept ${\rm M}_p = 1$ $-$ so for our purpose all the above parameters can be taken to be {\it dimensionless}\footnote{For example, if we denote $a_o \equiv {1\over \sqrt{\Lambda_o}}$ in \eqref{sissys}, then 
$\Lambda$ appearing in \eqref{akkim3} should actually be replaced by a dimensionless quantity ${\Lambda\over \Lambda_o}$. Similarly, ${\rm S}$ in \eqref{akkim} and \eqref{akkim3} is a dimensionful quantity and it's dimension is cancelled by $g^{1/\alpha}$ on the RHS of \eqref{akkim3}. ${\cal A}$ has the same dimension as $g^{-1/\alpha}$, so the combination 
${\cal A} {\rm S}^\alpha$ is dimensionless. The remaining two parameters 
${\rm V}$ and $\overline{\bf k}_{\rm IR}$ are dimensionful quantities, but we have already arranged to cancel the dimensions because both arise from the definition of the volume element in the Fourier space and we have taken ${\rm M}_p = 1$ (see footnote \ref{yaya}). Once we control the dimensions in the aforementioned way, we can give meaning to {\it large} and {\it small} values for the parameters.}.) Now since the LHS of \eqref{akkim3} is small and positive, the parameters $(a, b, \bar{c}, d)$ need to satisfy the following condition:
\begin{figure}[h]
\centering
\begin{tabular}{c}
\includegraphics[width=5in]{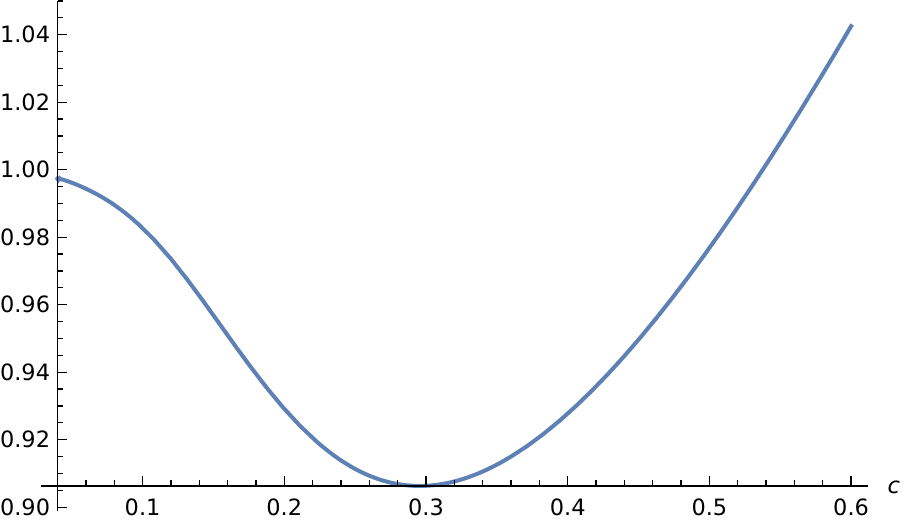}
\end{tabular}
\caption[]{Plot of $\left({c\over \mathbb{I}_{c, 2}}\right)^{3/4}$ versus $c \equiv (g{\cal A})^{1/2}$ where $\mathbb{I}_{c, 2}$ is the principal value integral given in \eqref{akkim00} for $\alpha = 2$. Observe that for $c \to 0$, ${c\over \mathbb{I}_{c, 2}} \to 1$ and the cosmological constant $\Lambda$ from \eqref{akkim00} takes the specific value of \eqref{akkim000}.}
\label{lambda1}
\end{figure} 
\bg\label{holliday}
3b  - 4\alpha d > 0, \nd
for $(\alpha, m^\ast_l) \in \mathbb{Z}_+$. The above inequality can be easily satisfied by taking $b$ large enough, {\it i.e.} by taking a sufficiently small coupling constant. Note the absence of $\bar{c}$ in \eqref{holliday} resulting from our choice $b_a = {\rm V}\overline{k}_{\rm IR}^{\nu(k_{\rm IR}) - 4} \equiv {{\rm V}\sqrt{2\pi/3}\over \Gamma\left(-{5/3}\right)}$. If we keep $b_a$ unfixed, then there could be a term proportional to $\bar{c}$ to \eqref{holliday}. For such a generic case, and by rearranging \eqref{akkim3}, the expression for the cosmological constant may be given by the following:
\bg\label{akkim0}
\boxed{\Lambda = \left({2{\rm V} g^{1/\alpha}\over b_a}\right)^{3/4} 
{\Lambda_0 \over \left(\PV \int\limits_0^\infty d{\rm S} ~{\rm exp}\left(-{{\rm S}\over g^{1/\alpha}}\right) {1\over 1 - {\cal A}{\rm S}^\alpha}\right)^{3/4}}}\nd
where $\Lambda_0 \equiv a^{-2}_0$ with $a_0$ as in \eqref{sissys}, is inserted in to take care of the dimensions; and we choose $b_a \equiv {{\rm V}\sqrt{2\pi/3}\over \Gamma\left(-{5/3}\right)}$. The above is an exact expression for the cosmological constant $\Lambda$ that appears from Borel resumming the Gevrey-$\alpha$ series and should be compared with the expression for $\Lambda$ in \eqref{hathway16}. Note that the volume ${\rm V}$ dependence cancels out, thus the result matches with the cosmological constant got in \eqref{camilahawk} using dominant nodal diagrams. We can rewrite \eqref{akkim0} in a slightly simpler and suggestive way by using the parameter 
$c \equiv ({\cal A}g)^{1/\alpha}$, that we had used earlier, as:
\bg\label{akkim00}
\Lambda = \left({2{\rm V}\Lambda^{4/3}_0\over b_a}\right)^{3/4} 
{c^{3/4}\over \left(\PV \int\limits_0^\infty du~{e^{-u/c}\over 1 - u^\alpha}\right)^{3/4}} = \left({2{\rm V}\Lambda^{4/3}_0\over b_a}\right)^{3/4} \left({c\over \mathbb{I}_{c,\alpha}}\right)^{3/4}, \nd
where $\mathbb{I}_{c,\alpha}$ is the principal value integral from the LHS
of \eqref{akkim00}. For our case, since both $({\cal A}, g) < 0$, we expect $c < 1$. Looking at {\Su Figures \ref{lambda1}, \ref{lambda2}} 
and 
{\Su \ref{lambdap}}, where we plot the behavior of ${c\over \mathbb{I}_{c,\alpha}}$ from \eqref{akkim00} for $\alpha = 2, 3, 4$ respectively, we see that in the limit $c \to 0$, ${c\over \mathbb{I}_{c,\alpha}} \to 1$ irrespective of the choice of $\alpha$. (For more details, see \eqref{brent111} and the proof for Proposition 2 therein.) This suggests that $\Lambda$ takes a definite value of:
\bg\label{akkim000}
\Lambda = \Lambda_0\left({2\Gamma(-5/3)\over \sqrt{2\pi/3}}\right)^{3/4}
\left({c\over \mathbb{I}_{c,\alpha}}\right)^{3/4} + 
{\cal O}\left(e^{-1/c}\right), \nd
which may be made small\footnote{From the fact that the cosmological constant $\Lambda$ may be determined for a given choice of $\alpha$ by the minima of the curves in {\Su Figures \ref{lambda1}, \ref{lambda2}} 
and 
{\Su \ref{lambdap}}. These numerical plots suggest that the minima are always {\it smaller} than 1 (in units that we took here) even if we increase the values for $\alpha$. We will not explore further on the consequence of this and a more detailed study will be performed elsewhere.} using the parametric range from \eqref{watts} and taking $\Lambda_0 = {\cal O}(1)$. 
However a detailed comparison with \eqref{hathway16} looks difficult at this stage because, {\Su one}, \eqref{akkim0} has many unknown parameters, namely $({\rm V}, g, b_a, {\rm \Lambda}_0)$ that need to be fixed 
first\footnote{The parameter ${\rm V}$ is not a problem as it is eliminated.} from appropriate physical arguments before a prediction for $\Lambda$ could be made. And {\Su two}, our analysis with three scalar fields is still a toy model, whereas \eqref{hathway16} is 
the actual picture with 256 degrees of freedom (additionally the latter uses the bulk Schwinger-Dyson's equations). 
Nevertheless the fact that such a simple model as ours produces a definite formula for $\Lambda$ suggests strongly that representing four-dimensional de Sitter space as a Glauber-Sudarshan state 
might be a step towards the right direction.

\subsubsection{Solutions to \eqref{maycamil3} and displacements of interacting vacua \label{sec4.6.3}}

\begin{figure}[h]
\centering
\begin{tabular}{c}
\includegraphics[width=4in]{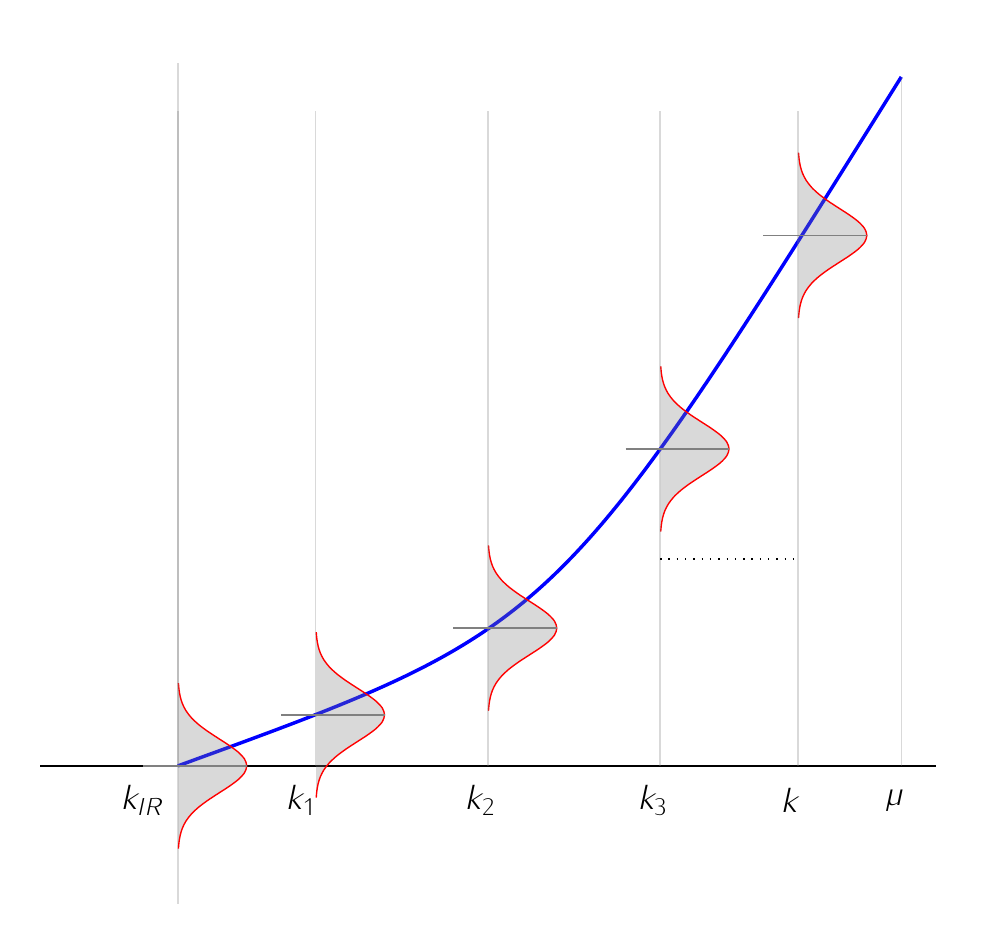}
\end{tabular}
\caption[]{The Glauber-Sudarshan states associated with $\alpha(k) \equiv {\rm V} \overline{\alpha}(k) = {1\over 2} k^{11/3}$ with $k_{\rm IR} \le k \le \mu$ in the configuration space. This is part of the full form for $\overline\alpha(k)$ given in \eqref{professor} with $\nu(k) \approx {11\over 3}$ from \eqref{mylenecake}.}
\label{GSfigure}
\end{figure}

The last thing remaining is to look for solutions of \eqref{maycamil3}, in other words, determine the values of $b_m$. However, in light of \eqref{yvesrocherme}, \eqref{evaBcaku} and \eqref{beatudi}, \eqref{maycamil3} may be re-written in the following way:

{\footnotesize
\bg\label{maycamil5}
\sum_{m = 0}^\infty b_m {\partial^p\over \partial k_0^p} 
\left({{\bf H}_m(k_0 + \overline\kappa_{\rm IR})\over 
\left(\overline{\bf k}^2_{\rm IR} - (k_0 + \overline\kappa_{\rm IR})^2\right)^{2-\nu(k)/2}}\right)
\Bigg\vert_{\kappa_{\rm IR}- \overline{\kappa}_{\rm IR}}^{\mu_0 - \overline\kappa_{\rm IR}} = -{\left({5\over 3}\right)! \over \left({5\over 3}-p\right)!}
~k_0^{{5\over 3} - p}\Bigg\vert_{\kappa_{\rm IR}- \overline{\kappa}_{\rm IR}}^{\mu_0 - \overline{\kappa}_{\rm IR}}, \nd}
where $\nu(k)$ is given by \eqref{mylenecake}. The above set of equations may be simplified slightly by taking $\overline\kappa_{\rm IR} = 0$, but not much. There are an infinite set of equations in \eqref{maycamil5}, and the only way to solve them is to find some hierarchy of scales (say by controlling the set of parameters $(\overline{\bf k}_{\rm IR}, \mu_0, \kappa_{\rm IR})$). Even then, the number of unknowns and the number of equations can be arbitrarily large. How do we know that the system has solutions? 
The answer, despite the apparent technical challenges, is rather obvious. To see this let us retrace the line of thoughts that led us to the set of equations in \eqref{maycamil5}. The vanishing of the perturbative contributions led us to impose the vanishing of the integrand in 
\eqref{600ween}. This gave us the following constraint:
\bg\label{601ween}
{\overline\alpha(k)\over a(k)} - 
{k^{\nu(k)}\over {\rm V}}\left({\overline{\alpha}^2(k)\over a^2(k)} + {1\over 2a(k)}\right) + {\cal O}\left({1\over {\rm V}^2}\right) = 0, \nd
with $\nu(k)$ given by \eqref{mylenecake}. Such a constraint immediately fixes $\overline\alpha(k)$ as in \eqref{evaBcaku}, which we may re-express in a slightly more suggestive way as a scaling relation of the form:
\bg\label{professor}
\boxed{\overline\alpha(k) = {\alpha(k)\over {\rm V}} = {k^{\nu(k)}\over 2{\rm V}} + {\cal O}\left({1\over {\rm V}^2}\right)} \nd
where the Glauber-Sudarshan states associated with $\alpha(k) = {k^{11/3}\over 2}$ are shown in {\Su Figure \ref{GSfigure}}. This in fact should suffice because the scaling of $k^{11/3}$ and the functional form for 
$\chi(k)$ are directly related to the $b_m$ coefficients because of the 
ans\"atze \eqref{greek}. This shows that, no matter how many $b_m$ coefficients we take, the set of equations in \eqref{maycamil5} {\it should} have finite solutions. Two quick consistency checks may be readily performed. The expectation vale of $\varphi_1$ after Borel resumming takes the form \eqref{arosecame3}:

{\tiny
\bg\label{arosavrile1}
 \langle\varphi_1\rangle_{\overline\sigma} = {\bf Re}~\delta\mathbb{T} + {1\over {\rm V}g^{1/\alpha}}
\int_0^\infty d{\rm S} ~{\rm exp}\left(-{{\rm S}\over g^{1/\alpha}}\right) {1\over 1 - {\cal A}{\rm S}^\alpha} 
\int_{k_{\rm IR}}^\mu d^{11}k \left(
{\overline{\alpha}^2(k)\over a^2(k)} + {1\over 2a(k)}\right) k^{\nu(k)}~ {\bf Re}\left(\psi_k({\rm X})~e^{-i(k_0 - \overline\kappa_{\rm IR})t}\right),\nonumber\\ \nd}
where $\delta\mathbb{T} = 0$ from \eqref{601ween}, and the volume term, together with the first integral, may be identified with the inverse of the cosmological constant from \eqref{akkim}. Since the exact form for 
$\overline\alpha(k)$ has a volume suppression as in \eqref{evaBcaku}, the term ${\overline\alpha^2(k)\over a^2(k)}$ is sub-dominant, and the second integral takes the form:
\bg\label{gorgina}
\int_{k_{\rm IR}}^\mu d^{11}k~{k^{\nu(k)}\over 2a(k)}~{\bf Re}\left(\psi_k({\rm X})~e^{-i(k_0 - \overline\kappa_{\rm IR})t}\right), \nd
with the same $\nu(k)$ as in \eqref{mylenecake}. On the other hand, from the last line of \eqref{arosecame3}, we can also rewrite the expectation value of $\varphi_1$ in the following way:

{\scriptsize
\bg\label{campbeel}
\langle\varphi_1\rangle_{\overline\sigma} = {\bf Re}~\delta\mathbb{T} + {1\over g^{1/\alpha}}
\int_0^\infty d{\rm S} ~{\rm exp}\left(-{{\rm S}\over g^{1/\alpha}}\right) {1\over 1 - {\cal A}{\rm S}^\alpha} 
\int_{k_{\rm IR}}^\mu d^{11}k ~{\overline\alpha(k)\over a(k)}~
{\bf Re}\left(\psi_k({\rm X})~e^{-i(k_0 - \overline\kappa_{\rm IR})t}\right), \nd}
using \eqref{601ween}, which doesn't have a volume suppression. However plugging in the exact form for $\overline\alpha(k)$ it is easy to see that the volume part makes the first integral transform to the inverse of the cosmological constant, and the second integral takes the form \eqref{gorgina}. These two ways of expressing the expectation value are exactly the same, showing the consistency of the set-up which in turn arose from the following upper bound:
\bg\label{yvesrochermey}
\sum\limits_{m_l({\rm N})}{\cal A}^{\rm N}\left[m_l({\rm N})\right] k^{m_l({\rm N})} \le {\cal A}^{\rm N} k^{\nu(k)}, ~~~~~ \forall {\rm N} \in \mathbb{Z}_+ \nd 
where it is assumed that the bound could be saturated once the full spectrum of fields (that include 44 components of gravitons, 84 components of the three-form fluxes and 128 components of the Rarita-Schwinger fermions) are taken into account. 

\section{Discussions and conclusions \label{disco}}

Constructing de Sitter space in string theory has proven to be a formidable task. Although there are arguments both for and against the possibility of having de Sitter space in string theory, what can be unambiguously stated is that exploring this has given us a remarkable insight into different aspects of mathematical physics. For instance, moduli stabilization in type IIB string compactifications have led to the understanding of how non-perturbative quantum effects are essential to lift the flat directions corresponding to some of the geometric moduli. Extensive investigations of both the KKLT and LVS scenarios, especially from the full ten-dimensional picture, have led to unveiling of other attributes of both Calabi-Yau and non-K\"ahler manifolds as well as of generic flux compactifications \cite{DRS}. 

In recent work, some of us have proposed that four-dimensional dS should be interpreted as a Glauber-Sudarshan state living in the Hilbert space of a eleven-dimensional supersymmetric Minkowski background. Without going into the merits of such an approach, which solves many problems of the current paradigms \cite{coherbeta, coherbeta2}, in this paper we have instead focused on two mathematical aspects which are relatively less familiar to physicists and which are essential in our approach. 

Firstly, we have introduced a general formalism for doing quantum field theory for a shifted \textit{interacting} vacuum, \textit{i.e.} the Glauber-Sudarshan state. And secondly, we have devised new mathematical procedures to sum up the amplitudes of certain one-point functions coming from the aforementioned field theory.
As emphasized several times in the main text, the calculations can be best controlled in a path integral formalism. As the acute reader would recall, our main idea is that the expectation value of the metric operator, when taken in the appropriate Glauber-Sudarshan state, leads to a de Sitter metric (with a time-dependent internal space). This, of course, was already shown to be the case, in our previous work \cite{coherbeta, coherbeta2}, \textit{at the tree level}\footnote{A more accurate statement would be that, while the computations in \cite{coherbeta} and \cite{coherbeta2} were performed at the tree-level, the full answer was only expected after all the quantum corrections were inserted in. This is most clearly evident from equations (6.6) and (6.7) of \cite{coherbeta2}.}. What we have done in this paper is to extend the analysis to include all higher order quantum corrections, coming from interactions, within a path integral approach. This, in full glory, would imply doing field theory for $44$ graviton components, $84$ flux components and $128$ fermionic fields. Instead of attempting to deal with this somewhat intractable problem, in this work we chose to deal with three fields  instead -- one representative sample for each of the above (a fermionic condensate for the last case). These three fields together represent the labels for constructing our Glauber-Sudarshan state as $(\overline\alpha(k), \overline\beta(k), \overline\gamma(k))$ respectively. 

The {\Su first} part of our quest was to answer how can one evaluate the path integral for a three-field theory, when constructed around a Glauber-Sudarshan state instead of the usual (Fock) vacuum. Although a coherent state is simply considered as a shift operator acting on the vacuum of the free theory, we have always emphasized why one must apply the shift operator on the interacting vacuum in M-theory (indeed, this is why we call our state `Glauber-Sudarshan' to emphatically announce its departure from usual coherent states). Given that we were \textit{not} dealing with a displacement operator in the free theory, it was quickly shown that nodal diagrams were the analogues of Feynman diagrams in this case. In other words, we have shown how the shifted vacuum structures of the path-integral of the three fields is most easily decomposed as collections of these so-called nodal diagrams. We reiterate that this part of the analysis was simply extending the previously known computations of path integrals in QFT when computed around a shifted vacuum by including all sorts of quantum corrections. One of the main advantages of using these nodal diagrams is to be able to deal with the momenta (and, in particular, momentum conservation) for the interacting diagrams. Another facet of our analysis illustrated how simple contributions which would necessarily vanish for Feynman diagrams, due to the presence of  Gaussian wave-functions, are now non-trivial when considering a nodal diagram, as a consequence of using a Glauber-Sudarshan state. 

The {\Su second}, and most significant, new mathematical tool used in our analysis is the Borel-\'Ecalle resummation and the resurgent trans-series of the path integral of the three field system. As shown in section \ref{borel}, the nodal diagrams mentioned above, have a factorial growth, which is denoted by $\alpha \geq 2$ for a three-field system. The factorial growth exhibited by the nodal diagrams, as one goes to higher orders in coupling constants, is the key reason why we find the expectation value of the metric to be an asymptotic series and not a convergent one. At this point, the reader might ask the following question -- if indeed most perturbative expansions in QFT are known to be asymptotic, what is the novelty in our case? The difference is precisely in the physical interpretation of what we are calculating in this case as opposed to standard QFTs underlying particle physics. Standard expressions of scattering amplitudes, say in perturbative QED, can be formally a divergent series and yet, for the purposes of the precision involved in everyday experiments, it is sufficient to consider calculations up to terms which are well in the perturbative regime and do not ``see'' the effects of the factorial growth. In other words, although mathematically unsatisfactory, one can restrict oneself to perturbative computations and not worry about the asymptotic nature of the full series. In our case, we are indeed evaluating the metric of space-time itself. This is not a computational quantity to be verified from experiments. Rather, we must have an \textit{exact} answer in order for it to be meaningful. If, for instance, we are able to show that one finds four-dimensional de Sitter space in M-theory only in the perturbative regime, and the expansion of the expectation value of the metric stops being meaningful at a certain order in perturbation theory, then we cannot conclude anything about the validity of such a solution. Therefore, our results tie in two things nicely -- \textit{Not only is it necessary to understand non-perturbative physics in order to construct de Sitter in string theory, but also perturbative quantum corrections are insufficient to describe such a solution.}

We show that the factorial growth of the expectation value of the metric is  of the type of {\bf Gevrey-$\alpha$ series}.  Indeed, the fact that we consider $3$-bosonic scalar fields leads to $\alpha \geq 2$, and we expect $\alpha$ to be much greater for full M-theory, going beyond our representative toy model. We show the pitfalls of naively `Borel-resumming' the asymptotic series coming from the nodal diagrams -- this can lead to all sorts of unphysical factors appearing in the non-perturbative contributions. However, a correct resummation of the Gevrey-$\alpha$ series (we have left $\alpha$ undetermined in most of the text to account for full M-theory), leads to a closed form expression for the cosmological constant (see \eqref{akkim0}). There are several subtleties which we have encountered on the way -- finding the dominant diagrams taking into consideration the denominator of the path integral and the fact that nodal diagrams are different from their Feynman ones and requiring an infrared cutoff for the momenta integrals. The second requirement is not a strong one -- gravitational theories have a long history of UV/IR coupling leading to IR cutoffs emerging from UV ones. And since we were always considering a Wilsonian EFT of M-theory, the Planck scale acts as a natural UV-cutoff in the action.

The careful analyses of the dominant diagrams require us to cancel a contribution coming from a term which can be regarded as a tadpole term for the Glauber-Sudarshan state (it is linear in $\overline{\alpha}(k)$) at the tree level. This leads to a surprising consequence for the allowed values of $\overline{\alpha}(k)$ (outlined in \eqref{mariec}). Although there can be two solutions, only one of them makes sense in the large volume limit\footnote{Perhaps the other solution, which is better defined for small volume ${\rm V}$, is the relevant one for constructing inflation in M-theory. We have not considered this question in detail here and leave this for future work.}, and yet, does not over-constrain the system (given the consistency relation \eqref{maycamil3}). To ensure that \eqref{maycamil3} and \eqref{mariec} can be simultaneously satisfied, we consider an ans\"atze of the form in \eqref{professor}. However, having done this, we find an interesting result. The value of the cosmological constant in this model {\it can} be a small one (suppressed due to the various factors appearing in \eqref{akkim000}). Such a conclusion is borne out from the fact that the displacement, that describes the Glauber-Sudarshan state from the vacuum in the configuration space, is very small ($\propto {\rm V}^{-1}$). In other words, although we do not find any way to construct 
four-dimensional de Sitter space in M-theory as a {\it vacuum} configuration, the Glauber-Sudarshan state describing it is nevertheless very close to it in a precise mathematical sense. Of course, this is only an indication (borne out by some numerical solutions shown later on) and to rigorously show that indeed smaller values of the cosmological constant are preferred in this theory, we would have to define a good measure on the space of solutions of our model. This we leave for future work.

However, the most important finding of our analysis is that the non-perturbative contribution, coming from the Borel resummation, which is (inversely) proportional to the cosmological constant turns out to be a \textbf{positive} quantity. This is rigorously proved in section \ref{sec4.5} and is the most striking of all our results. What we had set out to do is to construct de Sitter space as a coherent state on top of a supersymmetric Minkowski vacua. Even though we had shown previously how this can be achieved at the tree level, how the quantum corrections (especially the non-perturbative ones) contribute to the path integral, determining the value of the cosmological constant, was totally out of our control. It could have easily been the case that the non-perturbative effect was to essentially say that no positive cosmological constant was allowed and thus rule out our construction. Instead, we have shown that the principal value of the Borel-resummed non-perturbative result is to give a non-negative result and thus putting a lower bound on the value of the cosmological constant. Once again, the observation that the factorial growth $\alpha \geq 2$ plays a crucial role in this conclusion.

Finally, we have shown in section \ref{sec4.6} how this result matches qualitatively with previous findings of the cosmological constant derived solely from the quantum corrections, using the Schwinger-Dyson equations. Although an exact matching between the two approaches is beyond the scope of this paper, the main finding of both these approaches is that perturbative quantum corrections are not sufficient to give a positive cosmological constant. Although this was already shown in \cite{desitter2}, this approach using path integrals and Borel-\'Ecalle resummation makes this point more explicit. Some numerical solutions for the cosmological constant indeed show that the cosmological constant is small for small values of the coefficient $\mathcal{A}$ which controls the polynomial growth, and for small value of the coupling $g$, realizing the formal arguments made earlier.

To summarize, in this work we have induced two new mathematical techniques to realize de Sitter space a Glauber-Sudarshan state in M-theory. The first one is to use \textit{nodal diagrams} to evaluate path integrals about a displaced (interacting) vacuum. And the second is to apply the theory of Borel-\'Ecalle resummations to find non-perturbative contributions to expectation values coming from such path integrals. Although this analysis requires some new and sophisticated mathematical tools, it leads to two interesting results:
\begin{enumerate}
    \item The cosmological constant \eqref{akkim0}, coming from such non-perturbative resummations, is shown to be explicitly positive for a large number of fields (denoted by the value of $\alpha$) with the possibility that it can be very small.
    \item The Glauber-Sudarshan state \eqref{professor}, that represents the four-dimensional de Sitter space, has a small displacement from the vacuum configuration. This small difference from being a vacuum configuration is solely responsible for the system to overcome all the no-go and the swampland criteria. 
\end{enumerate}
Although our main aim was to investigate the validity of de Sitter space in Type IIB theory from its M-theory dual, it will be short-sighted to not mention the usefulness the methods elaborated in this paper might find in other applications in QFT. Indeed, the main techniques used are independent of M-theory, and we expect that they would find use in many other systems in physics.

\section*{Acknowledgements:} 

We would like to thank Maxim Emelin, Ori Ganor and Mithat Unsal for many helpful discussions and exchanges. KD will also like to thank Veronica Errasti-Di\'ez for some help with the figures. The work of SB is supported in part by the Higgs Fellowship. The work of KD, MMF, BK and VM is supported in part by a Discovery Grant from the Natural Sciences and
Engineering Research Council of Canada (NSERC). BP is supported by a Discovery Grant from NSERC; a Research Support for New
Academics grant from the Fonds de recherche du Qu\'ebec Nature et
technologies (FQRNT); and a Faculty Startup Grant from McGill University. For the purpose of open access, the authors have applied a Creative Commons Attribution (CC BY) licence to any Author Accepted Manuscript version arising from this submission.

\vskip.6in

\hskip1.5in {\it Everything begins and ends at exactly the right time and place.}

\hskip1.8in $-$ Peter Weir's ``{\Su Picnic at Hanging Rock}" (1975)

%\newpage


\vskip.6in

\begin{thebibliography}{99}

%\cite{Dunne:2012ae}
\bibitem{unsal}
G.~V.~Dunne and M.~\"Unsal,
``Resurgence and Trans-series in Quantum Field Theory: The CP${}^{{\rm N}-1}$ Model,''
JHEP \textbf{11}, 170 (2012), [arXiv:1210.2423 [hep-th]]; 
``Generating nonperturbative physics from perturbation theory,''
Phys. Rev. D \textbf{89}, no.4, 041701 (2014), 
[arXiv:1306.4405 [hep-th]]; 
G.~Basar, G.~V.~Dunne and M.~\"Unsal,
``Resurgence theory, ghost-instantons, and analytic continuation of path integrals,''
JHEP \textbf{10}, 041 (2013), [arXiv:1308.1108 [hep-th]].




%\cite{Dyson:1952tj}
\bibitem{dyson}
F.~J.~Dyson,
``Divergence of perturbation theory in quantum electrodynamics,''
Phys. Rev. \textbf{85}, 631-632 (1952).
%doi:10.1103/PhysRev.85.631
%621 citations counted in INSPIRE as of 11 Sep 2022


\bibitem{borelE}
E.~Borel, ``M\'emoire sur les s\'eries divergentes", Ann. Sci. \'Ec. Norm. Sup\'er., Series 3, {\bf 16}, 9–131 (1899).

%\cite{Shenker:1995xq}
\bibitem{shenker}
S.~H.~Shenker,
``Another length scale in string theory?,''
[arXiv:hep-th/9509132 [hep-th]].
%115 citations counted in INSPIRE as of 11 Oct 2022

%\cite{Polchinski:1995mt}
\bibitem{polchinski}
J.~Polchinski,
``Dirichlet Branes and Ramond-Ramond charges,''
Phys. Rev. Lett. \textbf{75}, 4724-4727 (1995)
[arXiv:hep-th/9510017 [hep-th]];
J.~Dai, R.~G.~Leigh and J.~Polchinski,
``New Connections Between String Theories,''
Mod. Phys. Lett. A \textbf{4}, 2073-2083 (1989)
%919 citations counted in INSPIRE as of 11 Oct 2022
%2852 citations counted in INSPIRE as of 11 Oct 2022


\bibitem{gevreyorig}
M.~Gevrey, ``Sur la nature analytique des solutions des \'equations aux d\'eriv\'ees partielles. Premier m\'emoire", {\it Annales scientifiques de l'\'Ecole Normale Sup\'erieure}, {\bf 35}, 129–190 (1918).

\bibitem{mittag}
G.~Mittag-Leffler, ``Sur la repr\'esentation arithm\'etique des fonctions analytiques d'une variable complexe", {\it Atti del IV Congresso Internazionale dei Matematici, Roma}; 6–11 (1908).


%\cite{Gukov:2016njj}
\bibitem{marino}
S.~Gukov, M.~Mari\~no and P.~Putrov,
``Resurgence in complex Chern-Simons theory,''
[arXiv:1605.07615 [hep-th]];
L.~Di Pietro, M.~Mari\~no, G.~Sberveglieri and M.~Serone,
``Resurgence and 1/N Expansion in Integrable Field Theories,''
JHEP \textbf{10}, 166 (2021)
[arXiv:2108.02647 [hep-th]].
%81 citations counted in INSPIRE as of 11 Oct 2022

%\cite{Emelin:2020buq}
\bibitem{Emelin}
M.~Emelin,
``Effective Theories as Truncated Trans-Series and Scale Separated Compactifications,''
JHEP \textbf{11}, 144 (2020)
[arXiv:2005.11421 [hep-th]].
%11 citations counted in INSPIRE as of 17 Oct 2022


\bibitem{KKLT}
S.~Kachru, R.~Kallosh, A.~D.~Linde and S.~P.~Trivedi,
``De Sitter vacua in string theory,''
Phys. Rev. D \textbf{68}, 046005 (2003)
[arXiv:hep-th/0301240 [hep-th]].
%2946 citations counted in INSPIRE as of 13 Mar 2021

\bibitem{swampland}
U.~H.~Danielsson and T.~Van Riet,
``What if string theory has no de Sitter vacua?,''
Int. J. Mod. Phys. D \textbf{27}, no.12, 1830007 (2018)
[arXiv:1804.01120 [hep-th]];\\
  G.~Obied, H.~Ooguri, L.~Spodyneiko and C.~Vafa,
  ``De Sitter Space and the Swampland,''
  arXiv:1806.08362 [hep-th];\\
  P.~Agrawal, G.~Obied, P.~J.~Steinhardt and C.~Vafa,
  ``On the Cosmological Implications of the String Swampland,''
  Phys.\ Lett.\ B {\bf 784}, 271 (2018)  
  [arXiv:1806.09718 [hep-th]];\\
 S.~K.~Garg and C.~Krishnan,
  ``Bounds on Slow Roll and the de Sitter Swampland,''
  arXiv:1807.05193 [hep-th];\\
  S.~K.~Garg and C.~Krishnan,
    ``Bounds on Slow Roll at the Boundary of the Landscape,''
  JHEP {\bf 1903}, 029 (2019)
   [arXiv:1810.09406 [hep-th]];\\
   H.~Ooguri, E.~Palti, G.~Shiu and C.~Vafa,
  ``Distance and de Sitter Conjectures on the Swampland,''
  Phys.\ Lett.\ B {\bf 788}, 180 (2019)  
  [arXiv:1810.05506 [hep-th]];\\
  D.~Andriot,
  ``On the de Sitter swampland criterion,''
  Phys.\ Lett.\ B {\bf 785}, 570 (2018)
  [arXiv:1806.10999 [hep-th]].  
  
 \bibitem{beno}
I.~Bena, G.~Giecold, M.~Grana, N.~Halmagyi and S.~Massai,
``The backreaction of anti-D3 branes on the Klebanov-Strassler geometry,''
JHEP \textbf{06}, 060 (2013)
[arXiv:1106.6165 [hep-th]]; I.~Bena, M.~Grana, S.~Kuperstein and S.~Massai,
``Anti-D3 Branes: Singular to the bitter end,''
Phys. Rev. D \textbf{87}, no.10, 106010 (2013)
[arXiv:1206.6369 [hep-th]].

%\cite{Bergshoeff:2015jxa}
\bibitem{bergshoeff}
E.~A.~Bergshoeff, K.~Dasgupta, R.~Kallosh, A.~Van Proeyen and T.~Wrase,
``$ \overline{\mathrm{D}3} $ and dS,''
JHEP \textbf{05}, 058 (2015)
[arXiv:1502.07627 [hep-th]];
K.~Dasgupta, M.~Emelin and E.~McDonough,
``Fermions on the antibrane: Higher order interactions and spontaneously broken supersymmetry,''
Phys. Rev. D \textbf{95}, no.2, 026003 (2017)
[arXiv:1601.03409 [hep-th]].
%144 citations counted in INSPIRE as of 17 Oct 2022



%\cite{Sethi:2017phn}
\bibitem{sothi}
S.~Sethi,
``Supersymmetry Breaking by Fluxes,''
JHEP \textbf{10}, 022 (2018)
[arXiv:1709.03554 [hep-th]].
%145 citations counted in INSPIRE as of 17 Oct 2022


%\cite{Denef:2018etk}
\bibitem{donof}
F.~Denef, A.~Hebecker and T.~Wrase,
``de Sitter swampland conjecture and the Higgs potential,''
Phys. Rev. D \textbf{98}, no.8, 086004 (2018)
[arXiv:1807.06581 [hep-th]].
%137 citations counted in INSPIRE as of 17 Oct 2022


\bibitem{evan}
K.~Dasgupta, M.~Emelin, E.~McDonough and R.~Tatar,
``Quantum Corrections and the de Sitter Swampland Conjecture,''
JHEP \textbf{01}, 145 (2019)
[arXiv:1808.07498 [hep-th]].
%53 citations counted in INSPIRE as of 03 Mar 2021

%\cite{Kachru:2018aqn}
\bibitem{kochru}
S.~Kachru and S.~P.~Trivedi,
``A comment on effective field theories of flux vacua,''
Fortsch. Phys. \textbf{67}, no.1-2, 1800086 (2019)
[arXiv:1808.08971 [hep-th]].
%91 citations counted in INSPIRE as of 17 Oct 2022


\bibitem{russo}
J.~G.~Russo and P.~K.~Townsend,
``Time-dependent compactification to de Sitter space: a no-go theorem,''
JHEP \textbf{06}, 097 (2019)
[arXiv:1904.11967 [hep-th]].


\bibitem{heliudson}
H.~Bernardo, S.~Brahma, K.~Dasgupta and R.~Tatar,
``Crisis on Infinite Earths: Short-lived de Sitter Vacua in the String Theory Landscape,''
JHEP \textbf{04}, 037 (2021)
[arXiv:2009.04504 [hep-th]];
``Purely nonperturbative AdS vacua and the swampland,''
Phys. Rev. D \textbf{104}, no.8, 086016 (2021)
[arXiv:2104.10186 [hep-th]].
%5 citations counted in INSPIRE as of 10 Apr 2021
 
%\cite{DallAgata:2022abm}
\bibitem{dologoto}
G.~Dall'Agata, M.~Emelin, F.~Farakos and M.~Morittu,
``Anti-brane uplift instability from goldstino condensation,''
JHEP \textbf{08}, 005 (2022)
[arXiv:2203.12636 [hep-th]];
M.~Emelin,
``Obstacles for dS in Supersymmetric Theories,''
[arXiv:2206.01603 [hep-th]].
%0 citations counted in INSPIRE as of 17 Oct 2022
%5 citations counted in INSPIRE as of 17 Oct 2022  
 
%\cite{Kallosh:2022fsc}
\bibitem{kalush}
R.~Kallosh, A.~Linde, T.~Wrase and Y.~Yamada,
``Goldstino condensation?,''
JHEP \textbf{08}, 166 (2022)
[arXiv:2206.04210 [hep-th]].
%1 citations counted in INSPIRE as of 17 Oct 2022 
  

\bibitem{tcc}
J.~Martin and R.~H.~Brandenberger,
``The Trans-Planckian problem of inflationary cosmology,''
Phys. Rev. D \textbf{63}, 123501 (2001)
[arXiv:hep-th/0005209 [hep-th]];\\
A.~Bedroya and C.~Vafa,
``Trans-Planckian Censorship and the Swampland,''
JHEP \textbf{09}, 123 (2020)
[arXiv:1909.11063 [hep-th]];\\
A.~Bedroya, R.~Brandenberger, M.~Loverde and C.~Vafa,
``Trans-Planckian Censorship and Inflationary Cosmology,''
Phys. Rev. D \textbf{101}, no.10, 103502 (2020)
[arXiv:1909.11106 [hep-th]];
S.~Brahma,
``Trans-Planckian censorship conjecture from the swampland distance conjecture,''
Phys. Rev. D \textbf{101}, no.4, 046013 (2020)
[arXiv:1910.12352 [hep-th]].


\bibitem{desitter2}
K.~Dasgupta, M.~Emelin, M.~M.~Faruk and R.~Tatar,
``de Sitter Vacua in the String Landscape,''
Nucl. Phys. B \textbf{969}, 115463 (2021)
[arXiv:1908.05288 [hep-th]];\\
``How a four-dimensional de Sitter solution remains outside the swampland,''
JHEP {\bf 07}, 109 (2021)
[arXiv:1911.02604 [hep-th]];\\
``de Sitter Vacua in the String landscape: La Petite Version,''
QTS2019
[arXiv:1911.12382 [hep-th]].


%\cite{Feynman:1963fq}
\bibitem{vernon}
R.~P.~Feynman and F.~L.~Vernon, Jr.,
``The Theory of a general quantum system interacting with a linear dissipative system,''
Annals Phys. \textbf{24}, 118-173 (1963).
%710 citations counted in INSPIRE as of 17 Oct 2022

%\cite{Agon:2014uxa}
\bibitem{balasub}
C.~Agon, V.~Balasubramanian, S.~Kasko and A.~Lawrence,
``Coarse Grained Quantum Dynamics,''
Phys. Rev. D \textbf{98}, no.2, 025019 (2018)
[arXiv:1412.3148 [hep-th]].
%37 citations counted in INSPIRE as of 22 Oct 2022

%\cite{Brahma:2022yxu}
\bibitem{brahma1}
S.~Brahma, A.~Berera and J.~Calder\'on-Figueroa,
``Quantum corrections to the primordial tensor spectrum: open EFTs \& Markovian decoupling of UV modes,''
JHEP \textbf{08}, 225 (2022)
[arXiv:2206.05797 [hep-th]].
%3 citations counted in INSPIRE as of 22 Oct 2022

%\cite{Colas:2022hlq}
\bibitem{colas1}
T.~Colas, J.~Grain and V.~Vennin,
``Benchmarking the cosmological master equations,''
[arXiv:2209.01929 [hep-th]].
%1 citations counted in INSPIRE as of 22 Oct 2022

%\cite{Burgess:2021luo}
\bibitem{burgess1}
C.~P.~Burgess, R.~Holman and G.~Kaplanek,
``Quantum Hotspots: Mean Fields, Open EFTs, Nonlocality and Decoherence Near Black Holes,''
Fortsch. Phys. \textbf{70}, no.4, 2200019 (2022)
[arXiv:2106.10804 [hep-th]];
C.~P.~Burgess, R.~Holman, G.~Kaplanek, J.~Martin and V.~Vennin,
``Minimal decoherence from inflation,''
[arXiv:2211.11046 [hep-th]].
%0 citations counted in INSPIRE as of 25 Nov 2022
%7 citations counted in INSPIRE as of 22 Oct 2022


%\cite{Bernardo:2021rul}
\bibitem{coherbeta2}
H.~Bernardo, S.~Brahma, K.~Dasgupta, M.~M.~Faruk and R.~Tatar,
``de Sitter Space as a Glauber-Sudarshan State: II,''
Fortsch. Phys. \textbf{69}, no.11-12, 2100131 (2021)
[arXiv:2108.08365 [hep-th]].
%2 citations counted in INSPIRE as of 12 Oct 2022


\bibitem{coherbeta}
S.~Brahma, K.~Dasgupta and R.~Tatar,
``Four-dimensional de Sitter space is a Glauber-Sudarshan state in string theory,''
JHEP {\bf 07}, 114 (2021)
[arXiv:2007.00786 [hep-th]];\\
``de Sitter Space as a Glauber-Sudarshan State,''
JHEP {\bf 02}, 104 (2021)
[arXiv:2007.11611 [hep-th]].





\bibitem{dvali}
G.~Dvali and C.~Gomez,
``Black Hole's Quantum N-Portrait,''
Fortsch. Phys. \textbf{61}, 742-767 (2013)
[arXiv:1112.3359 [hep-th]];\\
``Quantum Compositeness of Gravity: Black Holes, AdS and Inflation,''
JCAP \textbf{01}, 023 (2014)
[arXiv:1312.4795 [hep-th]];\\
G.~Dvali, C.~Gomez, R.~S.~Isermann, D.~Luest and S.~Stieberger,
``Black hole formation and classicalization in ultra-Planckian 2$\to$N scattering,''
Nucl. Phys. B \textbf{893}, 187-235 (2015)
[arXiv:1409.7405 [hep-th]];\\
G.~Dvali, C.~Gomez and S.~Zell,
``Quantum Break-Time of de Sitter,''
JCAP \textbf{06}, 028 (2017)
arXiv:1701.08776 [hep-th];\\
``Quantum Breaking Bound on de Sitter and Swampland,''
Fortsch. Phys. \textbf{67}, no.1-2, 1800094 (2019)
arXiv:1810.11002 [hep-th];\\
G.~Dvali and C.~Gomez,
``On Exclusion of Positive Cosmological Constant,''
Fortsch. Phys. \textbf{67}, no.1-2, 1800092 (2019)
[arXiv:1806.10877 [hep-th]];\\
G.~Dvali and C.~Gomez,
``Quantum Exclusion of Positive Cosmological Constant?,''
Annalen Phys. \textbf{528}, 68-73 (2016)
[arXiv:1412.8077 [hep-th]];\\
R.~Casadio, A.~Giugno and A.~Giusti,
``Corpuscular slow-roll inflation,''
Phys. Rev. D \textbf{97}, no.2, 024041 (2018)
[arXiv:1708.09736 [gr-qc]];\\
M.~Cadoni, R.~Casadio, A.~Giusti and M.~Tuveri,
``Emergence of a Dark Force in Corpuscular Gravity,''
Phys. Rev. D \textbf{97}, no.4, 044047 (2018)
[arXiv:1801.10374 [gr-qc]];\\
A.~Giusti,
``On the corpuscular theory of gravity,''
Int. J. Geom. Meth. Mod. Phys. \textbf{16}, no.03, 1930001 (2019).

\bibitem{maltz}
J.~Maltz and L.~Susskind,
``de Sitter Space as a Resonance,''
Phys. Rev. Lett. \textbf{118}, no.10, 101602 (2017)
[arXiv:1611.00360 [hep-th]].




\bibitem{GMN}
G.~W.~Gibbons,
``Aspects of supergravity theories,''
print-85-0061 (Cambridge);
J.~M.~Maldacena and C.~Nunez,
``Supergravity description of field theories on curved manifolds and a no go theorem,''
Int. J. Mod. Phys. A \textbf{16}, 822-855 (2001)
[arXiv:hep-th/0007018 [hep-th]];
G.~W.~Gibbons,
``Thoughts on tachyon cosmology,''
Class. Quant. Grav. \textbf{20}, S321-S346 (2003)
doi:10.1088/0264-9381/20/12/301
[arXiv:hep-th/0301117 [hep-th]];
K.~Dasgupta, R.~Gwyn, E.~McDonough, M.~Mia and R.~Tatar,
``de Sitter Vacua in Type IIB String Theory: Classical Solutions and Quantum Corrections,''
JHEP \textbf{07}, 054 (2014)
[arXiv:1402.5112 [hep-th]];
H.~Bernardo, S.~Brahma and M.~M.~Faruk,
``The inheritance of energy conditions: Revisiting no-go theorems in string compactifications,''
[arXiv:2208.09341 [hep-th]].



\bibitem{uvir}
A.~G.~Cohen, D.~B.~Kaplan and A.~E.~Nelson,
``Effective field theory, black holes, and the cosmological constant,''
Phys. Rev. Lett. \textbf{82}, 4971-4974 (1999)
[arXiv:hep-th/9803132 [hep-th]]; 
P.~Draper, I.~G.~Garcia and M.~Reece,
``Snowmass White Paper: Implications of Quantum Gravity for Particle Physics,''
[arXiv:2203.07624 [hep-ph]];
T.~W.~Kephart and H.~P\"as,
``UV/IR Mixing, Causal Diamonds and the Electroweak Hierarchy Problem,''
[arXiv:2209.03305 [hep-ph]].

\bibitem{dineseiberg}
M.~Dine and N.~Seiberg,
``Is the Superstring Weakly Coupled?,''
Phys. Lett. B \textbf{162}, 299-302 (1985).
%381 citations counted in INSPIRE as of 22 Dec 2022


\bibitem{DRS}
K.~Dasgupta, G.~Rajesh and S.~Sethi,
``M theory, orientifolds and G - flux,''
JHEP \textbf{08}, 023 (1999)
[arXiv:hep-th/9908088 [hep-th]];
S.~B.~Giddings, S.~Kachru and J.~Polchinski,
``Hierarchies from fluxes in string compactifications,''
Phys. Rev. D \textbf{66}, 106006 (2002)
[arXiv:hep-th/0105097 [hep-th]];
S.~Kachru, M.~B.~Schulz and S.~Trivedi,
``Moduli stabilization from fluxes in a simple IIB orientifold,''
JHEP \textbf{10}, 007 (2003)
[arXiv:hep-th/0201028 [hep-th]];
K.~Becker and K.~Dasgupta,
``Heterotic strings with torsion,''
JHEP \textbf{11}, 006 (2002)
[arXiv:hep-th/0209077 [hep-th]].

%\cite{Dasgupta:2018rtp}




\bibitem{gevrey}
J.~Martinet and J-P.~Ramis, ``Elementary acceleration and multisummability. I." Annales de l'I.H.P. Physique th\'eorique {\bf 54.4}, 331-401 (1991);
O.~Costin,
``Asymptotics and Borel summability," {\it Monographs and Surveys in Pure and Applied Mathematics 141}, Chapman and Hall Book (2009);
W.~Balser, ``From Divergent Power Series to Analytic Functions," {\it Lecture Notes in Mathematics}, {\bf vol 1582}, Springer, Berlin (1994).

\bibitem{dorigoni}
D.~Dorigoni,
``An Introduction to Resurgence, Trans-Series and Alien Calculus,''
Annals Phys. \textbf{409}, 167914 (2019)
[arXiv:1411.3585 [hep-th]].


\bibitem{ecalle}
J. \'Ecalle, 
``Les fonctions resurgentes," {\bf vol I - III}, Publ. Math. Orsay (1981).

\bibitem{hermite}
 W. N. Bailey, ``An integral representation for the product of two Hermite polynomials",
Journal London Math. Soc, {\bf 13} (1938), 202-203;
E. Feldheim, ``Quelques nouvelles relations pour les polynomes d'Hermite", Journal
London Math. Soc, {\bf 13} (1938), 22-29;
L. Carlitz, ``A Formula for the Product of Two Hermite Polynomials", 
Journal London Math. Soc, {\bf 32} (1957), 94-97.


\end{thebibliography}
\end{document}